\documentclass[review,10pt]{elsarticle}

\usepackage{lineno}
\modulolinenumbers[5]

\usepackage{hyperref} 
\hypersetup{ 
  colorlinks   = true, 
  urlcolor     = blue, 
  linkcolor    = blue, 
  citecolor    = blue, 
} 
\usepackage[fleqn]{amsmath}
\usepackage{amsfonts}
\usepackage{amssymb}
\usepackage{amsthm}

\usepackage{pifont}
\usepackage{natbib}
\usepackage{geometry}
\usepackage{graphicx}
\usepackage{txfonts}

\usepackage{setspace}
\usepackage[utf8]{inputenc}
\usepackage{caption}
\usepackage{subcaption}
\graphicspath{ {images/} }
\usepackage[usenames, dvipsnames]{color}
\usepackage{color,soul}
\setstcolor{red}

\biboptions{authoryear}

\begin{document}

\begin{frontmatter}

\title{Vorticity dynamics in transcritical liquid jet breakup}

\author{Jordi Poblador-Ibanez\fnref{myfootnote1}\corref{mycorrespondingauthor}}
\ead{poblador@uci.edu}
\author{William A. Sirignano\fnref{myfootnote3}}
\address{University of California, Irvine, CA 92697, United States}
\author{Fazle Hussain\fnref{myfootnote4}}
\address{Texas Tech University, Lubbock, TX 79409, United States}
\fntext[myfootnote1]{Junior Specialist, Department of Mechanical and Aerospace Engineering, UC Irvine.}
\fntext[myfootnote3]{Professor, Department of Mechanical and Aerospace Engineering, UC Irvine.}
\fntext[myfootnote4]{Professor, Department of Mechanical Engineering, Texas Tech University.}


\cortext[mycorrespondingauthor]{Corresponding author}


\begin{abstract}
Contrary to common assumptions, a transcritical domain exists during the early times of liquid hydrocarbon fuel injection at supercritical pressure. A sharp two-phase interface is sustained before substantial heating of the liquid. Thus, two-phase dynamics has been shown to drive the early three-dimensional deformation and atomisation. A recent study of a transcritical liquid jet shows distinct deformation features caused by interface thermodynamics, low surface tension, and intraphase diffusive mixing. In the present work, the vortex identification method \(\lambda_\rho\), which considers the fluid compressibility, is used to study the vortex dynamics in a cool liquid \textit{n}-decane transcritical jet surrounded by a hotter oxygen gaseous stream at supercritical pressures. The relationship between vortical structures and the liquid surface evolution is detailed, along with the vorticity generation mechanisms, including variable-density effects. The roles of hairpin and roller vortices in the early deformation of lobes, the layering and tearing of liquid sheets, and the formation of fuel-rich gaseous blobs are analysed. At these high pressures, enhanced intraphase mixing and ambient gas dissolution affect the local liquid structures (i.e., lobes). Thus, liquid breakup differs from classical sub-critical atomisation. Near the interface, liquid density and viscosity drop by up to 10\% and 70\%, respectively, and the liquid is more easily affected by the vortical motion (e.g., liquid sheets wrap around vortices). Despite the variable density, compressible vorticity generation terms are smaller than the vortex stretching and tilting. Layering traps and aligns the vortices along the streamwise direction while mitigating the generation of new rollers.
\end{abstract}


\end{frontmatter}


\setlength\abovedisplayshortskip{0pt}
\setlength\belowdisplayshortskip{-5pt}
\setlength\abovedisplayskip{-5pt}

\section{Introduction} 
\label{sec:introduction}

Optimisation of combustion efficiency and energy conversion calls for high-pressure combustion chambers. Diesel and gas turbines may operate in the range of 20 bar to 60 bar, while rocket engines can reach much higher pressures - between 70 bar and 200 bar. Typical fuels used in these applications are liquid hydrocarbon mixtures of \textit{n}-octane, \textit{n}-decane, and \textit{n}-dodecane, among other components, which have critical pressures around 20 bar. Therefore, operating conditions with near-critical or supercritical pressures for the fuel are common. Experimental studies at these extreme pressures reveal a thermodynamic transition where the distinction between liquid and gas is lost. The sharp phase interface is immersed in a variable-density layer with similar liquid and gas properties, and is rapidly distorted by hydrodynamic instabilities and turbulence~\citep{mayer1996propellant,h1998atomization,mayer2000injection,oschwald2006injection,chehroudi2012recent,falgout2016gas,crua2017transcritical}. In the past, a description was given of a sudden transition from a liquid to a gas-like supercritical state~\citep{spalding1959theory,rosner1967liquid}. However, the requirement that liquid and gas be in local thermodynamic equilibrium (LTE) at the interface provides evidence that a two-phase transcritical behaviour exists within a specific region of the mixture thermodynamic space. That is, the pressure is supercritical for the pure fluid but the temperature is below the mixture critical temperature~\citep{hsieh1991droplet,delplanque1993numerical,yang1994vaporization,delplanque1995transcritical,delplanque1996transcritical,poblador2018transient}. Above the fuel critical pressure, LTE enhances the solubility of the ambient gas into the liquid fuel, which causes a local change in mixture critical properties. Moreover, mixing layers with significant variations in the fluid properties develop in each phase~\citep{poblador2018transient,davis2019development,poblador2021selfsimilar}. \par

The phase-equilibrium assumption cannot apply beyond the mixture critical point where the two-phase interface transitions to a supercritical diffuse mixing. Other limitations also apply to the validity of LTE and have been discussed in the literature. LTE breaks down in scenarios where the interface presents a large thermal resistivity, for which the temperature jump across the phase non-equilibrium layer cannot be neglected~\citep{stierle2020selection}. Also,~\citet{dahms2013transition,dahms2015liquid,dahms2015non} and~\citet{dahms2016understanding} discuss and quantify the interface internal structure transition to the continuum domain at transcritical conditions. At temperatures near the mixture critical temperature and at very high pressures, the interfacial transition region may enter a continuum, with LTE no longer being a good modelling choice. It must be understood that the thermodynamic equilibrium of the continuous fluid is maintained throughout the layer; there is no question about the equation of state. However, the distinction between the two phases has disappeared and disallowed the use of the phase-equilibrium law. For interface temperatures sufficiently below the mixture critical temperature at a given pressure, LTE is well established. Mixing regions grow rapidly in the order of micrometers~\citep{poblador2018transient,davis2019development,poblador2021selfsimilar} while the thickness of the interfacial region remains in the nanoscale. Thus, the phase transition thickness becomes negligible, and the interface can be considered a sharp discontinuity with a jump in fluid properties. For practical purposes, the non-equilibrium layer of compressive shocks is treated as a discontinuity, despite its thickness being at least an order of magnitude greater than the phase non-equilibrium transition region. \par 

The transcritical behaviour explains why experimental observations have trouble capturing a two-phase environment. LTE drives the liquid and gas phases to be more alike near the interface. That is, the composition, density and viscosity of both fluids are similar (e.g., the gas phase becomes denser and the liquid viscosity drops to gas-like values)~\citep{yang2000modeling,poblador2018transient,davis2019development,poblador2021selfsimilar,poblador2021liquidjet,poblador2022temporal}. As a result, surface tension decreases substantially and vanishes at the mixture critical point. Therefore, the interface may experience a fast growth of short-wavelength surface perturbations, resulting in the early breakup into small ligaments and droplets and a mixing enhancement~\citep{poblador2021liquidjet}. \par

\begin{figure}
\centering
\begin{subfigure}{0.5\textwidth}
  \centering
  \includegraphics[width=1.0\linewidth]{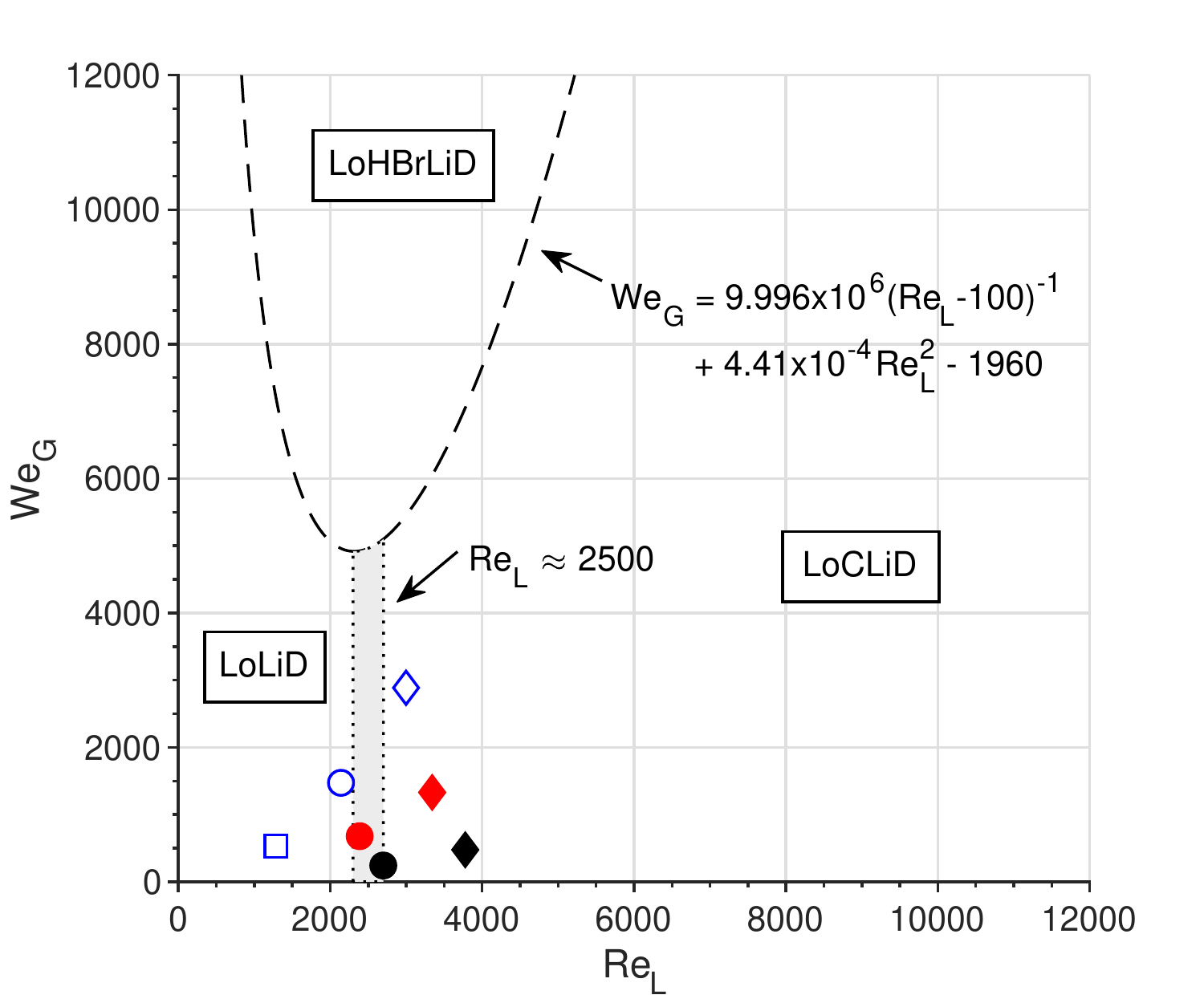}
  \caption{} 
  \label{subfig:Fig1a}
\end{subfigure}%
\begin{subfigure}{0.5\textwidth}
  \centering
  \includegraphics[width=1.0\linewidth]{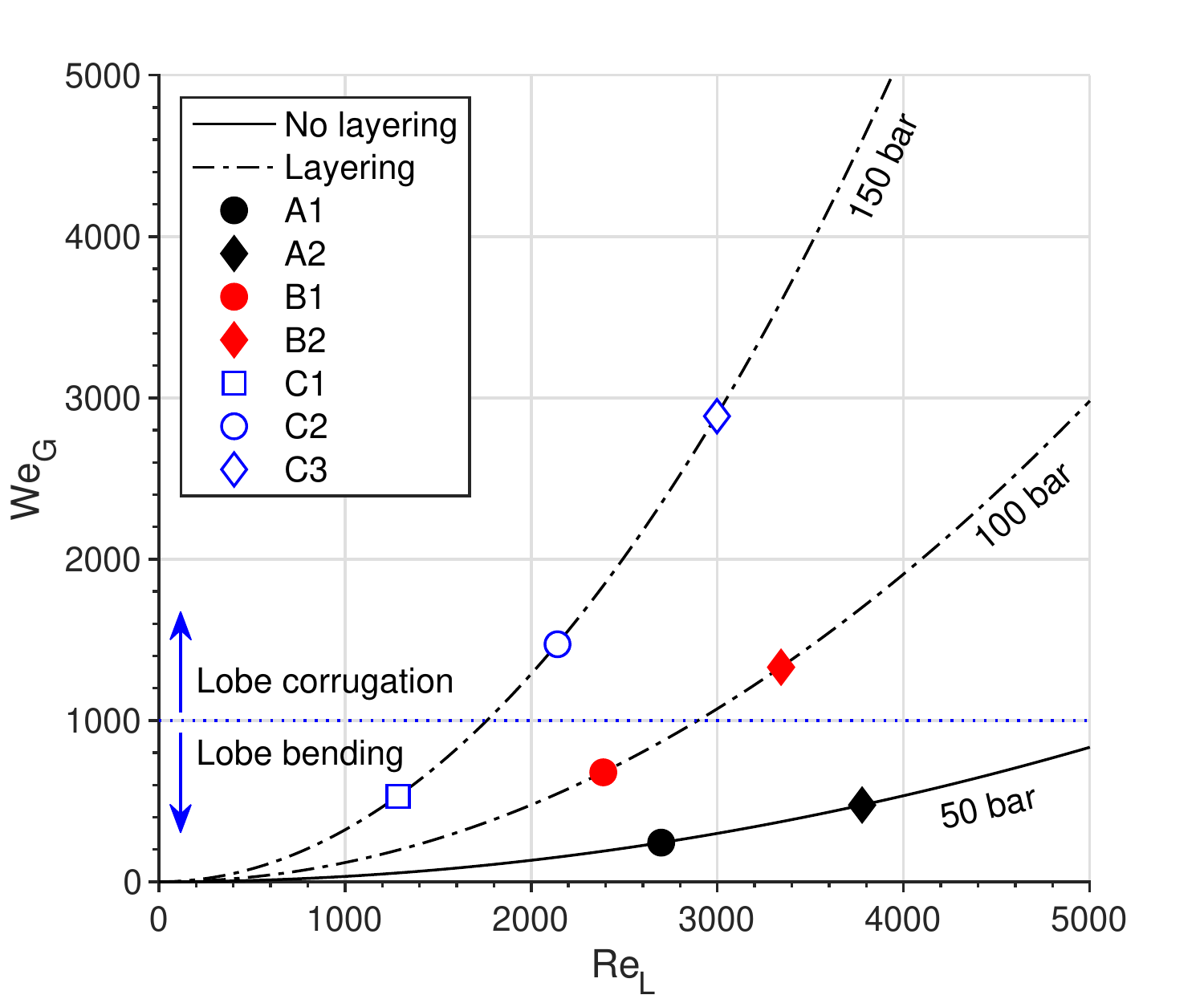}
  \caption{} 
  \label{subfig:Fig1b}
\end{subfigure}%
\caption{Classification of atomisation sub-domains and breakup features in a gas Weber number, \textit{We}\(_G\), vs liquid Reynolds number, \textit{Re}\(_L\), diagram. Also shown are the transcritical configurations from PS listed in table~\ref{tab:cases}. (a) incompressible framework described by~\citet{zandian2017planar}; and (b) transcritical work by PS.}
\label{fig:Fig1}
\end{figure}

Comprehending the early mixing process and atomisation in real-engine configurations prior to the eventual liquid mixture heating, vaporisation and possible shift to a supercritical state requires examining the fast surface deformation processes under transcritical conditions. Such analyses have been conducted within the limit of incompressible two-phase flows without interface thermodynamics and real-fluid modelling by~\citet{jarrahbashi2014vorticity},~\citet{jarrahbashi2016early} and~\citet{zandian2017planar,zandian2018understanding,zandian2019length,zandian2019vorticity}. \citet{zandian2017planar,zandian2018understanding} classify the injection of planar liquid jets using a liquid Reynolds number, \textit{Re}\(_L\) \(=(\rho_LUH)/\mu_L\), and a gas Weber number, \textit{We}\(_G\) \(=(\rho_GU^2H)/\sigma\), where \(\rho_L\) and \(\rho_G\) are the liquid and gas freestream densities, \(\mu_L\) the liquid viscosity, \(U\) the relative velocity between gas and liquid streams, \(H\) the jet thickness, and \(\sigma\) the surface-tension coefficient. Using \textit{We}\(_G\), the effects of the density ratio, \(\rho_G/\rho_L\), are embedded in the classification (note that other works define the density ratio as \(\rho_L/\rho_G\)). Three atomisation sub-domains are identified: (a) the Lobe-Ligament-Droplet (LoLiD) sub-domain characterised by low \textit{Re}\(_L\) and \textit{We}\(_G\) with the formation and stretching of lobes, which eventually break up into large droplets due to capillary instabilities; (b) the Lobe-Hole-Bridge-Ligament-Droplet (LoHBrLiD) sub-domain characterised by higher inertia forces compared to surface tension (i.e., higher gas density or lower surface-tension coefficient). Lobes are easily perforated by the gas phase, with holes expanding and forming bridges that eventually break off into ligaments and droplets; and (c) the Lobe-Corrugation-Ligament-Droplet (LoCLiD) similar to LoLiD, but where higher Reynolds numbers generate lobes that develop small-scale corrugations near the edge, forming smaller ligaments and droplets. Figure~\ref{subfig:Fig1a} summarises this classification in a \textit{We}\(_G\) vs \textit{Re}\(_L\) diagram, highlighting the transition between the sub-domains. Below \textit{We}\(_G<4900\), the LoLiD sub-domain transitions into the LoCLiD sub-domain around \textit{Re}\(_L\approx 2500\), whereas the boundary of the LoHBrLiD sub-domain is defined by a best-fit curve linking two different transition trends~\citep{zandian2017planar}. At low \textit{Re}\(_L\), the transitional boundary follows a hyperbolic relation (i.e., \textit{We}\(_G\sim\) \textit{Re}\(_L^{-1}\)), but at higher \textit{Re}\(_L\) a parabolic trend emerges defined by a modified Ohnesorge number, \textit{Oh}\(_m=\sqrt{\textit{We}_G}/\textit{Re}_L=\sqrt{\rho_G/\rho_L}\textit{Oh}\approx 0.021\), that includes the density ratio between gas and liquid. \par 

The results from these incompressible works indicate that very fast atomisation with the generation of small liquid structures occurs in real-engine configurations where gas inertia becomes important and surface-tension forces decrease. However, they fail to incorporate detailed physics and real-fluid thermodynamics during the atomisation of liquid fuels at engine-relevant conditions. Moreover, these works emphasise the parametric study by fixing the liquid density to 804 kg/m\(^3\), which is representative of heavy hydrocarbons, and varying \(U\), \(H\), \textit{Re}\(_L\), \textit{We}\(_G\), the density ratio and the viscosity ratio. As a result, some of the fluid properties for both liquid and gas might not be representative of the real fluids in typical combustion applications. \par 

The novelty in the analysis introduced by these authors is the inclusion of vorticity dynamics to explain the atomisation cascade process of liquid structures as a result of the interactions between hairpin vortices and Kelvin-Helmholtz (KH) vortices. Previous works have used vorticity dynamics to understand how vortex stretching, tilting and baroclinicity relate to the generation of three-dimensional perturbations leading to atomisation in incompressible flows (e.g., subcritical behaviour). A detailed literature review is provided in~\citet{jarrahbashi2014vorticity},~\citet{jarrahbashi2016early} and~\citet{zandian2018understanding}. \citet{zandian2019vorticity} extended the detailed analysis of vortex interactions to spatially-developing liquid jets submerged in a coaxial gas flow; more recently,~\citet{constante2021direct} have shown how the transition between the capillary-controlled and inertia-controlled regimes (i.e., low vs high \textit{We}\(_G\) and \textit{Re}\(_L\)) leads to streamwise vorticity generation and hairpin vortex formation. Additionally,~\citet{gao2022effect} have included phase change in their study of the atomisation of liquid round jets and have shown that droplet vaporisation has a substantial impact on the evolution and breakup of vortical structures. A reduction in droplets around the liquid core translates into a less turbulent flow field where vortex rings deform into strip-shaped or streamwise vortices that remain unperturbed for longer times. Vorticity dynamics are also considered in studies of jets injected into conditions ranging from trans- to supercritical environments. Usually, such works focus on the breakup of the primary vortical structures (e.g., rings or rollers) into smaller vortices as a measure of atomisation onset or to describe flow patterns~\citep{gnanaskandan2018side,lagarza2019large,wang2019three,koukouvinis2020high}. However, such studies neglect surface tension, and little focus is given to the complex interaction between vorticity and liquid atomisation. \par 

Following these works, \citet{poblador2022temporal}, hereinafter cited as PS, perform a temporal analysis of a three-dimensional planar liquid \textit{n}-decane jet submerged in a hotter gaseous oxygen stream at transcritical conditions. Various ambient pressures are considered well above the critical pressure of the liquid \textit{n}-decane. They show the influence of intraphase mixing, reduced surface tension and varying interface properties on the early deformation of the transcritical jet. A primary deformation mechanism arises at very high pressures (i.e., above 100 bar) as a direct result of the transcritical thermodynamics. The production of large ligaments and droplets is suppressed, and the liquid deforms in a gas-like manner with continuous stretching, folding, and layering of liquid sheets. Only localised mixing may cause the growth of short-wavelength perturbations that generate sub-micron ligaments and droplets. That is, certain liquid regions can become unstable if the liquid viscosity and surface tension drop. Apart from the layering mechanism, breakup patterns similar to those identified in the incompressible sub-domains LoHBrLiD or LoCLiD may occur at much lower \textit{Re}\(_L\) and \textit{We}\(_G\) as a direct result of the variation of fluid properties within each phase and the reduced surface tension as the ambient pressure increases. The shear across the initially perturbed phase interface generates early lobes that evolve differently depending on the problem configuration. For \textit{We}\(_G\) \(<1000\), a lobe bending mechanism occurs whereby lobes stretch, thin and easily bend toward the oxidizer stream before being perforated at very high pressures. In contrast, higher \textit{We}\(_G\) modify the deformation pattern with lobes corrugating around the streamwise direction before bursting into droplets, resembling a bag-breakup mechanism. In summary, higher pressures promote gas-like turbulent mixing with frequent hole formation. These transcritical effects observed in the \textit{n}-decane/oxygen mixture are shown in figure~\ref{subfig:Fig1b}, where the transition between lobe bending and corrugation mechanisms, as well as evidence of layering, are presented. For each pressure, a curve is obtained in the \textit{We}\(_G\) vs \textit{Re}\(_L\) diagram by fixing the properties of both fluids and the jet thickness, while varying the relative velocity \(U\). Note that figure~\ref{fig:Fig1} shows the classification of the cases analysed in PS (e.g., A1) in the \textit{We}\(_G\) vs \textit{Re}\(_L\) diagrams representative of the incompressible framework and the transcritical framework. These configurations are introduced in section~\ref{sec:description} and summarised in table~\ref{tab:cases}. A visualisation of the planar jet deformation process in case C2 is shown in figure~\ref{fig:Fig3}. \par 

The observations at transcritical conditions merit a vorticity dynamics study to explain how the deformation mechanisms shown in PS result from vortical motion. The gas-like behaviour of the liquid phase near the interface suggests that vortex-related surface deformation is stronger than seen in previous incompressible studies. For this study, we take the configurations of PS and post-process the data to identify the vortical features with the vortex identification method \(\lambda_\rho\) by~\citet{yao2018toward}. The \(\lambda_\rho\) method is a compressible extension of the incompressible \(\lambda_2\) method of~\citet{jeong1995identification} and it has been chosen in favour of another widely-used vortex identification method, the Q-criterion~\citep{jcr1988eddies}. The Q-criterion may be inconsistent for vortices subject to strong external strain, such as in liquid atomisation, and has an ambiguous extension to compressible flows. On the other hand, \(\lambda_2\) or \(\lambda_\rho\) are more suitable for this type of flows~\citep{kolavr2007vortex}. \par

This paper is structured as follows. First, section~\ref{sec:description} describes the transcritical configurations of PS (e.g., species, temperature, pressure, computational domain). Then, section~\ref{sec:modeling} summarises the physical modelling and numerical techniques used in the computations, for which a detailed description can be found in~\citet{pobladoribanez2021volumeoffluid}. The vortex identification method \(\lambda_\rho\) of~\citet{yao2018toward} is reviewed in section~\ref{sec:lambdarho}, and best practices are presented for its implementation in the post-processing of two-phase computations with a sharp interface. Section~\ref{sec:results} presents selected post-processing of the simulations of PS, focusing on the liquid deformation by identifying vortical structures, as well as on the vorticity generation mechanisms. Lastly, section~\ref{sec:conclusions} summarises this work's major findings and contributions. \par

\section{Flow Configuration} 
\label{sec:description}

This work analyses the computational data of the previously discussed numerical study by PS. The study considers a temporal analysis of a planar liquid \text{n}-decane jet initially at 450 K perturbed by a faster and hotter oxygen gaseous stream initially at 550 K. The hotter ambient gas provides enough energy to vaporise the fuel. Both species represent engines operating at high pressures with liquid hydrocarbon fuels injected into enriched air or pure oxygen. Under such conditions, both fluids might be supercritical away from the interface and the mixing layers. Therefore, the use of the terms ``liquid" and ``gas" to characterise each fluid might not be totally appropriate. However, we refer to the high-density, compressible fluid with \textit{n}-decane as the dominant component as liquid, whether it is subcritical or supercritical locally. Likewise, the lower density fluid with oxygen as the major species is described as gas over the transcritical domain. \par

\begin{figure}
\centering
\includegraphics[width=0.5\linewidth]{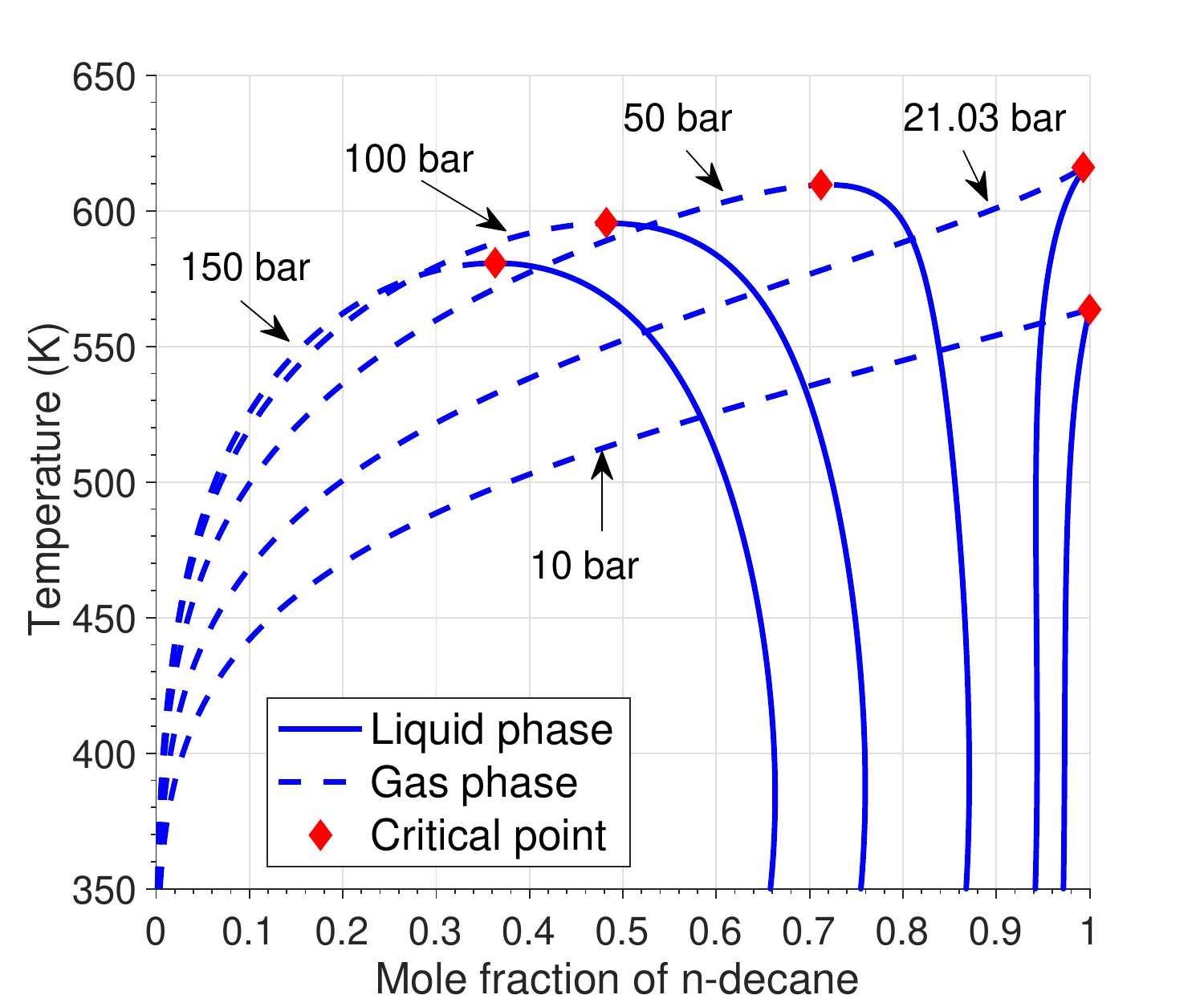}
\caption{Phase-equilibrium diagrams obtained using the SRK equation of state for the binary mixture of \textit{n}-decane and oxygen. The mixture composition in each phase is represented by the mole fraction of \textit{n}-decane as a function of temperature and pressure. The mixture critical point is shown. The figure is reproduced from PS.}
\label{fig:Fig2}
\end{figure}

Various values for the gas freestream velocity, \(u_G\), and ambient pressure, \(p_\text{amb}\), above the critical pressure of the liquid are considered (50, 100 and 150 bar). The critical pressures and critical temperatures of the analysed species are 21.03 bar and 617.70 K for \textit{n}-decane and 50.43 bar and 154.58 K for oxygen. Table~\ref{tab:cases} summarises the parameters for each configuration. All cases fall in a transcritical environment where a two-phase interface can be sustained with strong diffusive mixing surrounding it. For reference, figure~\ref{fig:Fig2} has been reproduced from PS to show the phase-equilibrium diagrams of the \textit{n}-decane/oxygen binary mixture at various pressures, where the domain of two-phase coexistence above the critical pressure of \textit{n}-decane is observed. \par

A reduced computational domain is considered to limit the computational cost. Symmetry is imposed on the centerline of the jet, periodic boundary conditions are used in the streamwise and spanwise directions, and outflow boundary conditions are applied at the top gaseous boundary away from the phase interface. The jet thickness is \(H=20\) \(\mu\)m (i.e., only the half-thickness of \(H'=10\) \(\mu\)m is included in the computational domain), and two initial streamwise (in \(x\)) and spanwise (in \(z\)) sinusoidal perturbations are superimposed on the jet's surface. The streamwise perturbation has a wavelength of 30 \(\mu\)m and an amplitude of 0.5 \(\mu\)m; the spanwise perturbation has a wavelength of 20 \(\mu\)m and an amplitude of 0.3 \(\mu\)m. The velocity field is initialised as \(\boldsymbol{u}=\big(u(y),0,0\big)\), where \(u(y)=(u_G/2)\big(\tanh{\big[6.5\times 10^{5}(y-H')\big]}+1\big)\), which distributes the streamwise velocity from 0 m/s in the liquid phase to \(u_G\) across the interface within a thin layer of about 6 \(\mu\)m. \par 

The initial instability is linear as the perturbation is sufficiently weak to let unstable waves evolve naturally. The perturbation along the spanwise direction is meant to accelerate the generation of three-dimensional flow structures. The initial wavelengths are chosen using estimates from~\citet{poblador2019axisymmetric} for an axisymmetric transcritical liquid jet. Guided by jet instability theory, the wavelengths are shorter than those used in incompressible studies where surface tension was generally stronger. Only one wavelength per direction is included in the numerical domain, with a size of \(L_x=L_y=30\) \(\mu\)m and \(L_z=20\) \(\mu\)m. The domain is discretised with a Cartesian uniform mesh with 450 x 450 x 300 cells (i.e., a uniform mesh size of \(\Delta=0.0667\) \(\mu\)m). \citet{pobladoribanez2021volumeoffluid} noted that this mesh resolution is adequate for our analysis. \par

\begin{table}
\begin{center}
\def~{\hphantom{0}}
\begin{tabular}{|r|r|r|r|r|r|r|r|r|r|} 
\multicolumn{1}{c}{} & 
\multicolumn{1}{c}{\(p_\text{amb}\) (bar)} & 
\multicolumn{1}{c}{\(u_G\) (m/s)} & \multicolumn{1}{c}{\(\rho_G\) (kg/m\(^3\))} &
\multicolumn{1}{c}{\(\rho_L\) (kg/m\(^3\))} &
\multicolumn{1}{c}{\(\mu_G\) (\(\mu\)Pa\(\cdot\)s)} &
\multicolumn{1}{c}{\(\mu_L\) (\(\mu\)Pa\(\cdot\)s)} &
\multicolumn{1}{c}{\(\sigma\) (mN/m)} &
\multicolumn{1}{c}{\textit{Re}\(_L\)} &
\multicolumn{1}{c}{\textit{We}\(_G\)}\\[3pt]
\multicolumn{1}{c}{A1} & 
\multicolumn{1}{c}{50} & 
\multicolumn{1}{c}{50} & 
\multicolumn{1}{c}{34.47} &
\multicolumn{1}{c}{615.18} &
\multicolumn{1}{c}{32.77} &
\multicolumn{1}{c}{228.01} &
\multicolumn{1}{c}{7.10} &
\multicolumn{1}{c}{2698} &
\multicolumn{1}{c}{243} \\
\multicolumn{1}{c}{A2} & 
\multicolumn{1}{c}{50} & 
\multicolumn{1}{c}{70} & 
\multicolumn{1}{c}{34.47} &
\multicolumn{1}{c}{615.18} &
\multicolumn{1}{c}{32.77} &
\multicolumn{1}{c}{228.01} &
\multicolumn{1}{c}{7.10} &
\multicolumn{1}{c}{3777} &
\multicolumn{1}{c}{476} \\
\multicolumn{1}{c}{B1} & 
\multicolumn{1}{c}{100} & 
\multicolumn{1}{c}{50} & 
\multicolumn{1}{c}{67.86} &
\multicolumn{1}{c}{632.59} &
\multicolumn{1}{c}{33.32} &
\multicolumn{1}{c}{265.14} &
\multicolumn{1}{c}{5.00} &
\multicolumn{1}{c}{2387} &
\multicolumn{1}{c}{679} \\
\multicolumn{1}{c}{B2} & 
\multicolumn{1}{c}{100} & 
\multicolumn{1}{c}{70} & 
\multicolumn{1}{c}{67.86} &
\multicolumn{1}{c}{632.59} &
\multicolumn{1}{c}{33.32} &
\multicolumn{1}{c}{265.14} &
\multicolumn{1}{c}{5.00} &
\multicolumn{1}{c}{3342} &
\multicolumn{1}{c}{1330} \\
\multicolumn{1}{c}{C1} & 
\multicolumn{1}{c}{150} & 
\multicolumn{1}{c}{30} & 
\multicolumn{1}{c}{100.14} &
\multicolumn{1}{c}{646.67} &
\multicolumn{1}{c}{34.02} &
\multicolumn{1}{c}{302.43} &
\multicolumn{1}{c}{3.40} &
\multicolumn{1}{c}{1285} &
\multicolumn{1}{c}{530} \\
\multicolumn{1}{c}{C2} & 
\multicolumn{1}{c}{150} & 
\multicolumn{1}{c}{50} & 
\multicolumn{1}{c}{100.14} &
\multicolumn{1}{c}{646.67} &
\multicolumn{1}{c}{34.02} &
\multicolumn{1}{c}{302.43} &
\multicolumn{1}{c}{3.40} &
\multicolumn{1}{c}{2141} &
\multicolumn{1}{c}{1473} \\
\multicolumn{1}{c}{C3} & 
\multicolumn{1}{c}{150} & 
\multicolumn{1}{c}{70} & 
\multicolumn{1}{c}{100.14} &
\multicolumn{1}{c}{646.67} &
\multicolumn{1}{c}{34.02} &
\multicolumn{1}{c}{302.43} &
\multicolumn{1}{c}{3.40} &
\multicolumn{1}{c}{2998} &
\multicolumn{1}{c}{2886} \\
\end{tabular}
\caption{List of analysed cases from PS using liquid \textit{n}-decane at 450 K and gaseous oxygen at 550 K. The subscripts \(G\) and \(L\) refer to freestream conditions for the gas and the liquid phases.}
\label{tab:cases}
\end{center}
\end{table}

Following the works by~\citet{zandian2017planar,zandian2018understanding}, table~\ref{tab:cases} presents the relevant freestream parameters to characterise each configuration in terms of \textit{Re}\(_L\) and \textit{We}\(_G\). Since interface properties vary in a transcritical jet, a value of \(\sigma\) characteristic of the interface before substantial deformation occurs is used. However, localised mixing has a strong influence on the characterisation of high-pressure atomisation problems, which raises the question whether \textit{Re}\(_L\) and \textit{We}\(_G\) alone are adequate to classify jet atomisation sub-domains in transcritical flows (PS). \par

\begin{figure}
\centering
\begin{subfigure}{0.5\textwidth}
  \centering
  \includegraphics[width=1\linewidth]{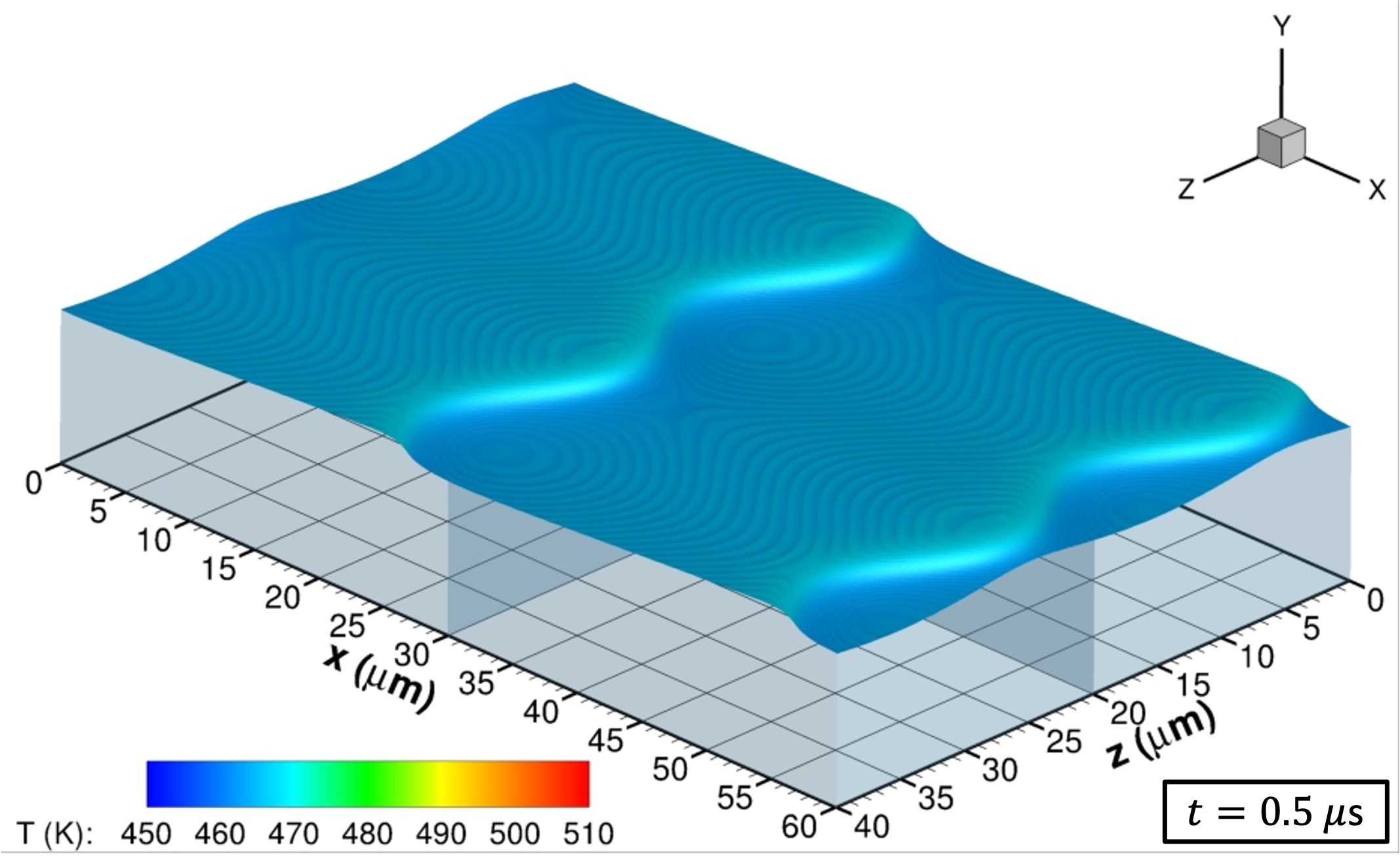}
  \label{subfig:Fig3a}
\end{subfigure}%
\begin{subfigure}{0.5\textwidth}
  \centering
  \includegraphics[width=1\linewidth]{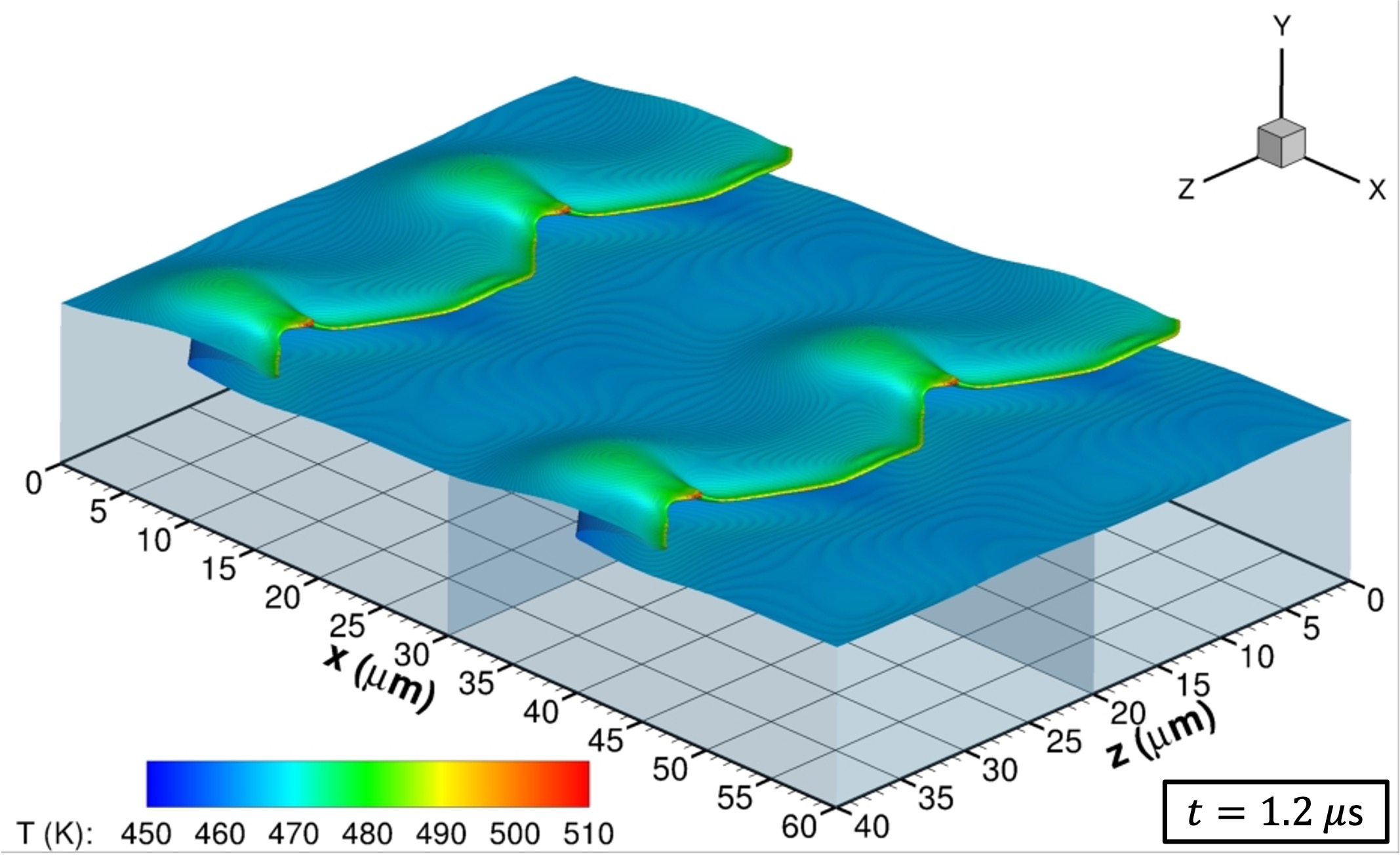}
  \label{subfig:Fig3b}
\end{subfigure}%
\\[-2ex]
\begin{subfigure}{0.5\textwidth}
  \centering
  \includegraphics[width=1\linewidth]{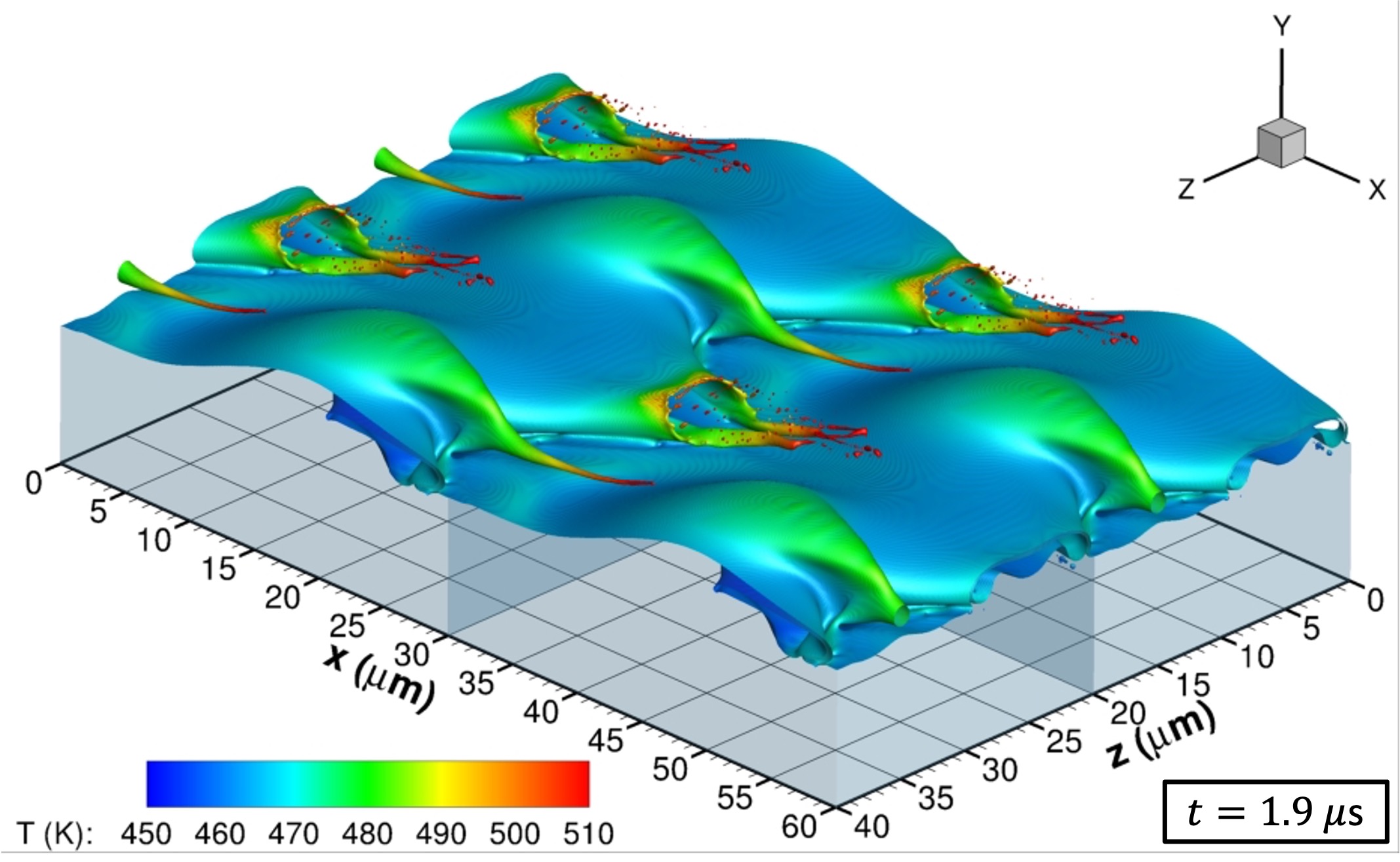}
  \label{subfig:Fig3c}
\end{subfigure}%
\begin{subfigure}{0.5\textwidth}
  \centering
  \includegraphics[width=1\linewidth]{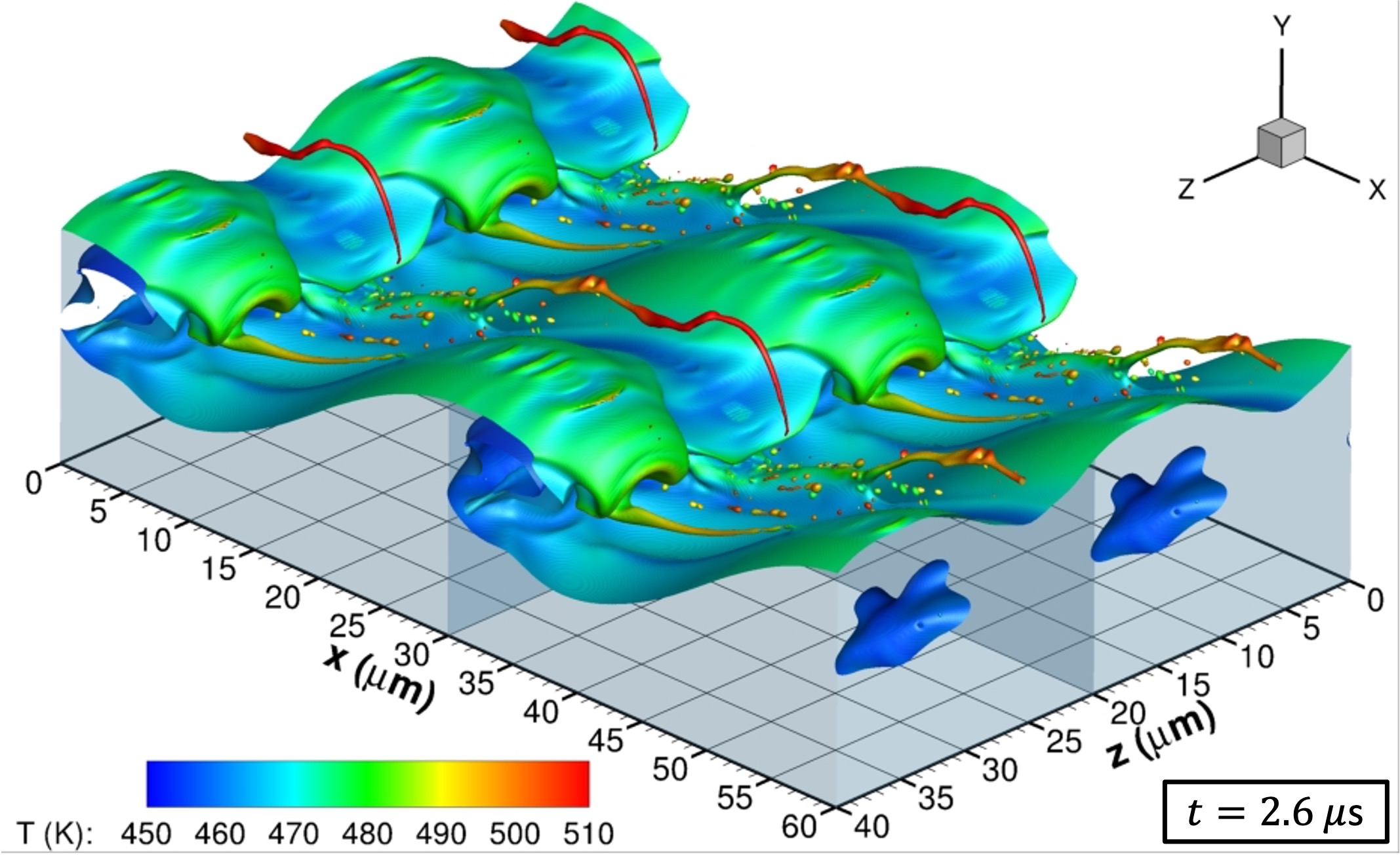}
  \label{subfig:Fig3d}
\end{subfigure}%
\\[-2ex]
\begin{subfigure}{0.5\textwidth}
  \centering
  \includegraphics[width=1\linewidth]{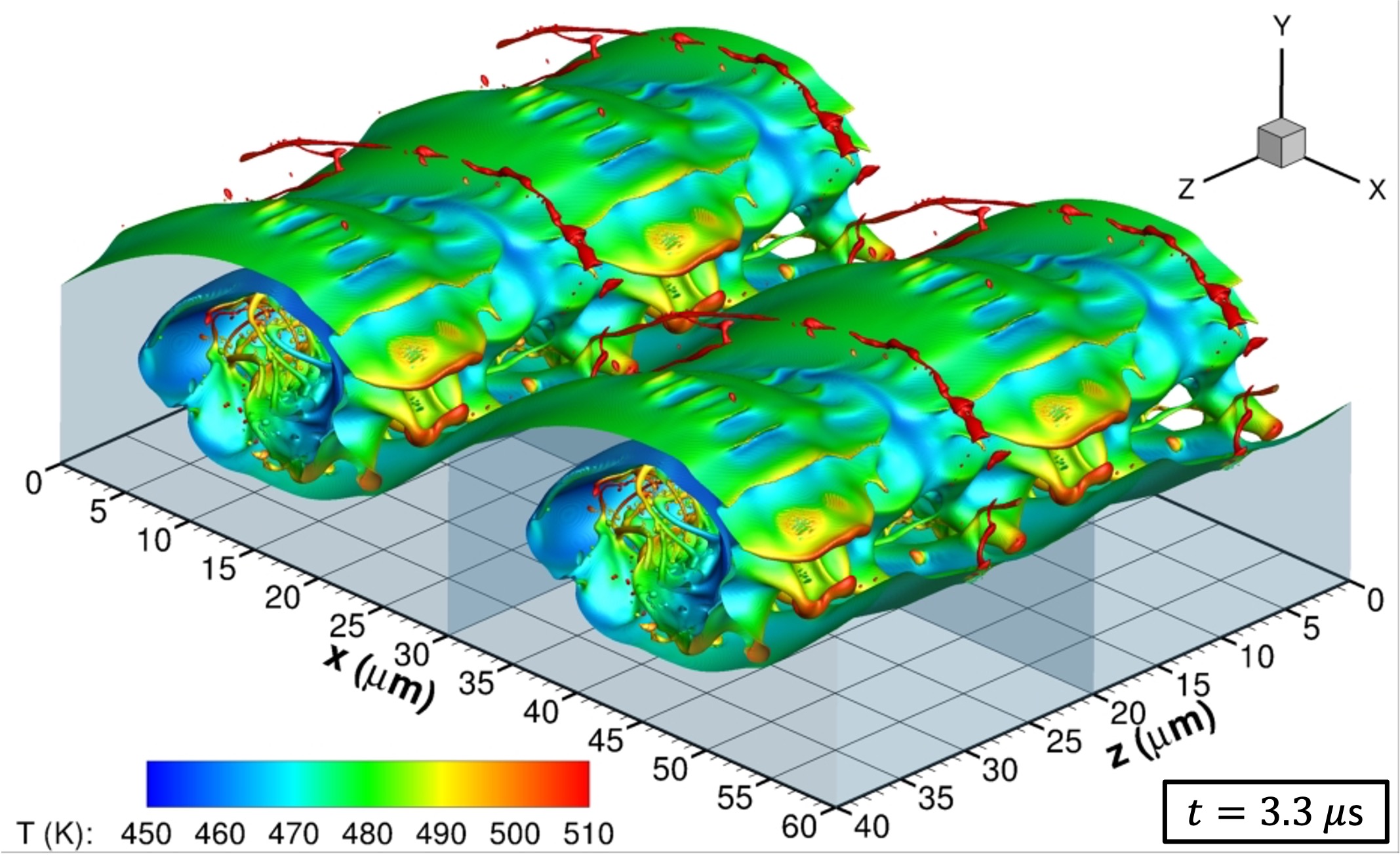}
  \label{subfig:Fig3e}
\end{subfigure}%
\begin{subfigure}{0.5\textwidth}
  \centering
  \includegraphics[width=1\linewidth]{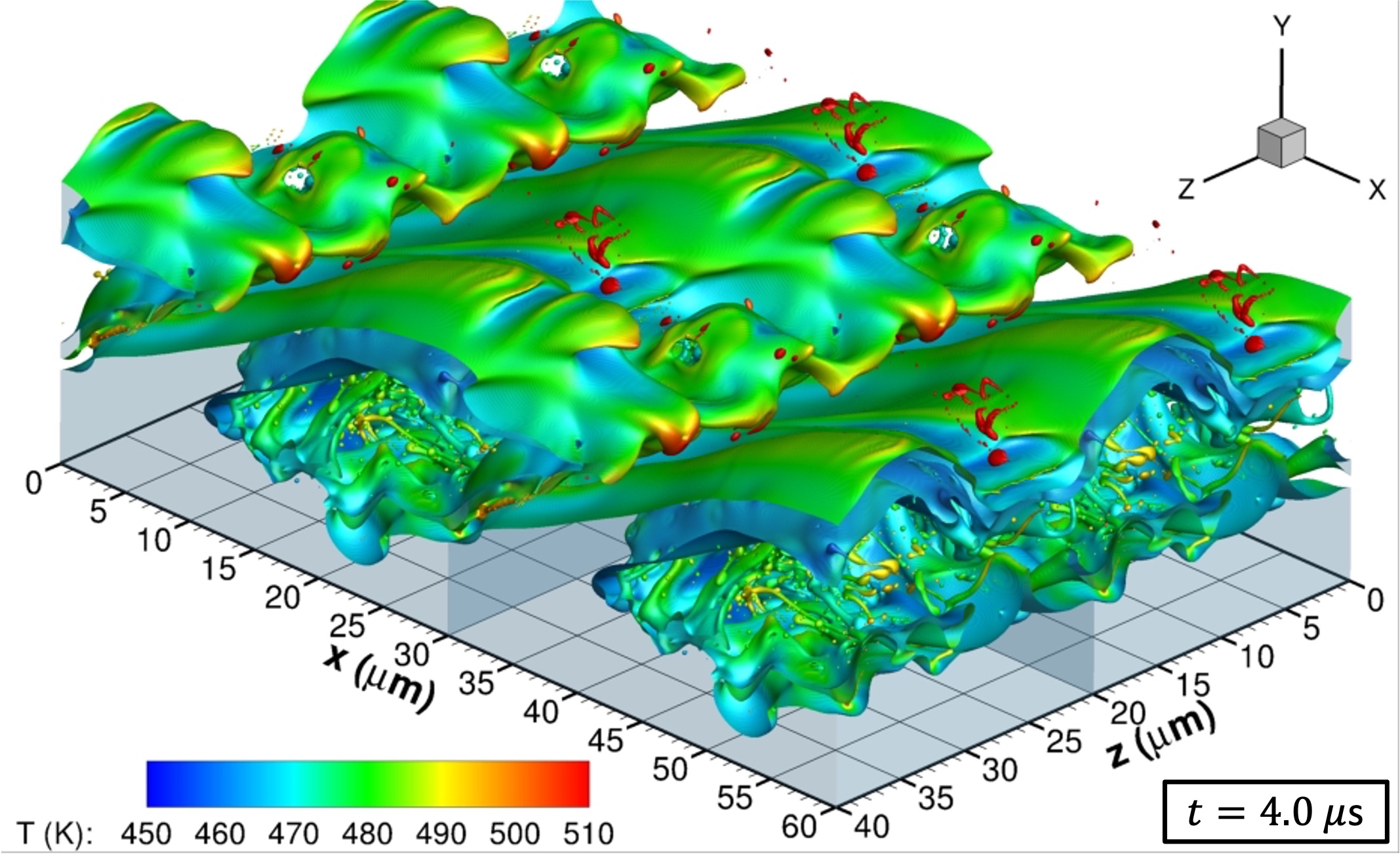}
  \label{subfig:Fig3f}
\end{subfigure}%
\caption{Temporal deformation of the planar jet configuration C2 analysed in PS. The liquid surface is identified by the isosurface with \(C=0.5\) and is coloured by the local temperature. The \(x\) and \(z\) periodicities are used to enlarge the computational domain.}
\label{fig:Fig3}
\end{figure}

Figure~\ref{fig:Fig3} shows an oblique view of the planar jet configuration from PS and the deformation process for case C2. For visualisation purposes, the assumed periodic behaviour is used to enlarge the domain in this and subsequent figures. The liquid-gas interface is coloured by its temperature to highlight the level of detail considered in these computations. Different local temperatures lead to a varying composition and fluid properties along the interface. The relation between equilibrium temperature and interface composition is seen in figure~\ref{fig:Fig2}. A more detailed description of the implications of this behaviour and how it is interpreted via vortex dynamics is provided in sections~\ref{subsec:lobe_bending},~\ref{subsec:lobe_crest_corrugation} and~\ref{subsec:layering}. \par

\section{Physical and Numerical Modelling}  
\label{sec:modeling}

A summary of the physical and numerical modelling used by PS is provided here. The physical modelling considered in the studies of non-reactive, transcritical liquid injection by~\citet{poblador2021liquidjet,poblador2022temporal} is based on a low-Mach-number formulation particularised for a binary mixture of a fuel, \(F\), and an oxidizer species, \(O\) (i.e., sum of mass fractions is \(Y_F+Y_O=1\)). Compressibility arises from species and thermal mixing at elevated pressures, but pressure variations in the thermodynamic model are neglected. The governing equations for mass~(\ref{eqn:cont}), momentum~(\ref{eqn:mom}), species transport~(\ref{eqn:spcont}) and energy~(\ref{eqn:energy}) are solved within each phase and satisfy the corresponding matching conditions across the moving liquid-gas interface~\citep{poblador2021liquidjet,poblador2022temporal,pobladoribanez2021volumeoffluid}. That is, mass, momentum and energy balance relations across the interface are ensured. In the low-Mach-number limit, the pressure field, \(p\), acts to correct the velocity field, \(\boldsymbol{u}=(u,v,w)\), to satisfy the volume dilatation obtained from the thermodynamics. In~(\ref{eqn:mom}), the viscous stress tensor is given by \(\tau=\mu [\boldsymbol{\nabla}\boldsymbol{u}+\boldsymbol{\nabla}\boldsymbol{u}^\text{T}-\frac{2}{3}(\boldsymbol{\nabla\cdot}\boldsymbol{u})I]\). 

\begin{equation}
\label{eqn:cont}
\frac{\partial \rho}{\partial t} + \boldsymbol{\nabla \cdot} (\rho\boldsymbol{u})=0
\end{equation}
\begin{equation}
\label{eqn:mom}
\frac{\partial}{\partial t}(\rho \boldsymbol{u})+\boldsymbol{\nabla \cdot} (\rho \boldsymbol{u}\boldsymbol{u}) = -\boldsymbol{\nabla} p + \boldsymbol{\nabla \cdot} \tau
\end{equation}
\begin{equation}
\label{eqn:spcont}
\frac{\partial}{\partial t}(\rho Y_O) + \boldsymbol{\nabla\cdot}(\rho Y_O \boldsymbol{u}) = \boldsymbol{\nabla \cdot} (\rho D_m \boldsymbol{\nabla} Y_O)
\end{equation}
\begin{equation}
\label{eqn:energy}
\frac{\partial}{\partial t}(\rho h) + \boldsymbol{\nabla\cdot}(\rho h \boldsymbol{u}) = \boldsymbol{\nabla \cdot} \bigg(\frac{\lambda}{c_p}\boldsymbol{\nabla} h \bigg) + \sum_{i=1}^{N=2} \boldsymbol{\nabla \cdot} \Bigg(\bigg[\rho D_m - \frac{\lambda}{c_p}\bigg]h_i \boldsymbol{\nabla} Y_i\Bigg)
\end{equation}

Although liquid atomisation is a problem involving a transition from laminar to turbulent flow, the formulation does not include turbulence modelling. A direct numerical approach suffices to analyse the early times of the liquid deformation cascade process, especially given the considered Reynolds numbers shown in table~\ref{tab:cases} and the grid resolution discussed in section~\ref{sec:description}. \par 

A volume-corrected Soave-Redlich-Kwong (SRK) equation of state~\citep{lin2006volumetric} is used in tandem with other models and correlations to evaluate fluid and transport properties. The equation of state provides the mixture density, \(\rho\), given the mixture temperature, composition and thermodynamic pressure (i.e., the ambient pressures reported in table 1). The concept of departure functions from the ideal state~\citep{poling2001properties} is implemented to evaluate the mixture enthalpy, \(h\), the partial enthalpy of species \(i\), \(h_i\), the mixture specific heat at constant pressure, \(c_p\), and the fugacity coefficient of species \(i\), \(\Phi_i\). The generalised multi-parameter correlations by~\citet{chung1988generalized} are used to evaluate the mixture viscosity, \(\mu\), and the mixture thermal conductivity, \(\lambda\), while a unified model for diffusion in non-ideal fluids is used to estimate the mass-diffusion coefficient, \(D_m\), for the binary mixture~\citep{leahy2007unified}. Lastly, the surface-tension coefficient is obtained from the Macleod-Sugden correlation as a function of the interface mixture properties and composition, as recommended in~\citet{poling2001properties}. This correlation provides the correct limit where \(\sigma \rightarrow 0\) at the mixture critical point. A necessary interface thermodynamic closure is given by LTE, defined by the fugacity of each species being equal on both sides of the interface~\citep{soave1972equilibrium,poling2001properties}. These interface relations define the interface state, such as the temperature and composition, the pressure jump, or the mass flux per unit area due to phase change, \(\dot{m}'\). Necessary thermodynamic relations based on the volume-corrected SRK equation of state are available in~\citet{davis2019development}. \par

An overview on the validity and limitations of the selected interface modelling for these transcritical flows has been provided in section~\ref{sec:introduction}. \citet{poblador2022temporal,pobladoribanez2021volumeoffluid} discuss in detail the validity of the methodology in configurations involving typical liquid hydrocarbon fuels and oxidizers. The expected interface thermodynamic states are below the mixture critical point as surface deformation and mixing occur. In a sufficiently hot environment, only small liquid structures (i.e., thin ligaments and droplets) might transition to a diffuse supercritical phase mixing, whereas the rest of the liquid core may be in a transcritical two-phase state for longer periods of time during the injection process. \par

The liquid-gas interface is captured using the Volume-of-Fluid (VOF) method described in more detail by~\citet{pobladoribanez2021volumeoffluid}, which accounts for liquid compressibility and mass exchange across the interface. This VOF method builds on the incompressible VOF model described in~\citet{baraldi2014mass}, which has been successfully used to study decaying isotropic turbulence in droplet-laden flows~\citep{dodd2014fast} and evaporating droplets in forced homogeneous isotropic turbulence in a low-Mach gas environment~\citep{dodd2021analysis}. \citet{pobladoribanez2021volumeoffluid} discuss the numerical approach to solve the governing equations. To summarise, the local interface equilibrium state is given by phase equilibrium and the mass and energy balances. Due to the simplified physical modelling, the interface solution is uncoupled from the momentum balance. An interface solver described in~\citet{poblador2018transient} is used together with a normal-probe technique to describe the interface locally. Then, the interface local solution is embedded in the numerical discretisation of~(\ref{eqn:spcont}) and~(\ref{eqn:energy}), which are solved to advance in time the mixture properties. \par 

Finally, a one-fluid approach (allowing for discontinuities in density and viscosity) is used to address the momentum equation and the pressure-velocity coupling for the low-Mach-number flow by solving a Poisson-type equation for the pressure field. The Continuum Surface Force (CSF) model~\citep{brackbill1992continuum,kothe1996volume,seric2018direct} is used to include the effects of surface tension as a localised body force in~(\ref{eqn:mom}) acting only at the interface, \(\boldsymbol{F_\sigma}\). The mixture density and viscosity are volume-averaged over a computational cell as \(\rho=\rho_g + (\rho_l-\rho_g)C\) and \(\mu=\mu_g + (\mu_l-\mu_g)C\), where \(C\) represents the cell's liquid volume fraction. Gas and liquid properties are given by the thermodynamic model as a function of composition, temperature and thermodynamic pressure at each cell. However, the interface properties are used in mixed cells (i.e., \(0<C<1\)) to simplify the computations. Following~\citet{dodd2014fast} and~\citet{dodd2021analysis}, a fast pressure solver based on a Fast Fourier Transform method or FFT is used~\citep{costa2018fft}. Similarly, the one-fluid velocity divergence is obtained by averaging each fluid's volume dilatation rate with \(C\) and by including the local volume expansion or contraction due to phase change

\begin{equation}
\label{eqn:veldiv}
\boldsymbol{\nabla \cdot}\boldsymbol{u}=-(1-C)\frac{1}{\rho_g}\frac{D\rho_g}{Dt}-C\frac{1}{\rho_l}\frac{D\rho_l}{Dt} + \dot{m}\bigg(\frac{1}{\rho_g}-\frac{1}{\rho_l}\bigg)
\end{equation}
\noindent
The material derivatives of \(\rho_g\) and \(\rho_l\) are obtained from thermodynamic relations based on the SRK equation of state and \(\dot{m} = \dot{m}'A_\Gamma/V_0\) is the mass flow per unit volume. \(\dot{m}'\) is the mass flux across the interface, which is negative when condensation occurs and is positive for vaporisation, \(A_\Gamma\) is the area of the interface plane crossing the cell, and \(V_0\) is the cell volume. The ratio \(A_\Gamma/V_0\) comes from the concept of surface-area density and is meant to activate phase-change effects only at interface cells where \(A_\Gamma\) is non-zero~\citep{palmore2019volume}. \par 

Additionally, an overview of known numerical issues that might appear with the proposed methodology and some solutions to mitigate them are discussed. Spurious currents around the interface caused by numerical inaccuracies are of special concern. These oscillations appear because of the limited smoothness of the discrete representation of a sharp interface using VOF and a lack of an exact interfacial pressure balance. Moreover, phase change is treated as a localised source term only active at the interface cells, which contributes further to the generation of spurious currents. Keeping track of these velocity oscillations is crucial in liquid atomisation computations, as the growth of surface instabilities that define the breakup cascade process must be of physical origin. For the present study, spurious currents introduce visualisation noise that affects post-processing techniques like the \(\lambda_\rho\) method. This issue is discussed in section~\ref{sec:lambdarho}. \par

Despite using a fast pressure solver, the numerical cost becomes prohibitive due to the resolution needed to capture the interface details, both its evolution and its thermodynamic state, and limit the numerical spurious currents around it. Moreover, atomisation involves a continuous generation of surface area. Therefore, the computational cost of resolving the interface local equilibrium state grows over time. For this reason, the analysed domain is small as described in section~\ref{sec:description}. Such problems are less severe in incompressible atomisation simulations.  \par

\section{The Vortex Identification Method}
\label{sec:lambdarho}

The vortex identification method \(\lambda_\rho\) proposed by~\citet{yao2018toward} is a direct extension of the \(\lambda_2\) vortex identification method for incompressible flows~\citep{jeong1995identification} commonly used in the literature. Here, a brief summary of this method and its implementation is provided. \par

\(\lambda_\rho\) aims to identify vortices by analysing local pressure minima in the flow field~\citep{jeong1995identification,yao2018toward}, which can only be determined in a two-dimensional plane or flow cross-section. Thus, a vortex is identified as a three-dimensional connected region following a particular set of pressure minima. The idea behind this search for a local pressure minimum comes from the expectation that the centrifugal force induced by fluid rotation (i.e., vorticity) causes a local drop in pressure. To find it, the gradient of the momentum equation (\ref{eqn:mom}) is analysed. Similarly to \(\lambda_2\), and using tensor notation, the gradient of (\ref{eqn:mom}) becomes

\begin{equation}
\label{eqn:phessian1}
\frac{D}{Dt}(\rho u_i)_{,j} + (\rho u_i)_{,k} u_{k,j} + (\rho \Theta u_i)_{,j} = -p_{,ij} + \tau_{ik,kj}
\end{equation}
\noindent
where \(\Theta=\boldsymbol{\nabla \cdot u}=u_{k,k}\) is the divergence of the velocity field or volume dilatation rate. The pressure Hessian, \(p_{,ij}\), contains information on the local curvature of the pressure field (i.e., second-order partial derivatives) and can be used to identify a pressure minimum. An equation for the pressure Hessian follows from the symmetric part of (\ref{eqn:phessian1}) as

\begin{equation}
\label{eqn:phessian2}
\frac{D S^{m}_{ji}}{Dt} - S^{\tau}_{ij} + S^{M}_{ij} + S^{\Theta}_{ij} = -p_{,ij}
\end{equation}
\noindent
with

\begin{equation}
\label{eqn:phessian3}
\begin{cases}
S^{m}_{ij} = \frac{1}{2}[(\rho u_i)_{,j} + (\rho u_j)_{,i}] \\
S^{\tau}_{ij} = \frac{1}{2}[\tau_{ik,kj} + \tau_{jk,ki}] \\
S^{M}_{ij} = \frac{1}{2}[(\rho u_i)_{,k} u_{k,j} + (\rho u_j)_{,k} u_{k,i}] \\
S^{\Theta}_{ij} = \frac{1}{2}[(\rho \Theta u_i)_{,j} + (\rho \Theta u_j)_{,i}]
\end{cases}
\end{equation}

A closer look to (\ref{eqn:phessian2}) shows that a pressure minimum may exist in the absence of a vortex since the unsteady fluid straining and viscous stress terms can create a pressure minimum. Therefore, a modified pressure Hessian, \(p_{,ij} \approx - S^{M}_{ij} - S^{\Theta}_{ij}\), is considered to focus only on the inertial effects that are usually linked to vortical motion. The eigenvalues of \(S^{M}_{ij} + S^{\Theta}_{ij}\) provide the necessary information to infer the existence of a local pressure minimum. That is, a vortex is a contiguous region with two negative eigenvalues of \(S^{M}_{ij} + S^{\Theta}_{ij}\) (i.e., two positive eigenvalues of \(p_{,ij}\)). In other words, the projection of the second-order partial derivatives of \(p\) in the eigenspace is the appropriate choice to represent the pressure variations. Ordering the tensor eigenvalues as \(\lambda_1\geq\lambda_2\geq\lambda_3\), the requirement is satisfied if \(\lambda_2<0\). This defines the incompressible vortex identification method \(\lambda_2\), which is replaced by \(\lambda_\rho\) in the compressible formulation to emphasise the inertial analysis once density is included. After finding the regions with an eigenvalue threshold value \(\lambda_{\rho,t}<0\), the isosurfaces with a constant negative value represent the vortices. Here, one assumes that the corresponding isosurface is aligned with the vorticity vector. In practice, this is true for a sufficiently large \(\lambda_{\rho,t}\). This value may be estimated from the problem scales defined in section~\ref{sec:description} as \(\lambda_{\rho,t}\sim-\rho(u_G/H)^2\), where \(\rho\) is the freestream density of the phase where we want to identify the vortices (i.e., liquid or gas). \par 

A direct connection between \(\lambda_2\) and \(\lambda_\rho\) follows from rewriting \(S^{M}_{ij}\)~\citep{yao2018toward} as

\begin{equation}
\label{eqn:phessian4}
S^{M}_{ij} = \rho(S_{ik}S_{kj}+\Omega_{ik}\Omega_{kj}) + \frac{\rho_{,k}}{2}(u_i u_{k,j} + u_j u_{k,i})
\end{equation}
\noindent
Here, the first term on the right-hand side is a density-weighted \(\lambda_2\), with \(S_{ij}\) and \(\Omega_{ij}\) being the symmetric and antisymmetric components, respectively, of the velocity gradient tensor, \(\boldsymbol{\nabla}\boldsymbol{u}\). Despite \(\lambda_2\) being seemingly a purely kinematic method, (\ref{eqn:phessian4}) highlights its dynamical nature in the limit of incompressible flows. The \(\lambda_\rho\) method can be applied to various compressible flows, including multi-phase flows~\citep{yao2018toward}. Similar to the viscous effects, the surface-tension force acting at the interface is neglected. Dimensional analysis shows that the viscous term in the momentum equation~(\ref{eqn:mom}) scales as \textit{Re}\(^{-1}\), while the body force representing the surface tension within the CSF framework scales with \textit{We}\(^{-1}\). For the study of the large-scale dynamics, both terms can be safely neglected since \(\textit{Re}\sim\mathcal{O}(10^3)\) and \(\textit{We}\sim\mathcal{O}(10^2)-\mathcal{O}(10^3)\) as seen in table~\ref{tab:cases}. \par 

\begin{figure}
\centering
\begin{subfigure}{0.33\textwidth}
  \centering
  \includegraphics[width=0.95\linewidth]{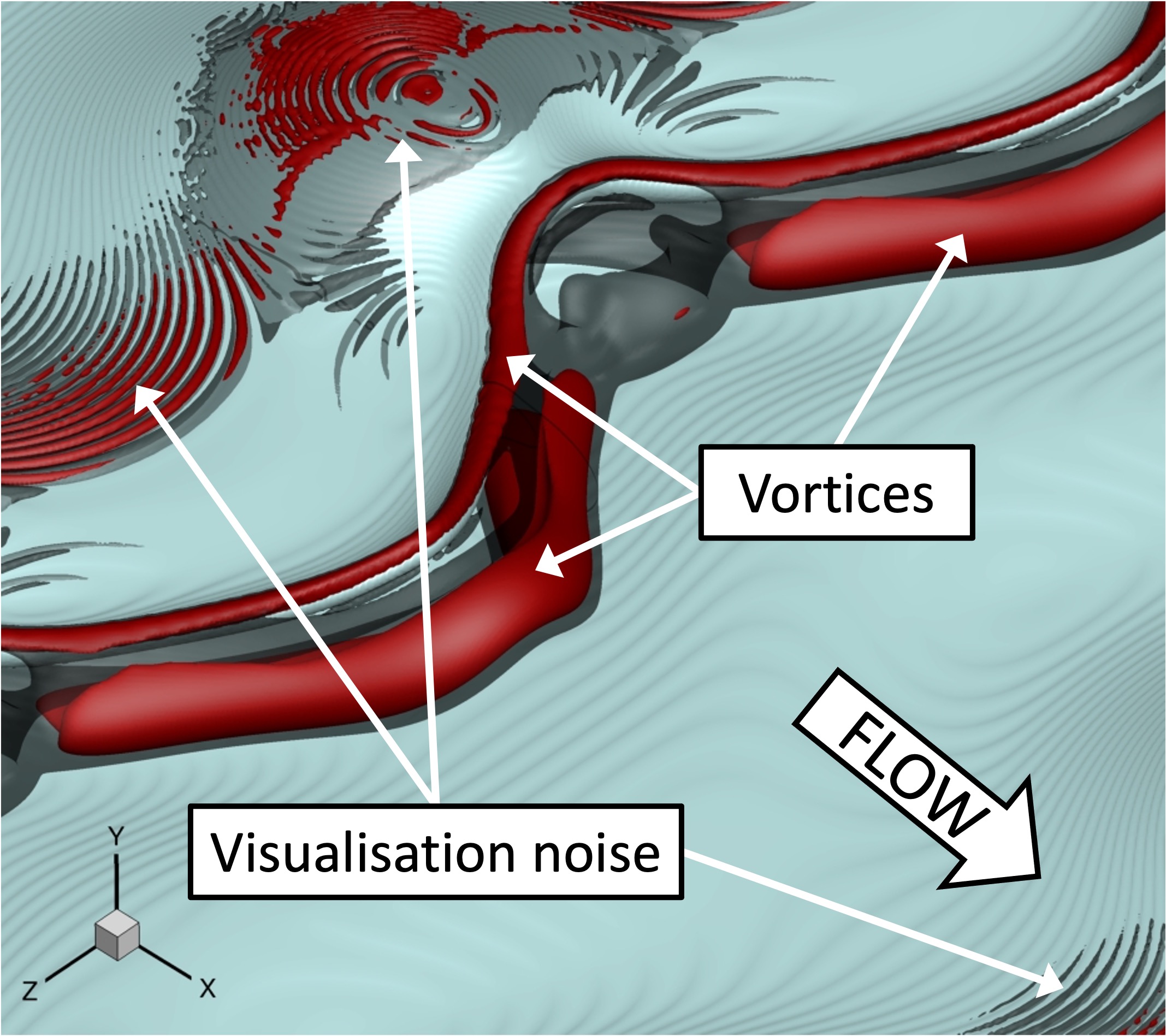}
  \caption{} 
  \label{subfig:Fig4a}
\end{subfigure}%
\begin{subfigure}{0.33\textwidth}
  \centering
  \includegraphics[width=0.95\linewidth]{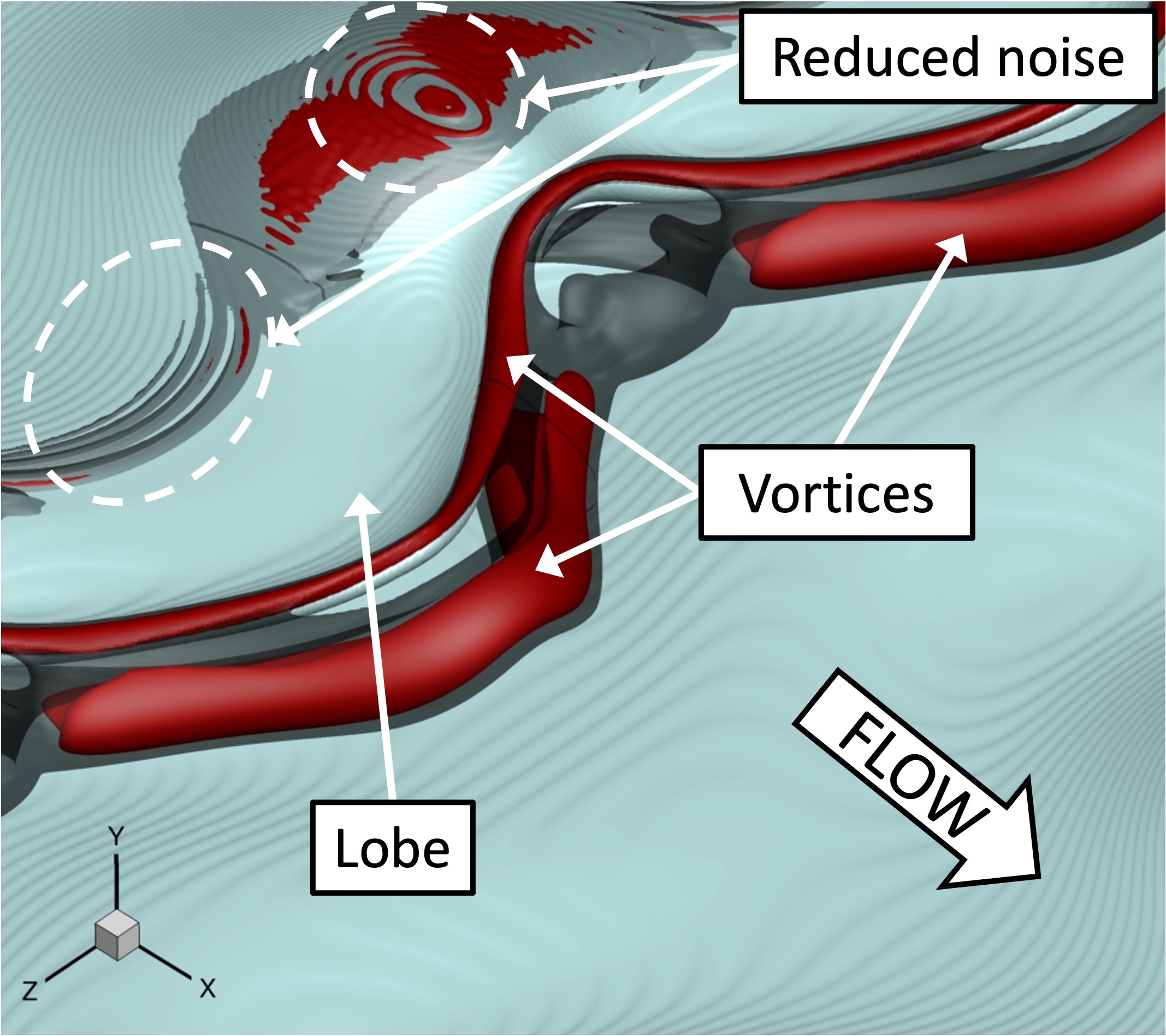}
  \caption{} 
  \label{subfig:Fig4b}
\end{subfigure}%
\begin{subfigure}{0.33\textwidth}
  \centering
  \includegraphics[width=0.95\linewidth]{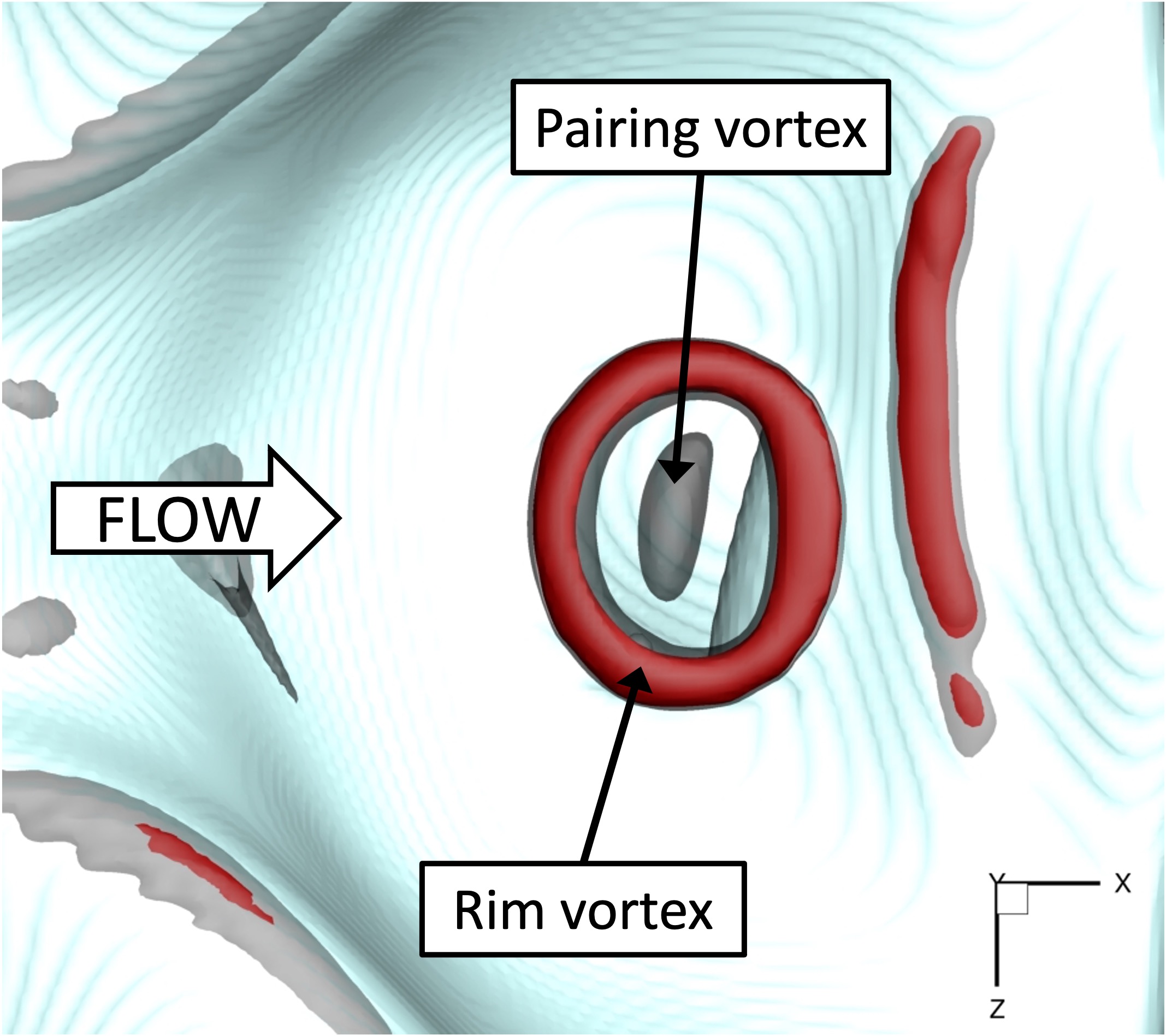}
  \caption{} 
  \label{subfig:Fig4c}
\end{subfigure}%
\caption{\(\lambda_\rho\) visualisation for case C1. The interface is identified by the blue isosurface with \(C=0.5\). The snapshot in 4a and 4b is at \(t=2.3\) \(\mu\)s and \(\lambda_{\rho,t} = -3\times 10^{15}\) (red) and \(\lambda_{\rho,t} = -1\times 10^{15}\) (translucid black). The snapshot 4c is at \(t=7.4\) \(\mu\)s and \(\lambda_{\rho,t} = -5\times 10^{15}\) (red) and \(\lambda_{\rho,t} = -2.5\times 10^{15}\) (translucid black). (a) noise generated by the sharp interface method without filtering the velocity field; (b) noise reduction by filtering the velocity field; and (c) loss of information caused by the choice of \(\lambda_{\rho,t}\).}
\label{fig:Fig4}
\end{figure}

Despite being a suitable vortex identification method for multi-phase flows, visualisation of \(\lambda_\rho\) might be affected by the discrete jumps of fluid properties and spurious currents around the interface emerging in sharp interface methods such as VOF. Their impact on the evolution of the liquid surface has been deemed negligible in the studied configurations (PS), but the calculation of \(\lambda_\rho\) magnifies the spurious currents. We refer to this issue as visualisation noise, which somewhat blurs vortex structures near the surface. Thus, it becomes essential to select a proper value of \(\lambda_{\rho,t}\). Visualisation noise is minimised during post-processing by considering phase-wise density gradients and locally averaging the velocity components, effectively filtering the velocity field with a spatial filter of \(\sim 2\Delta x\). The first action reduces the density gradient across the interface, while the filtered velocity smooths the spurious currents at the grid size level. After these steps, the physical vortices are still well captured, as evidenced in figure~\ref{fig:Fig4}. \par

Another consideration is that the \(\lambda_\rho\) method only identifies a vortex but does not provide information about its rotation (e.g., clockwise or counterclockwise). This information is extracted from the vorticity field, \(\boldsymbol{\omega}=\boldsymbol{\nabla\times u}=(\omega_x,\omega_y,\omega_z)\). Then, the magnitude of the eigenvalue can be understood as a local measure of the vorticity intensity. Therefore, some arbitrariness exists when representing a vortex with a specific value of \(\lambda_\rho\). Moreover, reduction of the visualisation noise demands a careful selection of the \(\lambda_\rho\) value. Small \(\lambda_\rho\) thresholds display more noise; thus, larger values are preferred. However, larger values might not capture all relevant structures. Here, we make an assessment to ensure the vortical structures are accurately captured by varying \(\lambda_\rho\) and addressing the changes in the vortex visualisation. \par

Figures~\ref{subfig:Fig4a} and~\ref{subfig:Fig4b} illustrate the visualisation noise problem for a snapshot of the 150-bar case C1 at \(t=2.3\) \(\mu\)s. \(\lambda_\rho\) has units of kg/(m\(^3\)s\(^2)\), but it is not specified in the text and figures for brevity. The unfiltered velocity field is used to evaluate \(\lambda_\rho\) in figure~\ref{subfig:Fig4a}, while the filtered velocity field is used in figure~\ref{subfig:Fig4b}. In both cases, the phase-wise density gradients are employed. The visualisation noise does not correspond to any physical vortical motion, and the filtering of the velocity field considerably reduces the amount of noise. Note how the vortices are still well captured after the filtering process. Then, the impact of the \(\lambda_{\rho,t}\) value is evident. The black translucid isosurface with \(\lambda_{\rho,t} = -1\times 10^{15}\) displays considerable noise on the liquid surface, but the red isosurface with \(\lambda_{\rho,t} = -3\times 10^{15}\) mitigates the problem while still capturing the stronger physical vortices everywhere else. The roller vortex or deformed KH vortex is seen in front of the lobes, as well as a vortex along their edges. Note that the choice of \(\lambda_{\rho,t} = -3\times 10^{15}\) erases the weaker vortical region between the two consecutive lobes, effectively ``breaking up" the vortex during visualisation. We know the roller must be a connected structure. Thus, it is easy to reconstruct the vortex accordingly. However, the loss of information can be critical as the liquid surface deforms and breaks into smaller structures. Figure~\ref{subfig:Fig4c} shows a later snapshot of case C1 at \(t=7.4\) \(\mu\)s. There, the liquid sheet is perforated, and new vortical structures emerge. With an isosurface represented by \(\lambda_{\rho,t} = -5\times 10^{15}\) the rim vortex following the edge of the hole is well captured, but not the pairing vortex that forms in the centre of the hole. This vortex can be captured with \(\lambda_{\rho,t} = -2.5\times 10^{15}\), but its existence is not evident and could easily be missed. \par

Note that the apparent waviness of the blue isosurface representing the liquid surface in figure~\ref{fig:Fig4} is a plotting issue. Since \(C\) is not a smooth function across the interface, the isosurface \(C=0.5\) might look like a stepwise function depending on the plotting software. For maximum misalignment of the surface with respect to the grid axes, the waviness is of the order of the mesh size. Further grid refinement mitigates the problem. Nonetheless, the numerical code uses a proper interface representation given by the VOF method (outlined in section~\ref{sec:modeling}), and the visualisation noise of \(\lambda_\rho\) is of numerical origin. This issue is apparent in figure~\ref{fig:Fig3} and in the figures presented in section~\ref{sec:results}. \par

\section{Results and Discussion}
\label{sec:results}

This section is structured as follows. First, some of the early deformation mechanisms discussed in PS are described in sections~\ref{subsec:lobe_bending} and~\ref{subsec:lobe_crest_corrugation}, showing the strong coupling between vortices in the gaseous phase and the variation of fluid properties within the transcritical liquid. Then, section~\ref{subsec:layering} highlights the impact of layering on the evolution of the vortical field, and section~\ref{subsec:fuelblobs} shows the interaction of vortices with fuel-rich gaseous blobs. Lastly, section~\ref{subsec:vorticity_generation} peers into the transcritical flow to describe the evolution of vortex structures and quantify the terms in the vorticity equation. \par

The major deformation mechanisms identified by PS are summarised in figure~\ref{subfig:Fig1b}. The analysed configurations presented in table~\ref{tab:cases} and visualised in figure~\ref{subfig:Fig1b} are classified as follows. Looking at the early deformation stages, cases A1, A2, B1 and C1 involve a lobe stretching, bending and perforation mechanism (section~\ref{subsec:lobe_bending}). Note that perforation may occur only at very high pressures. Thus, cases A1 and A2 show negligible hole formation. Then, cases B2, C2 and C3 fall within the lobe and crest corrugation mechanism (section~\ref{subsec:lobe_crest_corrugation}). For all cases with an ambient pressure of 100 bar and 150 bar (i.e., B and C cases), the layering of liquid sheets becomes the dominant deformation mechanism after some time (section~\ref{subsec:layering}). For reference, the Supplemental Material in PS shows the jet deformation for all cases that support the discussion that follows. \par 

In this study, a non-dimensional time, \(t^*=t/t_c=t\frac{u_G}{H}\), is used to compare between configurations. The characteristic time, \(t_c\), is defined from the jet thickness and the gas freestream velocity (see section~\ref{sec:description}). Due to limited computational resources, the simulations run only to \(t^*=15\). During this time, a fluid particle in the gas freestream crosses the computational domain ten times in all configurations. Instead, the perturbation wave immersed in the shear layer is estimated to travel across the domain at least five times. \par

\subsection{Lobe stretching, bending, and perforation}
\label{subsec:lobe_bending}

The early deformation pattern for \textit{We}\(_G\) \(<1000\) involves lobes stretching downstream into the gas phase above the liquid surface. For higher thermodynamic pressures, the lobes become thinner and their dynamics are easily influenced by the gas flow around them. In particular, they bend upward and are eventually perforated (e.g., cases B1 and C1). Here, case C1 is analysed in detail since it presents the strongest real-fluid thermodynamics and intraphase mixing effects. Figure~\ref{fig:Fig8} conveys the differences between cases A2, B1 and C1, all of which have a similar \textit{We}\(_G\). \par 

\begin{figure}
\centering
\begin{subfigure}{0.25\textwidth}
  \centering
  \includegraphics[width=1.0\linewidth]{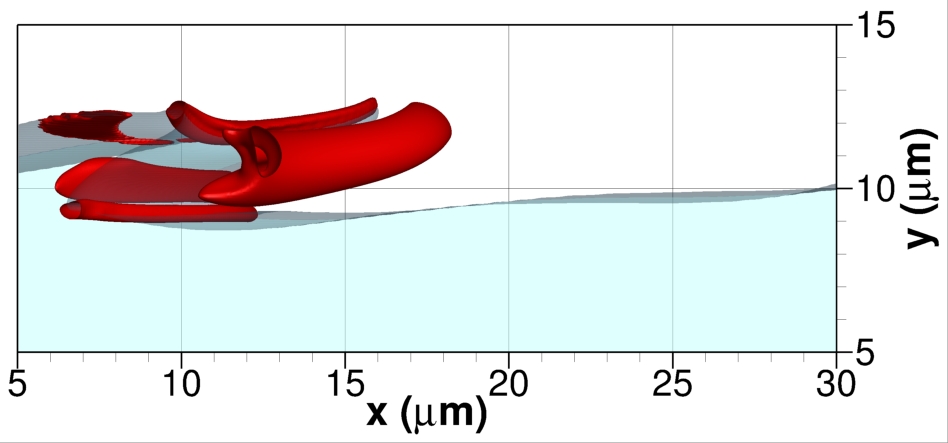}
  \label{subfig:Fig5a}
\end{subfigure}%
\begin{subfigure}{0.25\textwidth}
  \centering
  \includegraphics[width=1.0\linewidth]{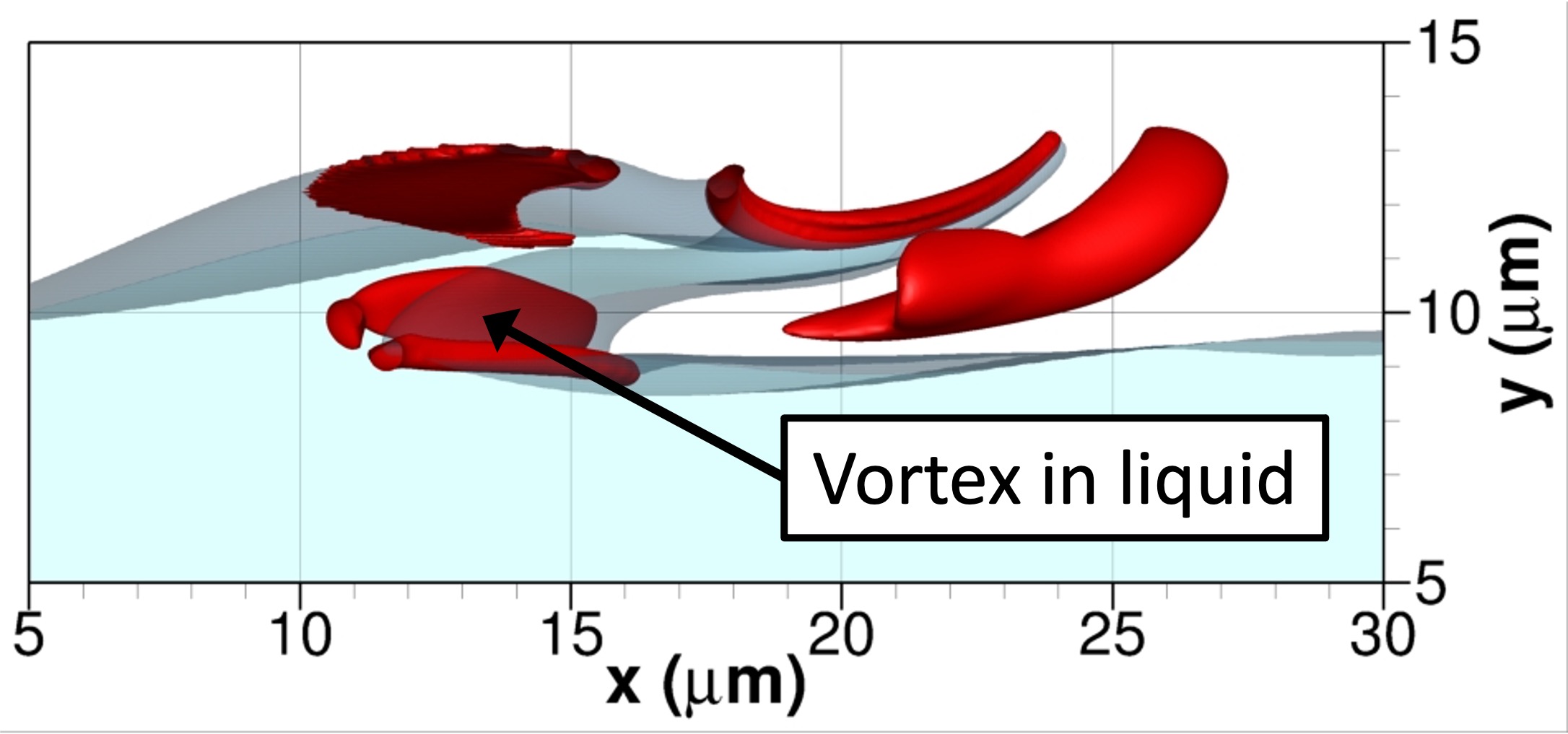}
  \label{subfig:Fig5b}
\end{subfigure}%
\begin{subfigure}{0.25\textwidth}
  \centering
  \includegraphics[width=1.0\linewidth]{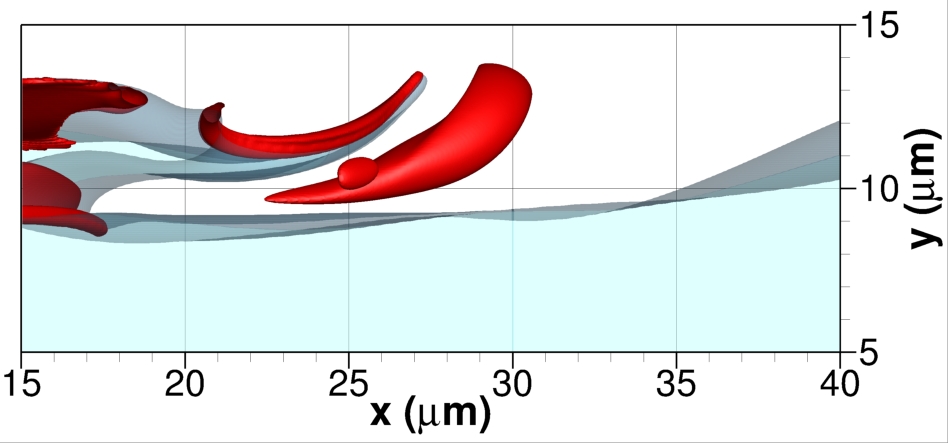}
  \label{subfig:Fig5c}
\end{subfigure}%
\begin{subfigure}{0.25\textwidth}
  \centering
  \includegraphics[width=1.0\linewidth]{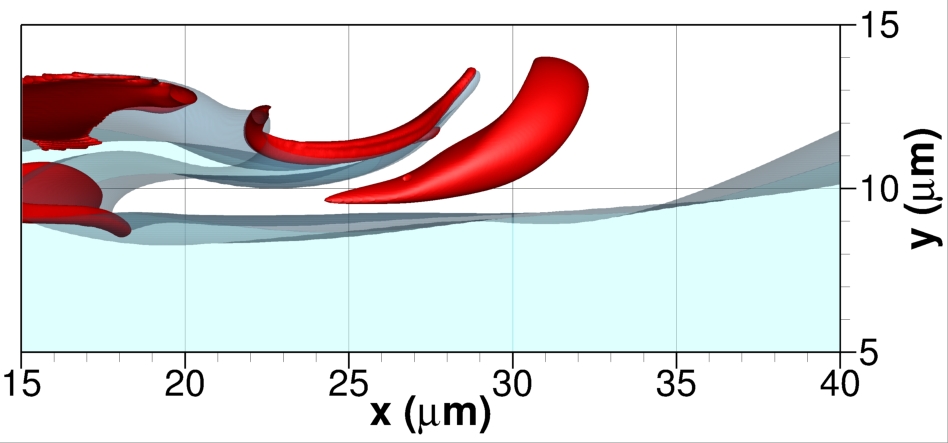}
  \label{subfig:Fig5d}
\end{subfigure}%
\\[-2.7ex]
\begin{subfigure}{0.25\textwidth}
  \centering
  \includegraphics[width=1.0\linewidth]{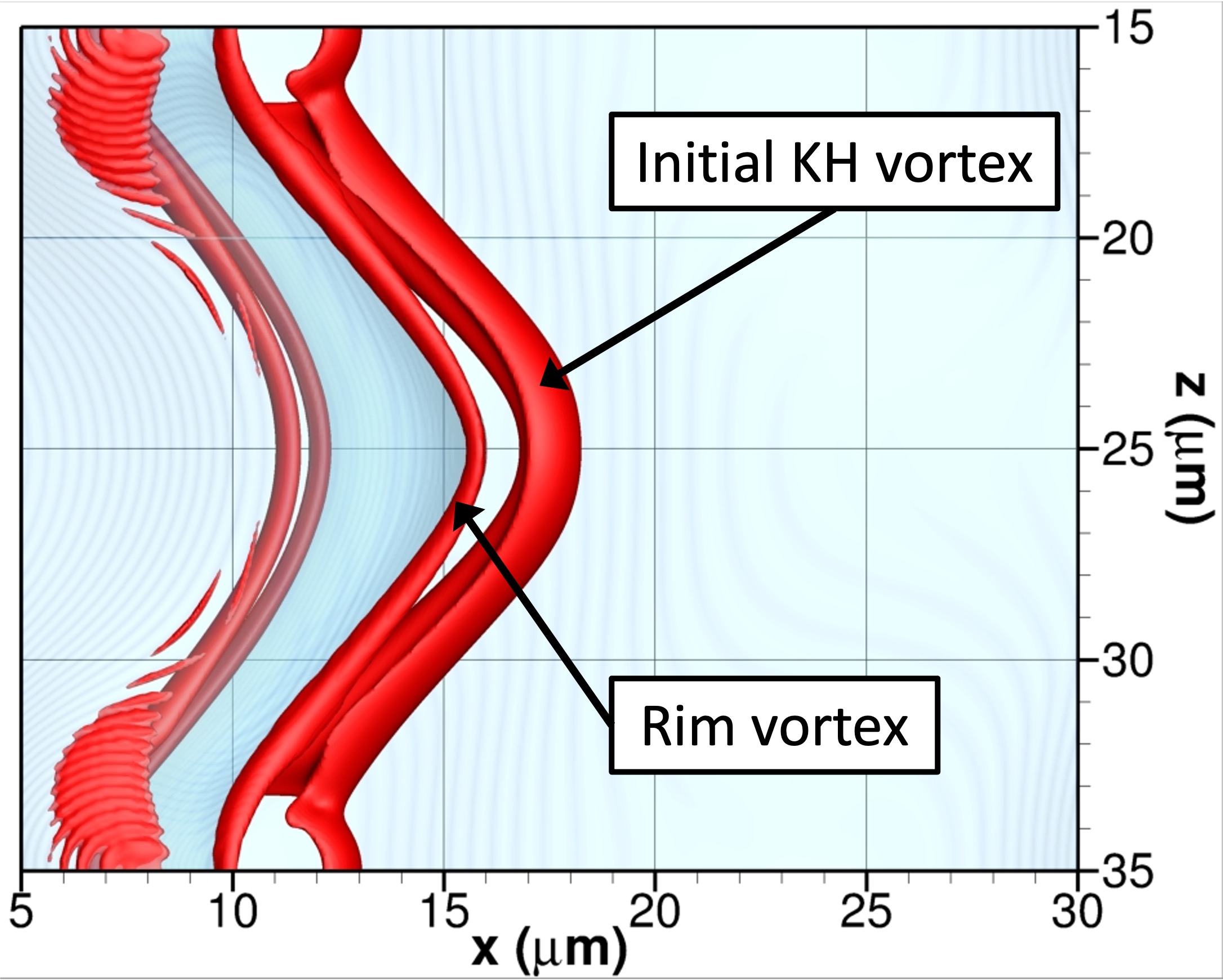}
  \caption{} 
  \label{subfig:Fig5e}
\end{subfigure}%
\begin{subfigure}{0.25\textwidth}
  \centering
  \includegraphics[width=1.0\linewidth]{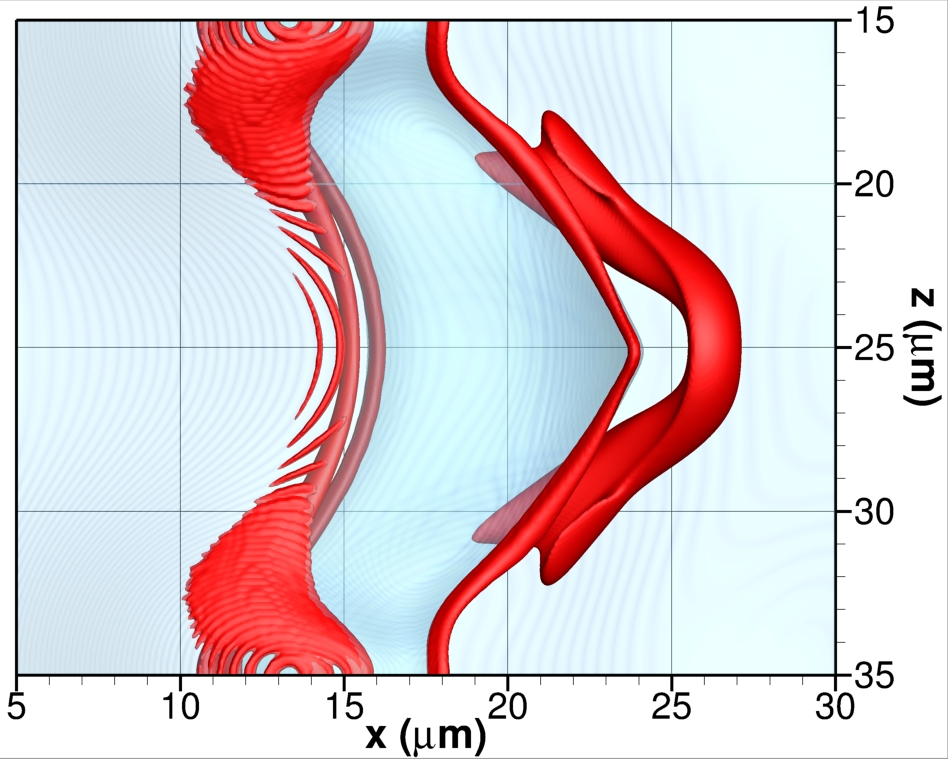}
  \caption{} 
  \label{subfig:Fig5f}
\end{subfigure}%
\begin{subfigure}{0.25\textwidth}
  \centering
  \includegraphics[width=1.0\linewidth]{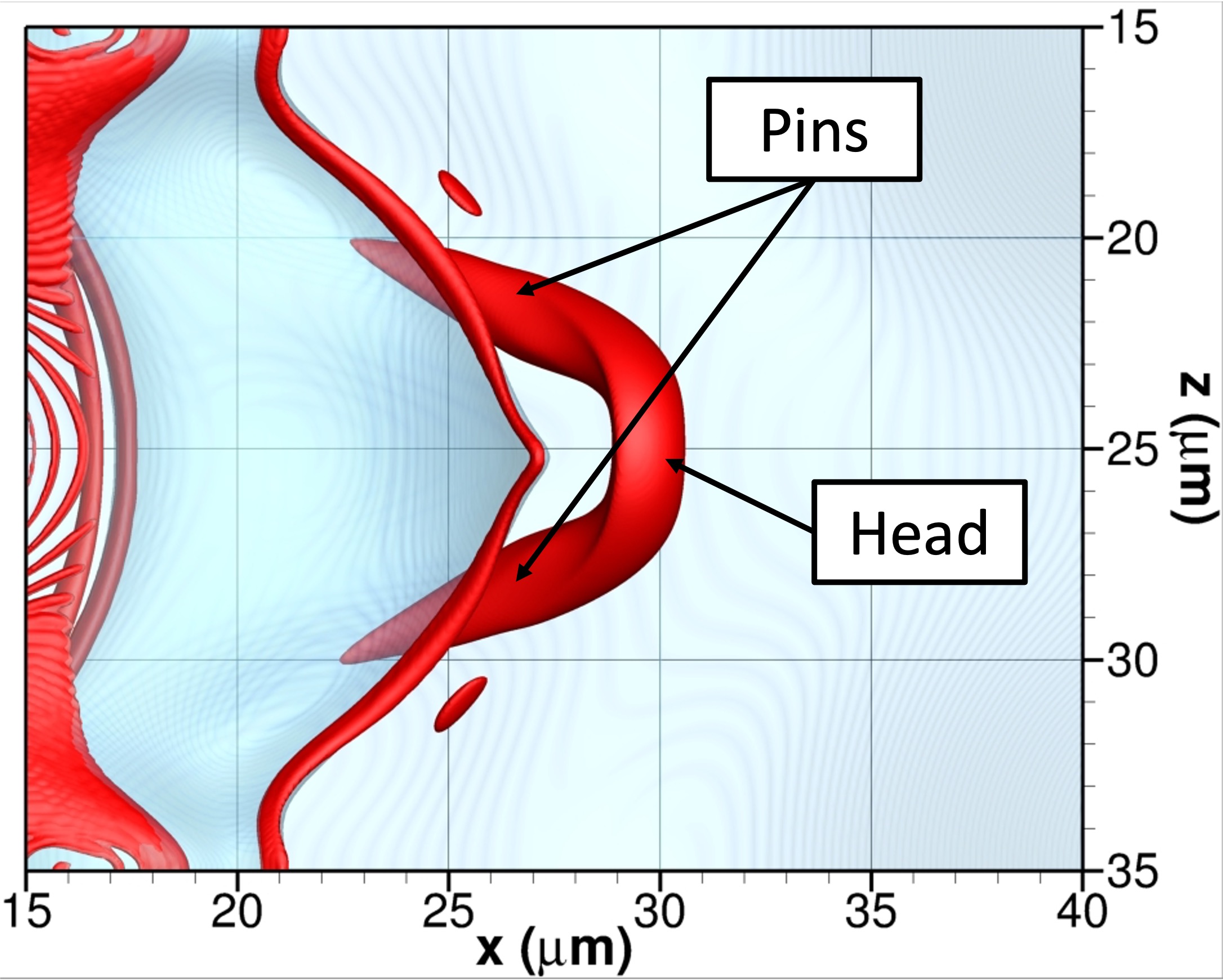}
  \caption{} 
  \label{subfig:Fig5g}
\end{subfigure}%
\begin{subfigure}{0.25\textwidth}
  \centering
  \includegraphics[width=1.0\linewidth]{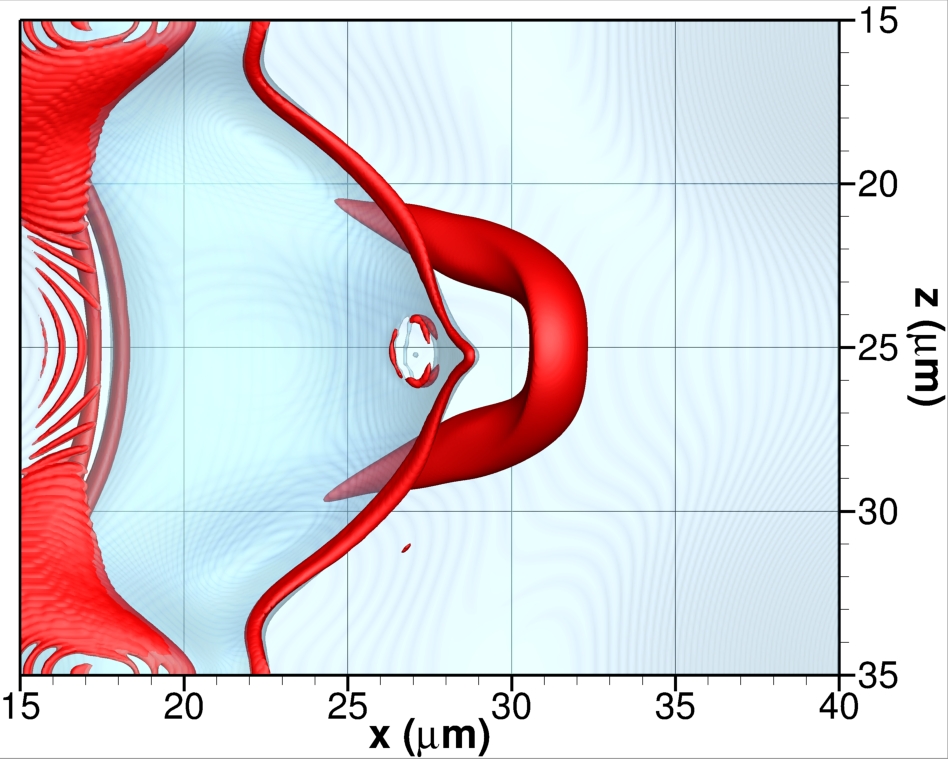}
  \caption{} 
  \label{subfig:Fig5h}
\end{subfigure}%
\caption{Vortex dynamics of the lobe extension, bending and perforation mechanism for case C1. The top figures show the side view at \(z=35\) \(\mu\)m and the bottom figures show the top view of the liquid. The liquid surface is identified by the translucid blue isosurface with \(C=0.5\) and the vortex structures are the red isosurface with \(\lambda_{\rho,t}=-2.5\times 10^{15}\). (a) \(t^*=3\); (b) \(t^*=3.75\); (c) \(t^*=4.05\); and (d) \(t^*=4.2\).}
\label{fig:Fig5}
\end{figure}

The evolution of the initial vortex (KH vortex) forming from the shear between the fluids is presented in figure~\ref{fig:Fig5}. This vortex is responsible for the early lobe deformation. \(\lambda_{\rho,t}=-2.5\times 10^{15}\) has been chosen to represent this vortex, as it is a good compromise between vortex identification and visualisation noise. Note that during the deformation of the vortex, some regions weaken and the magnitude of \(\lambda_\rho\) falls below the chosen threshold \(\lambda_{\rho,t}\). This roller vortex rotates clockwise (i.e., toward \(-z\)) as seen from the \(xy\) planes in figure~\ref{fig:Fig5}, and rapidly deforms into a hairpin vortex between \(t^*=3\) and \(t^*=4.2\). During this process, the vortex head is always located downstream of the lobe's tip. Initially, the vortex induces the stretching of the lobe in the streamwise direction, which is easily deformed due to the extreme transcritical environment at 150 bar. The dissolution of the ambient gas into the liquid phase and the heating decrease the liquid density and generate a liquid mixture with a gas-like viscosity near the interface (see figure 9 in PS). Thus, this fluid is easily perturbed by the dense and compressed gaseous phase at 150 bar. As the vortex pins (or legs) align with the streamwise direction, the vortex head moves upward into the oxidizer stream. As a consequence, the lobe stretches and bends sharply, facing the incoming gas before being perforated between \(t^*=4.05\) and \(t^*=4.2\). This combined motion of lobe and vortex is caused by the velocity induced by the counter-rotating vortex pins as the gas flows underneath them and into the lower side of the lobe. Such self-induction mechanism is described in~\citet{zandian2018understanding}. Two other vortices are identified: a vortex along the lobe's flank and a vortex within the liquid phase. Following the nomenclature presented in figure~\ref{fig:Fig3}, the first vortex is henceforth called rim vortex. A similar vortex is identified as a hairpin vortex in~\citet{zandian2018understanding} because of its shape. The other vortex in the liquid phase is identified via the density-weighting effect of \(\lambda_\rho\). Both vortices rotate clockwise when viewed toward \(-z\). \par 

\begin{figure}
\centering
\begin{subfigure}{0.33\textwidth}
  \centering
  \includegraphics[width=1.0\linewidth]{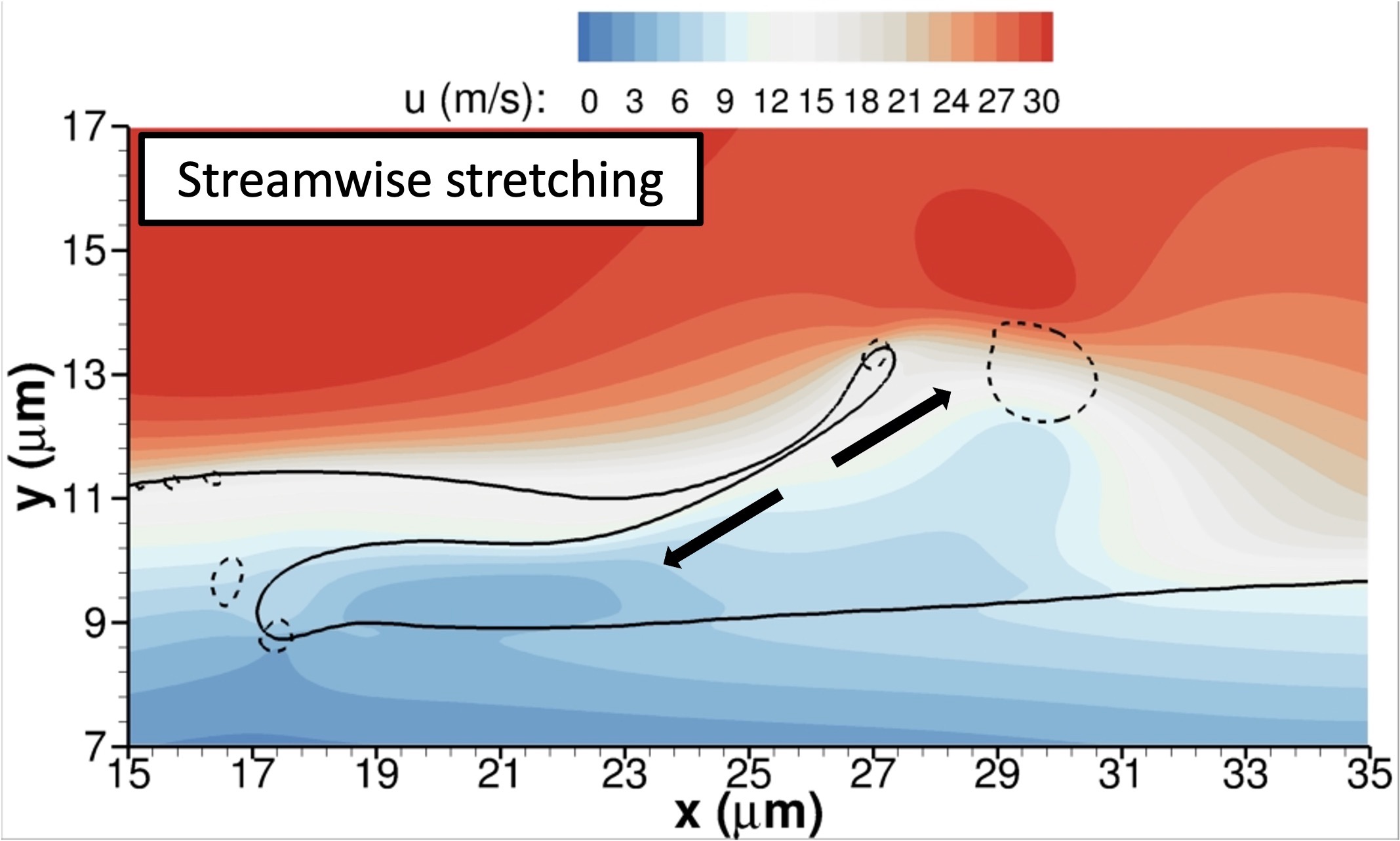}
  \caption{} 
  \label{subfig:Fig6a}
\end{subfigure}%
\begin{subfigure}{0.33\textwidth}
  \centering
  \includegraphics[width=1.0\linewidth]{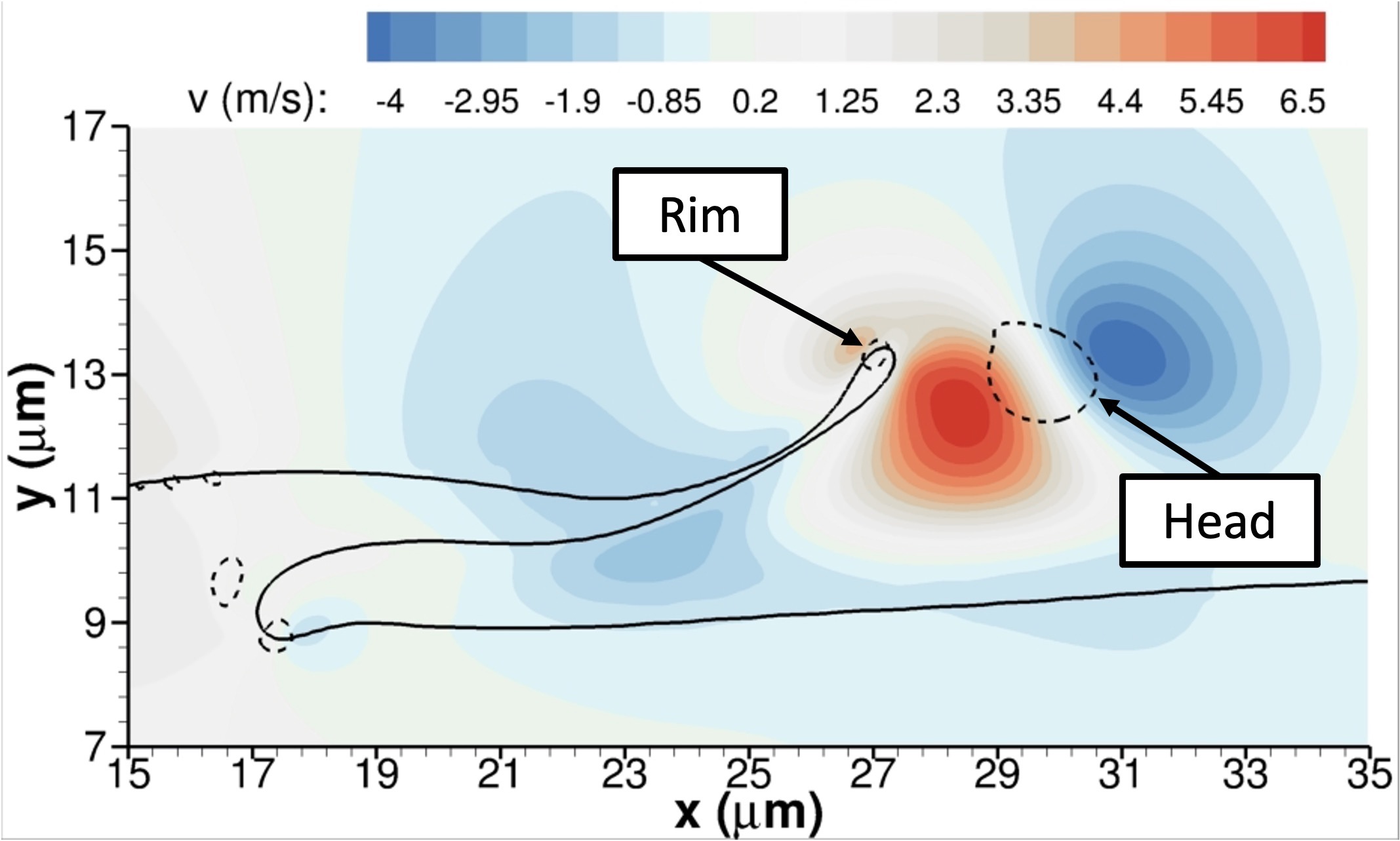}
  \caption{} 
  \label{subfig:Fig6b}
\end{subfigure}%
\begin{subfigure}{0.33\textwidth}
  \centering
  \includegraphics[width=1.0\linewidth]{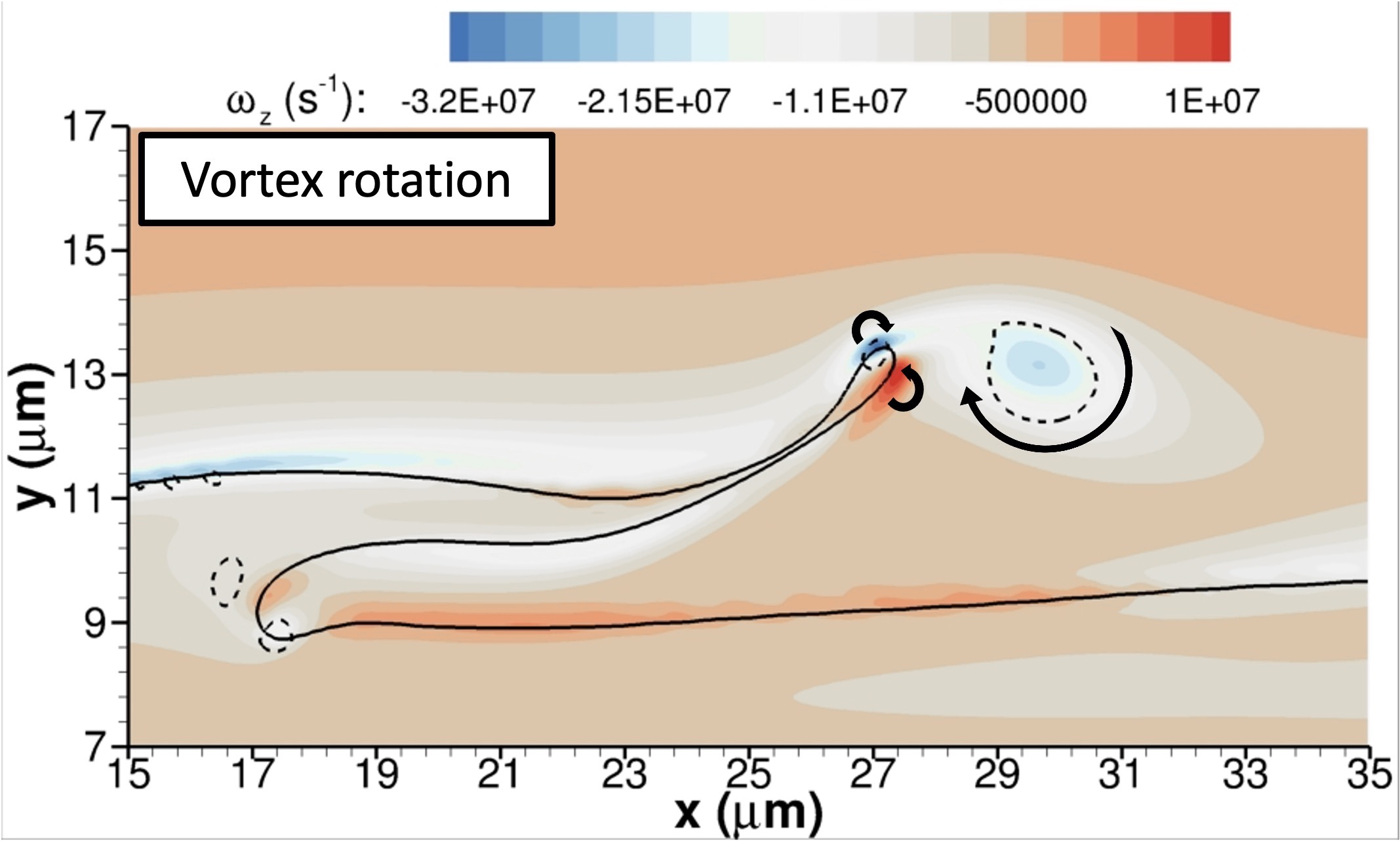}
  \caption{} 
  \label{subfig:Fig6c}
\end{subfigure}%
\\
\begin{subfigure}{0.33\textwidth}
  \centering
  \includegraphics[width=1.0\linewidth]{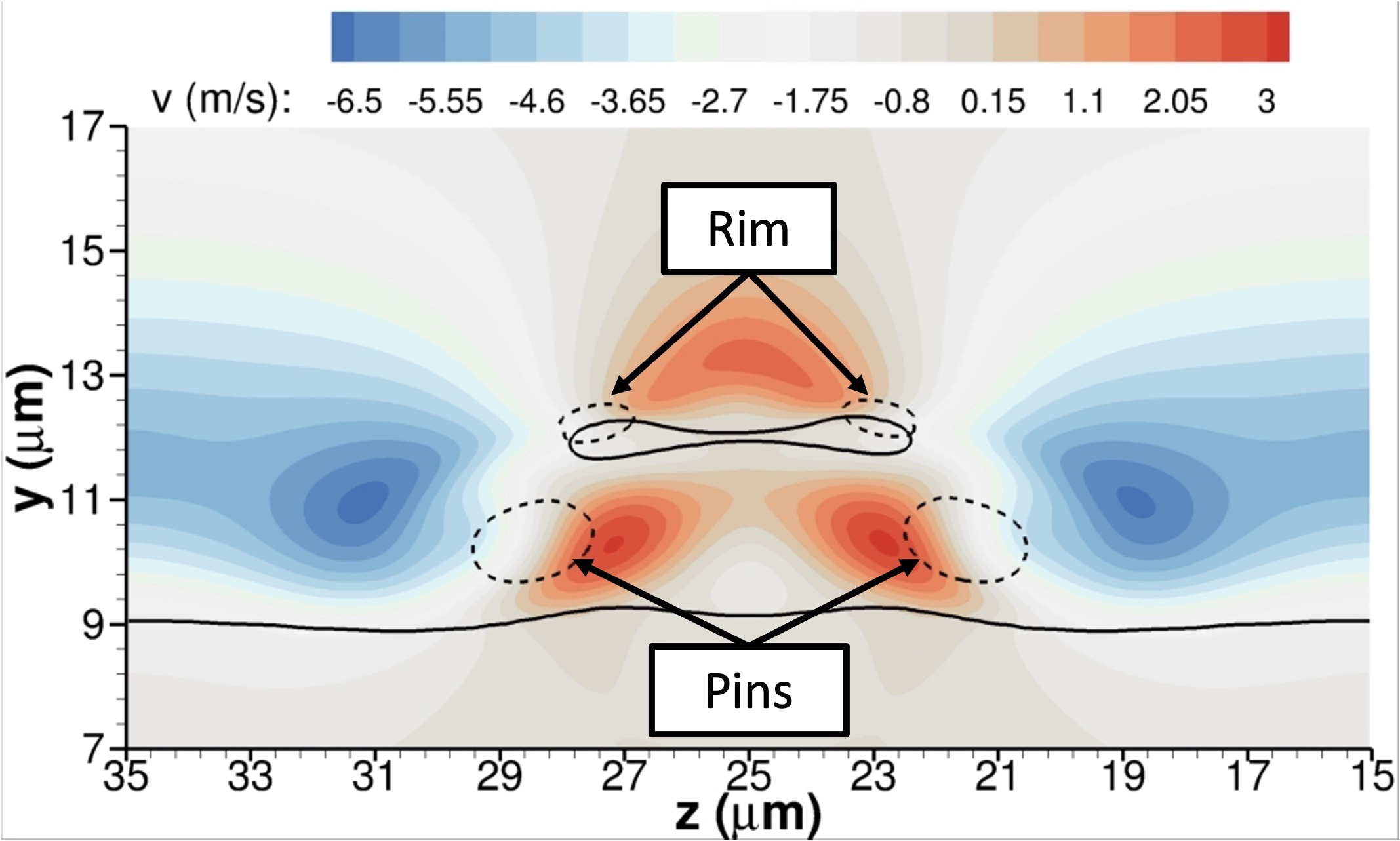}
  \caption{} 
  \label{subfig:Fig6d}
\end{subfigure}%
\begin{subfigure}{0.33\textwidth}
  \centering
  \includegraphics[width=1.0\linewidth]{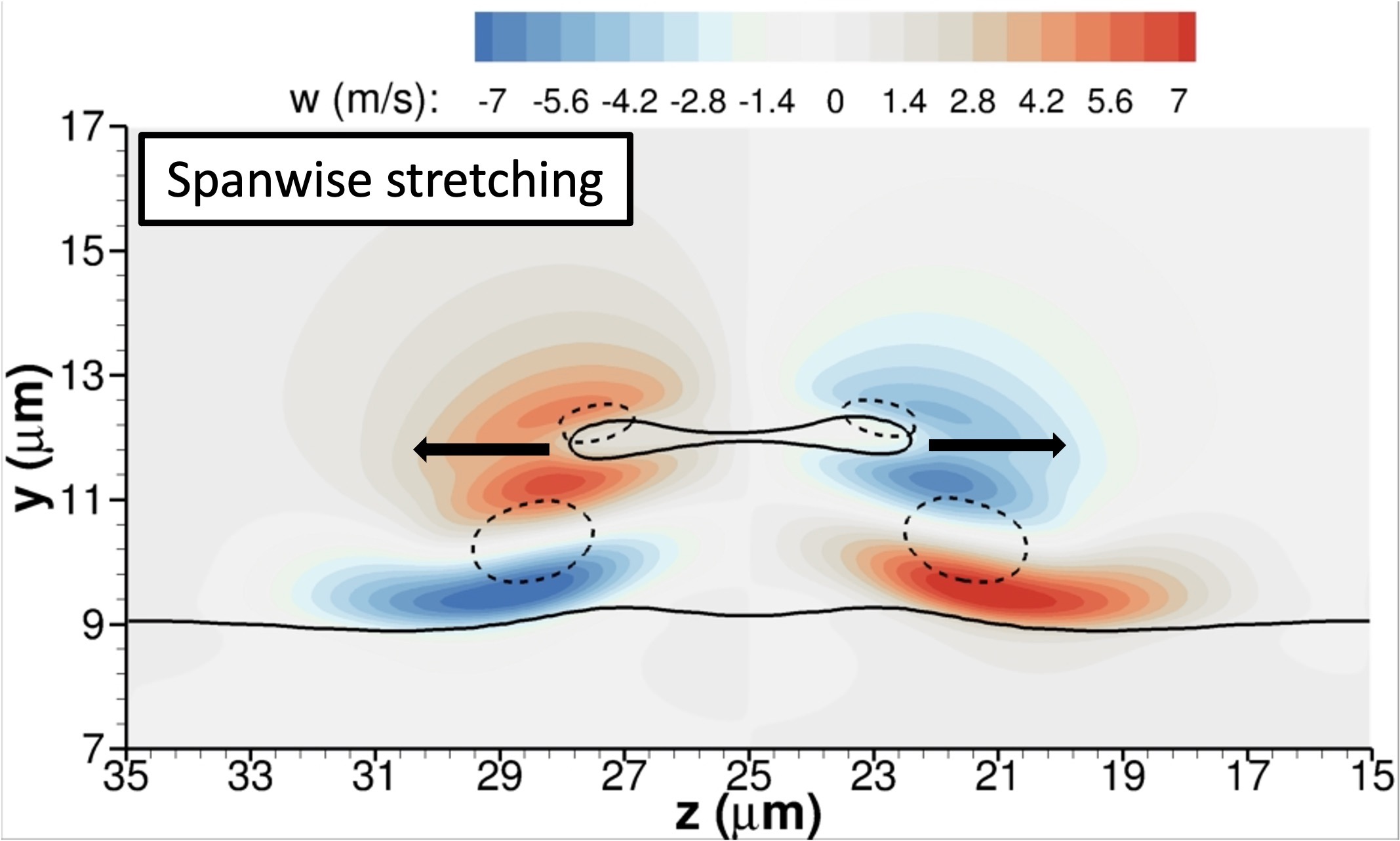}
  \caption{} 
  \label{subfig:Fig6e}
\end{subfigure}%
\begin{subfigure}{0.33\textwidth}
  \centering
  \includegraphics[width=1.0\linewidth]{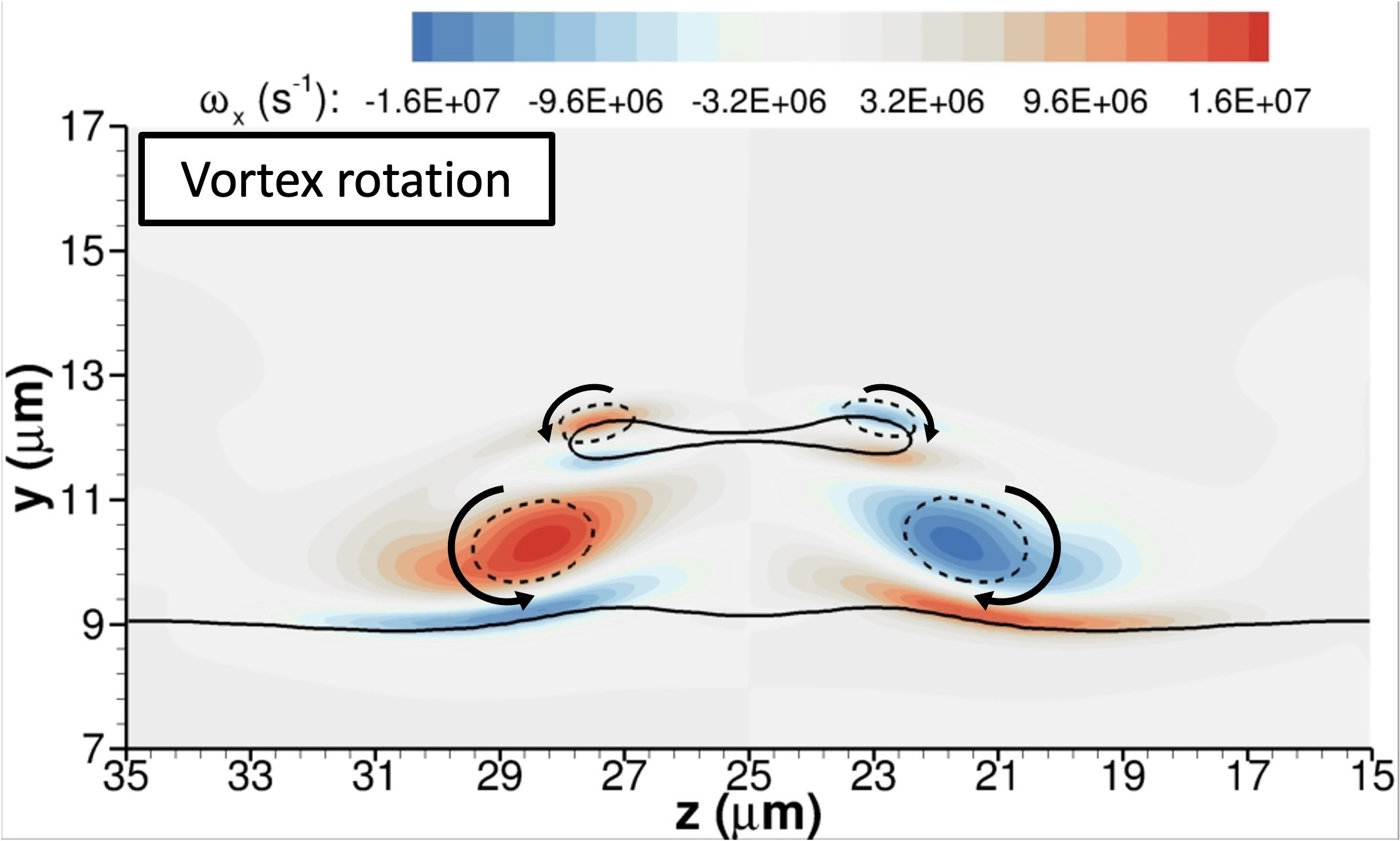}
  \caption{} 
  \label{subfig:Fig6f}
\end{subfigure}%
\caption{Vortex dynamics of the lobe stretching, bending and perforation mechanism for case C1 at \(t^*=4.05\). The interface is identified by the solid isocontour with \(C=0.5\) and the vortex cross-sections are identified by the dashed isocontour with \(\lambda_{\rho,t}=-2.5\times 10^{15}\). (a) \(u\) at \(z=25\) \(\mu\)m; (b) \(v\) at \(z=25\) \(\mu\)m; (c) \(\omega_z\) at \(z=25\) \(\mu\)m; (d) \(v\) at \(x=26\) \(\mu\)m; (e) \(w\) at \(x=26\) \(\mu\)m; and (f) \(\omega_x\) at \(x=26\) \(\mu\)m.}
\label{fig:Fig6}
\end{figure}

The velocity and vorticity fields explain the deformation process. Figure~\ref{fig:Fig6} presents contours of relevant components of the velocity and vorticity vectors in two flow cross-sections at \(t^*=4.05\). The \(xy\) plane at \(z=25\) \(\mu\)m and the \(yz\) plane at \(x=26\) \(\mu\)m cut the lobe near its tip. The initial evolution of the lobe looks spanwise symmetric across its centerline, but over time symmetry is broken and the liquid structures behave independently. The rotation of the vorticity field is highlighted for completeness. The vortex head is aligned with the spanwise direction, while the vorticity vector in the vortex pins is predominantly in the streamwise direction (see figure~\ref{fig:Fig5}). The vorticity around the rim vortex is also noteworthy. \par 

Figure~\ref{fig:Fig6} captures well the gas flow underneath the hairpin vortex and how it is ejected upward and toward the lobe, causing the described lobe-vortex deformation and hole formation. The hairpin vortex stretches and thins the lobe in both \(x\) and \(z\), leading to perforation. Liquid-phase stretching is enhanced at high pressures and defines a dominant deformation mechanism where liquid stretching and folding precedes the formation of liquid sheets or layers. This is a result of the low surface tension and the strong variation of liquid properties caused by the dissolution of the ambient gas (PS). \par

\begin{figure}
\centering
\begin{subfigure}{0.5\textwidth}
  \centering
  \includegraphics[width=1.0\linewidth]{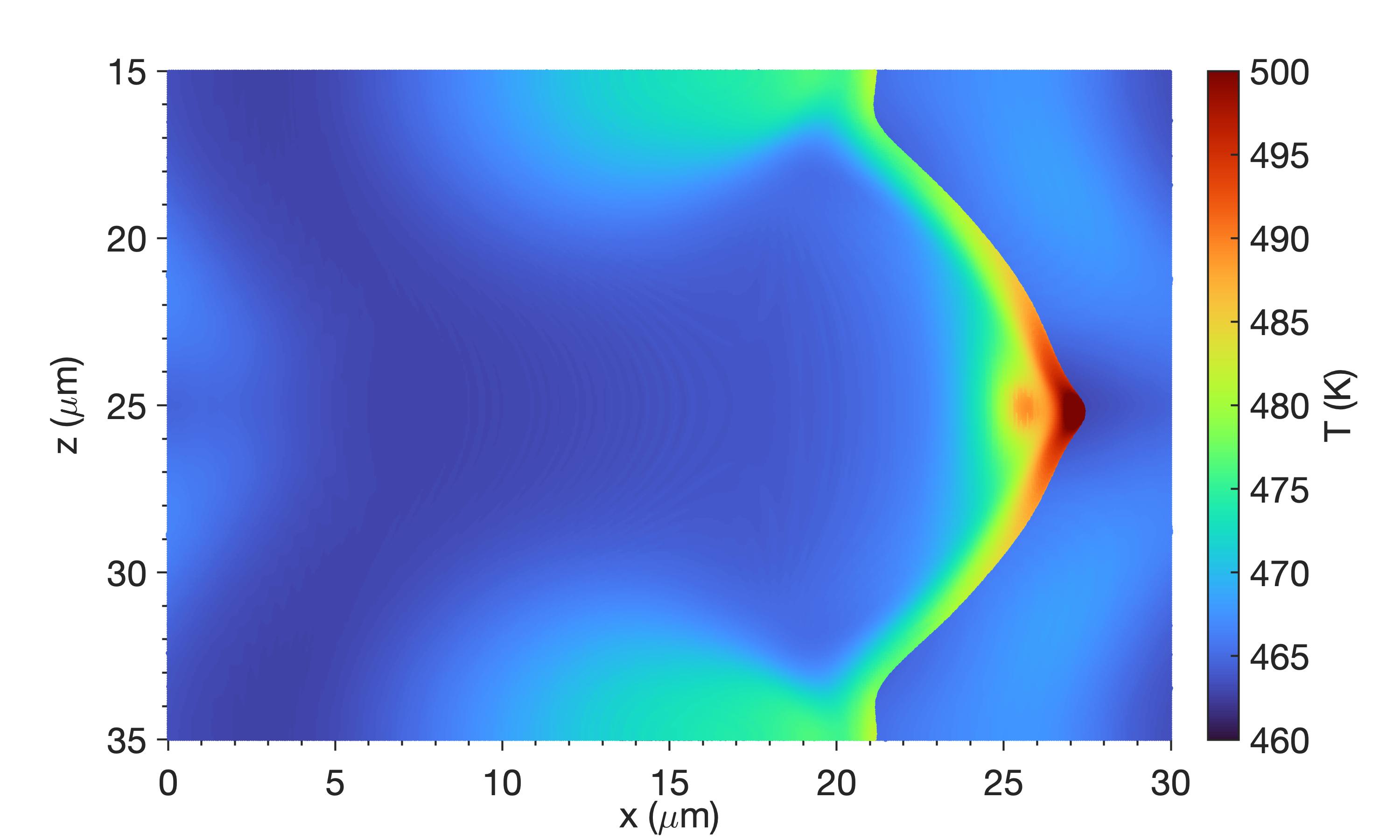}
  \caption{} 
  \label{subfig:Fig7a}
\end{subfigure}%
\begin{subfigure}{0.5\textwidth}
  \centering
  \includegraphics[width=1.0\linewidth]{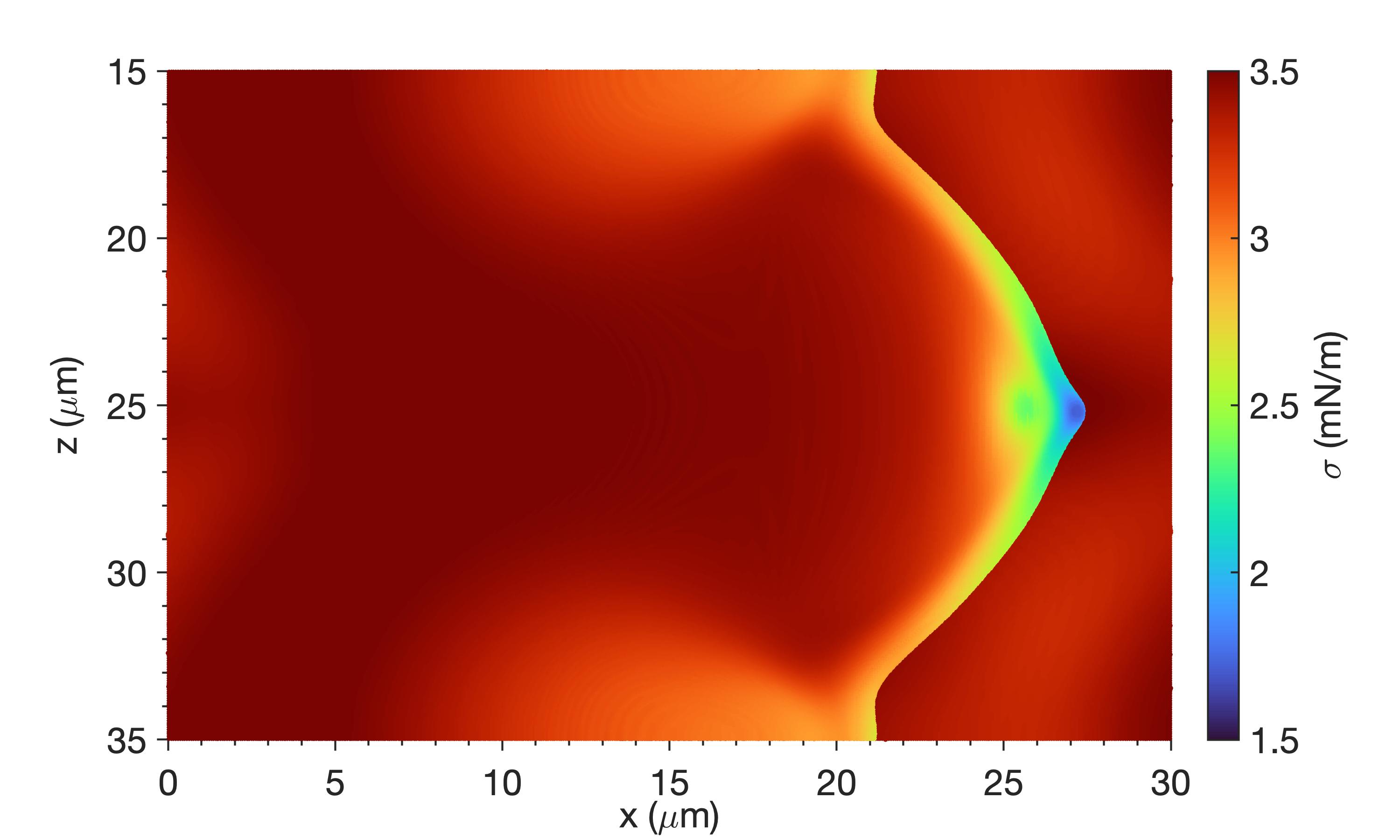}
  \caption{} 
  \label{subfig:Fig7b}
\end{subfigure}%
\\
\begin{subfigure}{0.5\textwidth}
  \centering
  \includegraphics[width=1.0\linewidth]{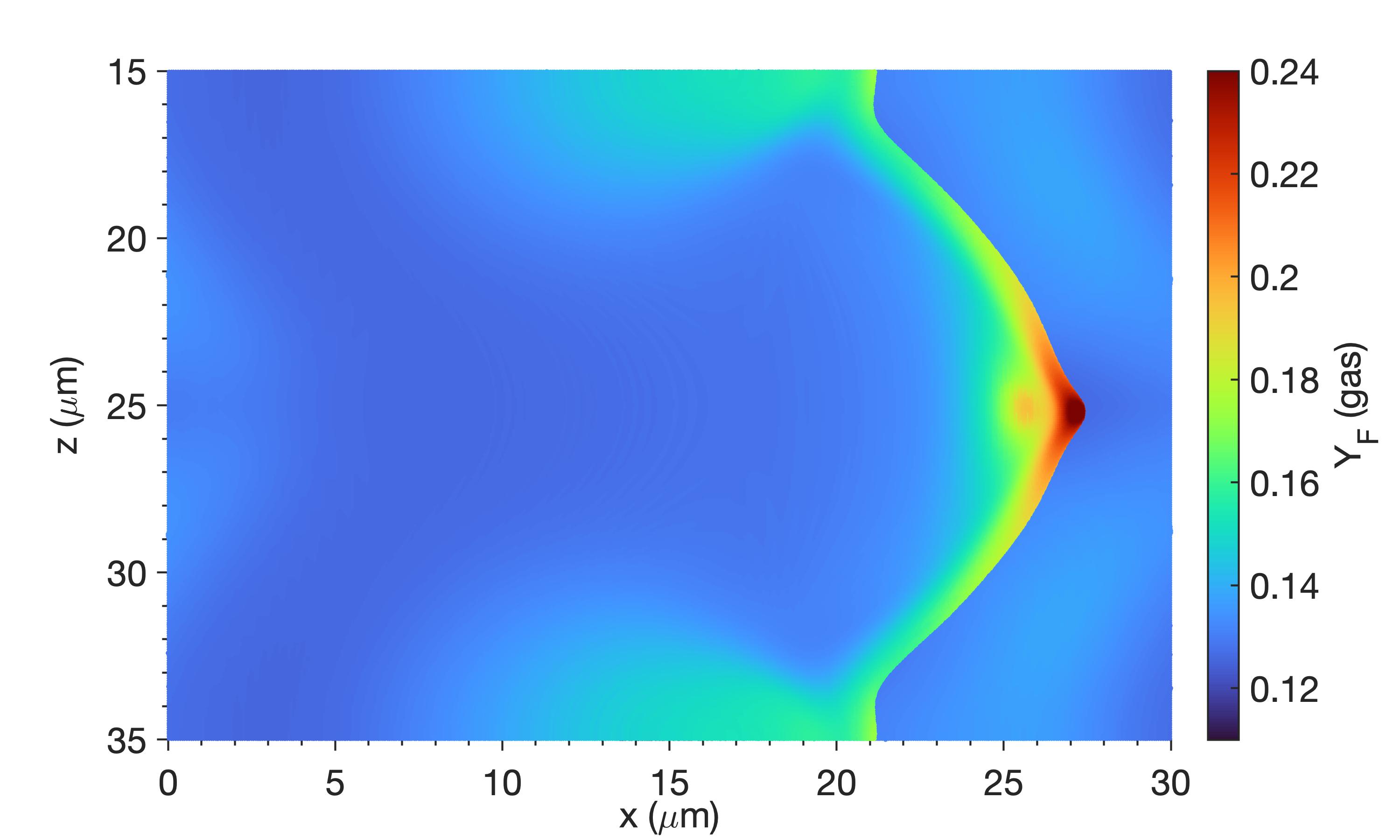}
  \caption{} 
  \label{subfig:Fig7c}
\end{subfigure}%
\begin{subfigure}{0.5\textwidth}
  \centering
  \includegraphics[width=1.0\linewidth]{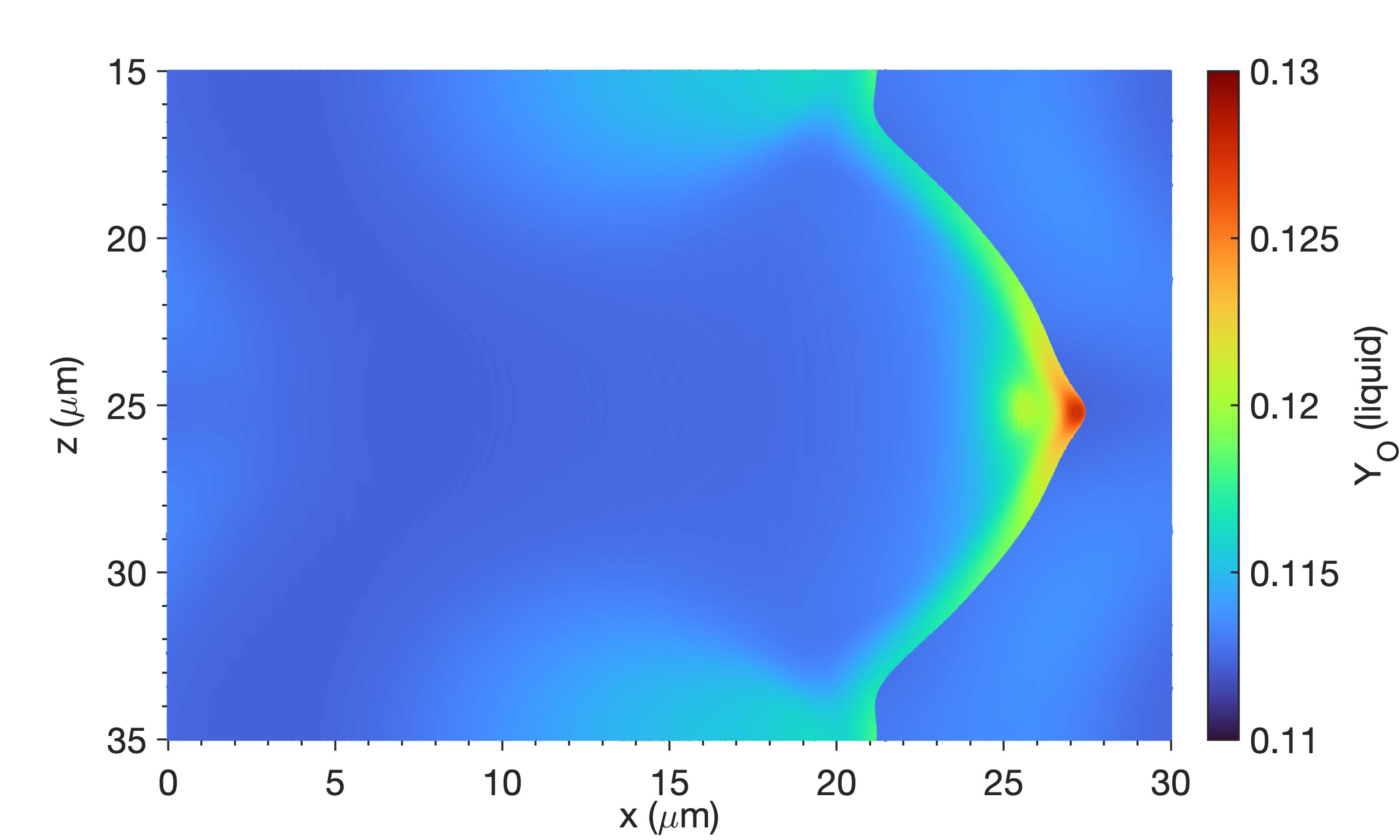}
  \caption{} 
  \label{subfig:Fig7d}
\end{subfigure}%
\caption{Solution of the local thermodynamic equilibrium along the lobe for case C1 at \(t^*=4.05\). A top view of the lobe from above the liquid surface is shown. The liquid surface is coloured by the local interface value for (a) temperature, \(T\); (b) surface-tension coefficient, \(\sigma\); (c) fuel vapour mass fraction, \(Y_F\); and (d) dissolved oxygen mass fraction, \(Y_O\).}
\label{fig:Fig7}
\end{figure}

\begin{figure}
\centering
\begin{subfigure}{0.33\textwidth}
  \centering
  \includegraphics[width=1.0\linewidth]{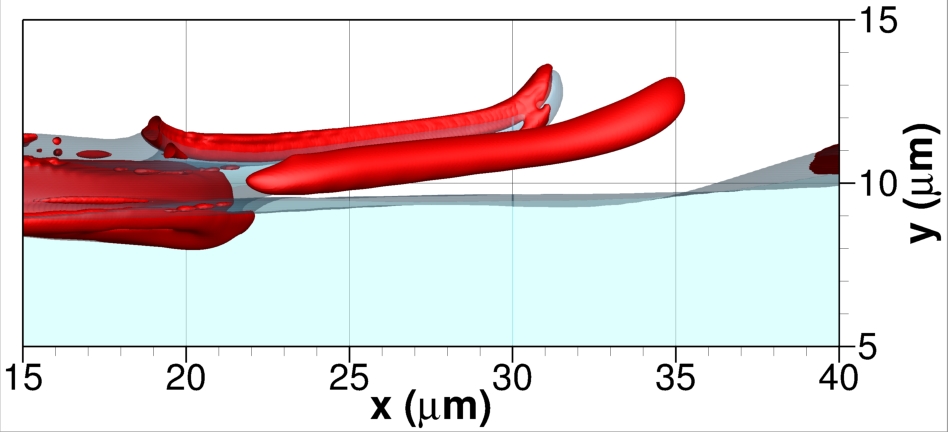}
  \label{subfig:Fig8a}
\end{subfigure}%
\begin{subfigure}{0.33\textwidth}
  \centering
  \includegraphics[width=1.0\linewidth]{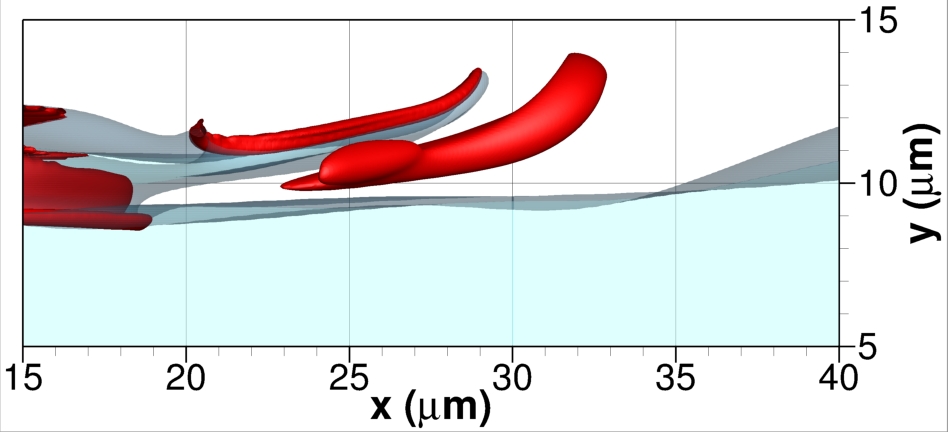}
  \label{subfig:Fig8b}
\end{subfigure}%
\begin{subfigure}{0.33\textwidth}
  \centering
  \includegraphics[width=1.0\linewidth]{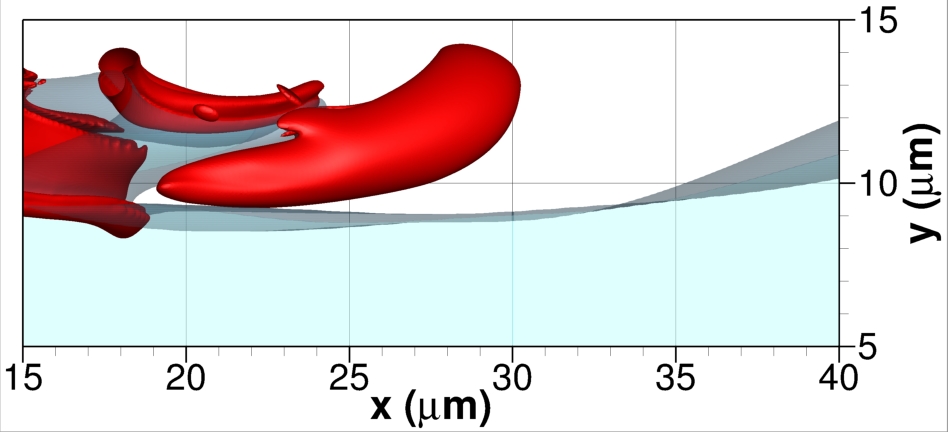}
  \label{subfig:Fig8c}
\end{subfigure}%
\\[-2.7ex]
\begin{subfigure}{0.33\textwidth}
  \centering
  \includegraphics[width=1.0\linewidth]{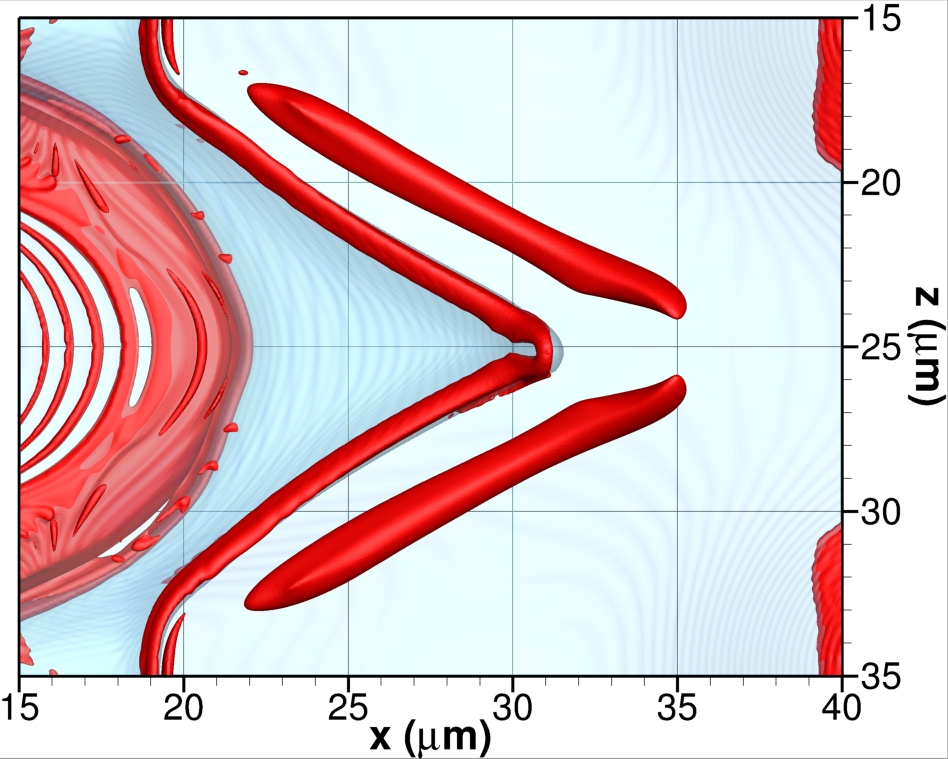}
  \caption{} 
  \label{subfig:Fig8d}
\end{subfigure}%
\begin{subfigure}{0.33\textwidth}
  \centering
  \includegraphics[width=1.0\linewidth]{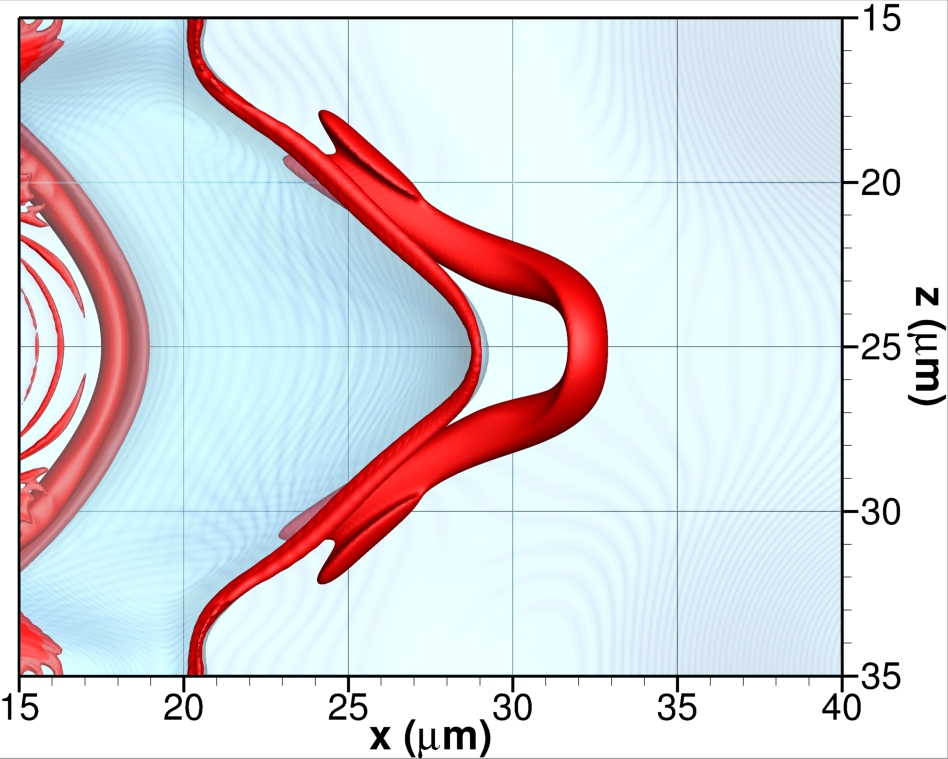}
  \caption{} 
  \label{subfig:Fig8e}
\end{subfigure}%
\begin{subfigure}{0.33\textwidth}
  \centering
  \includegraphics[width=1.0\linewidth]{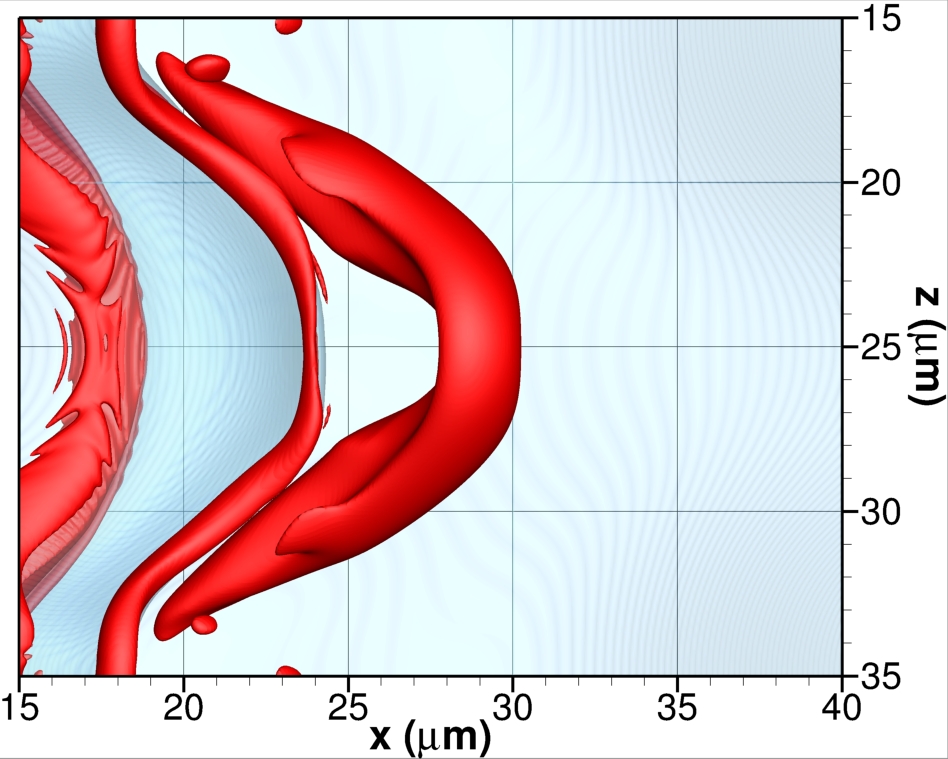}
  \caption{} 
  \label{subfig:Fig8f}
\end{subfigure}%
\caption{Pressure and mixing effects on the vorticity dynamics of the lobe extension, bending and perforation mechanism. The top figures show the side view at \(z=35\) \(\mu\)m and the bottom figures show the top view of the liquid. The liquid surface is identified by the translucid blue isosurface with \(C=0.5\) and the vortex structures are identified by the red isosurface of \(\lambda_\rho\). (a) case A2 at \(t^*=4.2\) with \(\lambda_{\rho,t}=-3\times 10^{15}\); (b) case B1 at \(t^*=4\) with \(\lambda_{\rho,t}=-5\times 10^{15}\); and (c) case C1i at \(t^*=4.05\) with \(\lambda_{\rho,t}=-1\times 10^{15}\).}
\label{fig:Fig8}
\end{figure}

The complex coupling between vorticity dynamics, thermodynamics and surface deformation is a key characteristic of transcritical flows. Not only the variation of fluid properties within each phase caused by mixing are determinant, but also the local interface state (e.g., surface tension, ambient gas dissolution, liquid vaporisation). Figure~\ref{fig:Fig7} shows the lobe at \(t^*=4.05\) from above the liquid surface. In each sub-figure, the surface is coloured by the local interface temperature, fuel mass fraction in the gas phase, oxygen mass fraction in the liquid phase, and surface-tension coefficient. Various features become immediately clear, such as the enhanced oxygen dissolution in the liquid phase at 150 bar or the strong correlation between surface temperature and composition given by phase equilibrium (see figure~\ref{fig:Fig2}). For the observed temperature range, the amount of oxygen dissolved in the liquid does not vary much with temperature. However, considerably more fuel is vaporised as the interface temperature increases. Altogether, higher local temperatures result in lower surface tension. Here, the vortical motion responsible for the lobe stretching and bending carries the lobe's tip toward hotter regions in the oxidizer stream. Thus, a temperature rise occurs. Moreover, locally thin liquid regions heat faster. That is, the heat coming from the hotter gas cannot diffuse into colder liquid regions. This effect is observed along the lobe's edge and at the perforation location. \par 

The local heating of the liquid enhances all the above-mentioned transcritical features (e.g., reduced liquid density and viscosity, denser gas), explaining the vorticity-induced enhanced lobe deformation. Additionally, PS show that, at very high pressures, the interface may undergo net condensation or net vaporisation depending on the local thermodynamic state and the mass and energy balances. Despite the constant fuel evaporation caused by the hotter gaseous environment, the dissolution of the ambient gas increases substantially. Consequently, there are interfacial regions where the liquid internal energy exceeds the gas internal energy, thereby resulting in net condensation~\citep{poblador2018transient}. For the analysed configuration, net condensation dominates at interface temperatures around 460 K to 470 K. At higher temperatures, \textit{n}-decane evaporates faster than the dissolution rate of oxygen (i.e., net vaporisation occurs). \par 

The changes in ambient or thermodynamic pressure affect strongly the evolution of the lobe and the initial roller vortex. Although the evolution of vorticity is coupled to the highly-varying properties of the transcritical fluid, we only focus on certain dynamical aspects and summarise the observed features. Figure~\ref{fig:Fig8} shows how cases A2, B1 and C1i evolve differently than C1 despite having similar \textit{We}\(_G\). Case C1i consists of an incompressible simulation without real-fluid thermodynamics and where the fluid properties are set equal to those in table~\ref{tab:cases} for case C1. The goal of this simulation is to understand the influence of liquid intraphase mixing in the deformation process. \par 

Figure~\ref{fig:Fig8} presents the liquid lobe and the deformed roller vortex right before the hole forms in cases B1 and C1 around \(t^*\approx 4\). The formation of a clear hairpin vortex is visible for the 100-bar and 150-bar cases at the chosen \(\lambda_{\rho,t}\) values. At 50 bar (i.e., case A2), the vortex is subject to higher velocities with the 70 m/s gas freestream velocity, resulting in rapid streamwise stretching and weakening of the vortex head compared to the cases with higher pressures and smaller velocities. How each vortex affects the lobe depends on the properties of the liquid phase and how they vary at a given ambient or thermodynamic pressure. At lower pressures, lobes are thicker and rounder because of higher surface tension. That is, smaller curvatures are required for similar pressure jumps across an interface. Moreover, less oxygen dissolves into the liquid, and the density and viscosity remain more similar to the pure liquid properties. Coupled with the lower gas inertia, the lobes do not deform as easily as at higher pressures. \par 

Further insights are obtained by looking at the relative position between the hairpin vortex and the lobe (see figures~\ref{fig:Fig5} and~\ref{fig:Fig8}). As pressure decreases, the relative distance between the vortex and the lobe increases. For case C1 at 150 bar, the lobe's tip is closer to the vortex head than for case B2 at 100 bar. Similarly, this distance increases between case B2 and case A2 at 50 bar. Disregarding the differences in gas freestream velocity between cases, the separation between the lobe's tip and the vortex head can be understood as a measure of the lobe's streamwise stretching (i.e., influenced by the vortex induced velocity). Similarly, the separation between the vortex pins and the lobe's lateral edges can be understood as a measure of the lobe's spanwise stretching. For case C1, the lobe easily overlaps the vortex pins, but it barely occurs for case B2. For case A2, the vortex pins and the lobe are separated. As a result, case A2 does not undergo hole formation while in cases B2 and C1 the lobe stretches enough to become sufficiently and be perforated. Figure~\ref{fig:Fig8} also illustrates how intraphase mixing affects the evolution of the lobe. Even though the hairpin vortex evolves similarly in cases C1 and C1i and the freestream fluid properties are the same, the lobe is not stretched in the same manner. The constant fluid properties across the lobe due to lack of ambient gas dissolution prevent the lobe's stretching and a hole cannot develop. \par

\subsection{Lobe and crest corrugation}
\label{subsec:lobe_crest_corrugation}

Another early deformation mechanism is observed for configurations with \textit{We}\(_G\) \(>1000\), which is characteristic of cases B2, C2 and C3 (i.e., above ambient pressures of 100 bar and 150 bar). Initially, the lobes extend over the liquid surface, similar to the deformation mechanism detailed in section~\ref{subsec:lobe_bending}. However, instead of experiencing a spanwise rotation, a streamwise corrugation of the lobe and the crest of the main perturbation is observed. The corrugation of the lobe generates a very thin liquid film that expands and bursts into droplets, similar to a bag-breakup mechanism. \par

\begin{figure}
\centering
\begin{subfigure}{0.33\textwidth}
  \centering
  \includegraphics[width=1.0\linewidth]{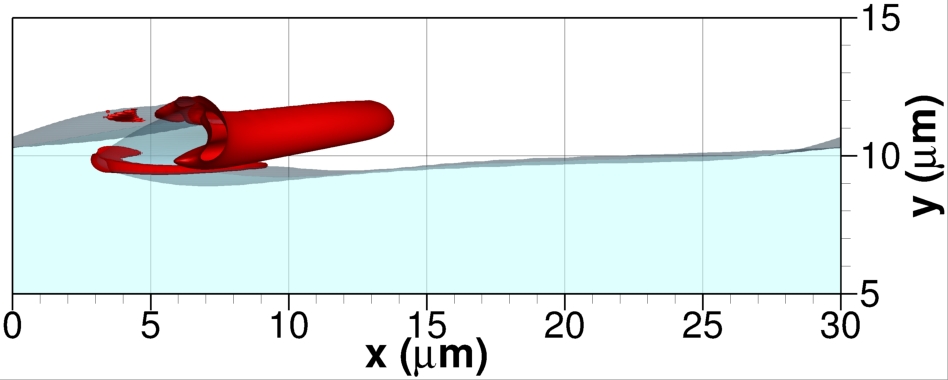}
  \label{subfig:Fig9a}
\end{subfigure}%
\begin{subfigure}{0.33\textwidth}
  \centering
  \includegraphics[width=1.0\linewidth]{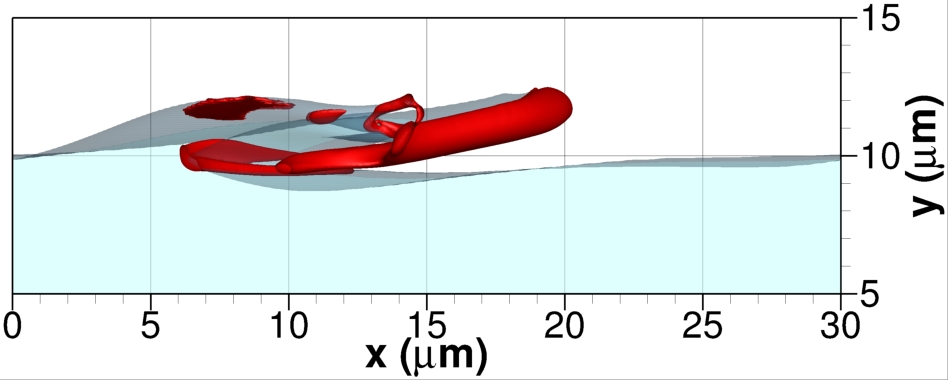}
  \label{subfig:Fig9b}
\end{subfigure}%
\begin{subfigure}{0.33\textwidth}
  \centering
  \includegraphics[width=1.0\linewidth]{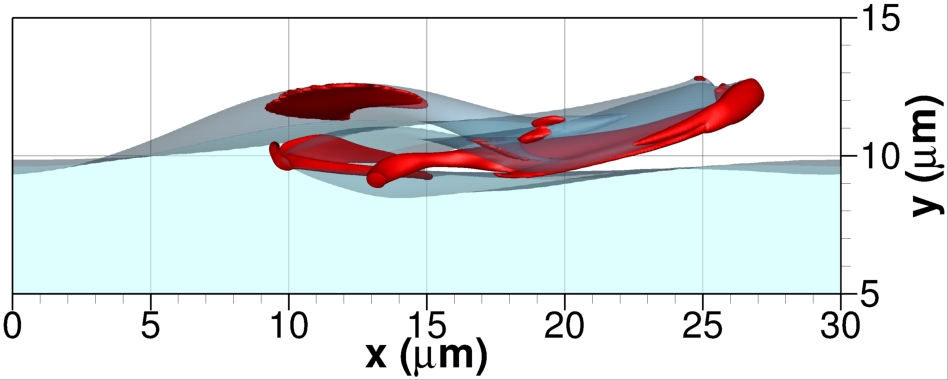}
  \label{subfig:Fig9c}
\end{subfigure}%
\\[-2.7ex]
\begin{subfigure}{0.33\textwidth}
  \centering
  \includegraphics[width=1.0\linewidth]{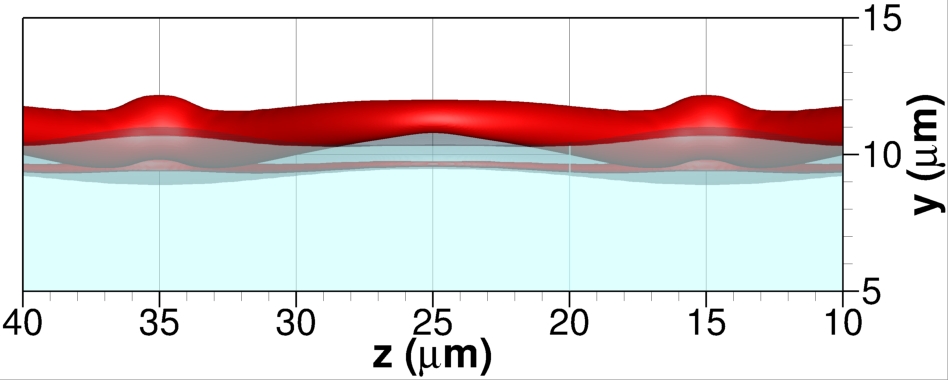}
  \label{subfig:Fig9d}
\end{subfigure}%
\begin{subfigure}{0.33\textwidth}
  \centering
  \includegraphics[width=1.0\linewidth]{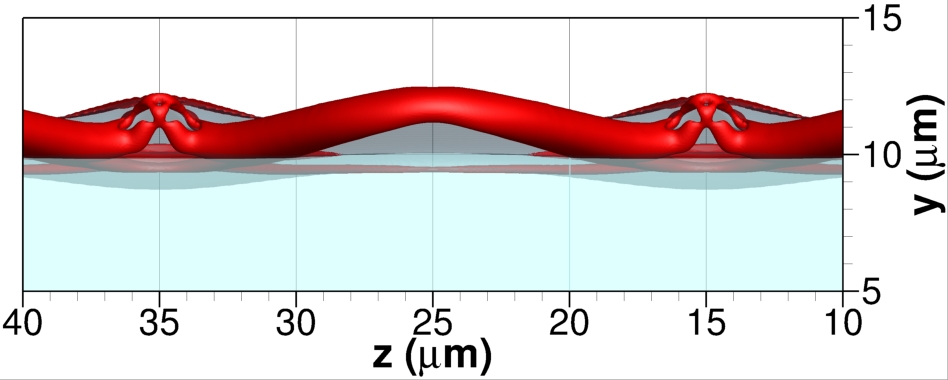}
  \label{subfig:Fig9e}
\end{subfigure}%
\begin{subfigure}{0.33\textwidth}
  \centering
  \includegraphics[width=1.0\linewidth]{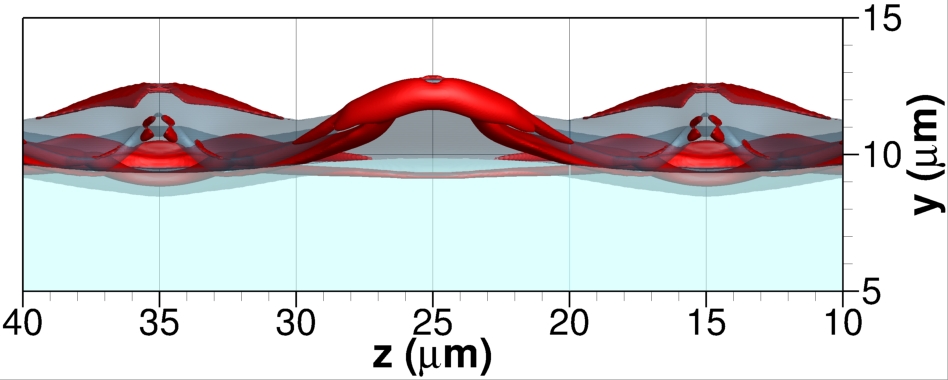}
  \label{subfig:Fig9f}
\end{subfigure}%
\\[-2.7ex]
\begin{subfigure}{0.33\textwidth}
  \centering
  \includegraphics[width=1.0\linewidth]{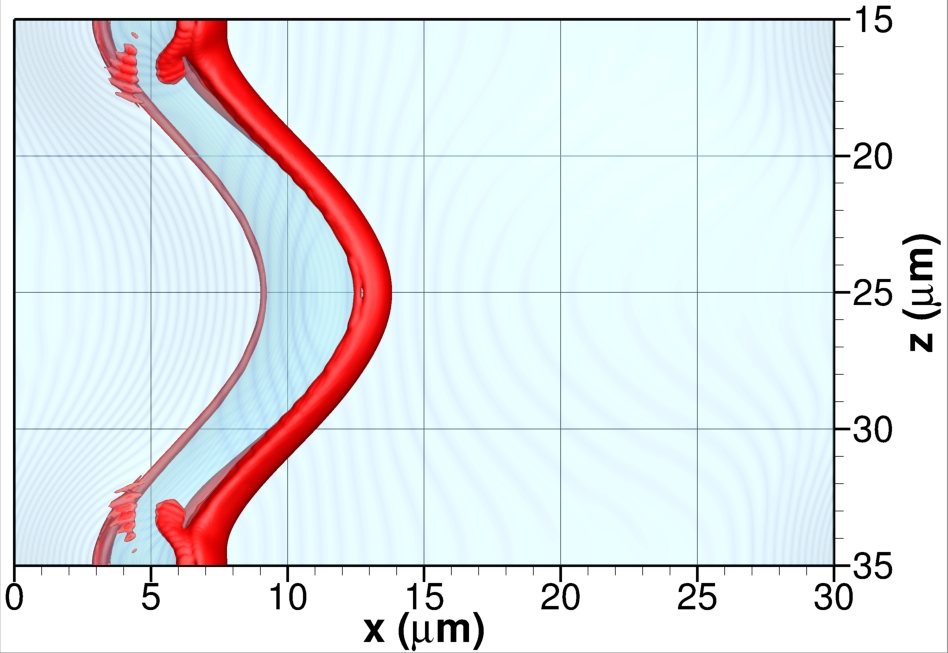}
  \caption{} 
  \label{subfig:Fig9g}
\end{subfigure}%
\begin{subfigure}{0.33\textwidth}
  \centering
  \includegraphics[width=1.0\linewidth]{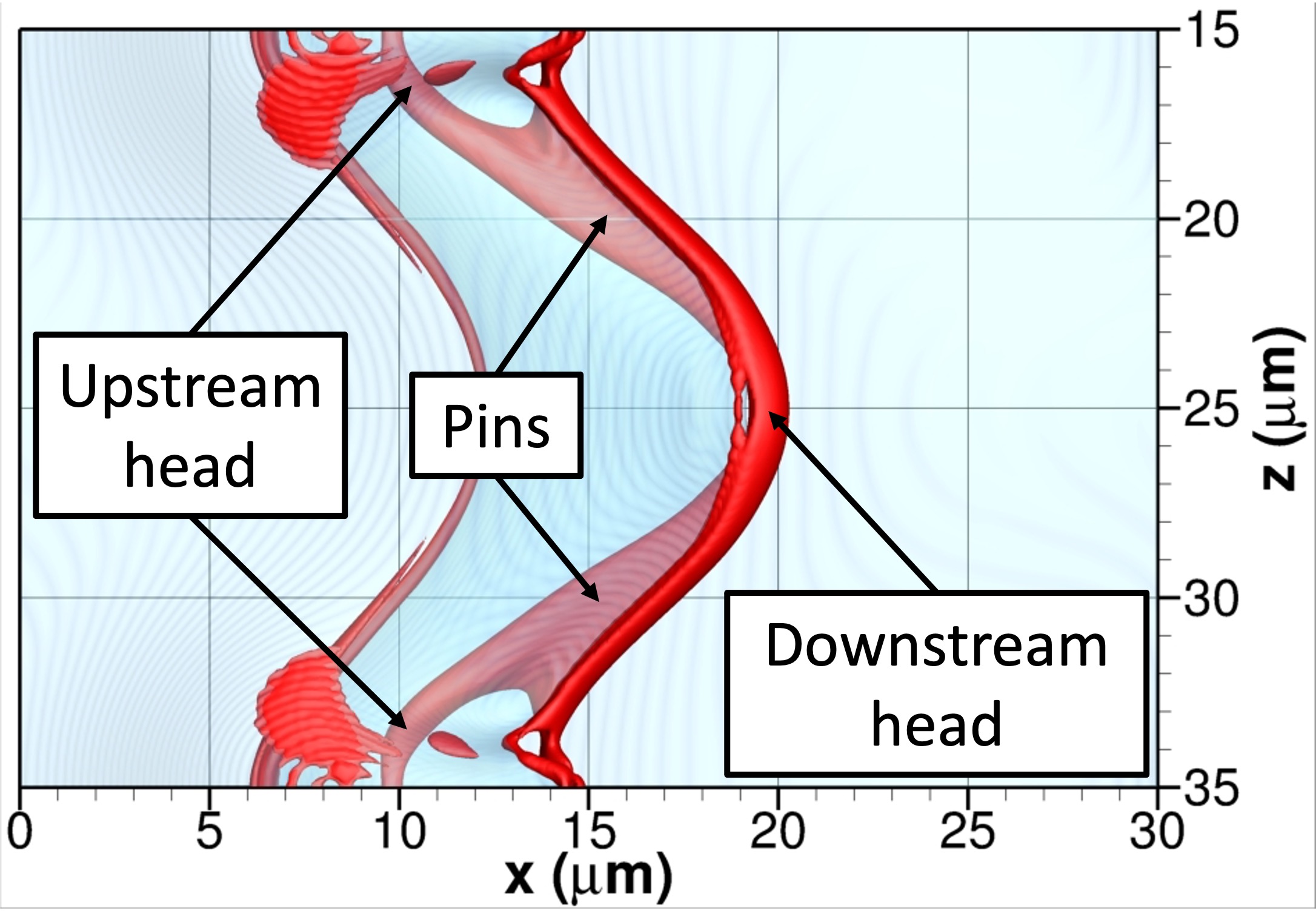}
  \caption{}
  \label{subfig:Fig9h}
\end{subfigure}%
\begin{subfigure}{0.33\textwidth}
  \centering
  \includegraphics[width=1.0\linewidth]{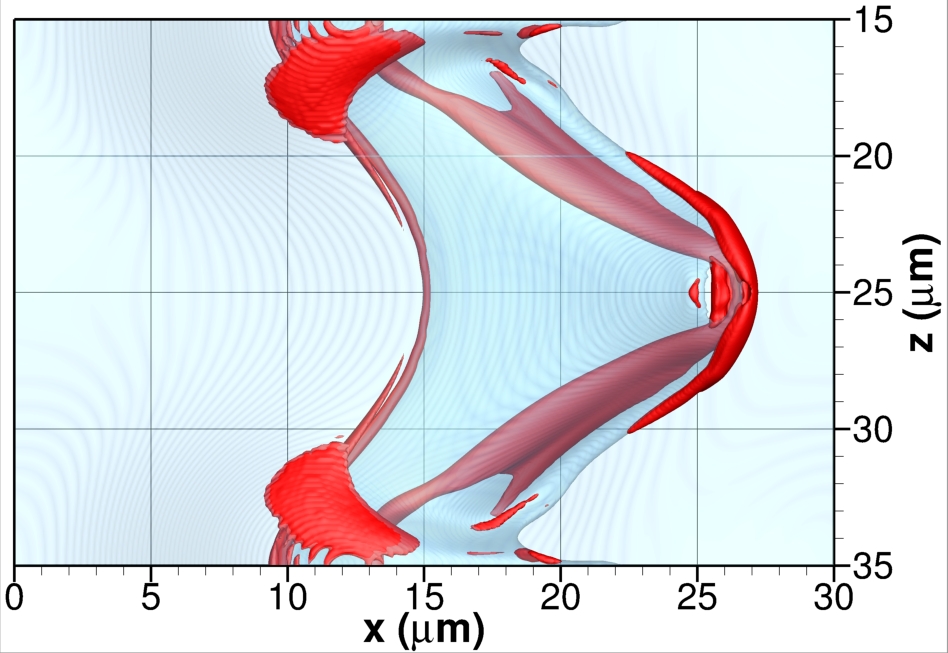}
  \caption{}
  \label{subfig:Fig9i}
\end{subfigure}%
\\
\begin{subfigure}{0.33\textwidth}
  \centering
  \includegraphics[width=1.0\linewidth]{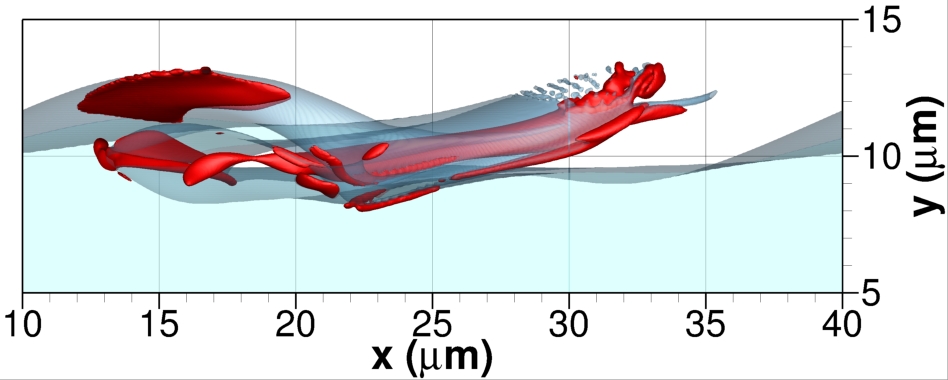}
  \label{subfig:Fig9j}
\end{subfigure}%
\begin{subfigure}{0.33\textwidth}
  \centering
  \includegraphics[width=1.0\linewidth]{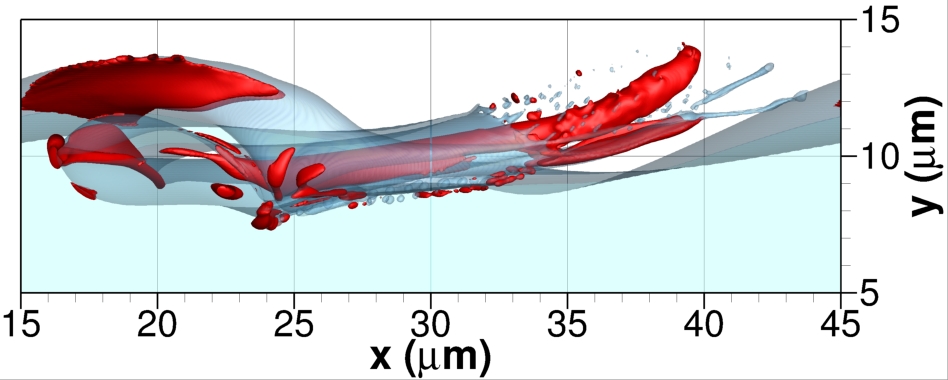}
  \label{subfig:Fig9k}
\end{subfigure}%
\begin{subfigure}{0.33\textwidth}
  \centering
  \includegraphics[width=1.0\linewidth]{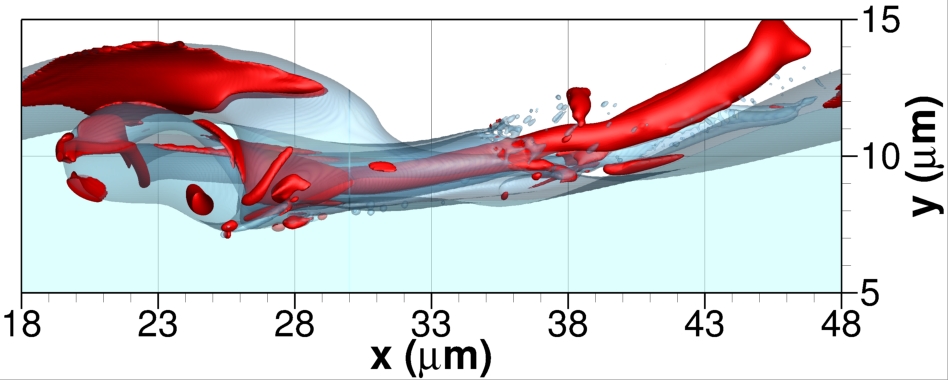}
  \label{subfig:Fig9l}
\end{subfigure}%
\\[-2.7ex]
\begin{subfigure}{0.33\textwidth}
  \centering
  \includegraphics[width=1.0\linewidth]{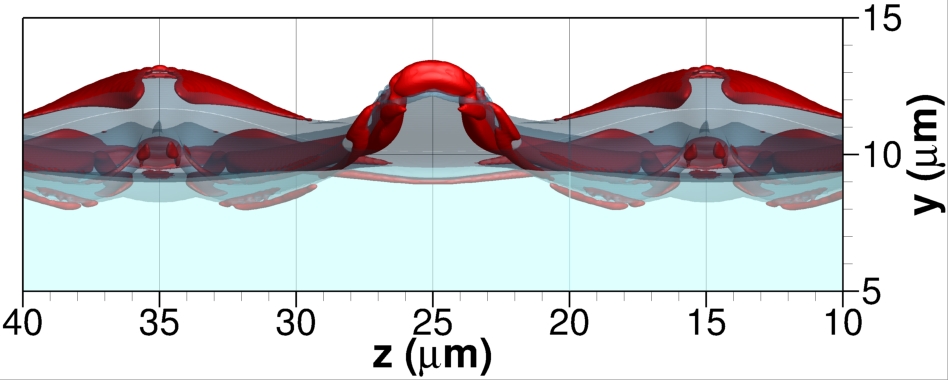}
  \label{subfig:Fig9m}
\end{subfigure}%
\begin{subfigure}{0.33\textwidth}
  \centering
  \includegraphics[width=1.0\linewidth]{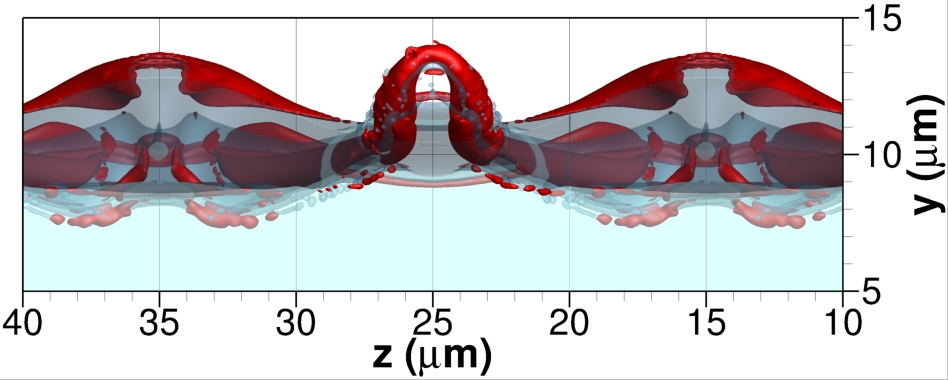}
  \label{subfig:Fig9n}
\end{subfigure}%
\begin{subfigure}{0.33\textwidth}
  \centering
  \includegraphics[width=1.0\linewidth]{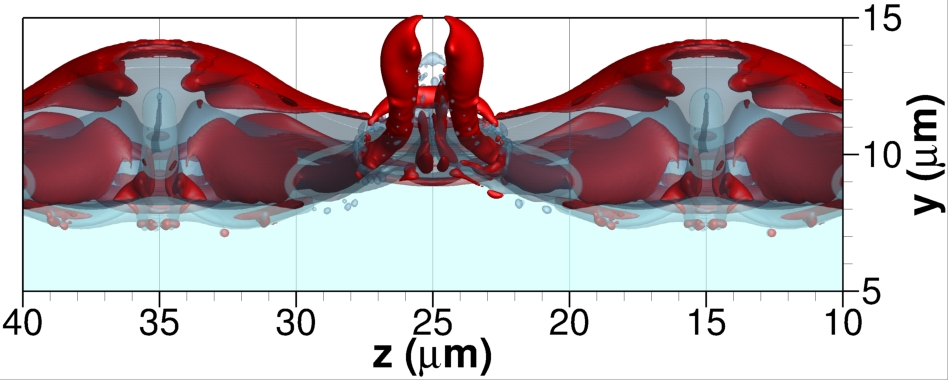}
  \label{subfig:Fig9o}
\end{subfigure}%
\\[-2.7ex]
\begin{subfigure}{0.33\textwidth}
  \centering
  \includegraphics[width=1.0\linewidth]{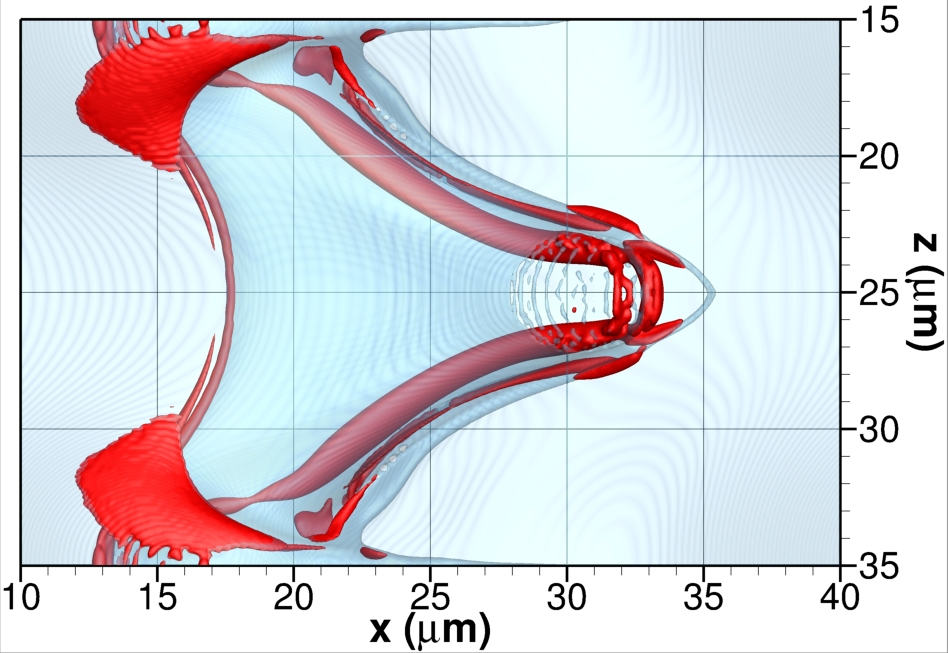}
  \caption{} 
  \label{subfig:Fig9p}
\end{subfigure}%
\begin{subfigure}{0.33\textwidth}
  \centering
  \includegraphics[width=1.0\linewidth]{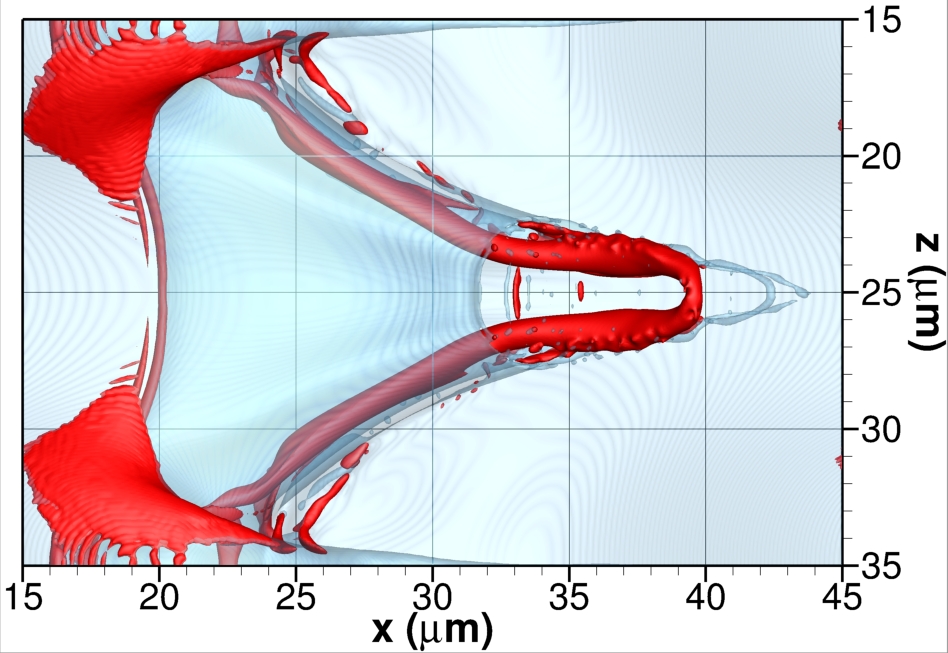}
  \caption{}
  \label{subfig:Fig9q}
\end{subfigure}%
\begin{subfigure}{0.33\textwidth}
  \centering
  \includegraphics[width=1.0\linewidth]{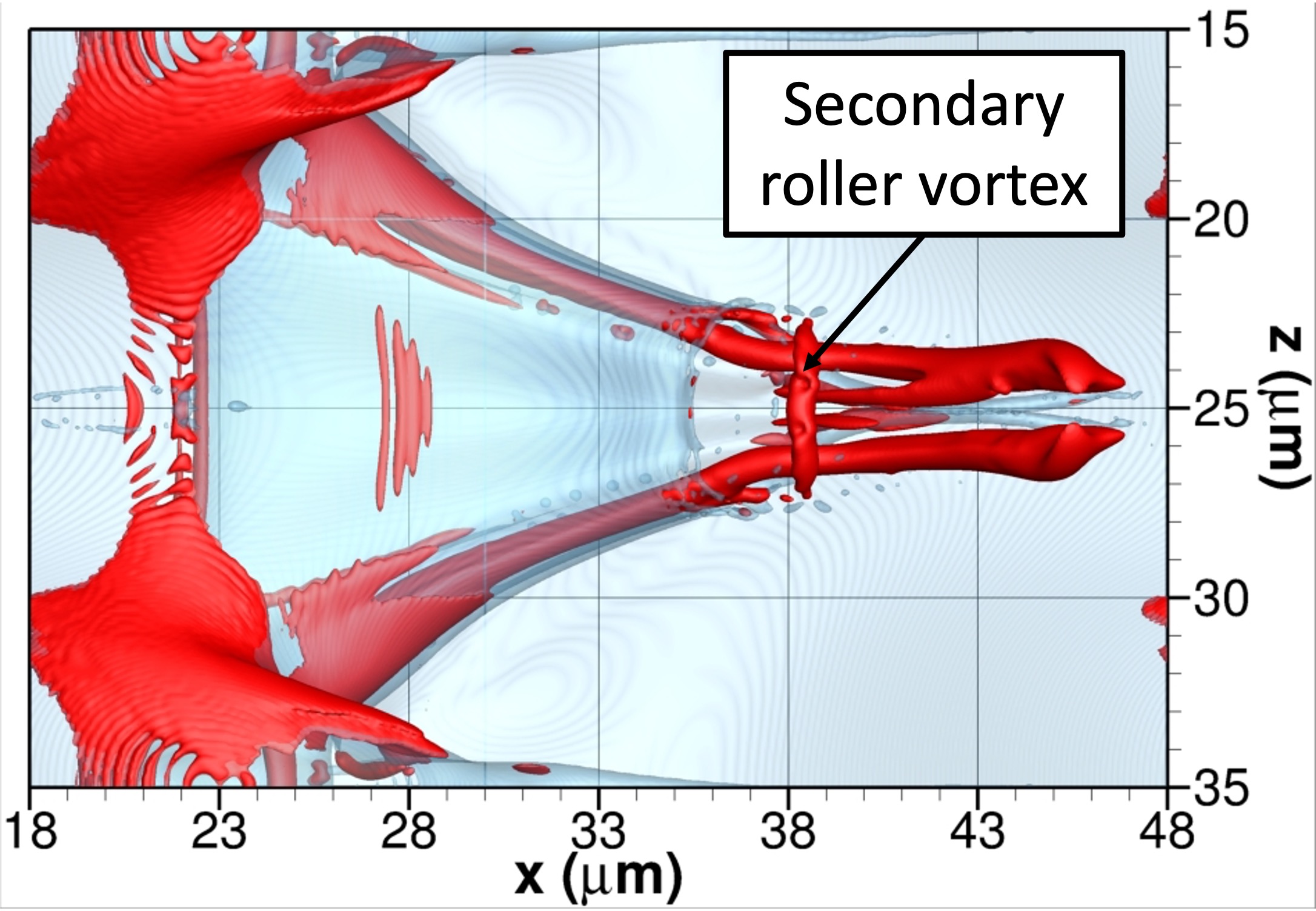}
  \caption{}
  \label{subfig:Fig9r}
\end{subfigure}%
\caption{Vortex dynamics of the lobe and crest corrugation mechanism for case C2. The top figures show the side view at \(z=35\) \(\mu\)m, the middle figures show the front view from the respective sub-figure domain, and the bottom figures show the top view of the liquid. The liquid surface is identified by the translucid blue isosurface with \(C=0.5\) and the vortex structures are identified by the red isosurface with \(\lambda_{\rho,t}=-9\times 10^{15}\). (a) \(t^*=2.5\); (b) \(t^*=3\); (c) \(t^*=3.5\); (d) \(t^*=4\); (e) \(t^*=4.5\); and (f) \(t^*=5\).}
\label{fig:Fig9}
\end{figure}

Case C2 is chosen for the analysis of this deformation mechanism. Similar to section~\ref{subsec:lobe_bending}, one of the 150-bar cases is chosen to illustrate the transcritical environment. Figure~\ref{fig:Fig9} presents the evolution of the vortical structures during the early times, which are identified using a value of \(\lambda_{\rho,t} = -9 \times 10^{15}\). Note that higher gas freestream velocities translate into higher vorticity as highlighted by the magnitude of \(\lambda_{\rho,t}\). The initial roller vortex, which rotates clockwise when seen toward \(-z\), also deforms into a hairpin vortex. However, now the vortex remains attached to the lobe's edge everywhere, as seen in figure~\ref{subfig:Fig9g}. This is consistent with the stronger vorticity inducing further lobe stretching than in case C1. Comparing against figure~\ref{fig:Fig5}, the rim vortex identified at lower \textit{We}\(_G\) is not observed (i.e., merges with the roller vortex). \par 

As the roller vortex deforms into a hairpin, the stronger stretching causes the vortex pins to move underneath the lobe between \(t^*=2.5\) and \(t^*=3.5\); or, equivalently, the gas-like liquid phase wraps around the vortex. Once the vortex pins are fully captured by the lobe, the spanwise stretching is enhanced and the gas flow underneath the lobe is intensified. This process rapidly inflates the lobe, which becomes a very thin sheet, and then bursts into droplets. The bursting occurs first near the lobe's tip, which is stretched both in \(z\) by the vortex pins and in \(x\) by the head of the hairpin vortex. That is, the thinning occurs similar to the results shown in figure~\ref{fig:Fig6} for case C1. After the gas perforates the lobe, the hole rapidly recedes under the action of surface tension. The strong recirculation in the region promotes the formation of tiny droplets leading to a highly perturbed flow. From \(t^*=3.5\) to \(t^*=5\), liquid ligaments and droplets accelerate into the oxidizer stream. The hairpin vortex stretches rapidly and additional vortical structures emerge. For instance, a small secondary roller vortex above the hairpin vortex is seen at \(t^*=5\). More details about the formation of this vortex are given in section~\ref{subsubsec:rollers}. \par 

\begin{figure}
\centering
\begin{subfigure}{0.33\textwidth}
  \centering
  \includegraphics[width=1.0\linewidth]{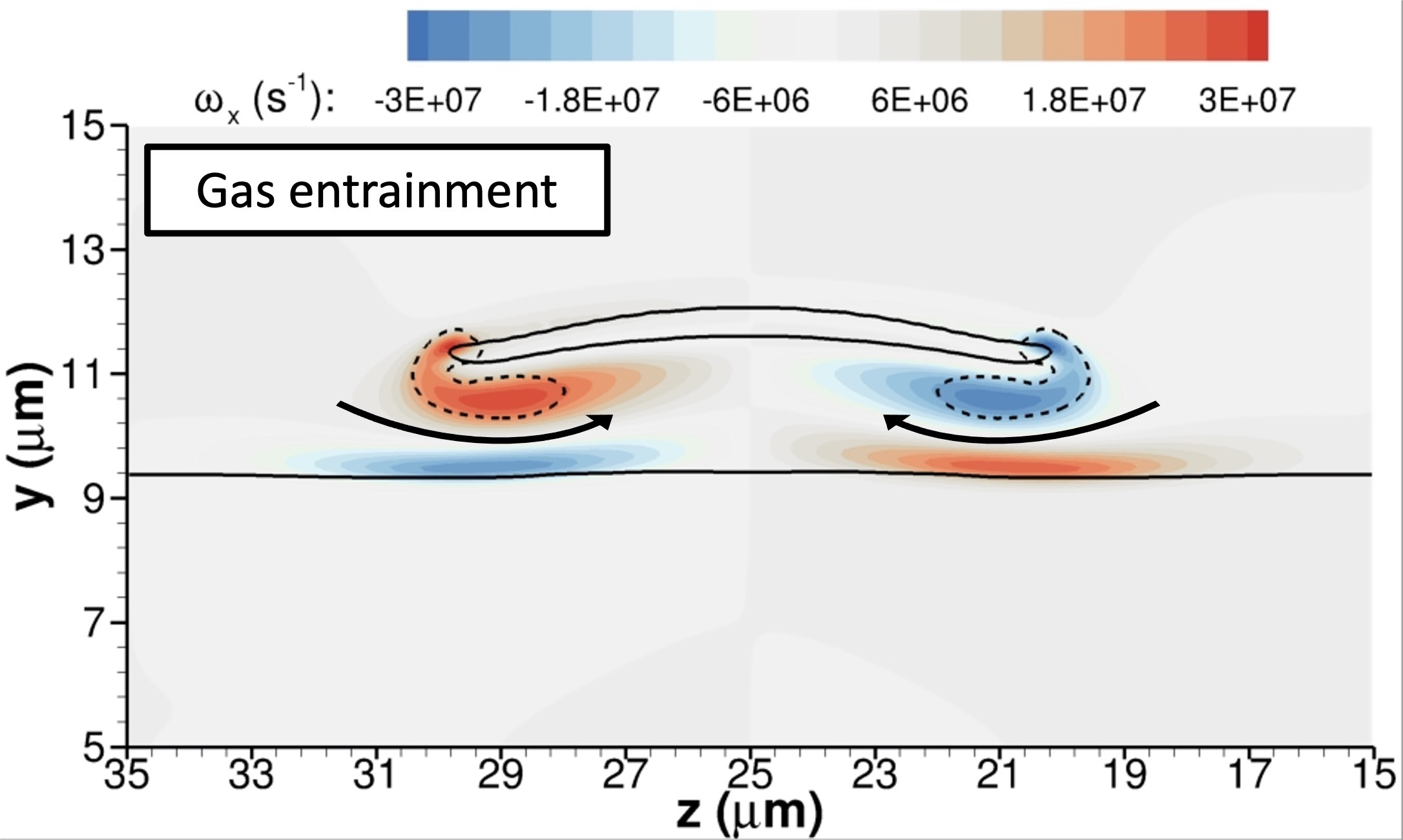}
  \caption{} 
  \label{subfig:Fig10a}
\end{subfigure}%
\begin{subfigure}{0.33\textwidth}
  \centering
  \includegraphics[width=1.0\linewidth]{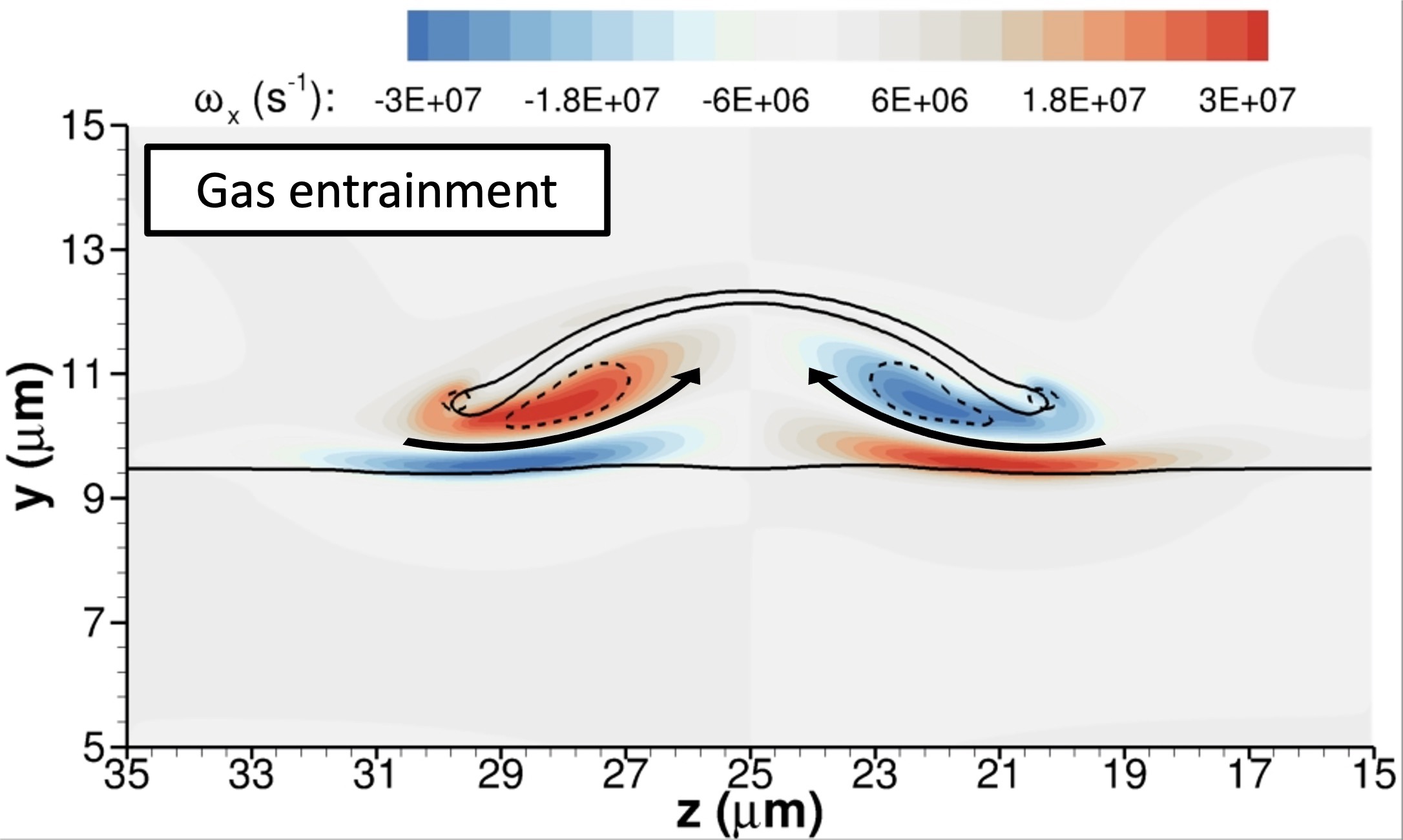}
  \caption{}
  \label{subfig:Fig10b}
\end{subfigure}%
\begin{subfigure}{0.33\textwidth}
  \centering
  \includegraphics[width=1.0\linewidth]{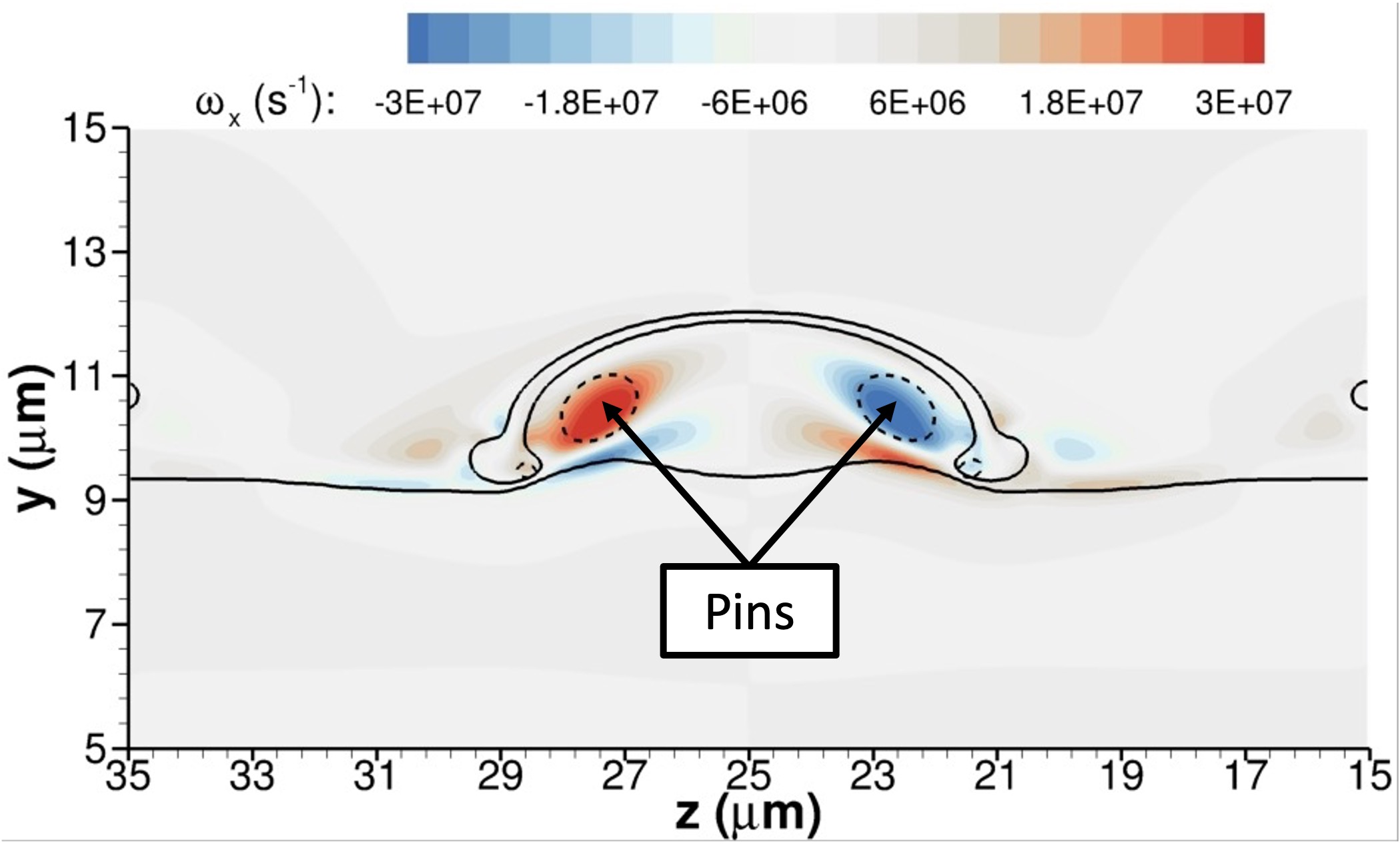}
  \caption{}
  \label{subfig:Fig10c}
\end{subfigure}%
\\
\begin{subfigure}{0.33\textwidth}
  \centering
  \includegraphics[width=1.0\linewidth]{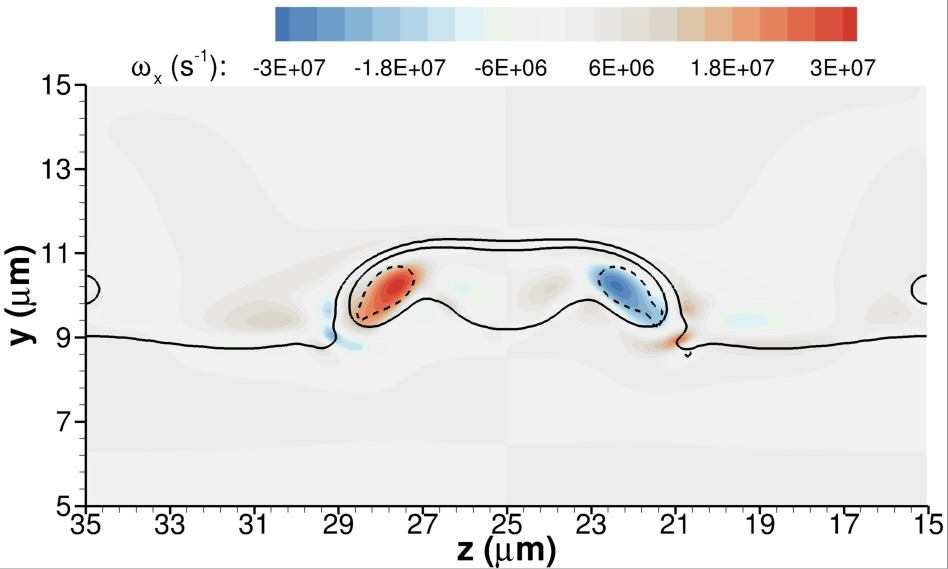}
  \caption{} 
  \label{subfig:Fig10d}
\end{subfigure}%
\begin{subfigure}{0.33\textwidth}
  \centering
  \includegraphics[width=1.0\linewidth]{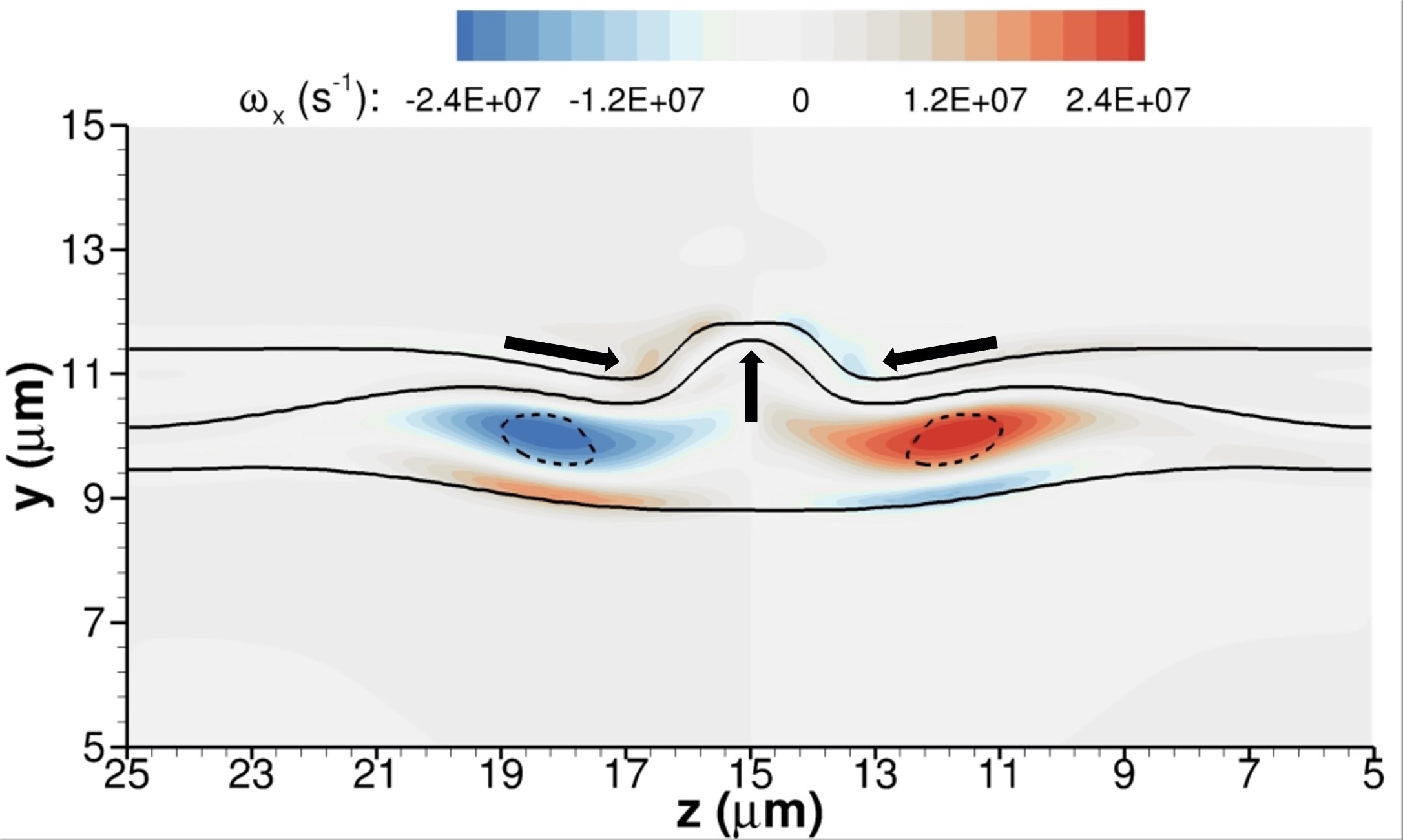}
  \caption{}
  \label{subfig:Fig10e}
\end{subfigure}%
\begin{subfigure}{0.33\textwidth}
  \centering
  \includegraphics[width=1.0\linewidth]{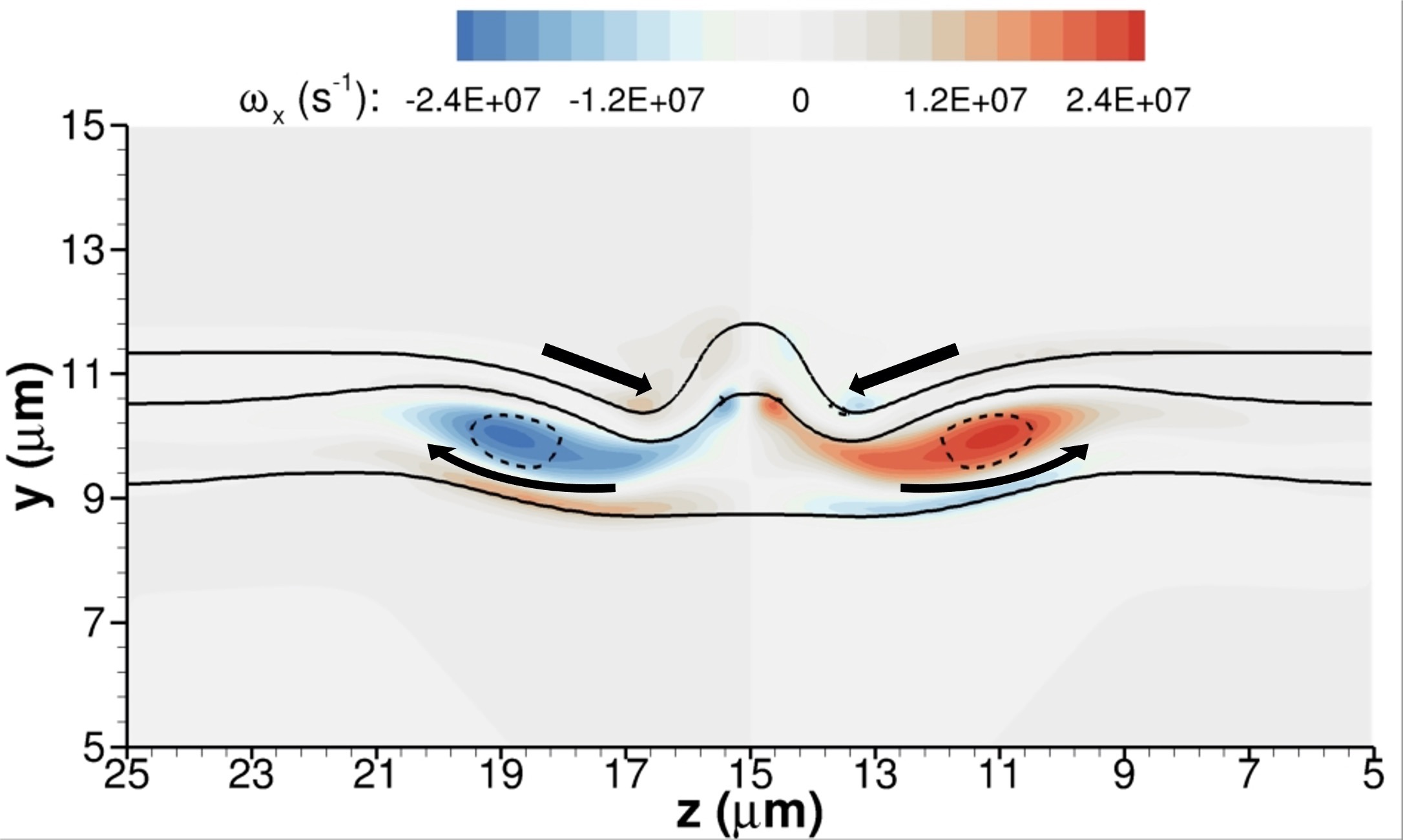}
  \caption{}
  \label{subfig:Fig10f}
\end{subfigure}%
\caption{Vortex dynamics of the lobe corrugation mechanism for case C2. The contours of \(\omega_x\) are shown on \(yz\) planes at various \(x\) following the lobe and the perturbation crest over time. The interface is identified by the solid isocontour with \(C=0.5\) and the vortex cross-sections are identified by the dashed isocontour with \(\lambda_{\rho,t}=-9\times 10^{15}\). (a) lobe at \(x=17\) \(\mu\)m and \(t^*=3\); (b) lobe at \(x=23\) \(\mu\)m and \(t^*=3.5\); (c) lobe at \(x=27\) \(\mu\)m and \(t^*=4\); (d) lobe at \(x=30\) \(\mu\)m and \(t^*=4.5\); (e) crest at \(x=12.5\) \(\mu\)m and \(t^*=3\); and (f) crest at \(x=17\) \(\mu\)m and \(t^*=3.5\).}
\label{fig:Fig10}
\end{figure}

The deformation process is visualised in figures~\ref{subfig:Fig10a}-\ref{subfig:Fig10d} where contours of \(\omega_x\) are shown in \(yz\) planes at various times and streamwise locations. As the liquid phase wraps around the vortex pins, the gap between the lobe's sides and the liquid surface below is reduced, accelerating the gas entrainment and intensifying the vortex streamwise strength. Additionally, some vorticity appears under the vortex pins with the opposite rotation. Previous incompressible works by~\citet{jarrahbashi2016early} and~\citet{zandian2018understanding} do not show such a vortex-capturing mechanism. Lobes overlapping vortices are observed, but these vortices are secondary. Therefore, the results presented here differ substantially and the capturing process of the initial roller vortex is attributed to the transcritical environment. The sides of the lobe eventually merge with the liquid below. This ends the lateral gas entrainment and traps the vortex pins in the gaseous cavity underneath the lobe. The remaining vortical motion is then responsible for lifting the liquid surface below and flattening the lobe as continuous spanwise stretching occurs. \par 

\begin{figure}
\centering
\includegraphics[width=0.7\linewidth]{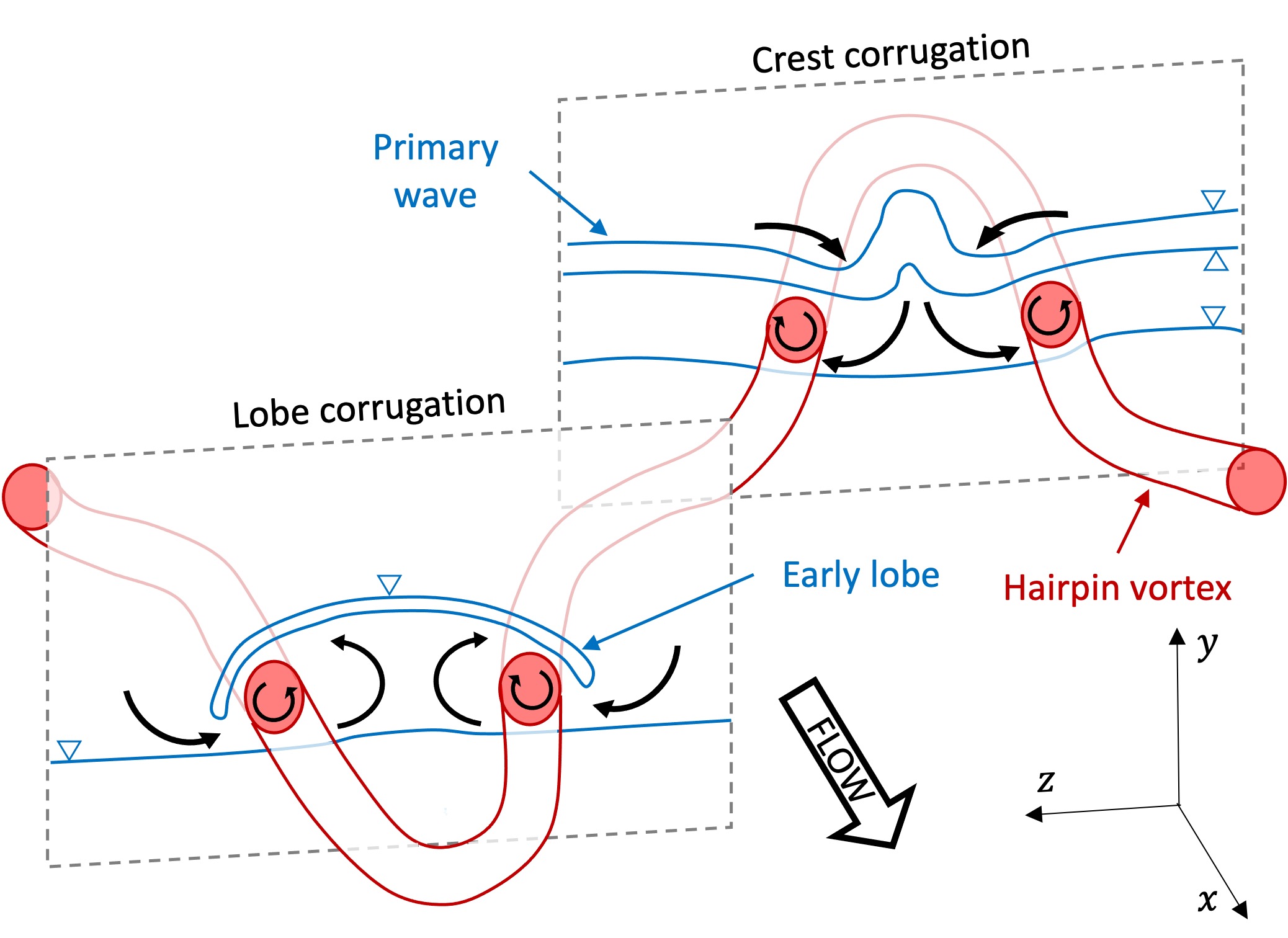}
\caption{Sketch of the lobe and crest corrugation mechanisms and generated by the deformed initial roller or hairpin vortex.}
\label{fig:Fig11}
\end{figure}

Figures~\ref{subfig:Fig10e} and~\ref{subfig:Fig10f} show the crest corrugation mechanism, which is caused by the induced flow from the upstream-facing side of the hairpin vortex. The upstream head of the hairpin vortex can be observed in figure~\ref{subfig:Fig9h} under the liquid wave at \(z=15\) \(\mu\)m or \(z=35\) \(\mu\)m between \(x=10\) \(\mu\)m and \(x=15\) \(\mu\)m. Vorticity deforms the thicker liquid sheet in that region. First, the induced velocity field brings together the sides of the liquid sheet. This displacement corrugates or folds the liquid around \(z=15\) \(\mu\)m, much like a sheet of paper folds when you bring together the two opposite edges. The deformation process continues with the liquid sheet weakly wrapping around the vortex pins, ejecting the gas in between the vortical structures, the liquid-core surface, and the liquid sheet. Altogether, the wave crest corrugates, presenting a characteristic nose-like shape highlighted in PS. Figure~\ref{fig:Fig11} presents a sketch of the lobe and crest corrugation mechanisms to emphasise the main features and how both deformations are directly caused by the hairpin vortex. The lobe corrugation sketch is representative of the numerical solution for case C2 depicted in figure~\ref{subfig:Fig10b}, while the crest corrugation sketch follows figure~\ref{subfig:Fig10f}. \par

The crest corrugation mechanism is clearly observed in cases B2, C2 and C3. It is also apparent in other analysed configurations. The main reason why this mechanism becomes more relevant at high ambient pressures is directly linked to the increased liquid layer formation at highly transcritical conditions. As previously discussed, the thickness of the initial lobes depends on the surface tension and the dissolution of the ambient gas. Similarly, the liquid sheet near the wave crest becomes thinner at higher pressures. Thus, the thinner layers in the 100-bar and 150-bar configurations are affected more by the deformation mechanism discussed here. \par

\begin{figure}
\centering
\begin{subfigure}{0.5\textwidth}
  \centering
  \includegraphics[width=1.0\linewidth]{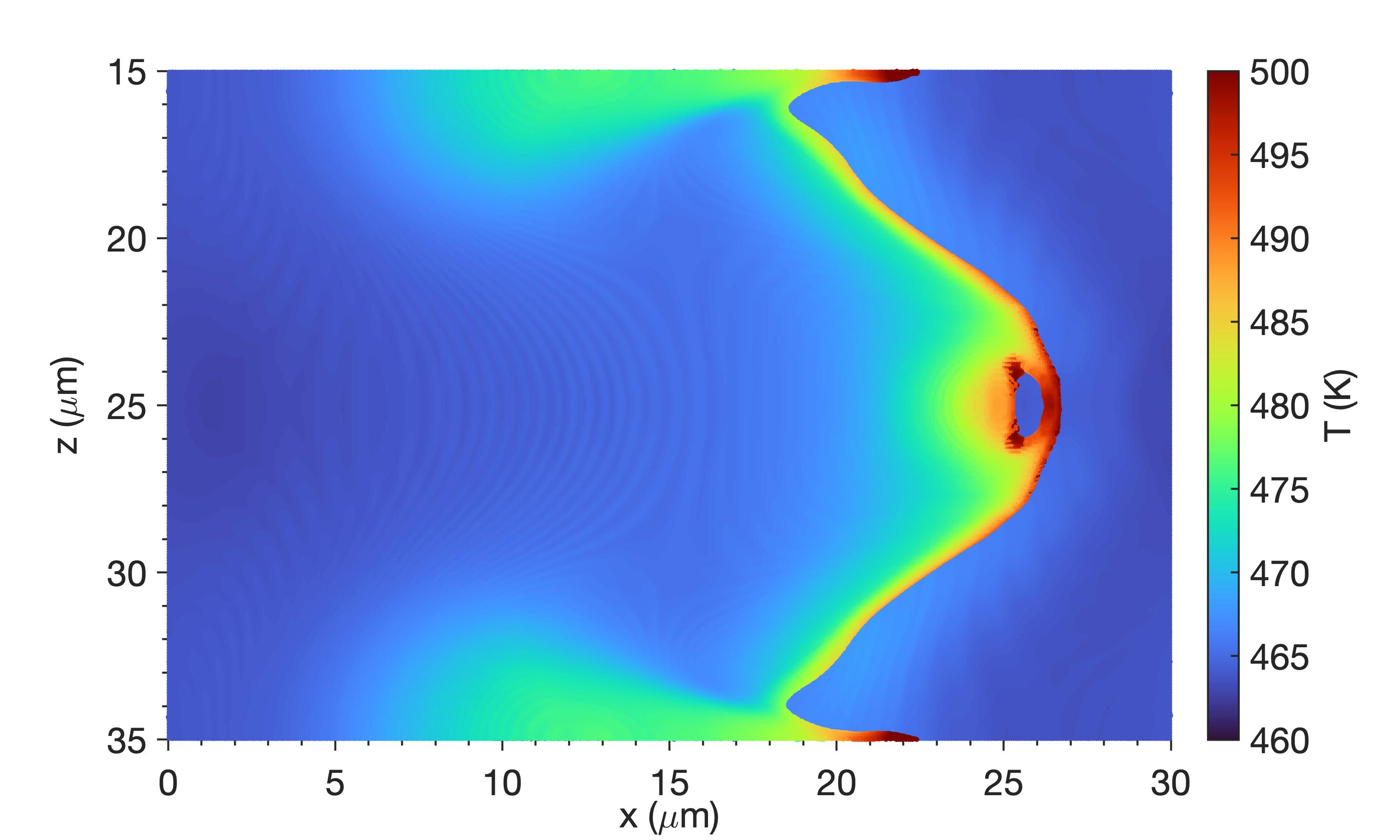}
  \caption{} 
  \label{subfig:Fig12a}
\end{subfigure}%
\begin{subfigure}{0.5\textwidth}
  \centering
  \includegraphics[width=1.0\linewidth]{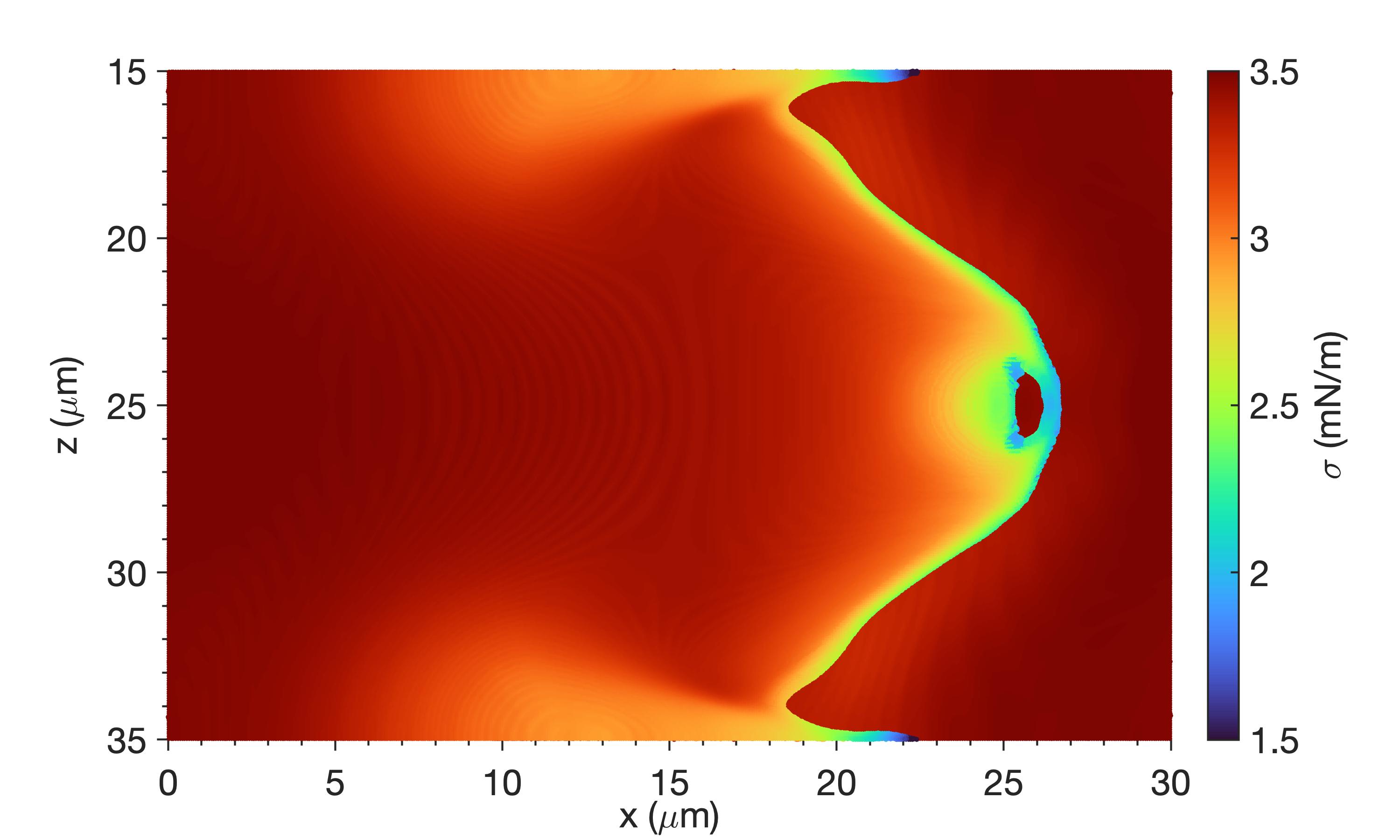}
  \caption{} 
  \label{subfig:Fig12b}
\end{subfigure}%
\caption{Solution of the local thermodynamic equilibrium along the lobe for case C2 at \(t^*=3.5\). A top view from an \(xz\) plane located above the liquid surface is provided. The liquid surface is coloured by the local interface value for (a) temperature, \(T\); and (b) surface-tension coefficient, \(\sigma\).}
\label{fig:Fig12}
\end{figure}

As done in section~\ref{subsec:lobe_bending}, figure~\ref{fig:Fig12} presents the lobe in case C2 at \(t^*=3.5\) viewed from an \(xz\) plane above the liquid surface, showcasing the local thermodynamic equilibrium solution at the interface. Compared to case C1 with a lower freestream velocity of 30 m/s, further lobe stretching and thinning are observed. This results in a faster heating of the liquid, a higher surface temperature, and a lower surface-tension coefficient. Therefore, the edge of the lobe in case C1 is rounder than in case C2. The change in composition as the interface temperature increases is not shown here, and the reader is referred to figures~\ref{fig:Fig2} and~\ref{fig:Fig7} for more details. \par 

Other features are observed. Once the lobe bursts, smaller liquid structures are generated that heat very fast, resulting in an increased vaporisation of the fuel. Additionally, the corrugation of the perturbation crest submerges it into a hotter gaseous region, significantly raising the surface temperature compared to the cooler liquid below. Since crest corrugation is less pronounced in case C1, such a large temperature increase is not observed in figure~\ref{fig:Fig7}. The heating of the perturbation crest and subsequent decrease in surface tension might cause a substantial change in the surface dynamics (PS). As the crest submerges into the hotter gas, it also moves into a faster gaseous stream. This increase in relative velocity between liquid and gas, coupled with the decrease in \(\sigma\), results in the emergence of short wavelength surface instabilities that quickly grow, leading to ligament shredding and the formation of small droplets (PS). \par

\subsection{Formation of liquid sheets or layering}
\label{subsec:layering}

Similar to the deformation cascade process that the liquid jet undergoes, vortices form, deform, and break up over time. Understanding this process is the key since it defines the evolution of the liquid surface, its atomisation, and the mixing within each phase. For instance, vorticity is directly responsible for the formation of high temperature regions and fuel-rich blobs in the gas phase. Therefore, we examine the evolution of vortex structures to identify regions of vortex generation and how these vortices evolve toward smaller scales. In particular, we focus on the impact of the layering of liquid sheets characteristic of very high-pressure transcritical environments. \par 

\begin{figure}
\centering
\begin{subfigure}{0.33\textwidth}
  \centering
  \includegraphics[width=1.0\linewidth]{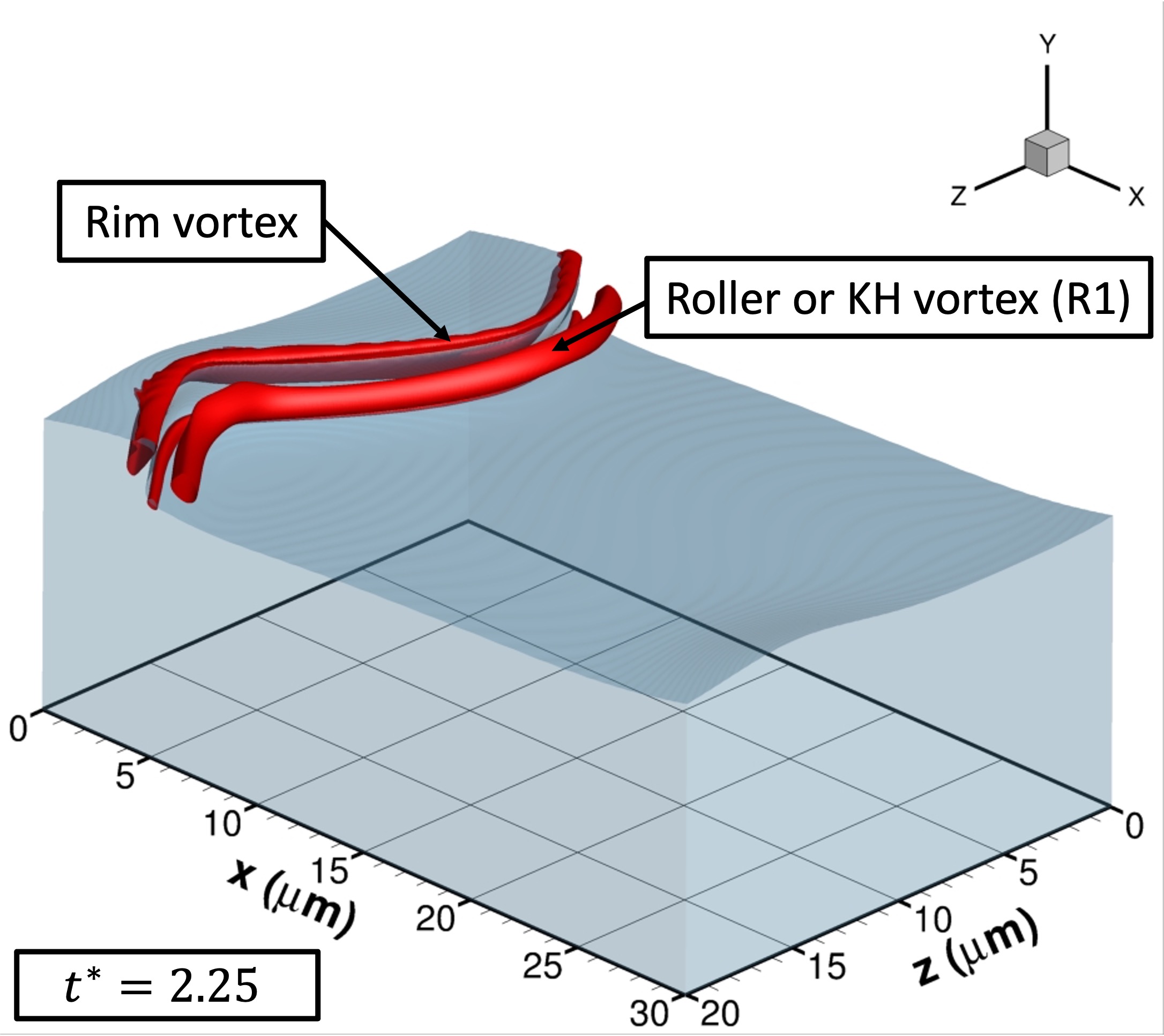}
  \label{subfig:Fig13a}
\end{subfigure}%
\begin{subfigure}{0.33\textwidth}
  \centering
  \includegraphics[width=1.0\linewidth]{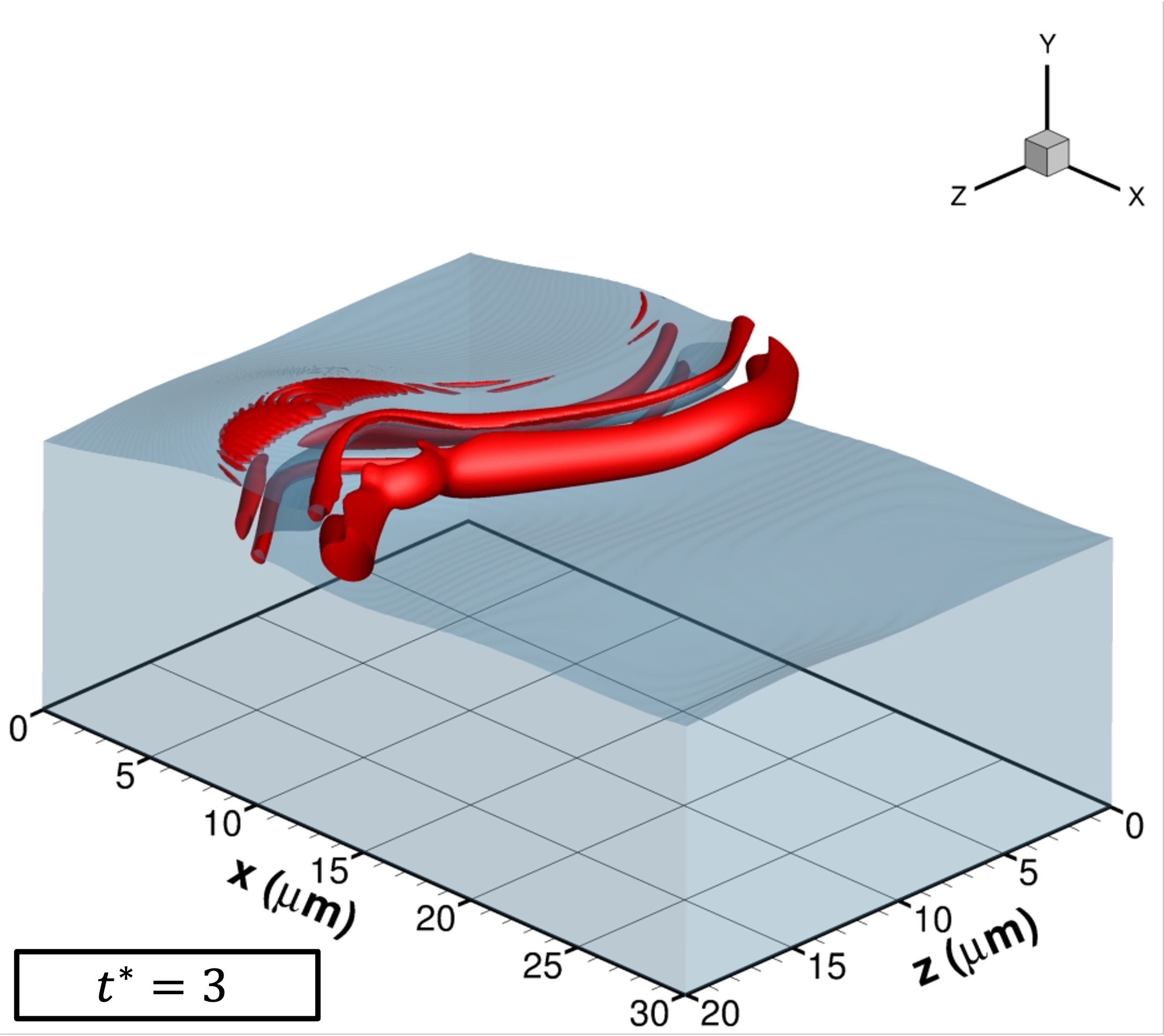}
  \label{subfig:Fig13b}
\end{subfigure}%
\begin{subfigure}{0.33\textwidth}
  \centering
  \includegraphics[width=1.0\linewidth]{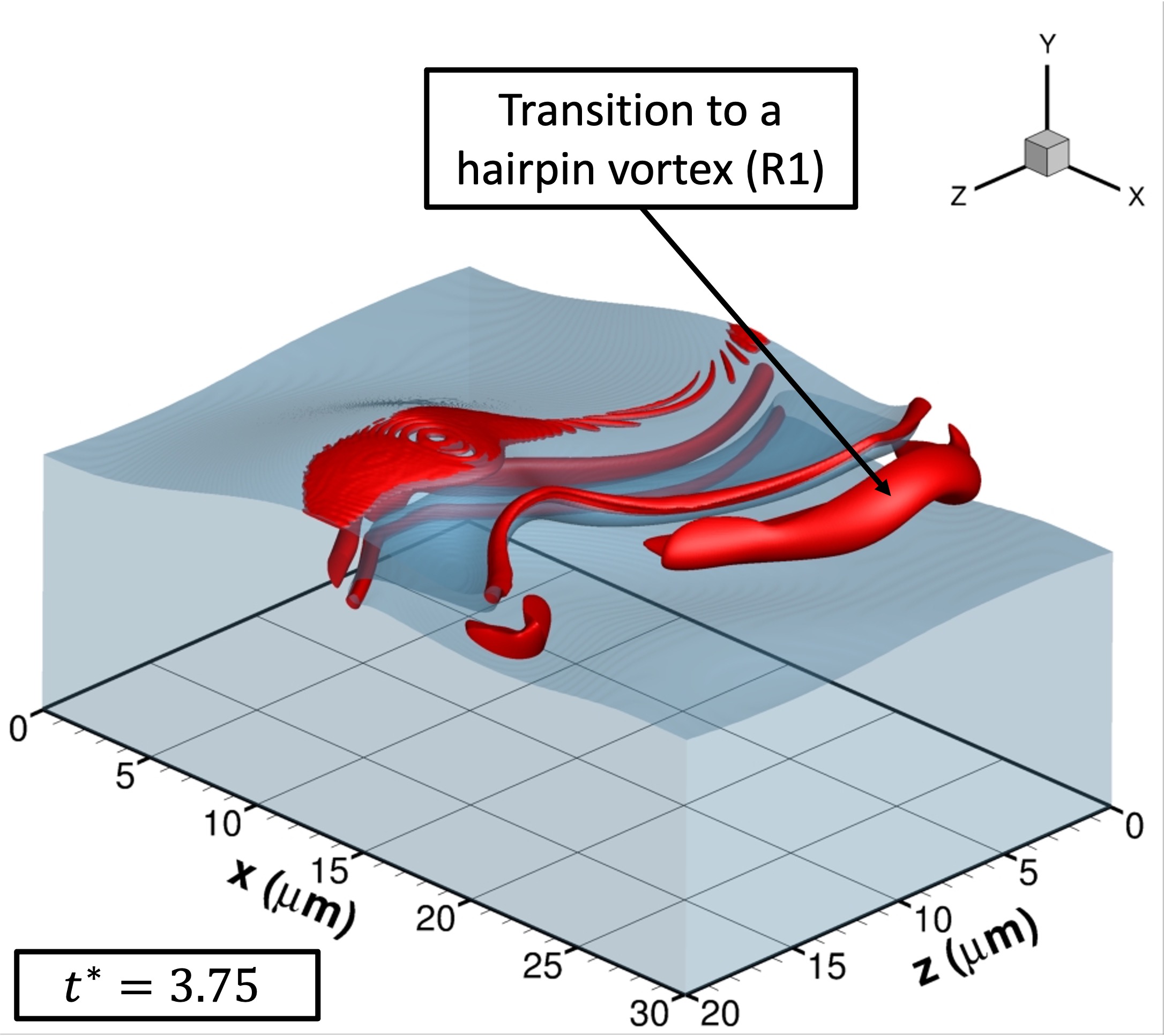}
  \label{subfig:Fig13c}
\end{subfigure}%
\\[-2ex]
\begin{subfigure}{0.33\textwidth}
  \centering
  \includegraphics[width=1.0\linewidth]{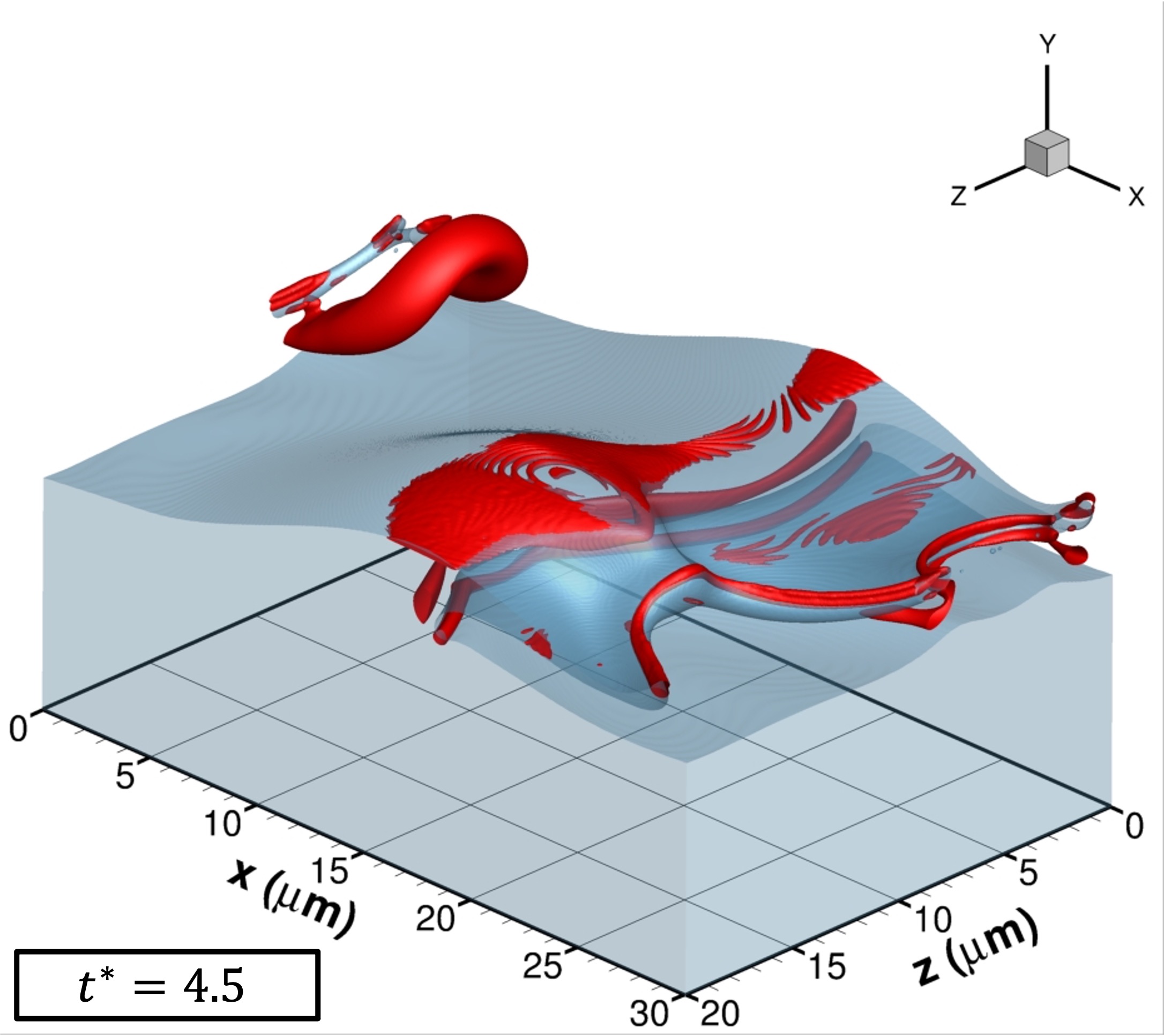}
  \label{subfig:Fig13d}
\end{subfigure}%
\begin{subfigure}{0.33\textwidth}
  \centering
  \includegraphics[width=1.0\linewidth]{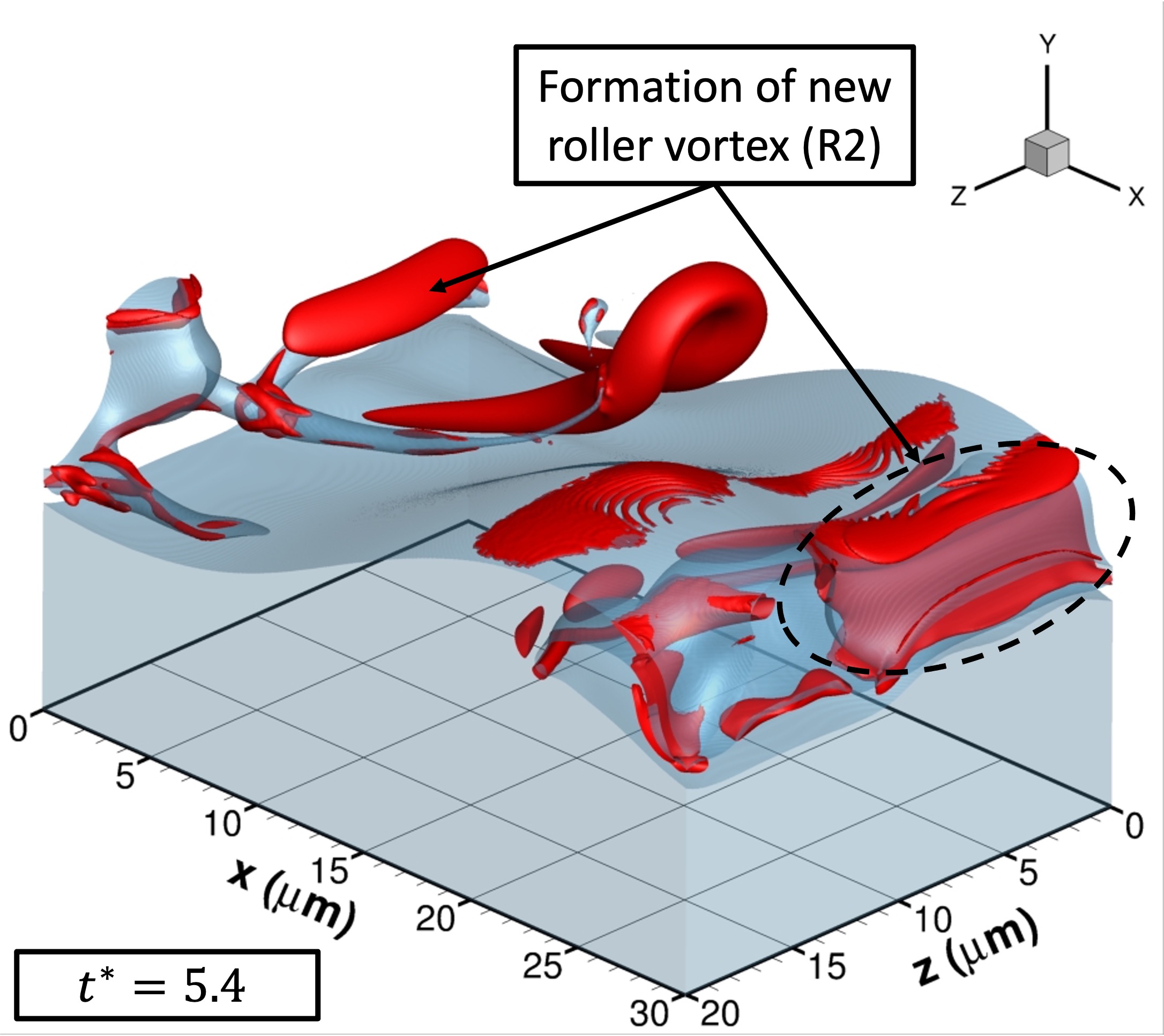}
  \label{subfig:Fig13e}
\end{subfigure}%
\begin{subfigure}{0.33\textwidth}
  \centering
  \includegraphics[width=1.0\linewidth]{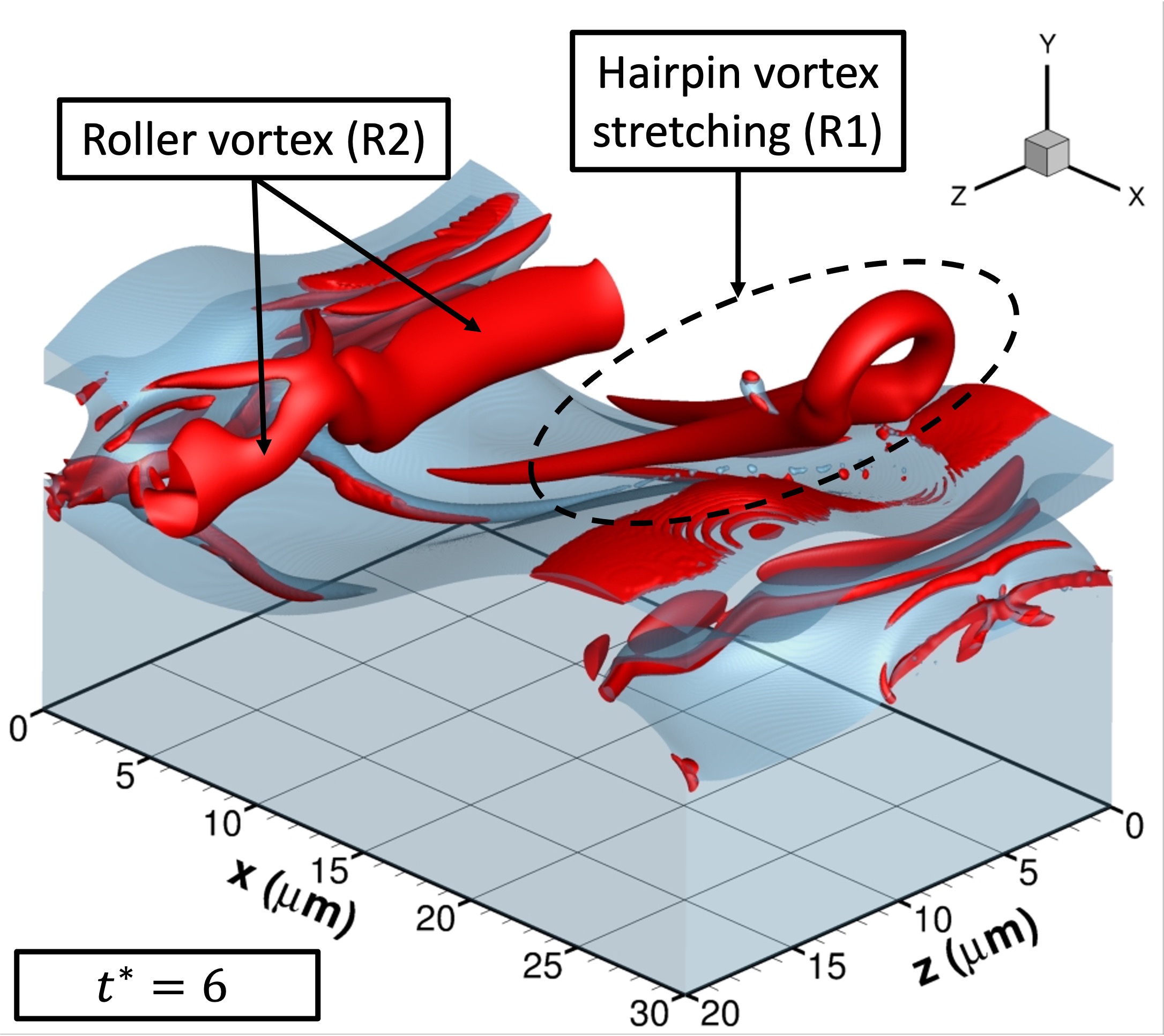}
  \label{subfig:Fig13f}
\end{subfigure}%
\\[-2ex]
\begin{subfigure}{0.33\textwidth}
  \centering
  \includegraphics[width=1.0\linewidth]{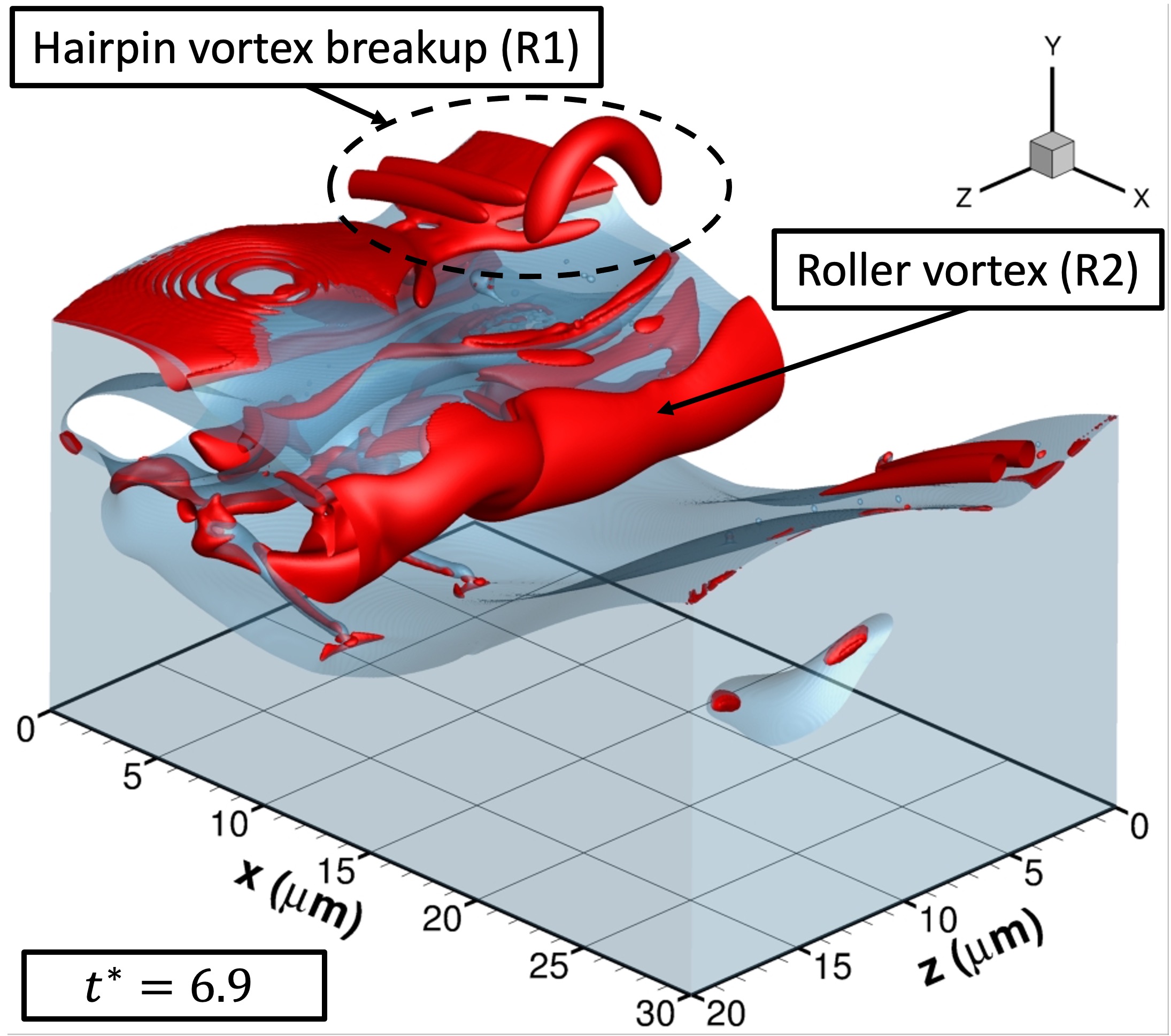}
  \label{subfig:Fig13g}
\end{subfigure}%
\begin{subfigure}{0.33\textwidth}
  \centering
  \includegraphics[width=1.0\linewidth]{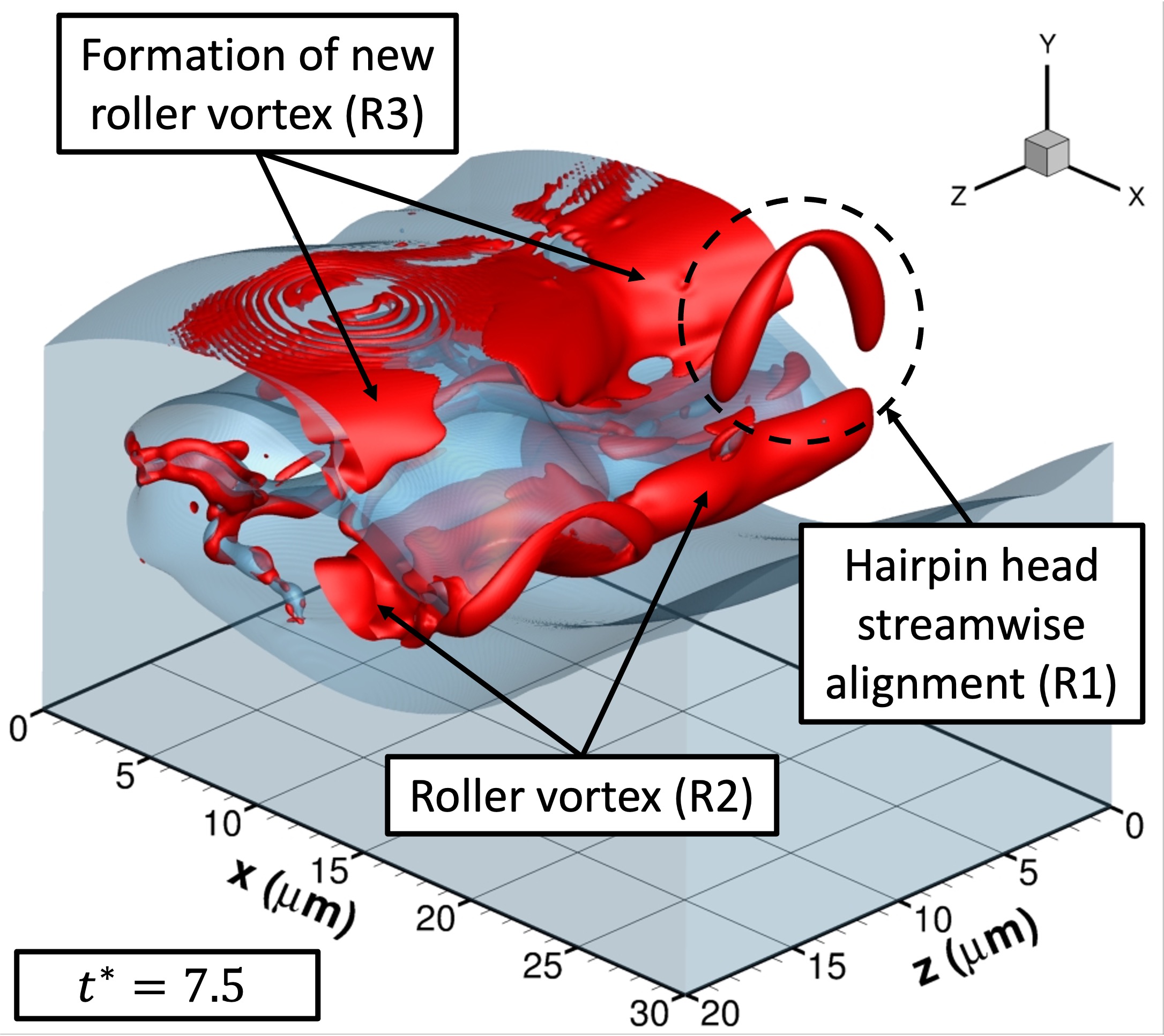}
  \label{subfig:Fig13h}
\end{subfigure}%
\begin{subfigure}{0.33\textwidth}
  \centering
  \includegraphics[width=1.0\linewidth]{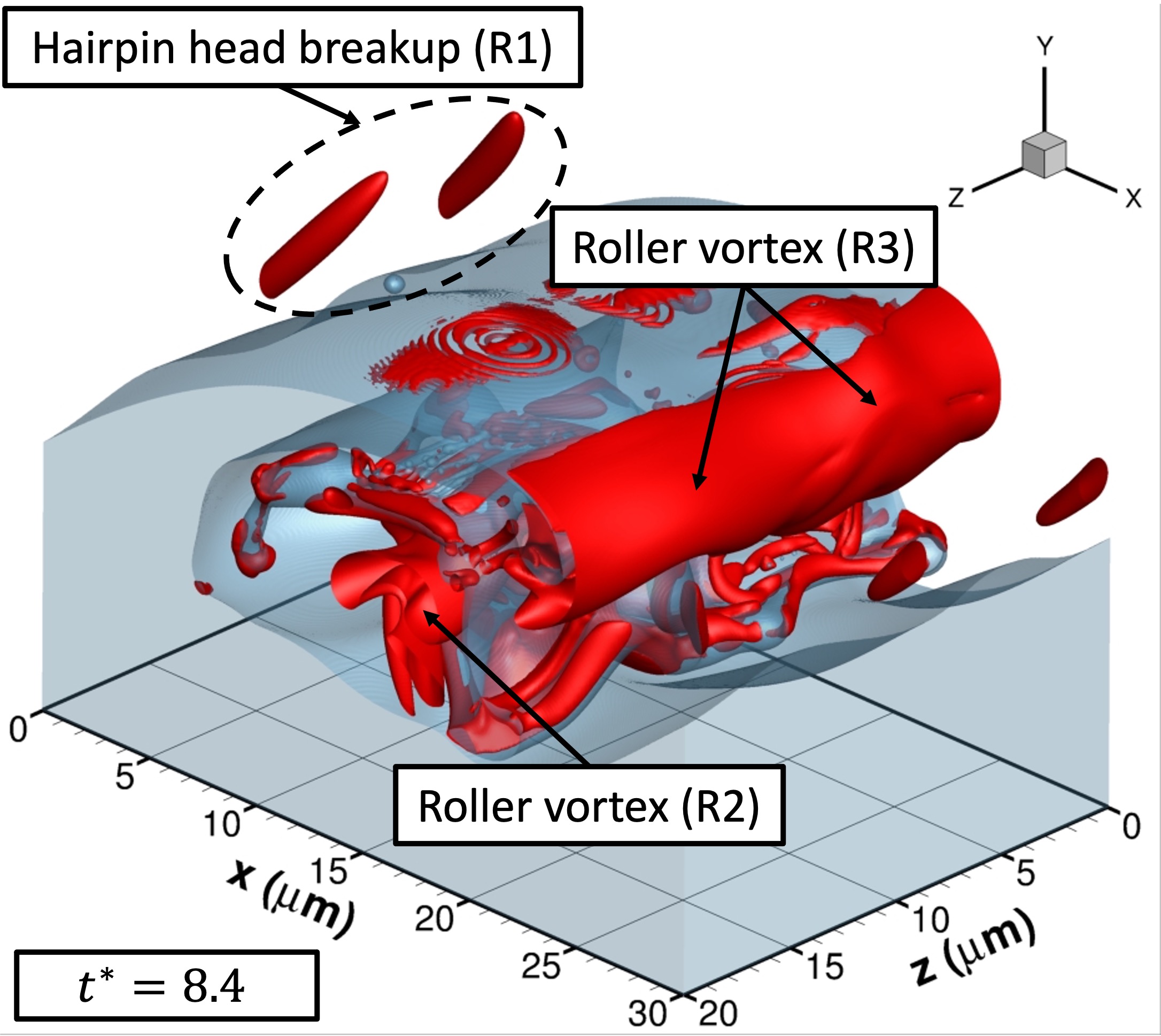}
  \label{subfig:Fig13i}
\end{subfigure}%
\\[-2ex]
\begin{subfigure}{0.33\textwidth}
  \centering
  \includegraphics[width=1.0\linewidth]{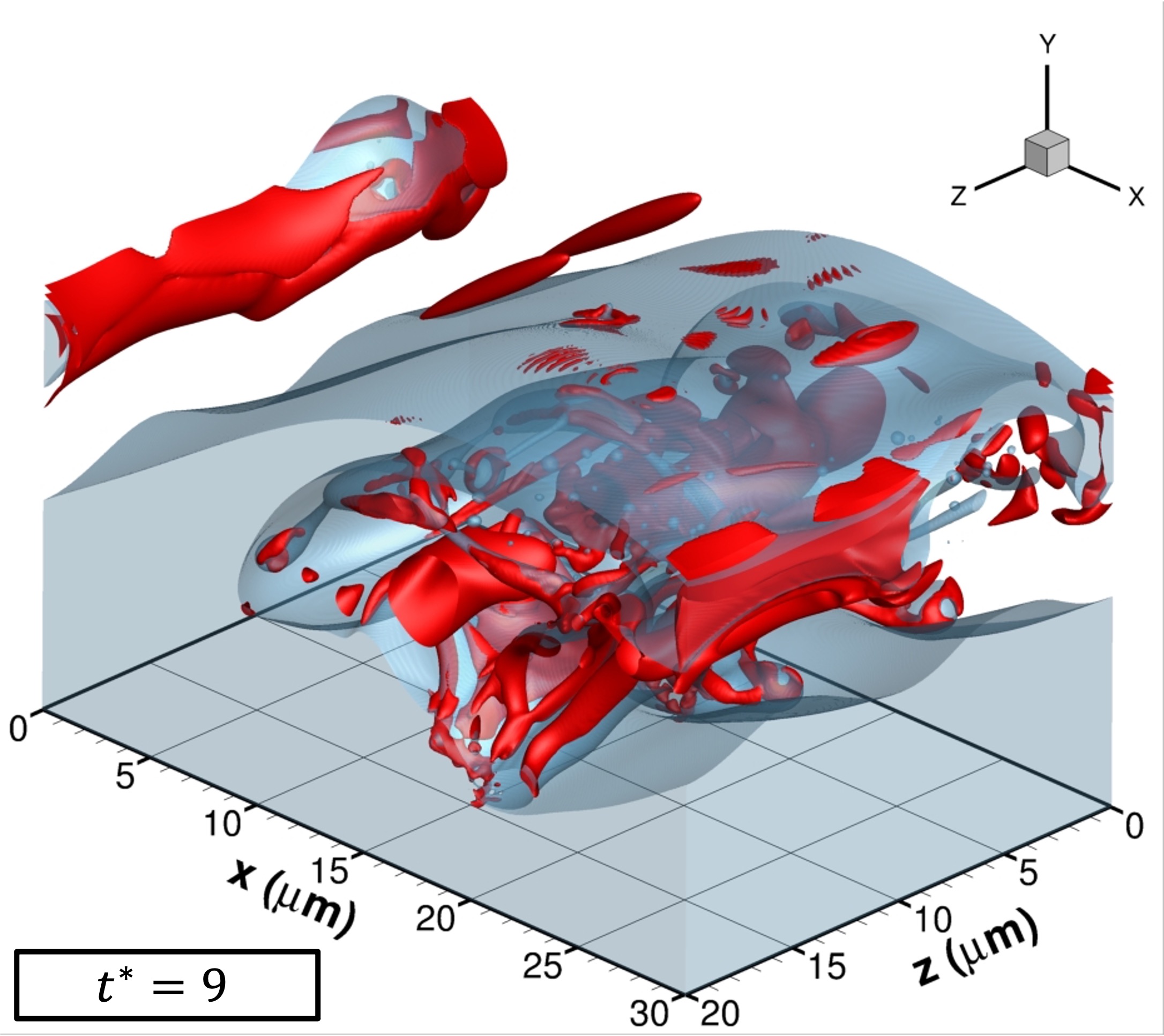}
  \label{subfig:Fig13j}
\end{subfigure}%
\begin{subfigure}{0.33\textwidth}
  \centering
  \includegraphics[width=1.0\linewidth]{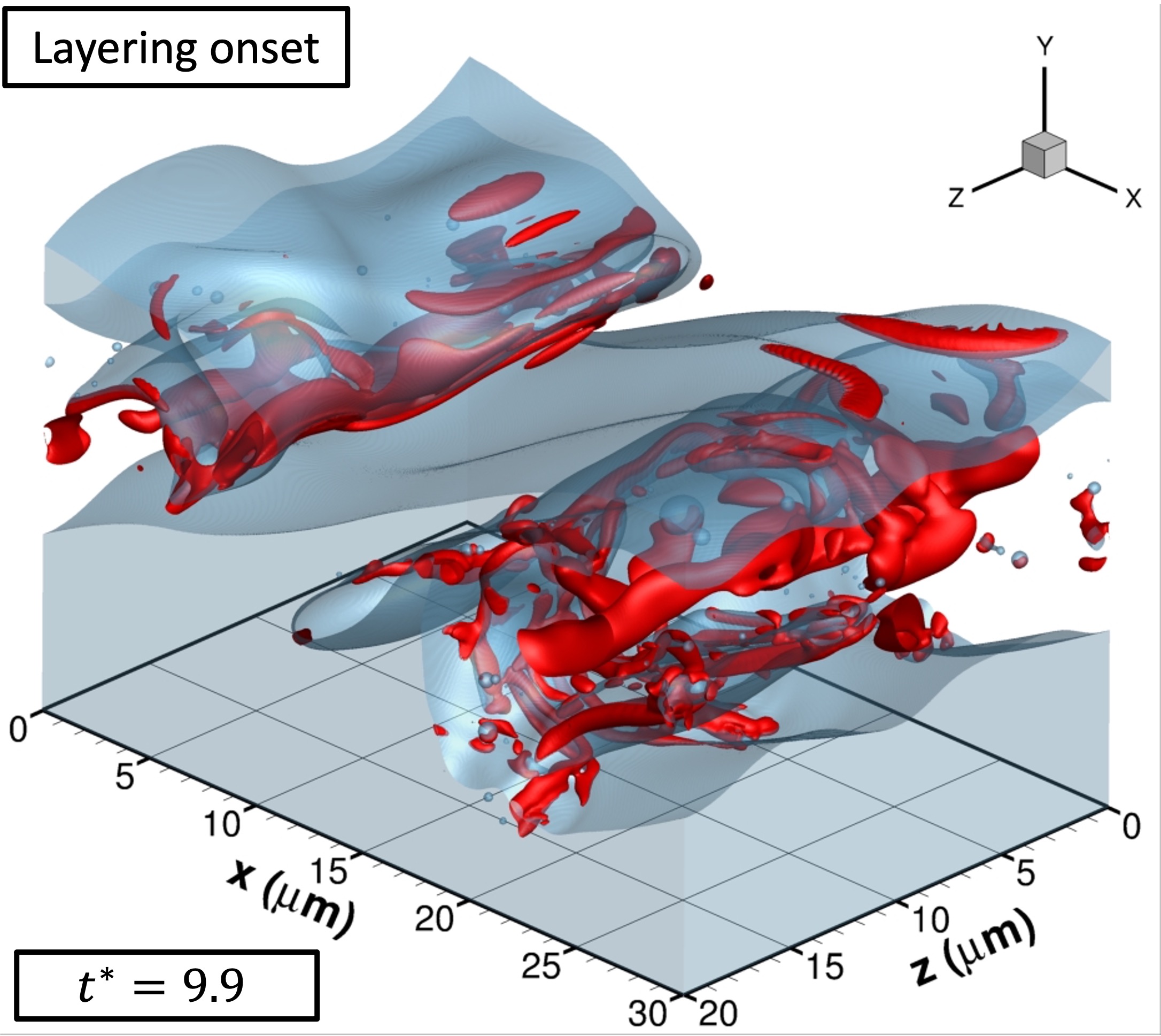}
  \label{subfig:Fig13k}
\end{subfigure}%
\begin{subfigure}{0.33\textwidth}
  \centering
  \includegraphics[width=1.0\linewidth]{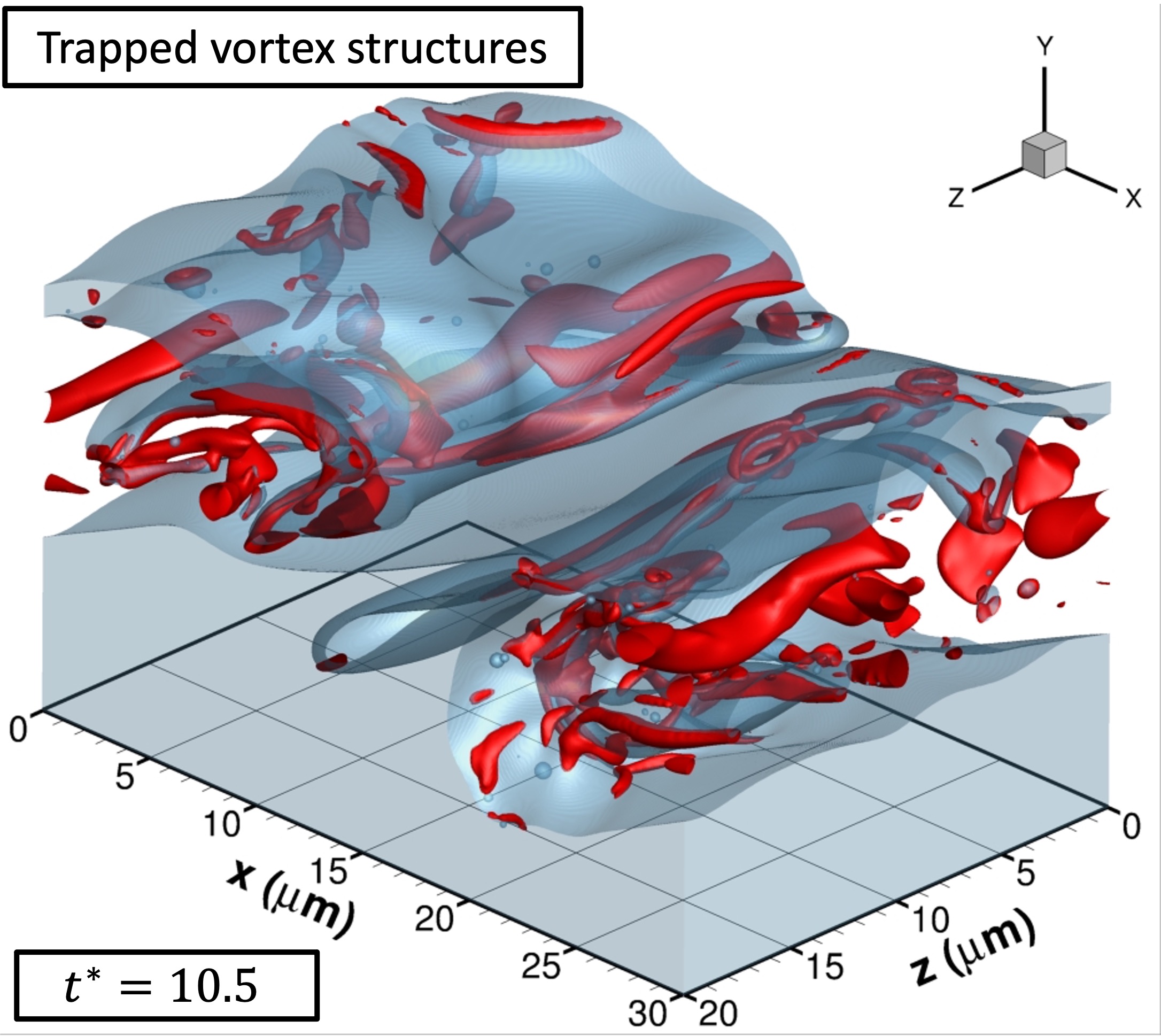}
  \label{subfig:Fig13l}
\end{subfigure}%
\caption{Evolution of vortex structures for case C1. An oblique view shows the liquid surface identified by the translucid blue isosurface with \(C=0.5\); the vortex structures are identified by the red isosurface with \(\lambda_{\rho,t}=-2.5\times 10^{15}\).}
\label{fig:Fig13}
\end{figure}

\begin{figure}
\centering
\begin{subfigure}{0.33\textwidth}
  \centering
  \includegraphics[width=1.0\linewidth]{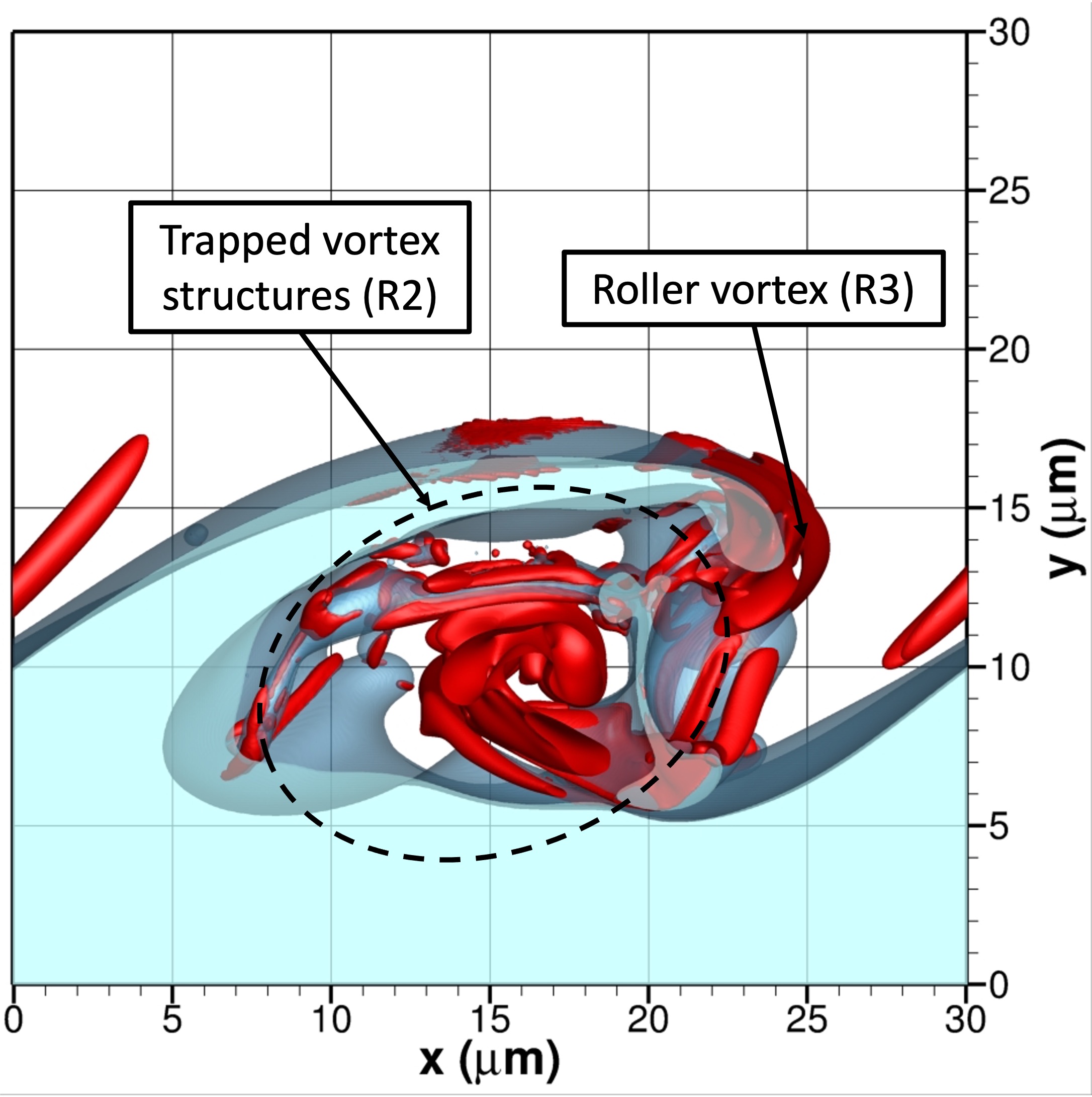}
  \caption{} 
  \label{subfig:Fig14a}
\end{subfigure}%
\begin{subfigure}{0.33\textwidth}
  \centering
  \includegraphics[width=0.72\linewidth]{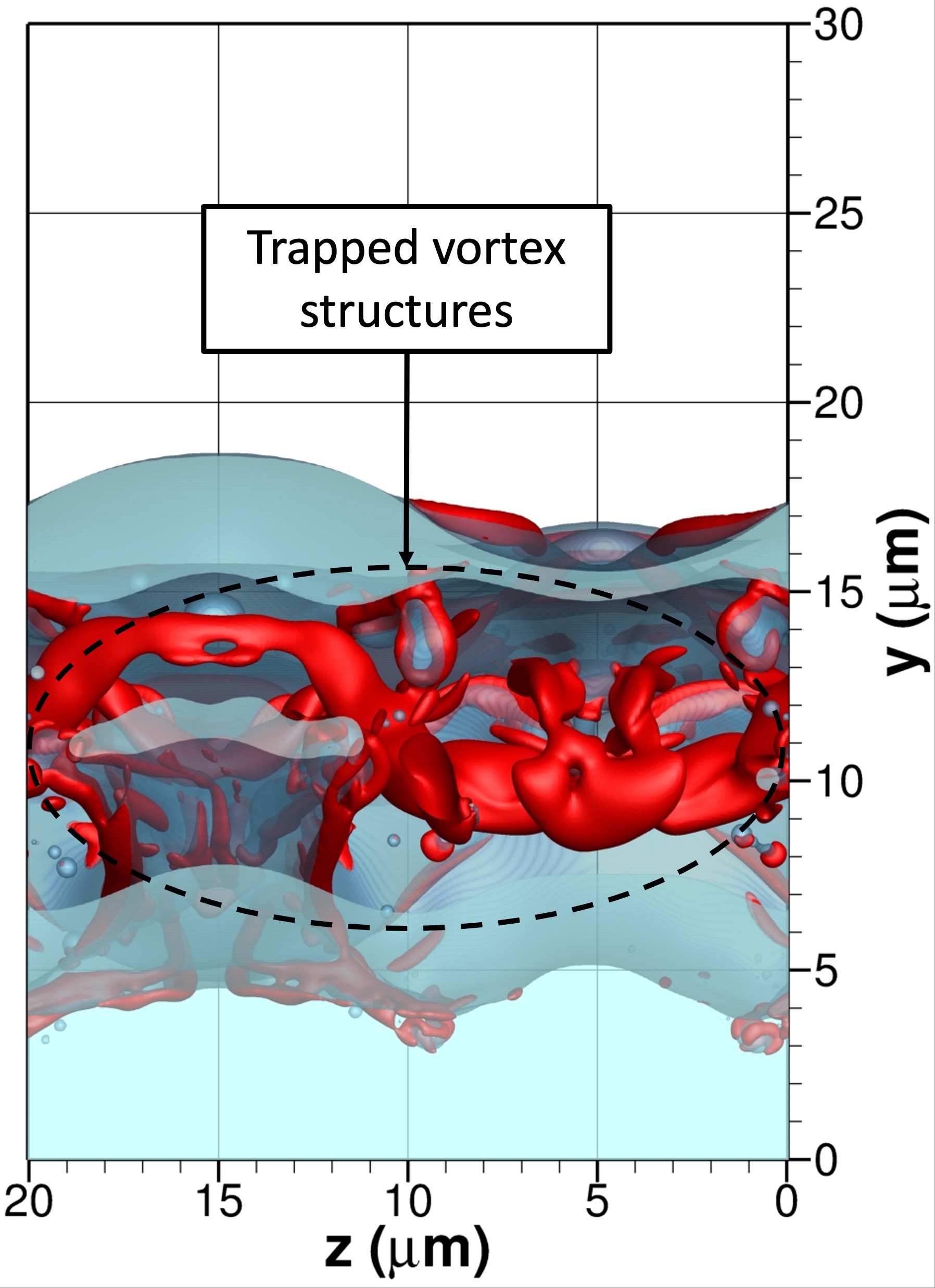}
  \caption{}
  \label{subfig:Fig14b}
\end{subfigure}%
\begin{subfigure}{0.33\textwidth}
  \centering
  \includegraphics[width=1.0\linewidth]{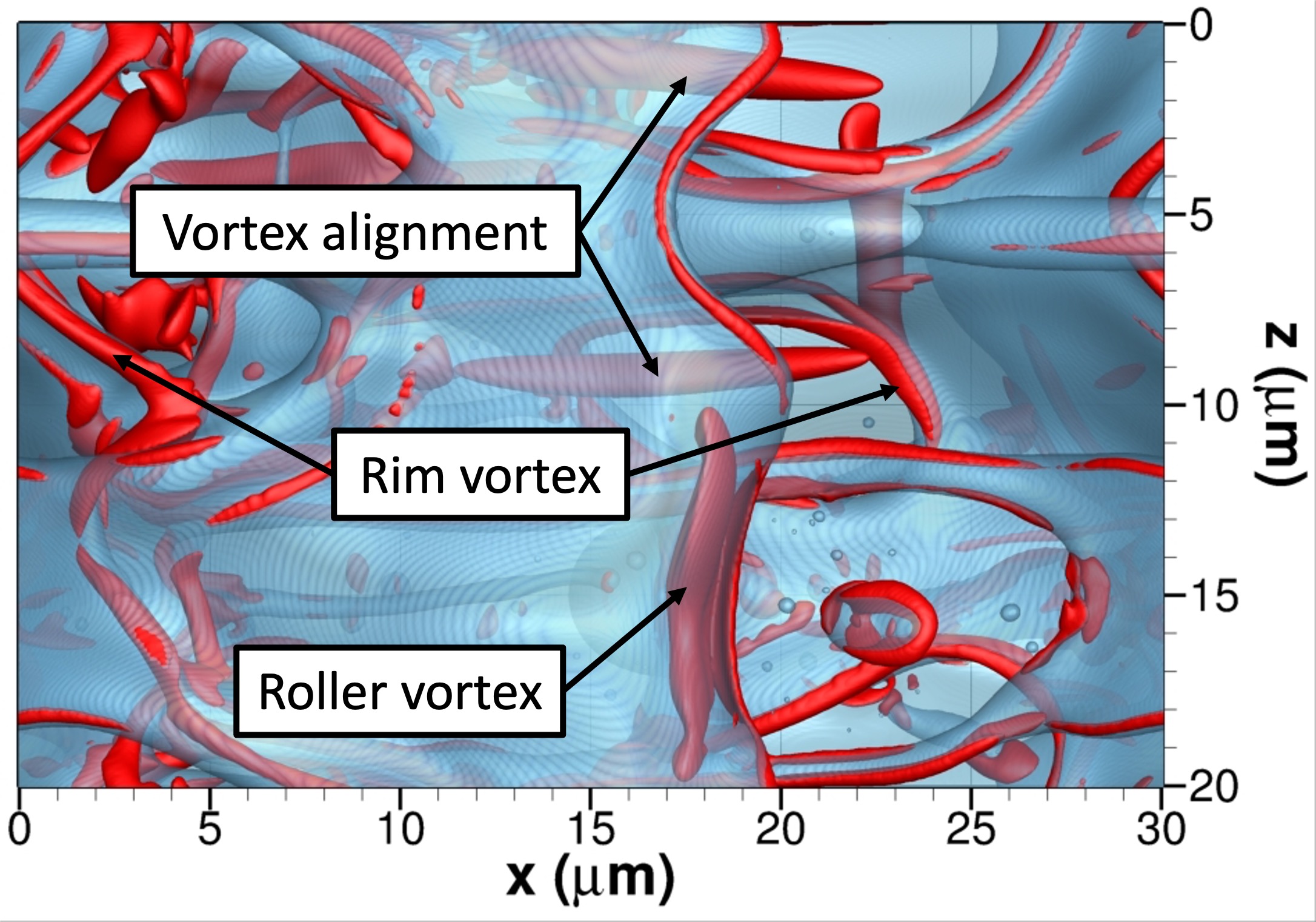}
  \caption{}
  \label{subfig:Fig14c}
\end{subfigure}%
\caption{Evolution of vortex structures for case C1. 14a shows a side view at \(z=20\) \(\mu\)m, 14b a front view at \(x=30\) \(\mu\)m, and 14c a top view from above the liquid surface. Each snapshot corresponds to a different \(t^*\). The liquid surface is identified by the translucid blue isosurface with \(C=0.5\) and the vortex structures are identified by the red isosurface with \(\lambda_{\rho,t}=-2.5\times 10^{15}\). (a) \(t^*=8.4\); (b) \(t^*=9.9\); and (c) \(t^*=15\).}
\label{fig:Fig14}
\end{figure}

Major events in the evolution of the vorticity field for case C1 are presented in figures~\ref{fig:Fig13} and~\ref{fig:Fig14}. The 150-bar lower velocity configuration has been chosen to clearly show the layering. In the following, we describe the major differences between case C1 and higher velocity cases. Initially, the shear between liquid and gas generates a roller or KH vortex along the perturbed surface, downstream of the lobe (see figure~\ref{fig:Fig13} at \(t^*=2.25\)). We name this vortex structure R1 (i.e., the first roller). Usually, a distinct rim vortex also appears along the lobe's edge. The roller vortex evolves into a hairpin vortex, whose impact on the early lobe deformation has been described in sections~\ref{subsec:lobe_bending} and~\ref{subsec:lobe_crest_corrugation}. R1 is advected into the oxidizer stream, subject to continuous stretching and streamwise alignment (i.e., \(\omega_x\) generation). Eventually, the hairpin head breaks from the hairpin legs, and then it reorients along the streamwise direction before breaking again. This vortex breakup process must be understood as the weakening of the local vorticity; thus, the value of \(\lambda_{\rho,t}\) cannot capture the complete vortex structure anymore. The subsequent deformation of R1 can be observed in figure~\ref{fig:Fig13} from \(t^*=2.25\) to \(t^*=8.4\). \par 

Over time, new roller vortices form downstream of the growing perturbation. As seen in figure~\ref{fig:Fig13}, vortex R2 forms between \(t^*=5.4\) and \(t^*=6\), and vortex R3 forms between \(t^*=7.5\) and \(t^*=8.4\). This vortex formation can be understood as a vortex pairing process. During the growth of the surface perturbation, the vortex sheet between liquid and gas detaches at intervals. The resulting spanwise vorticity downstream of the wave's edge tends to concentrate, forming the vortex tubes or rollers we identify as R2 and R3. Compared to R1, these new roller vortices remain very close to the liquid surface and further contribute to the liquid stretching that leads to the folding and layering of the liquid sheets. The wave growth and subsequent surface area growth enhance the heat and oxygen mass transfer into the liquid, accelerating the features that facilitate the liquid phase stretching. The rapid wave growth also causes strong gas entrainment, which swallows the roller vortices underneath the wave (see figures~\ref{subfig:Fig14a} and~\ref{subfig:Fig14b}). Such strong gas entrainment can be visualised at \(t^*=8.4\), where the roller vortex is stretching under the liquid and resembles more a vortex sheet. \par 

\begin{figure}
\centering
\begin{subfigure}{0.5\textwidth}
  \centering
  \includegraphics[width=0.9\linewidth]{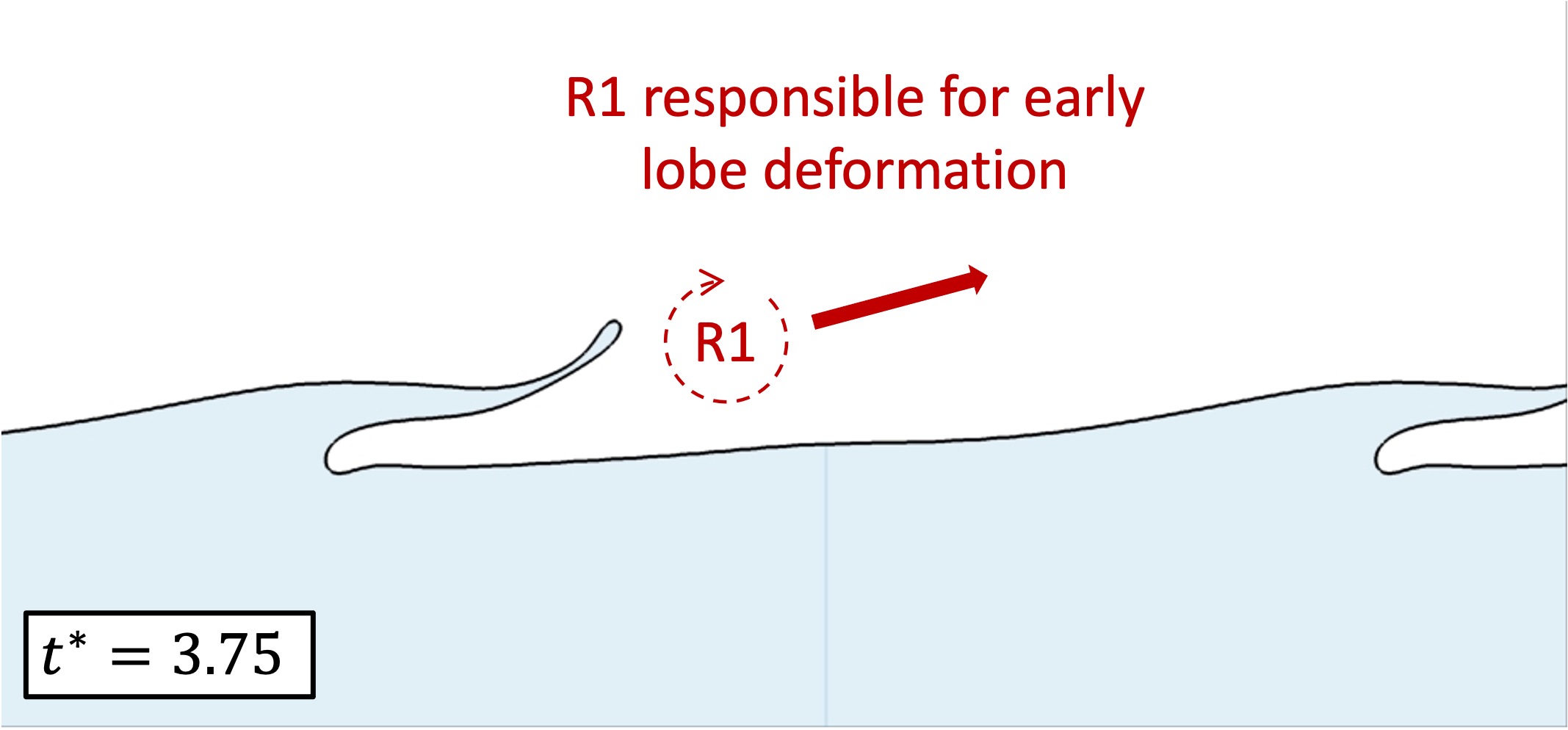}
  \label{subfig:Fig15a}
\end{subfigure}%
\begin{subfigure}{0.5\textwidth}
  \centering
  \includegraphics[width=0.9\linewidth]{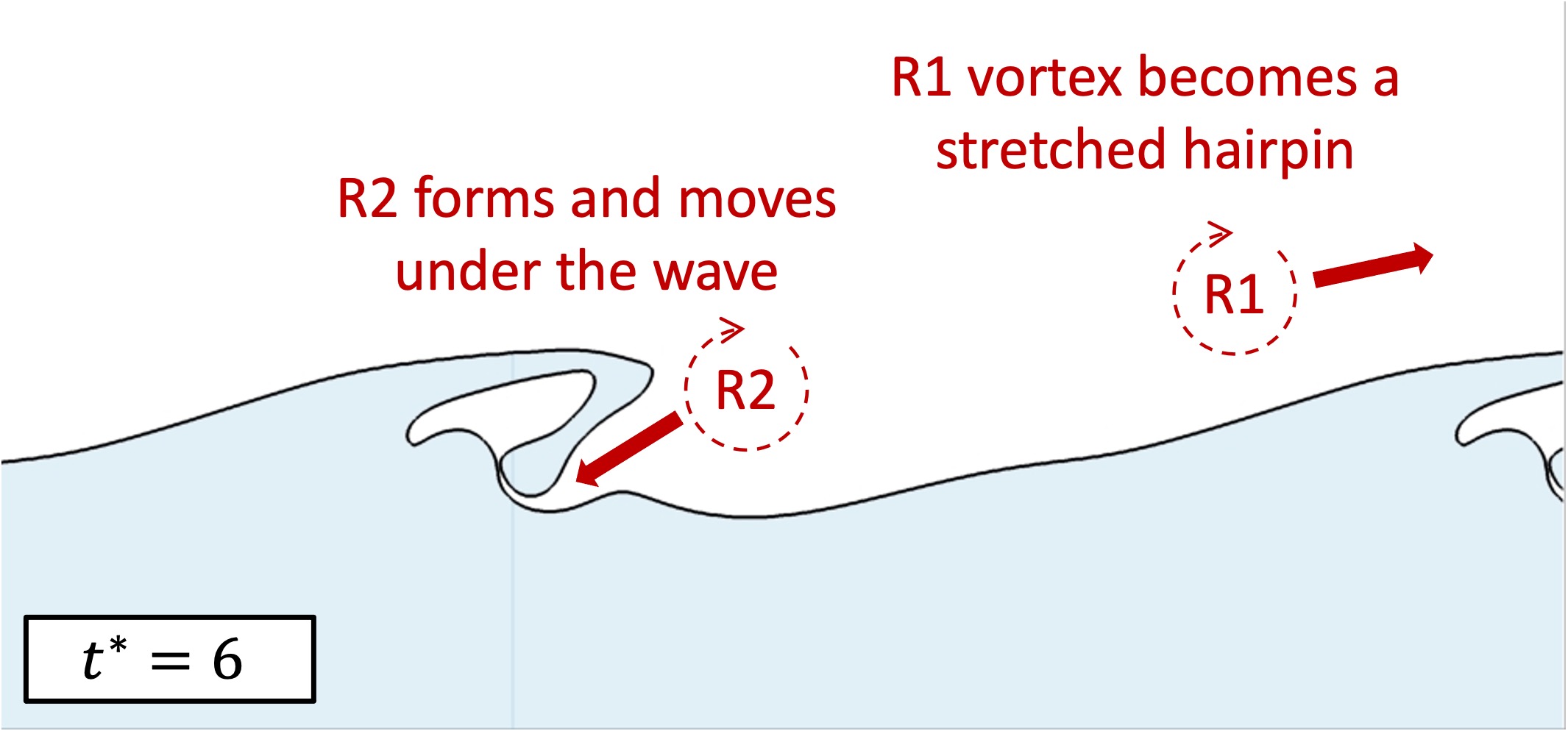}
  \label{subfig:Fig15b}
\end{subfigure}%
\\[3ex]
\begin{subfigure}{0.5\textwidth}
  \centering
  \includegraphics[width=0.9\linewidth]{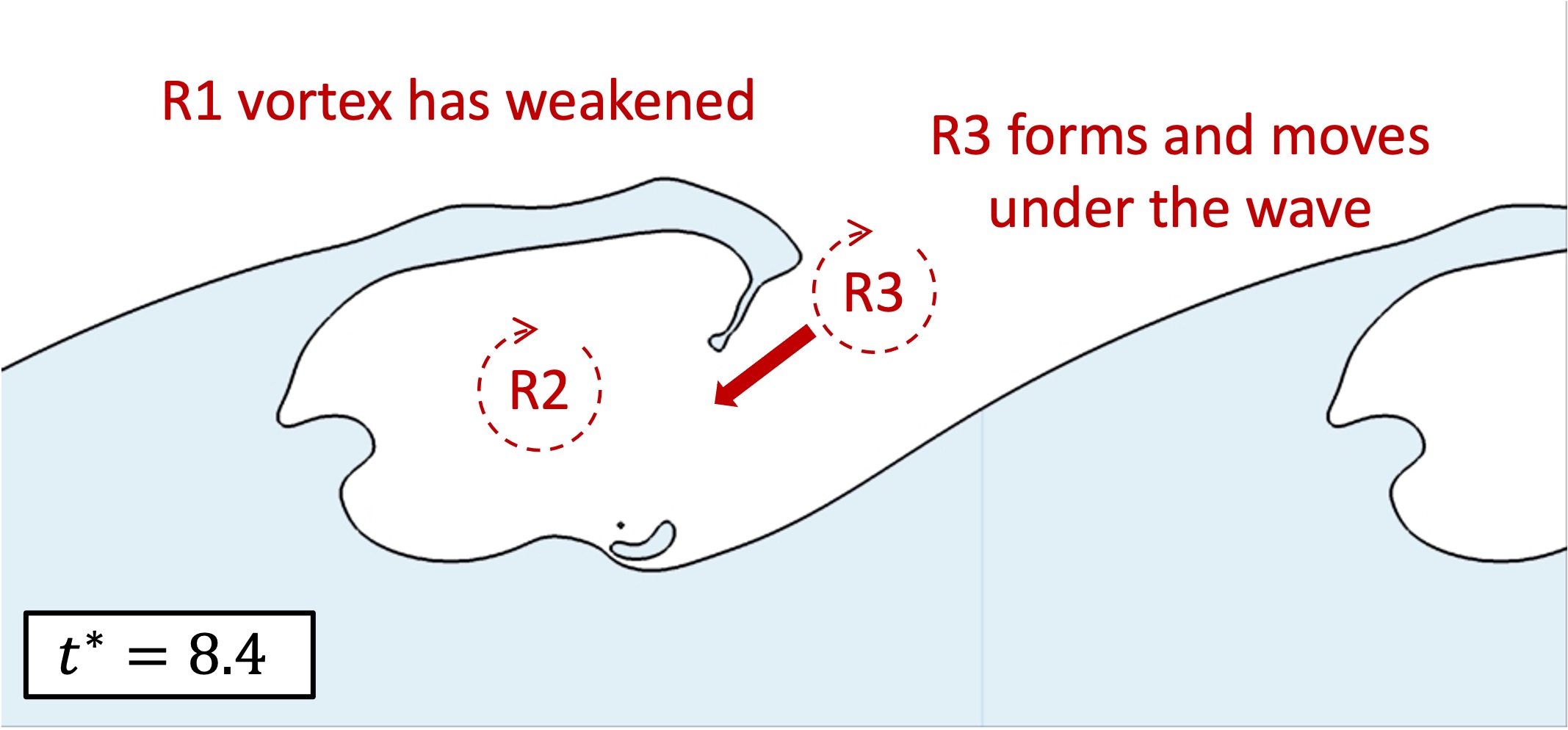}
  \label{subfig:Fig15c}
\end{subfigure}%
\begin{subfigure}{0.5\textwidth}
  \centering
  \includegraphics[width=0.9\linewidth]{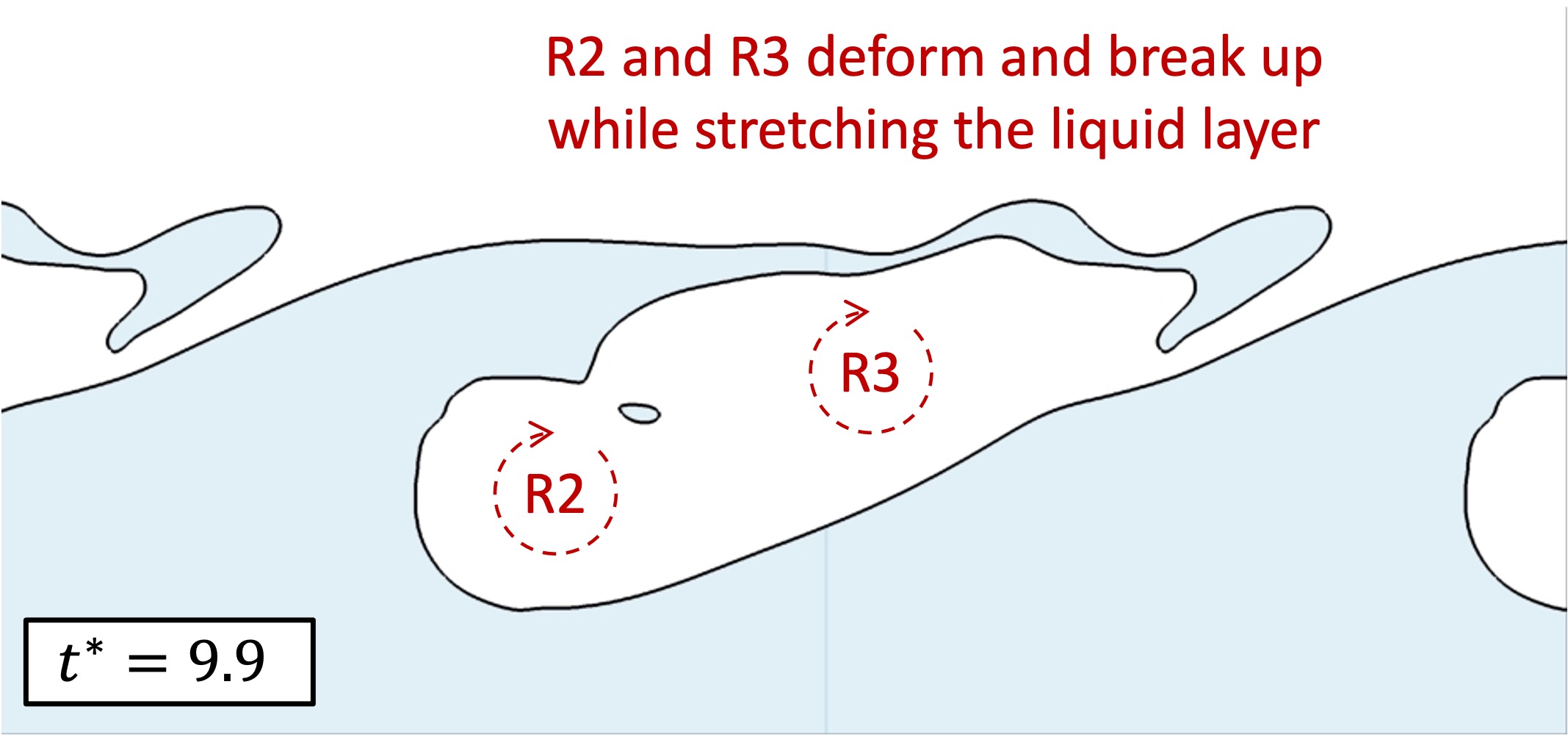}
  \label{subfig:Fig15d}
\end{subfigure}%
\caption{Vortex capturing mechanism in case C1 and how it leads to the layering of liquid sheets. The interface is represented by the isocontour of \(C=0.5\) along \(xy\) planes at \(z=5\) \(\mu\)m.}
\label{fig:Fig15}
\end{figure}

This vortex-capturing mechanism defines the local formation of small liquid sheets, ligaments and droplets under the liquid wave and, most importantly, explains the layering mechanism that dominates the transcritical liquid jet deformation. Figure~\ref{fig:Fig15} shows this process, highlighting the formation of the various roller vortices (i.e., R1, R2 and R3). After the onset of layering (i.e., for \(t^*>10\)), generation of strong vortices above the liquid layers is reduced, and most of the vortices visualised with \(\lambda_{\rho,t}=-2.5\times 10^{15}\) are confined within the various liquid layers (see figure~\ref{subfig:Fig14c}). \par

\begin{figure}
\centering
\begin{subfigure}{0.33\textwidth}
  \centering
  \includegraphics[width=0.98\linewidth]{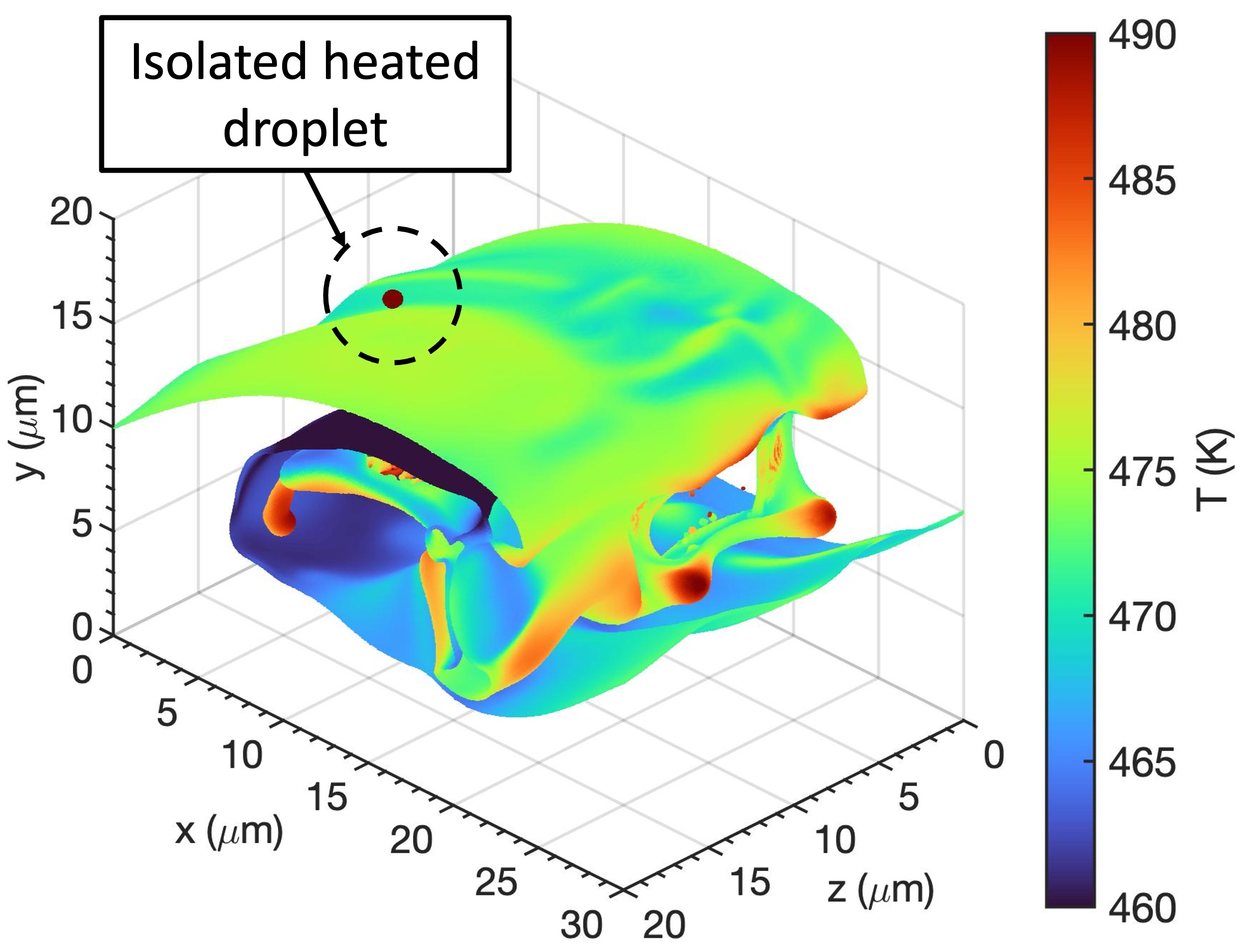}
  \label{subfig:Fig16a}
\end{subfigure}%
\begin{subfigure}{0.33\textwidth}
  \centering
  \includegraphics[width=0.98\linewidth]{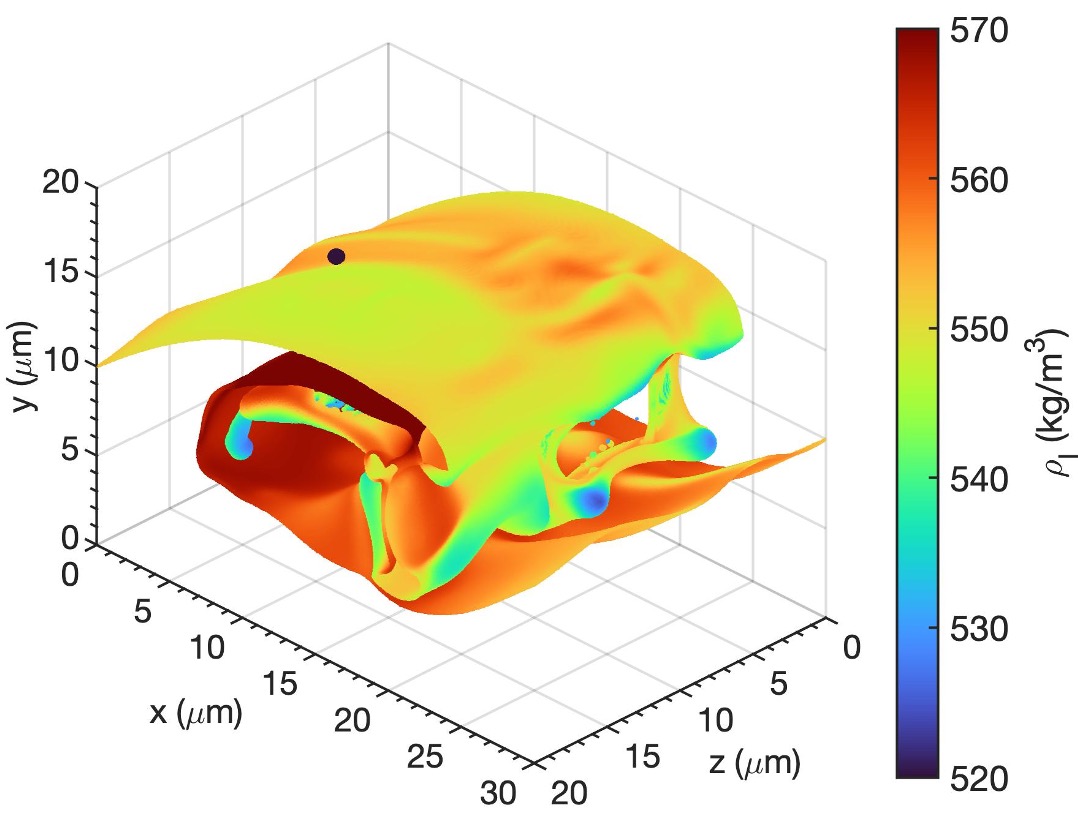}
  \label{subfig:Fig16b}
\end{subfigure}%
\begin{subfigure}{0.33\textwidth}
  \centering
  \includegraphics[width=0.98\linewidth]{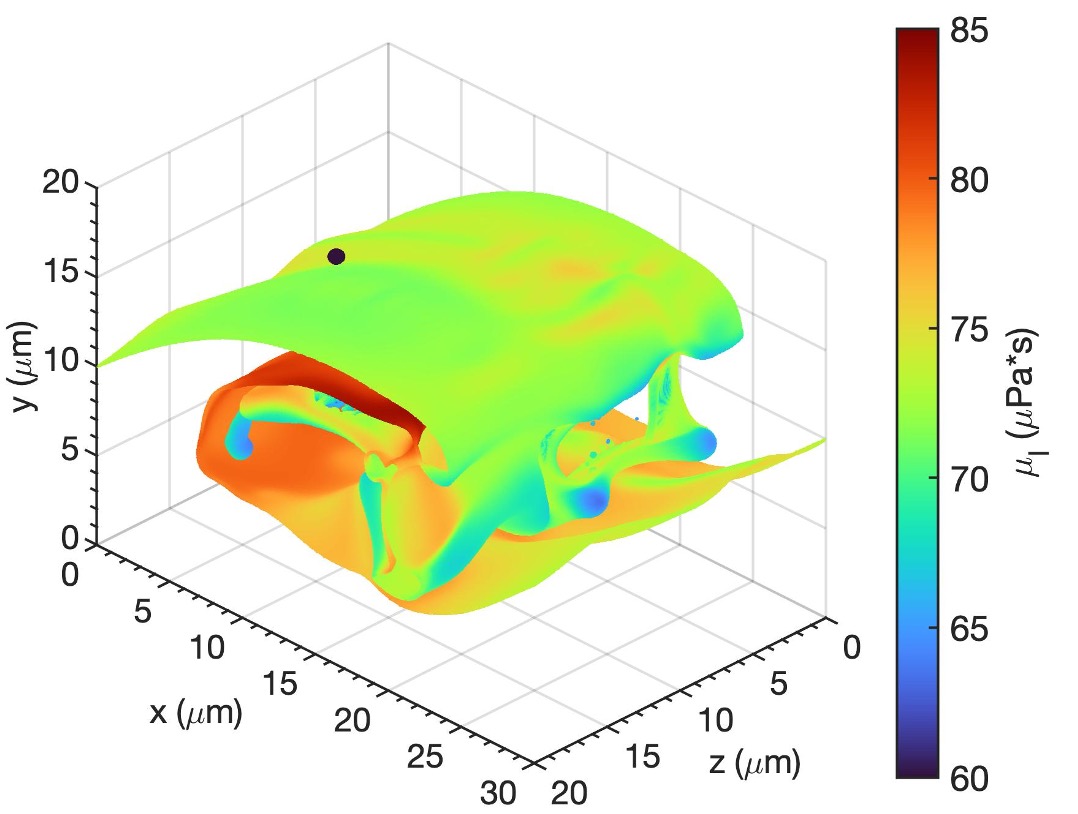}
  \label{subfig:Fig16c}
\end{subfigure}%
\\[1ex]
\begin{subfigure}{0.33\textwidth}
  \centering
  \includegraphics[width=0.95\linewidth]{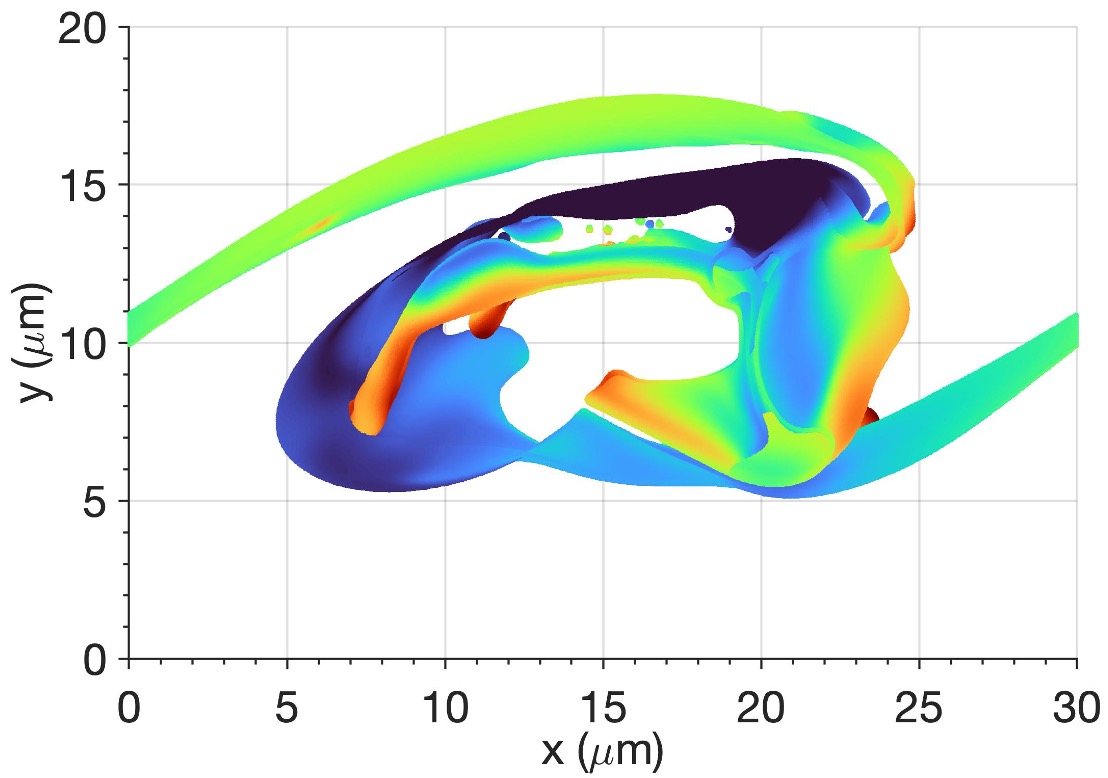}
  \caption{} 
  \label{subfig:Fig16d}
\end{subfigure}%
\begin{subfigure}{0.33\textwidth}
  \centering
  \includegraphics[width=0.95\linewidth]{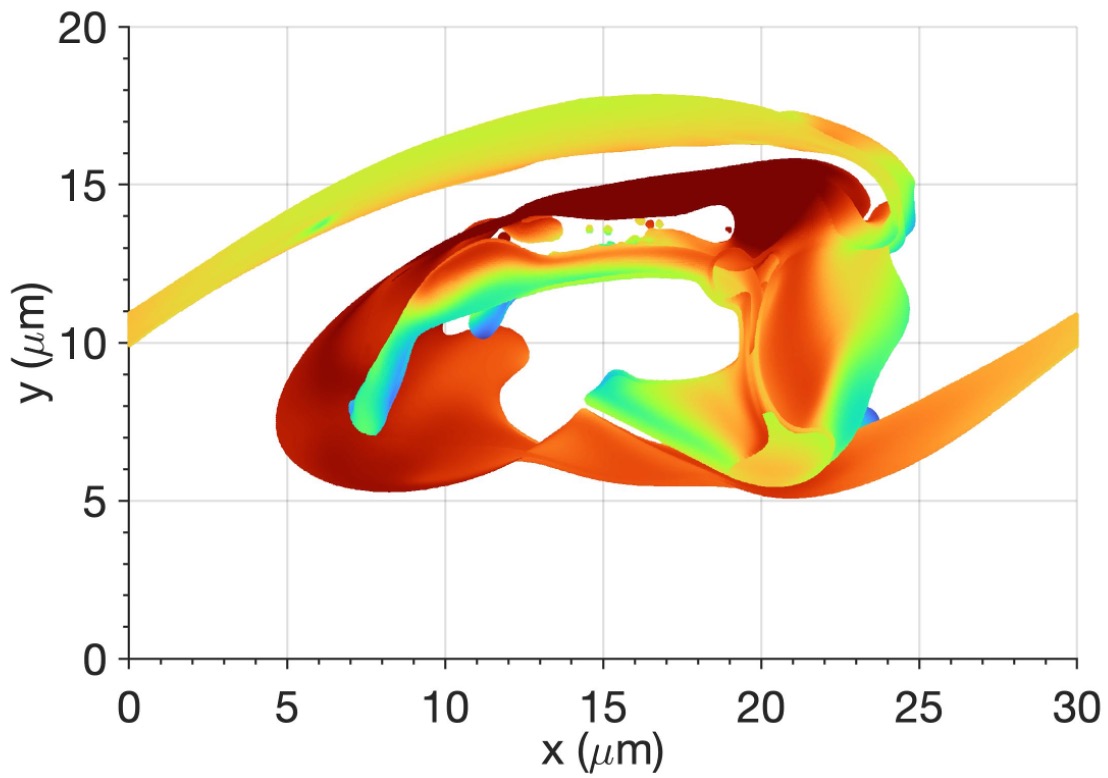}
  \caption{} 
  \label{subfig:Fig16e}
\end{subfigure}%
\begin{subfigure}{0.33\textwidth}
  \centering
  \includegraphics[width=0.95\linewidth]{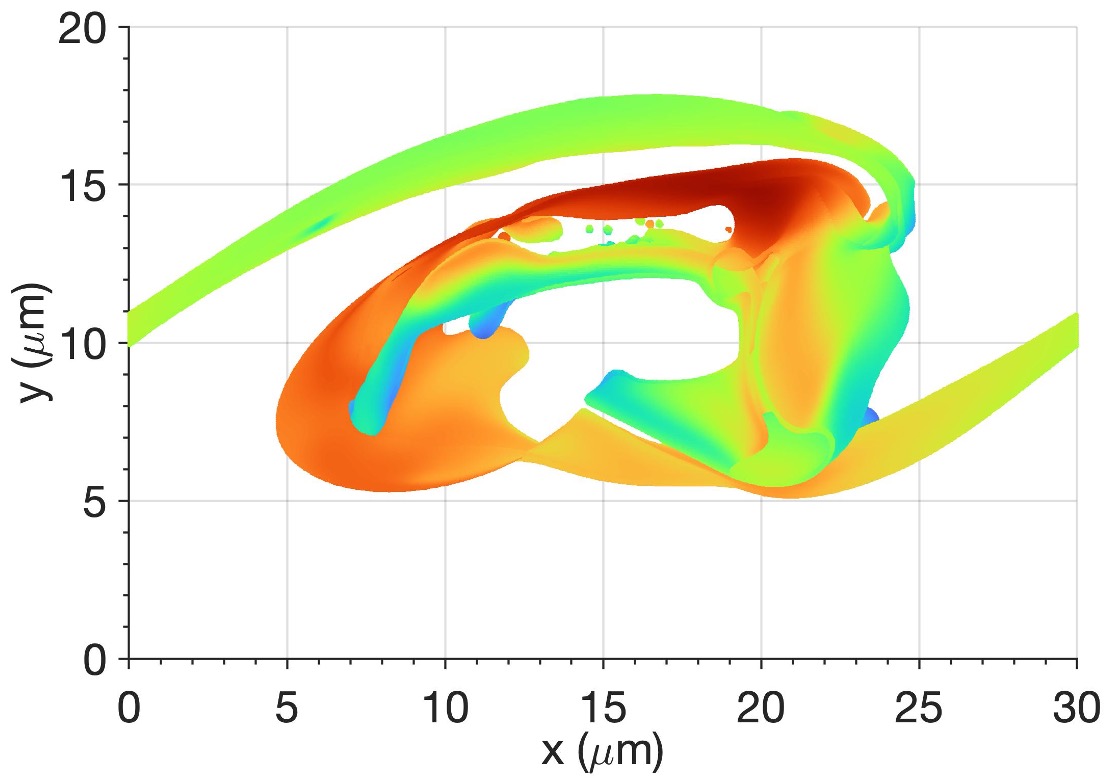}
  \caption{} 
  \label{subfig:Fig16f}
\end{subfigure}%
\caption{Solution of the local thermodynamic equilibrium along the liquid for case C1 at \(t^*=8.4\). A three-dimensional view and a side view from an \(xy\) plane located at \(z=20\) \(\mu\)m are provided. The liquid surface is coloured by the local interface value for (a) temperature, \(T\); (b) liquid density, \(\rho_l\); and (c) liquid viscosity, \(\mu_l\).}
\label{fig:Fig16}
\end{figure}

At this deformation stage, the liquid-gas mixing mechanisms are more complex, having a larger variability in interface properties and fluid properties within each phase. Figure~\ref{fig:Fig16} shows the interface temperature, liquid density and liquid viscosity for case C1 at \(t^*=8.4\) before layering becomes dominant. During the growth of the main wave, the liquid region facing the hotter oxidizer stream is heated, raising the interface temperature. Once the roller vortices stretch the wave and cause its rolling, hotter liquid blobs move underneath the wave due to gas entrainment. As seen in figure~\ref{subfig:Fig16d}, various liquid structures (e.g., sheets, ligaments and droplets) at different temperatures start to interact with the colder liquid in the wave trough. This non-uniform temperature's effect on the interface composition and surface-tension coefficient was presented in figures~\ref{fig:Fig2},~\ref{fig:Fig7} and~\ref{fig:Fig12}. The liquid density and viscosity in figure~\ref{fig:Fig16} emphasise the variation of dynamical properties along the liquid surface and, consequently, within the liquid phase. Note that due to the dissolved oxygen at 150 bar, the liquid density and viscosity in the cold interface already differ substantially from the freestream properties detailed in table~\ref{tab:cases}. \par 

An extreme case of the heating effects on smaller liquid structures is seen in figure~\ref{fig:Fig16}. An isolated droplet is immersed in the hotter gas above the liquid, formed during the early lobe breakup described in section~\ref{subsec:lobe_bending}. Due to the smaller liquid volume, the surface temperature increases substantially (i.e., above 500 K). The increased temperature results in more fuel vapour and more dissolved oxygen than in other liquid regions. Thus, the liquid density, viscosity and surface-tension coefficient are substantially smaller than anywhere else in the liquid. \par 

Some common vorticity patterns are observed in the layering deformation regime. The tearing of the liquid sheets and frequent hole formation may result in the formation of rim vortices, similar to the rim vortex identified during the early lobe deformation in case C1. As the hole expands, these vortices tend to weaken. Moreover, additional roller vortices are occasionally generated, whose importance is limited but continuing to contribute to the layering process. However, layering mitigates the gas entrainment into the two-phase mixture and flow recirculation overall, with little spanwise vorticity generation. As shown in PS, layering tends to uniformise the momentum mixing layer and the velocity field. Thus, the remaining vortices within the liquid sheets tend to align with the streamwise direction. \par 

All configurations at 100 bar and 150 bar show the layering mechanism with a qualitative description similar to that discussed in previous lines. Nonetheless, the interface is more easily perturbed at higher velocities, with short-wavelength perturbations causing frequent ligament shredding and droplet formation. This behaviour is discussed in section~\ref{subsec:lobe_crest_corrugation} and in more detail in PS. This scenario is critical in cases B2, C2 and C3. The short-wavelength perturbations generate smaller roller vortices that are responsible for the formation and stretching of smaller lobes. Eventually, these lobes break up into numerous ligaments and droplets, hence multi-phase turbulence increases and flow visualisation becomes difficult for the purposes of this study. As the smaller ligaments and droplets heat faster, the variations in surface properties become larger. \par 

Although not shown here, the long-term evolution of vortex structures in the flow at 50 bar (i.e., cases A1 and A2) resembles more that of subcritical atomisation. Layering occurs rarely and the liquid phase cannot be so easily stretched due to the higher surface tension and the mitigation of intraphase mixing (PS). The vortex structures in the gas phase do not weaken as fast as in the higher-pressure cases, which emphasises the following: (a) layering is a deformation process that reduces vorticity generation by stratifying the flow, (b) a gaseous phase with lower density and viscosity reduces viscous diffusion and dissipation, and (c) other vorticity generation mechanisms, such as baroclinicity, may play a key role in the lower ambient pressure. \par

\subsection{Formation of fuel-rich gas blobs}
\label{subsec:fuelblobs}

The fuel-oxidizer mixing at trans- and supercritical conditions is commonly assumed to be driven by mass diffusion and the rapid formation of a nearly homogeneous mixture beneficial for combustion purposes~\citep{mayer1996propellant,roy2010experimental,hickey2013supercritical}. That is, the fluid is characterised by a gas-like diffusivity and the mixing front displays less irregularities, promoting a more uniform combustion. Such behaviour is key for combustion stability and a cleaner fuel burning. \par

However, the computations by PS show that the emergence of fuel-rich gaseous blobs is common during the early atomisation stages where a two-phase regime is sustained, even at pressures up to 150 bar with \textit{n}-decane as a fuel. These exist due to the non-uniform fuel vaporisation along the surface and the breakup of liquid structures. Similar to the liquid deformation, the fluid dynamics (i.e., vorticity) also determine the evolution of these blobs, rather than mass diffusion alone. Thus, the dimension and shape of these blobs relate to the size of the liquid structures (i.e., microns) and the vorticity. Because the results presented by PS do not extend beyond a few microseconds, it is unclear whether these fuel-rich pockets diffuse quickly. Nonetheless, the existence of multitude of cool liquid structures which take time to vaporise suggests that fuel-rich blobs will appear as long as the liquid mixture remains transcritical. \par 

\begin{figure}
\centering
\begin{subfigure}{0.66\textwidth}
  \centering
  \includegraphics[width=0.9\linewidth]{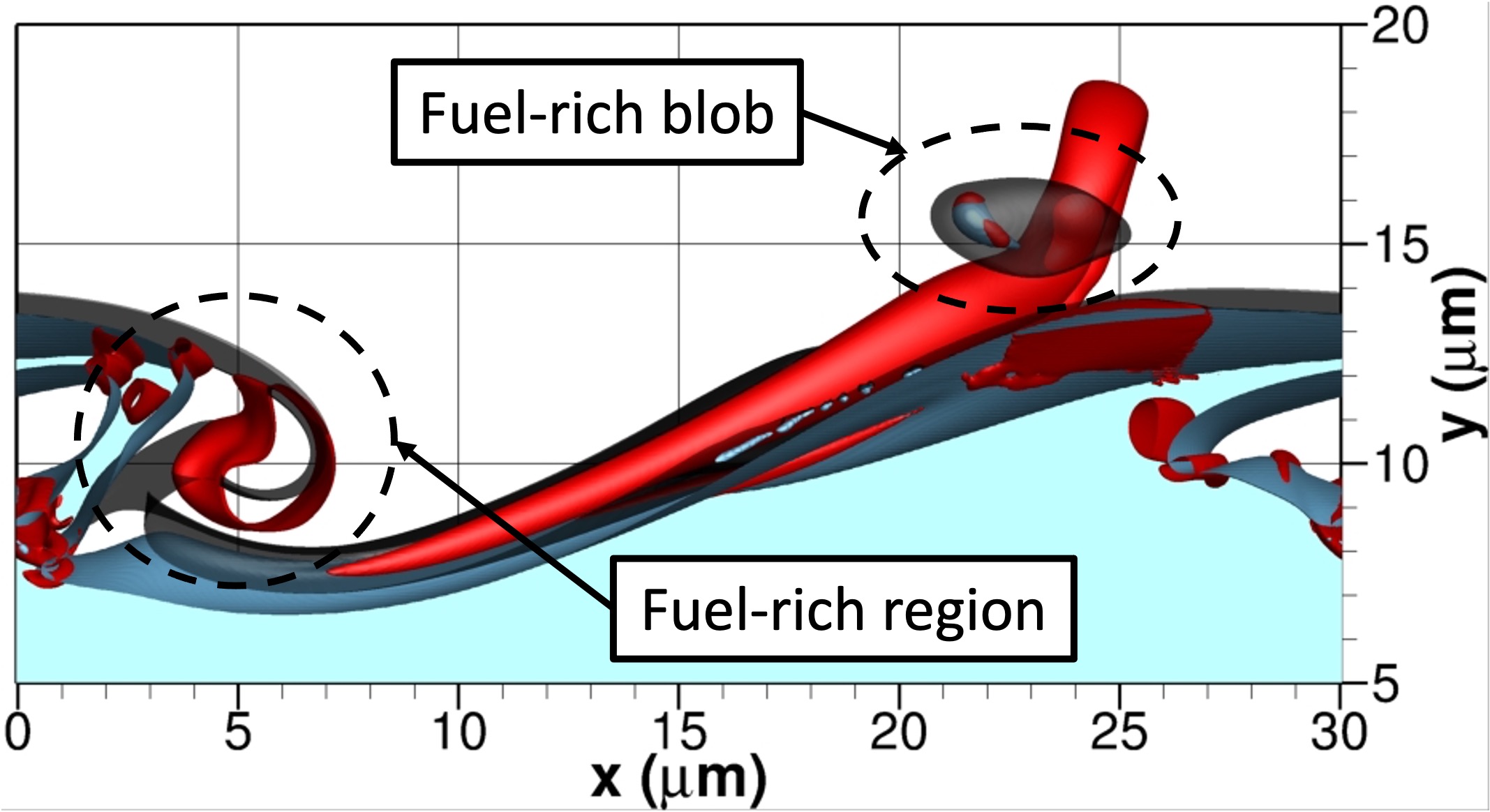}
  \label{subfig:Fig17a}
\end{subfigure}%
\begin{subfigure}{0.33\textwidth}
  \centering
  \includegraphics[width=0.75\linewidth]{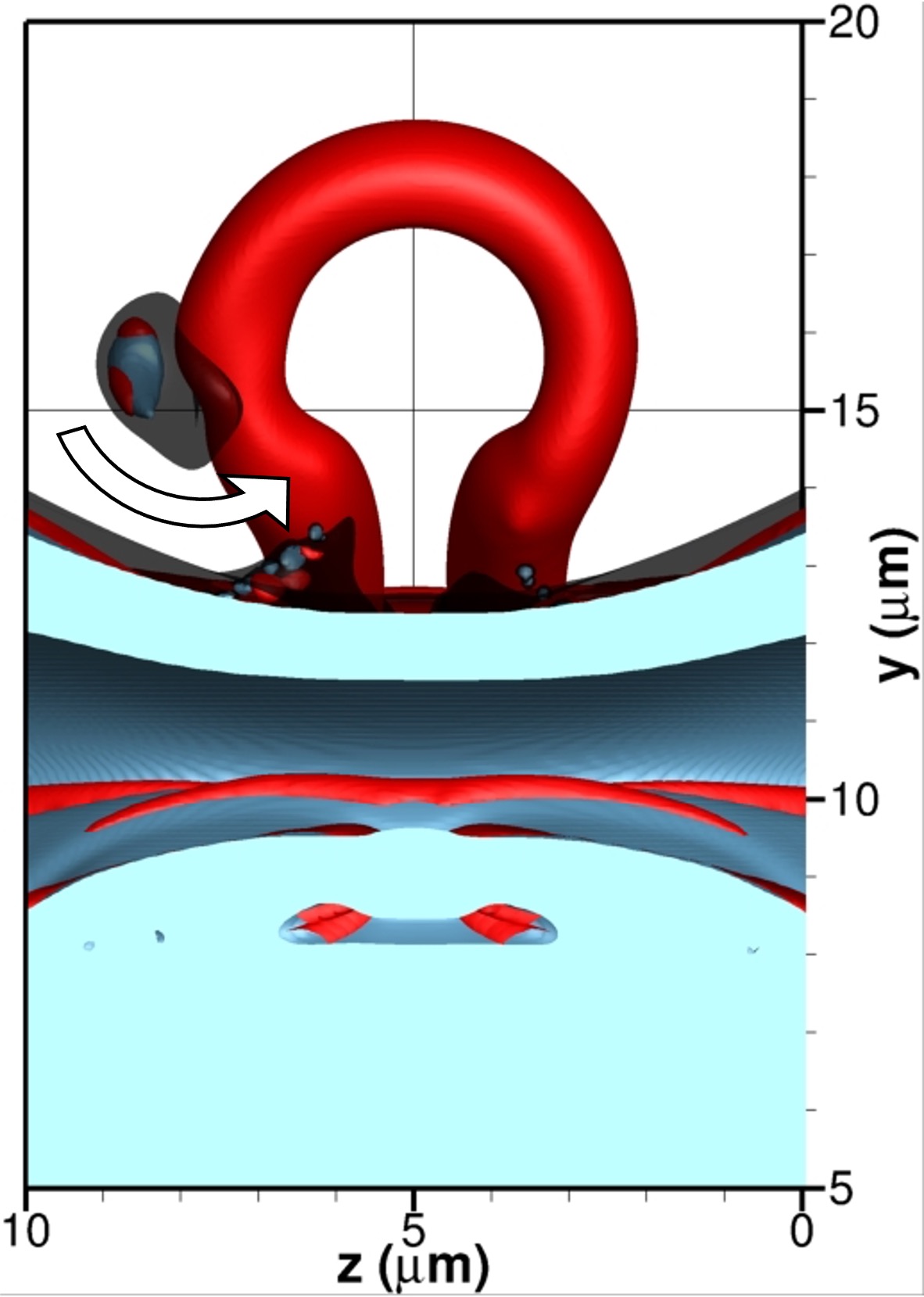}
  \label{subfig:Fig17b}
\end{subfigure}%
\\[2ex]
\begin{subfigure}{1.0\textwidth}
  \centering
  \includegraphics[width=0.6\linewidth]{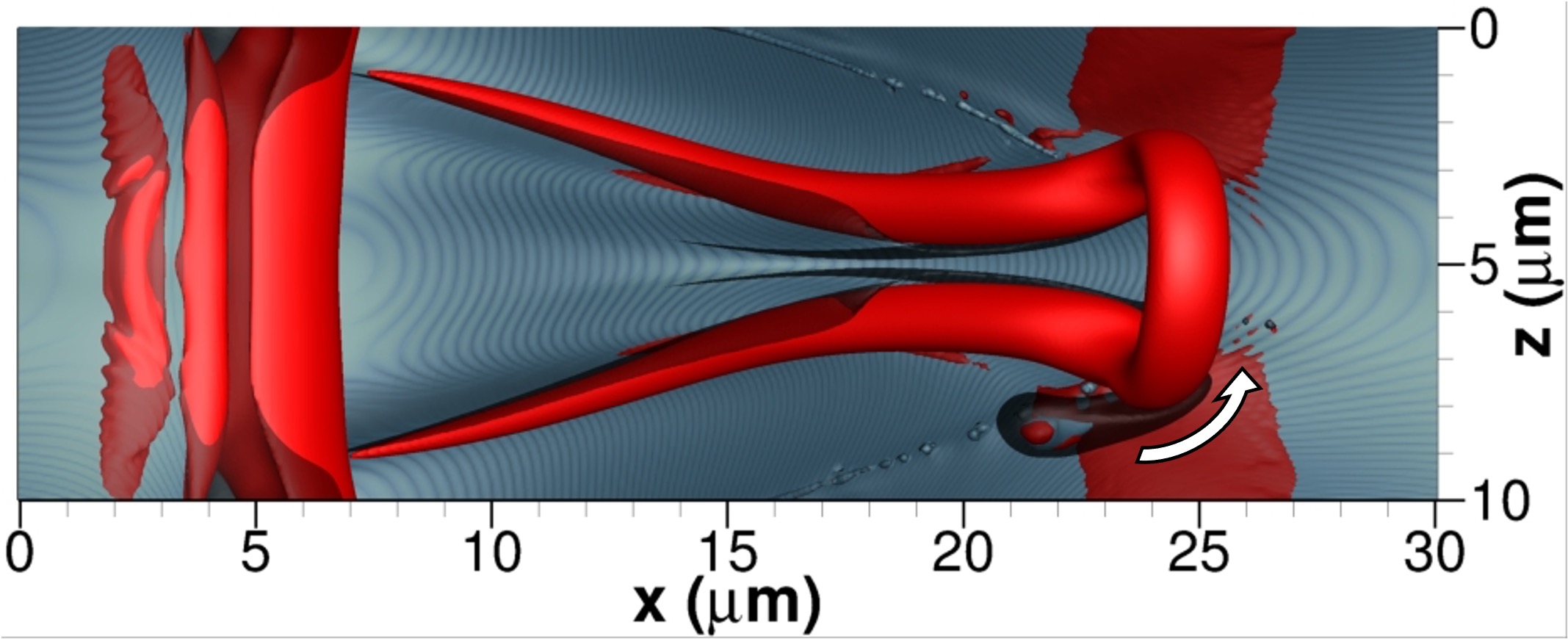}
  \label{subfig:Fig17c}
\end{subfigure}%
\caption{Interaction between fuel-rich gaseous blobs and vortex structures for case C1 at \(t^*=6\). The liquid surface is identified by the blue isosurface with \(C=0.5\), the vortex structures are identified by the red isosurface with \(\lambda_{\rho,t}=-2.5\times 10^{15}\), and the fuel-rich regions are identified by the translucid black isosurface with \(Y_F=0.05\).}
\label{fig:Fig17}
\end{figure}

\begin{figure}
\centering
\begin{subfigure}{0.5\textwidth}
  \centering
  \includegraphics[width=0.95\linewidth]{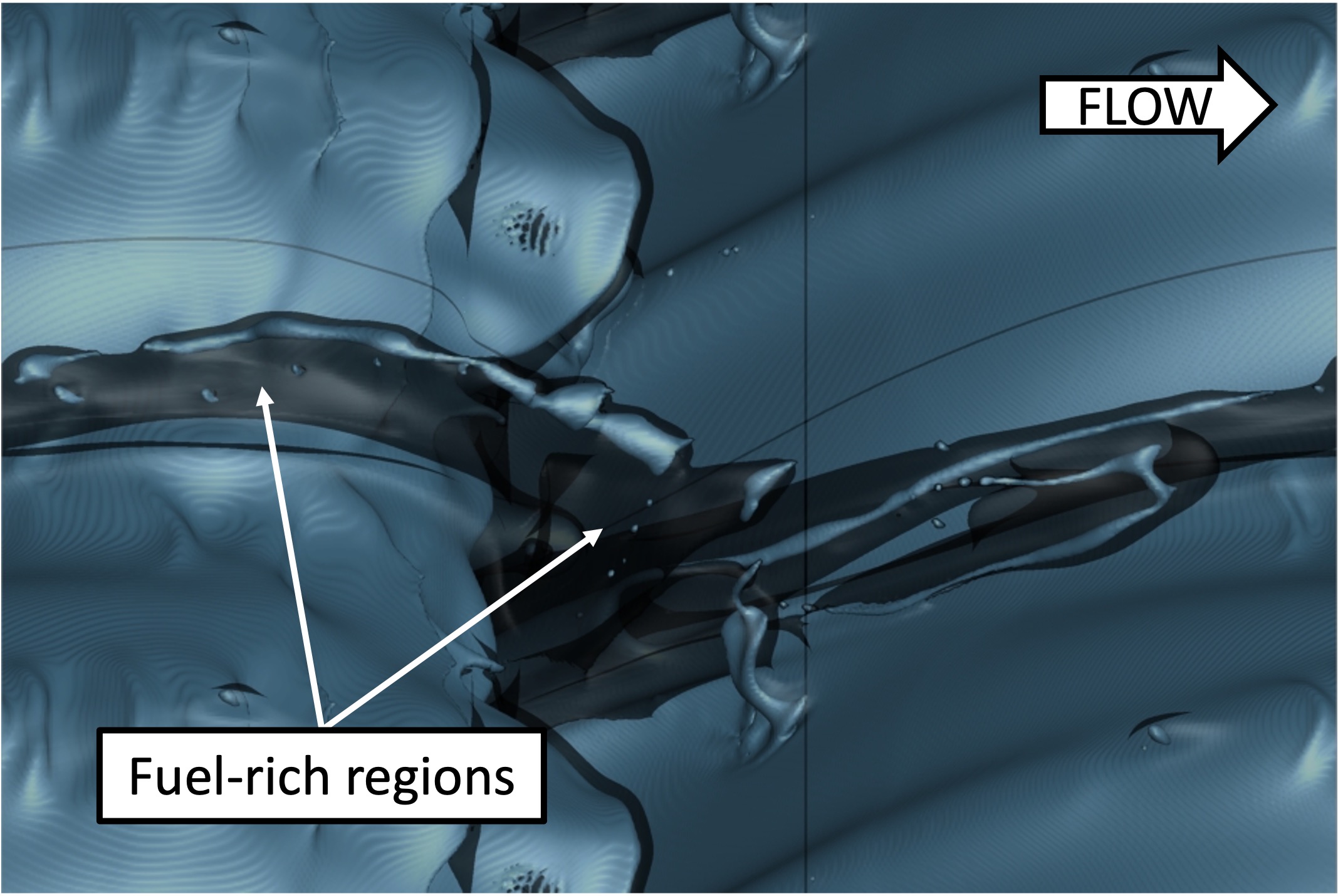}
  \caption{} 
  \label{subfig:Fig18a}
\end{subfigure}%
\begin{subfigure}{0.5\textwidth}
  \centering
  \includegraphics[width=0.95\linewidth]{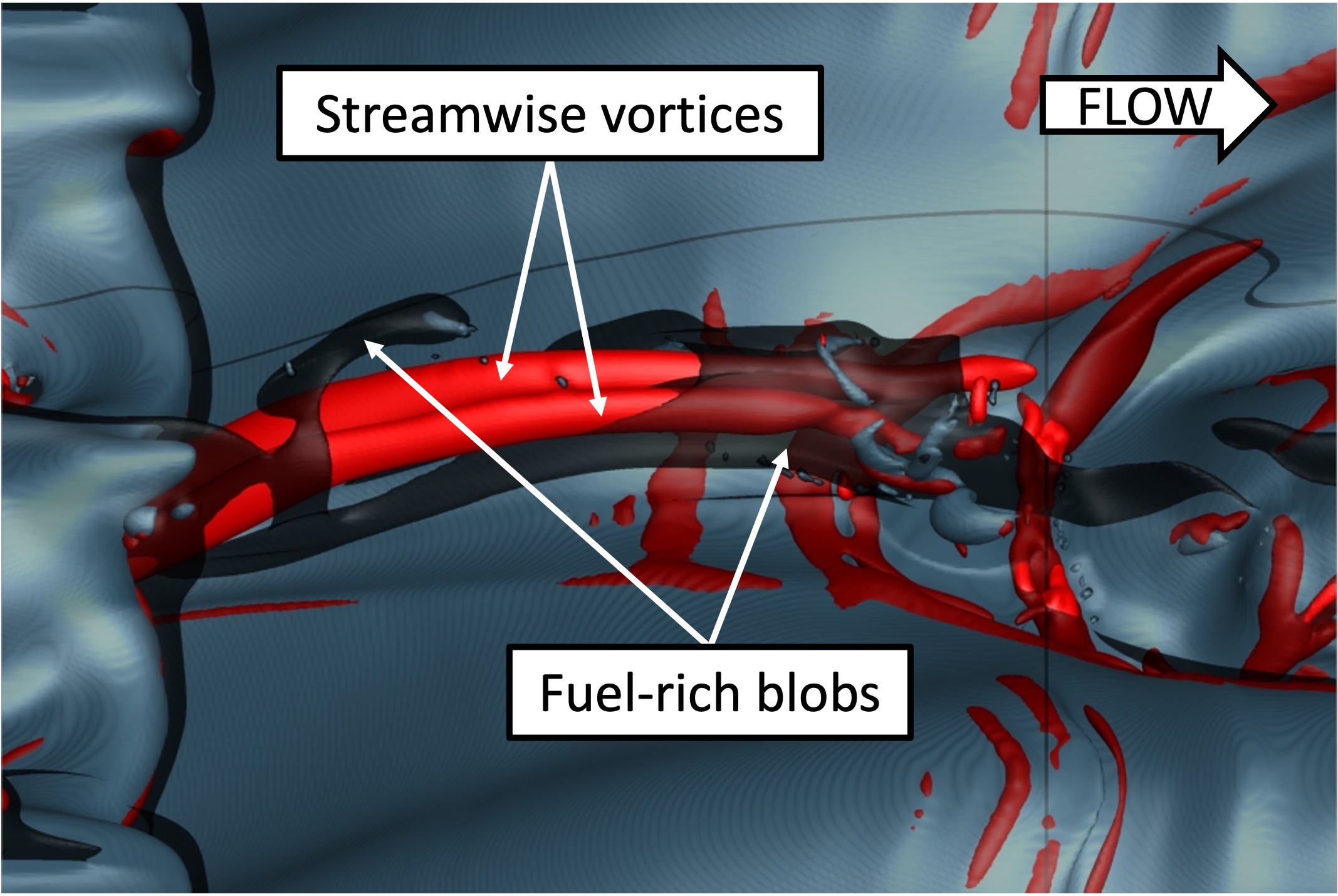}
  \caption{} 
  \label{subfig:Fig18b}
\end{subfigure}%
\caption{Interaction between fuel-rich gaseous blobs and vortex structures for case C2. The liquid surface is identified by the blue isosurface with \(C=0.5\), the vortex structures are identified by the red isosurface with \(\lambda_{\rho,t}=-9\times 10^{15}\), and the fuel-rich regions are identified by the translucid black isosurface with \(Y_F=0.05\). (a) \(t^*=8.25\); and (b) \(t^*=10\)}
\label{fig:Fig18}
\end{figure}

Some examples of these fuel-rich regions are seen in figure~\ref{fig:Fig17} for case C1 and in figure~\ref{fig:Fig18} for case C2, which depict the isosurface of the fuel mass fraction with \(Y_F=0.05\). First, figure~\ref{fig:Fig17} shows a fuel-rich blob around the isolated vaporising droplet at \(t^*=6\), while the roller vortex R2 (see figure~\ref{fig:Fig14}) promotes more fuel vaporisation along the edge of the growing wave. In particular, the blob is influenced strongly by the fluid dynamics. It presents a streamwise-elongated shape, and easily wraps around the head of the hairpin vortex R1 because of the induced flow. Then, figure~\ref{subfig:Fig18a} highlights the fuel-rich regions at \(t^*=8.25\) (i.e., \(t=3.3\) \(\mu\)s for case C2) around the train of liquid structures seen evolving in figure~\ref{fig:Fig3}. Later, these blobs interact with the vortices similar to figure~\ref{fig:Fig17} by wrapping around them (see figure~\ref{subfig:Fig18b} at \(t^*=10\)). Another example of fuel-rich and fuel-lean regions is figure~\ref{fig:Fig27} discussing the vorticity and mixing within the liquid layers, which shows large variations of the fuel vapour concentration. \par 

Although we only show the isosurface with \(Y_F=0.05\), the blobs typically have a much higher concentration of fuel vapour inside, especially when they surround broken liquid structures like those depicted in figures~\ref{fig:Fig17} and \ref{fig:Fig18}. These small droplets and ligaments are observed to reach temperatures of 520 K, resulting in a fuel vapour concentration up to \(Y_F=0.33\) around them. This is also highlighted by the phase-equilibrium diagrams shown in figure~\ref{fig:Fig2}. \par 

\subsection{Vorticity generation in the transcritical flow}
\label{subsec:vorticity_generation}

Sections~\ref{subsec:lobe_bending} to~\ref{subsec:fuelblobs} detailed the deformation of vortical structures and how their interaction with the transcritical liquid mixture defines the fuel-oxidizer mixing and atomisation processes. For example, the growth of three-dimensional instabilities during the atomisation of the liquid is directly related to the generation of streamwise vorticity, as described in previous works such as~\citet{jarrahbashi2014vorticity,jarrahbashi2016early,zandian2017planar,zandian2018understanding}. We impose an initial sinusoidal perturbation along the spanwise direction with a wavelength equal to the domain size (i.e., \(L_z=20\) \(\mu\)m), but toward the end of our computations (e.g., figure~\ref{subfig:Fig14c}) two or more spanwise wavelengths are seen over the same distance (PS). This process occurs because the initial roller vortex deforms into a hairpin where the pins align with \(x\) (i.e., \(\omega_x\) is generated). Thus, \(\omega_x\) perturbs the surface and generates additional waves in \(z\). Vortex tilting and streamwise alignment occur frequently (e.g., see figures~\ref{fig:Fig13} and~\ref{fig:Fig14}), defining both the liquid deformation and the fuel-oxidizer mixing (e.g., deformation patterns, fuel-rich gas blobs). \par 

The vorticity equation is obtained by taking the curl of the momentum equation~(\ref{eqn:mom}). The compressible form of the vorticity equation becomes

\begin{equation}
\label{eqn:vorticity}
\frac{D\boldsymbol{\omega}}{Dt}=(\boldsymbol{\omega\cdot\nabla})\boldsymbol{u}-\boldsymbol{\omega}(\boldsymbol{\nabla\cdot u})+\frac{\boldsymbol{\nabla}\rho\boldsymbol{\times\nabla} p}{\rho^2} + \boldsymbol{\nabla \times} \bigg(\frac{1}{\rho}\boldsymbol{\nabla\cdot\tau}\bigg)+\boldsymbol{\nabla\times F_\sigma}
\end{equation}
\noindent
where \(\boldsymbol{F_\sigma}\) is the body force term used to model surface tension (i.e., CSF approach). In the analysis that follows, the surface tension term and the viscous term \(\boldsymbol{\nabla \times} \bigg(\frac{1}{\rho}\boldsymbol{\nabla\cdot\tau}\bigg)\) are neglected~\citep{zandian2017planar}. This approximation is also done for the \(\lambda_\rho\) post-processing, as discussed in section~\ref{sec:lambdarho}. Hence, only the vortex stretching and tilting term, \((\boldsymbol{\omega\cdot\nabla})\boldsymbol{u}\), the vortex stretching due to volume dilatation, \(-\boldsymbol{\omega}(\boldsymbol{\nabla\cdot u})\), and the baroclinic term, \(\frac{\boldsymbol{\nabla}\rho\boldsymbol{\times\nabla} p}{\rho^2}\), are analysed to study the vorticity of the large-scale dynamics. At smaller scales, viscosity may become important. Since~(\ref{eqn:vorticity}) is a vector equation, each component of the vorticity field is given by

\begin{equation}
\label{eqn:vorticityX}
\frac{D\omega_x}{Dt} = \omega_x\frac{\partial u}{\partial x} + \omega_y\frac{\partial u}{\partial y} + \omega_z\frac{\partial u}{\partial z} - \omega_x (\boldsymbol{\nabla \cdot u}) + \frac{1}{\rho^2}\bigg(\frac{\partial \rho}{\partial y}\frac{\partial p}{\partial z} - \frac{\partial \rho}{\partial z}\frac{\partial p}{\partial y}\bigg)
\end{equation}
\begin{equation}
\label{eqn:vorticityY}
\frac{D\omega_y}{Dt} = \omega_x\frac{\partial v}{\partial x} + \omega_y\frac{\partial v}{\partial y} + \omega_z\frac{\partial v}{\partial z} - \omega_y (\boldsymbol{\nabla \cdot u}) + \frac{1}{\rho^2}\bigg(\frac{\partial \rho}{\partial z}\frac{\partial p}{\partial x} - \frac{\partial \rho}{\partial x}\frac{\partial p}{\partial z}\bigg)
\end{equation}
\begin{equation}
\label{eqn:vorticityZ}
\frac{D\omega_z}{Dt} = \omega_x\frac{\partial w}{\partial x} + \omega_y\frac{\partial w}{\partial y} + \omega_z\frac{\partial w}{\partial z} - \omega_z (\boldsymbol{\nabla \cdot u}) + \frac{1}{\rho^2}\bigg(\frac{\partial \rho}{\partial x}\frac{\partial p}{\partial y} - \frac{\partial \rho}{\partial y}\frac{\partial p}{\partial x}\bigg)
\end{equation}

Fluid compressibility and variable density are important for vortex generation in transcritical flows. Vortices submerged in the mixing layers are subject to density and pressure gradients. Volume dilatation affects vorticity, and the baroclinicity now extends beyond the interfacial region. Nonetheless, baroclinicity might be negligible compared to other terms in dense fluids due to the scaling with \(\rho^{-2}\)~\citep{zandian2017planar,zandian2018understanding}, such as in the liquid phase or the gas phase at very high pressures. \par 

Similar to the \(\lambda_\rho\) post-processing, the filtered velocity field (see section~\ref{sec:lambdarho}) is used to quantify vorticity and the relevant terms in equations~(\ref{eqn:vorticity}),~(\ref{eqn:vorticityX}),~(\ref{eqn:vorticityY}) and~(\ref{eqn:vorticityZ}) to minimise the influence of spurious currents around the liquid-gas interface. This filtering process does not affect significantly the results shown in this section; however, the differences in \(\boldsymbol{\nabla\cdot u}\) become important in the incompressible case C1i shown in figure~\ref{fig:Fig8} (i.e., filtered velocity field is not divergence free). Although we could gain more insight into the vorticity behaviour by just looking at the full terms in (\ref{eqn:vorticity}), splitting the vorticity vector equation into components allows us to focus on the specific vorticity generation mechanisms relevant to atomisation (e.g., streamwise vorticity). \par

In the discussion that follows, we refer to streamwise (\(\partial/\partial x\)), transverse (\(\partial/\partial y\)) and spanwise (\(\partial/\partial z\)) terms to describe the vortex stretching and tilting for each vorticity component. Stretching occurs along the direction of the analysed vorticity component (e.g., streamwise stretching for \(\omega_x\)) and tilting takes places in the other two orthogonal directions (e.g., transverse and spanwise tilting for \(\omega_x\)). The vortex stretching and tilting in (\ref{eqn:vorticity}) has three stretching components and six tilting terms (i.e., two per direction). Here, we define \(\dot{\omega}_{x\rightarrow x}=\omega_x\frac{\partial u}{\partial x}\), \(\dot{\omega}_{y\rightarrow y}=\omega_y\frac{\partial v}{\partial y}\) and \(\dot{\omega}_{z\rightarrow z}=\omega_z\frac{\partial w}{\partial z}\) as the three vortex stretching contributions to vorticity generation. Then, transverse tilting refers to the reorientation of transverse vorticity or \(\omega_y\). The two terms are defined as \(\dot{\omega}_{y\rightarrow x}=\omega_y\frac{\partial u}{\partial y}\) and \(\dot{\omega}_{y\rightarrow z}=\omega_y\frac{\partial w}{\partial y}\). Similarly, streamwise tilting terms are \(\dot{\omega}_{x\rightarrow y}=\omega_x\frac{\partial v}{\partial x}\) and \(\dot{\omega}_{x\rightarrow z}=\omega_x\frac{\partial w}{\partial x}\), and spanwise tilting terms are \(\dot{\omega}_{z\rightarrow x}=\omega_z\frac{\partial u}{\partial z}\) and \(\dot{\omega}_{z\rightarrow y}=\omega_z\frac{\partial v}{\partial z}\). \par 

The following sections present a detailed description of the vorticity generation mechanisms of some of the most relevant features observed in sections \ref{subsec:lobe_bending}, \ref{subsec:lobe_crest_corrugation} and \ref{subsec:layering}. Section~\ref{subsubsec:hairpin} focuses on the \(\omega_x\) generation that explains the alignment with \(x\) of the initial hairpin vortex legs. Then, section~\ref{subsubsec:rollers} describes the \(\omega_z\) generation mechanisms of the secondary roller vortex shown in figure~\ref{fig:Fig9}. Lastly, section~\ref{subsubsec:layers} addresses the early stages of the layered flow to understand the sources of \(\omega_x\) that promote the streamwise alignment of vortical structures trapped between the liquid layers. \par

\subsubsection{Stretching of the hairpin vortices}
\label{subsubsec:hairpin}

The initial velocity distribution described in section~\ref{sec:description} results in a vortex sheet with only spanwise vorticity (\(\omega_z<0\)), and a single non-zero velocity gradient (\(\partial u/\partial y >0\)). With the initial interface displacement, the major source of \(\omega_x\) at the start of the simulation is the baroclinic term. After some time, vorticity is generated in the other directions and \(\partial u/\partial x\) and \(\partial u/\partial z\) are, in general, non-zero. Given the relative strength of \(\omega_z\) and \(\partial u/\partial y\) with respect to the other quantities, \(\dot{\omega}_{y\rightarrow x}\) and \(\dot{\omega}_{z\rightarrow x}\) become important \(\omega_x\) generators. After the initial transient, \citet{zandian2017planar,zandian2018understanding} show that, for low density ratios (i.e., \(\rho_G/\rho_L\)), the streamwise vorticity generation mechanism responsible for the formation of hairpin vortices in the gas phase is the baroclinic term. As the density ratio increases (i.e., the gas phase becomes denser), the relative importance of the baroclinic term diminishes due to its \(\rho^{-2}\) scaling in favour of \(\dot{\omega}_{x\rightarrow x}\). In general, \(\dot{\omega}_{z\rightarrow x}\) and \(\dot{\omega}_{y\rightarrow x}\) are stronger than \(\dot{\omega}_{x\rightarrow x}\), but they tend to cancel each other. A similar behaviour is observed in round jets~\citep{jarrahbashi2014vorticity}. \par 

\begin{figure}
\centering
\begin{subfigure}{0.5\textwidth}
  \centering
  \includegraphics[width=1.0\linewidth]{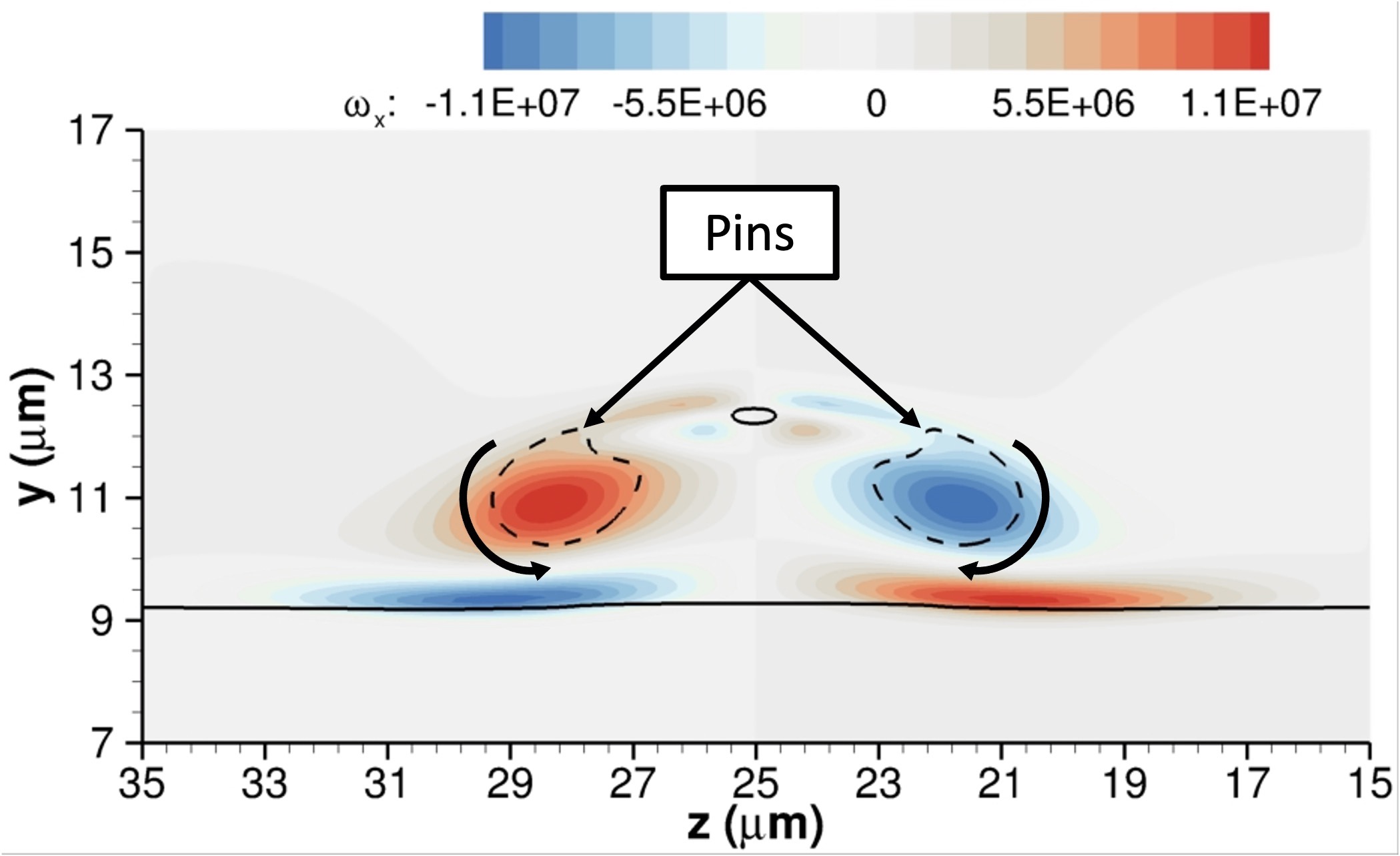}
  \caption{} 
  \label{subfig:Fig19a}
\end{subfigure}%
\begin{subfigure}{0.5\textwidth}
  \centering
  \includegraphics[width=1.0\linewidth]{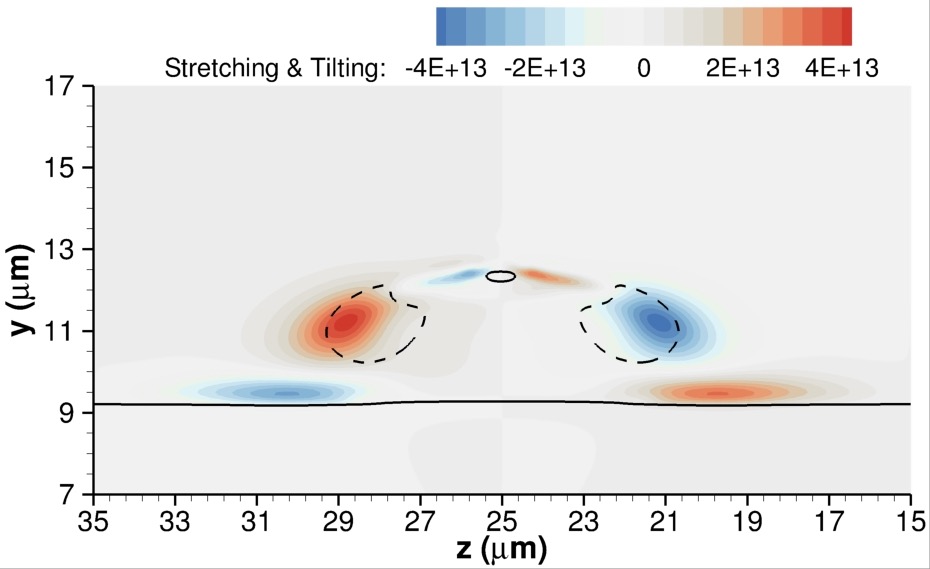}
  \caption{} 
  \label{subfig:Fig19b}
\end{subfigure}%
\\
\begin{subfigure}{0.5\textwidth}
  \centering
  \includegraphics[width=1.0\linewidth]{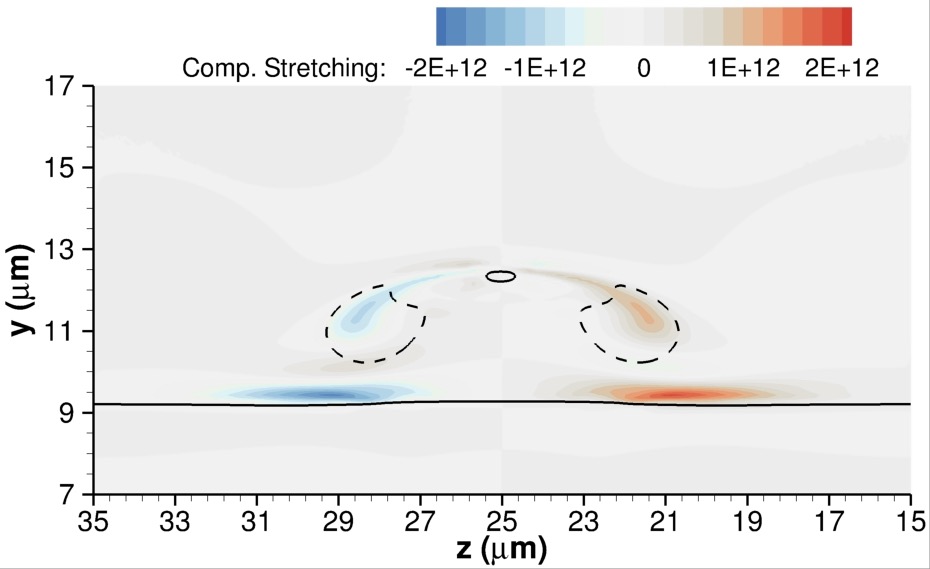}
  \caption{} 
  \label{subfig:Fig19c}
\end{subfigure}%
\begin{subfigure}{0.5\textwidth}
  \centering
  \includegraphics[width=1.0\linewidth]{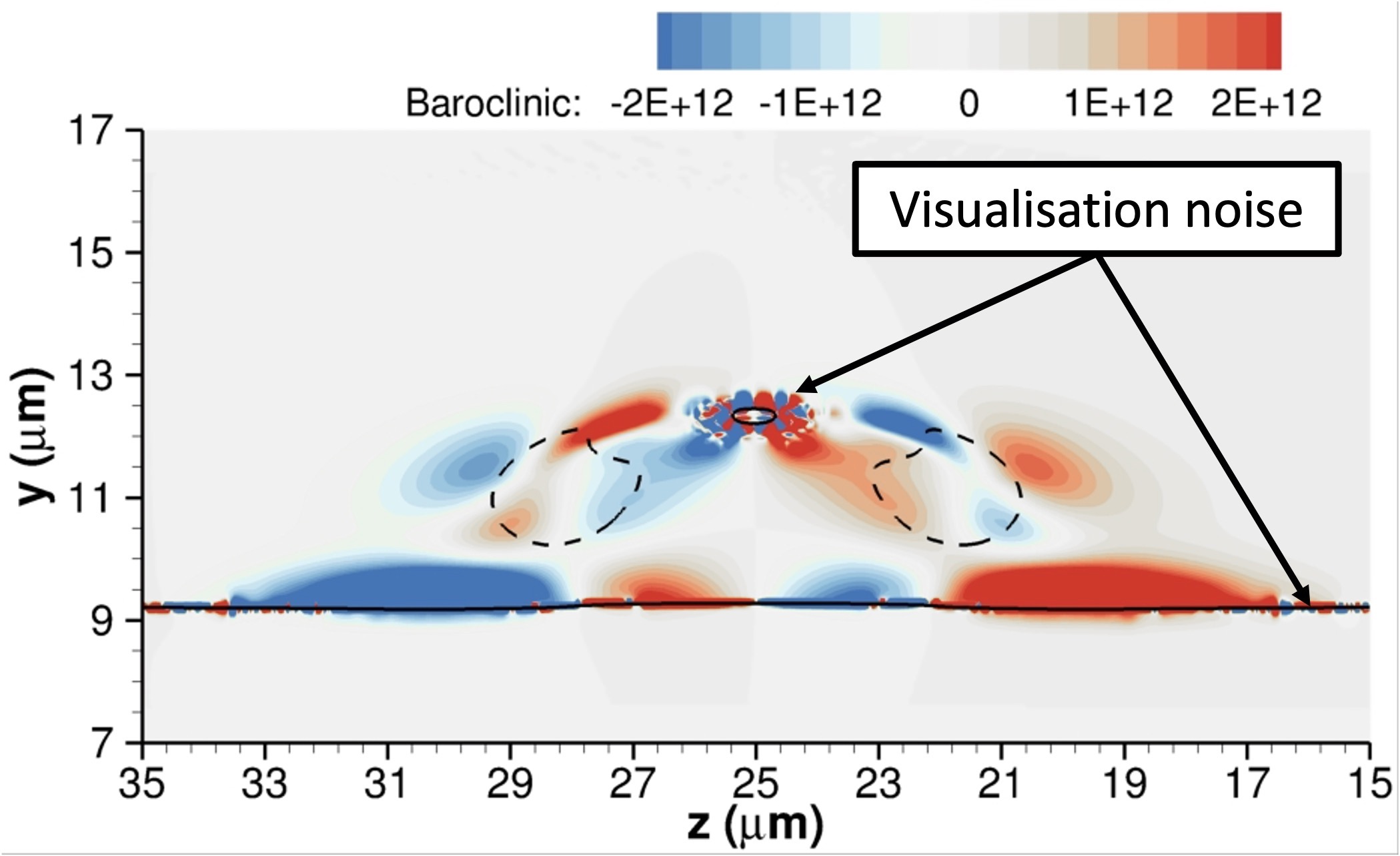}
  \caption{} 
  \label{subfig:Fig19d}
\end{subfigure}%
\caption{\(\omega_x\) generation terms during the stretching of the initial hairpin vortex for case C1 at \(t^*=3\). The front view from a \(yz\) plane at \(x=16\) \(\mu\)m is shown. The interface is identified by the solid isocontour with \(C=0.5\), and the cut vortex structures are identified by the dashed isocontour with \(\lambda_{\rho,t}=-2.5\times 10^{15}\). (a) \(\omega_x\); (b) vortex stretching and tilting; (c) compressible stretching; and (d) baroclinic term.}
\label{fig:Fig19}
\end{figure}

This description explains the initial deformation of the KH roller vortex into a hairpin; thus, let us look at \(t^*>2\) to understand the additional stretching the vortex pins undergo. \citet{zandian2017planar,zandian2018understanding} visualised the baroclinic term because their computations used a more diffuse interface capturing method (i.e., Level-Set). The density gradient between liquid and gas extends a few cells beyond the interface, allowing a smoother visualisation. However, our sharp interface VOF method concentrates the density gradient at the interface cells. Together with the pressure oscillations near the interface due to the generation of spurious currents, the baroclinic contribution of the two-phase flow cannot be properly visualised during the early times (see figure~\ref{subfig:Fig19d}). Once the mixing layers grow sufficiently, we can capture the baroclinic term. \par 

To understand the continuous stretching of the hairpin vortex pins, figure~\ref{fig:Fig19} shows the streamwise vorticity and its generation terms from (\ref{eqn:vorticityX}) for case C1 at \(t^*=3\) in a \(yz\) plane at \(x=16\) \(\mu\)m. The \(\omega_x\) contours in figure~\ref{subfig:Fig19a} resemble those in figure~\ref{subfig:Fig6f}, which depicts \(\omega_x\) at a later time. The vorticity vector is predominantly aligned with \(x\) inside the vortex pins. Another region of strong \(\omega_x\) appears between the hairpin and the liquid below. Despite the variable density in the transcritical flow, the vortex stretching and tilting terms are at least an order of magnitude larger than the compressible stretching or baroclinic terms. This characteristic repeats for all analysed cases. In our study, lower pressures are associated with larger characteristic velocities and milder density variations across the mixing regions. Thus, vortex stretching and tilting terms continue to dominate, even if the lower density in the gas phase favours a stronger baroclinic term. \par 

\begin{figure}
\centering
\begin{subfigure}{0.33\textwidth}
  \centering
  \includegraphics[width=1.0\linewidth]{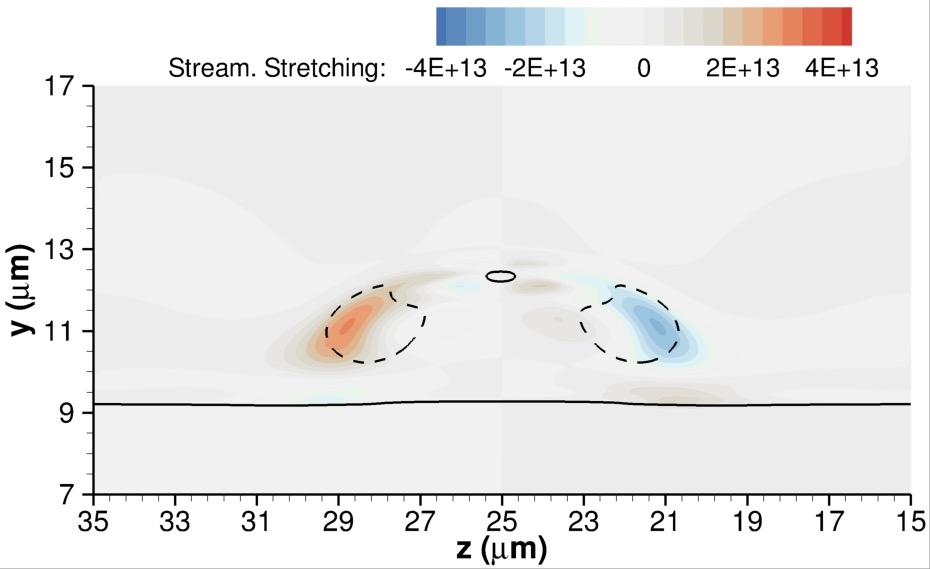}
  \caption{} 
  \label{subfig:Fig20a}
\end{subfigure}%
\begin{subfigure}{0.33\textwidth}
  \centering
  \includegraphics[width=1.0\linewidth]{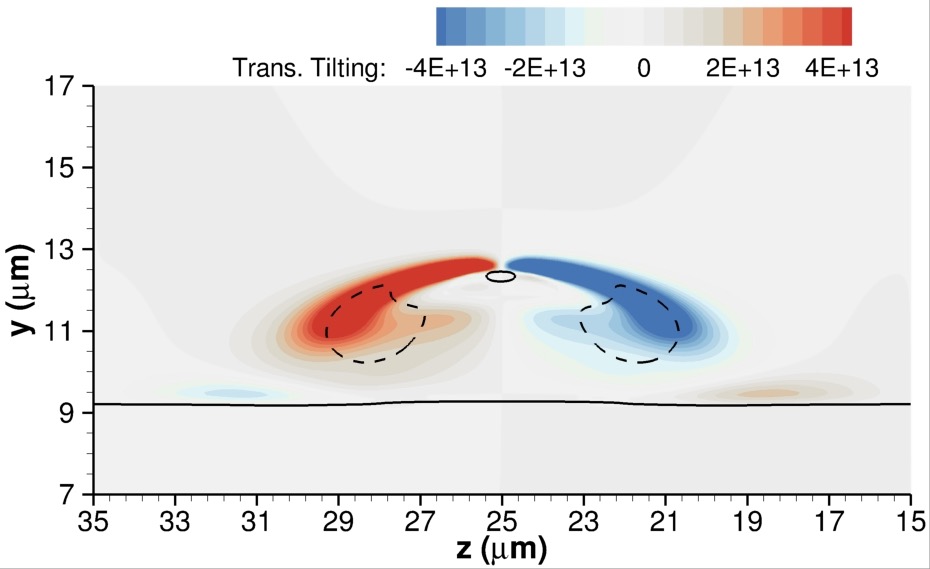}
  \caption{} 
  \label{subfig:Fig20b}
\end{subfigure}%
\begin{subfigure}{0.33\textwidth}
  \centering
  \includegraphics[width=1.0\linewidth]{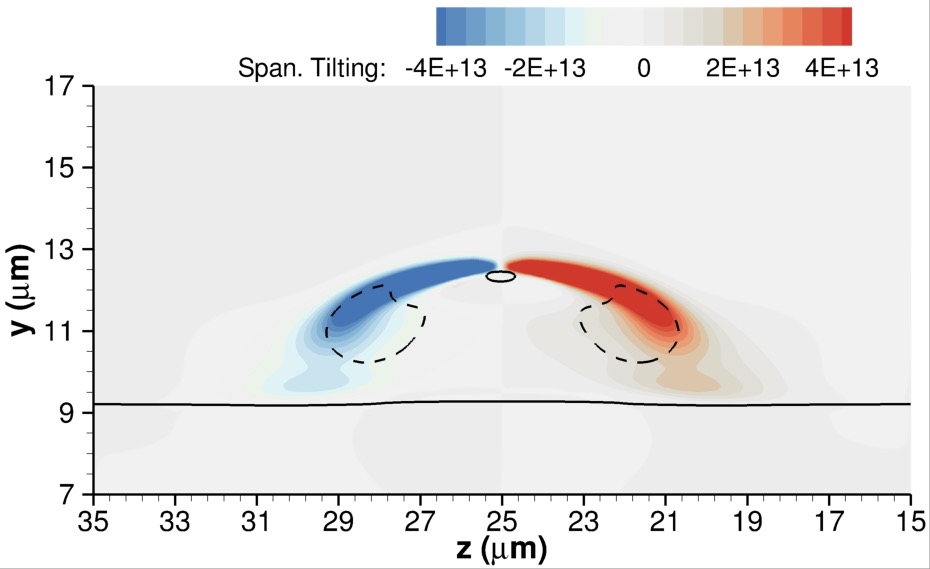}
  \caption{} 
  \label{subfig:Fig20c}
\end{subfigure}%
\caption{\(\omega_x\) generation terms during the stretching of the initial hairpin vortex for case C1 at \(t^*=3\). The front view from a \(yz\) plane at \(x=16\) \(\mu\)m is shown. The interface is identified by the solid isocontour with \(C=0.5\), and the cut vortex structures are identified by the dashed isocontour with \(\lambda_{\rho,t}=-2.5\times 10^{15}\). (a) \(\dot{\omega}_{x\rightarrow x}\); (b) \(\dot{\omega}_{y\rightarrow x}\); and (c) \(\dot{\omega}_{z\rightarrow x}\).}
\label{fig:Fig20}
\end{figure}

The combined effect of vortex stretching and tilting increases the \(\omega_x\) magnitude in the vortex pins, effectively stretching the vortical structure. Compressible stretching opposes this deformation, but it has a weak contribution. Baroclinicity inside the vortex pins is not strong. Figure~\ref{fig:Fig20} presents the individual components of the vortex stretching and tilting terms from (\ref{eqn:vorticityX}), and shows the partial cancellation of \(\dot{\omega}_{y\rightarrow x}\) and \(\dot{\omega}_{z\rightarrow x}\) reported in~\citet{zandian2017planar,zandian2018understanding}. A closer look into figure~\ref{subfig:Fig19b} reveals that \(\dot{\omega}_{x\rightarrow x}\) is relatively weak and does not justify the strong streamwise vorticity generation inside the vortex pins. Despite the partial cancellation, \(\dot{\omega}_{y\rightarrow x}\) is larger than \(\dot{\omega}_{z\rightarrow x}\) in some regions near the vortex pins, strengthening the combined effect of stretching and tilting. Although not presented here, a slice through \(x=26\) \(\mu\)m at \(t^*=4.05\) shows that \(\omega_x\) inside the vortex pins is still strong. Nonetheless, none of the terms of the streamwise vorticity equation significantly increase or decrease \(\omega_x\) inside the vortex. \par 

\begin{figure}
\centering
\begin{subfigure}{0.5\textwidth}
  \centering
  \includegraphics[width=1.0\linewidth]{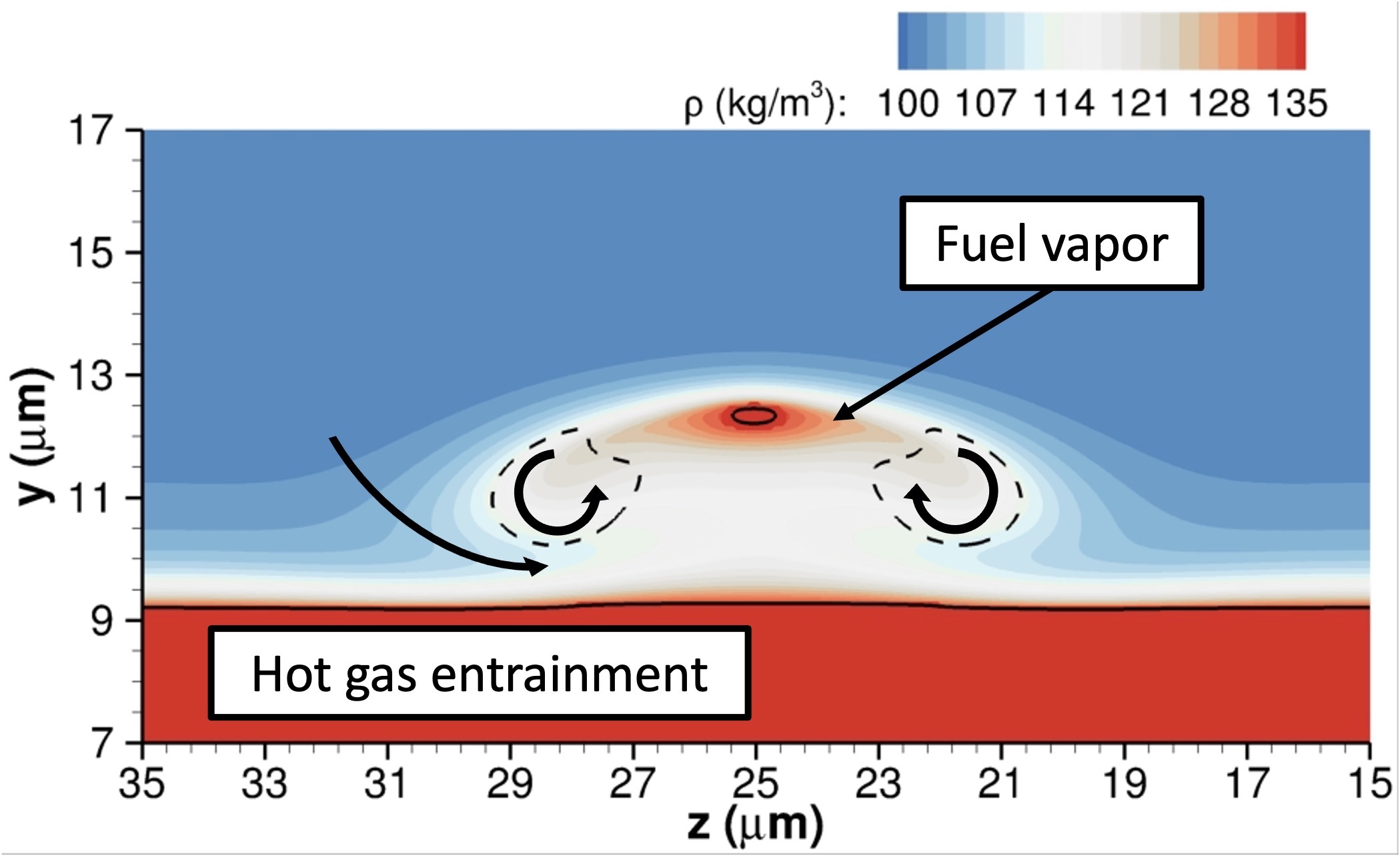}
  \caption{} 
  \label{subfig:Fig21a}
\end{subfigure}%
\begin{subfigure}{0.5\textwidth}
  \centering
  \includegraphics[width=1.0\linewidth]{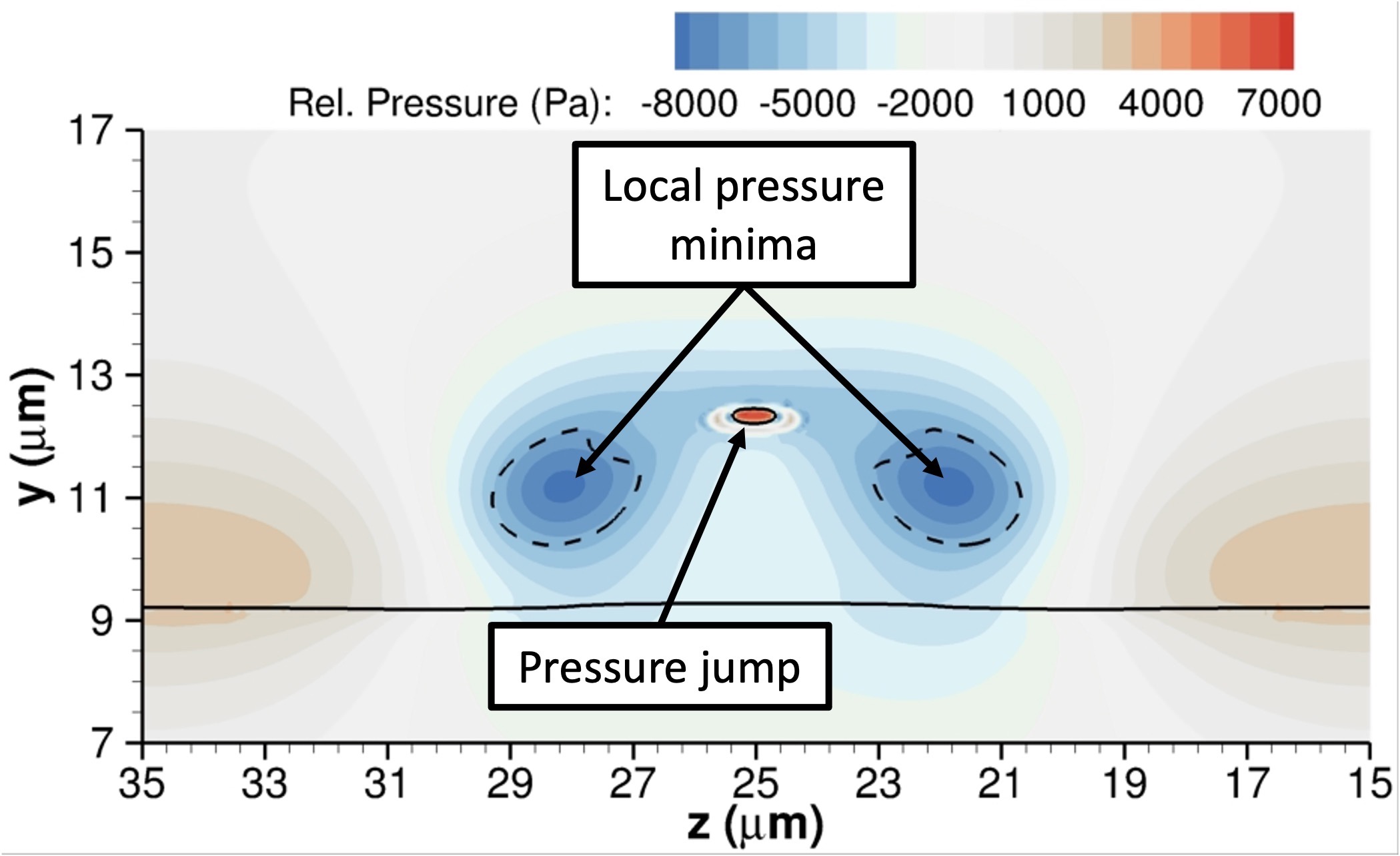}
  \caption{} 
  \label{subfig:Fig21b}
\end{subfigure}%
\\
\begin{subfigure}{0.5\textwidth}
  \centering
  \includegraphics[width=1.0\linewidth]{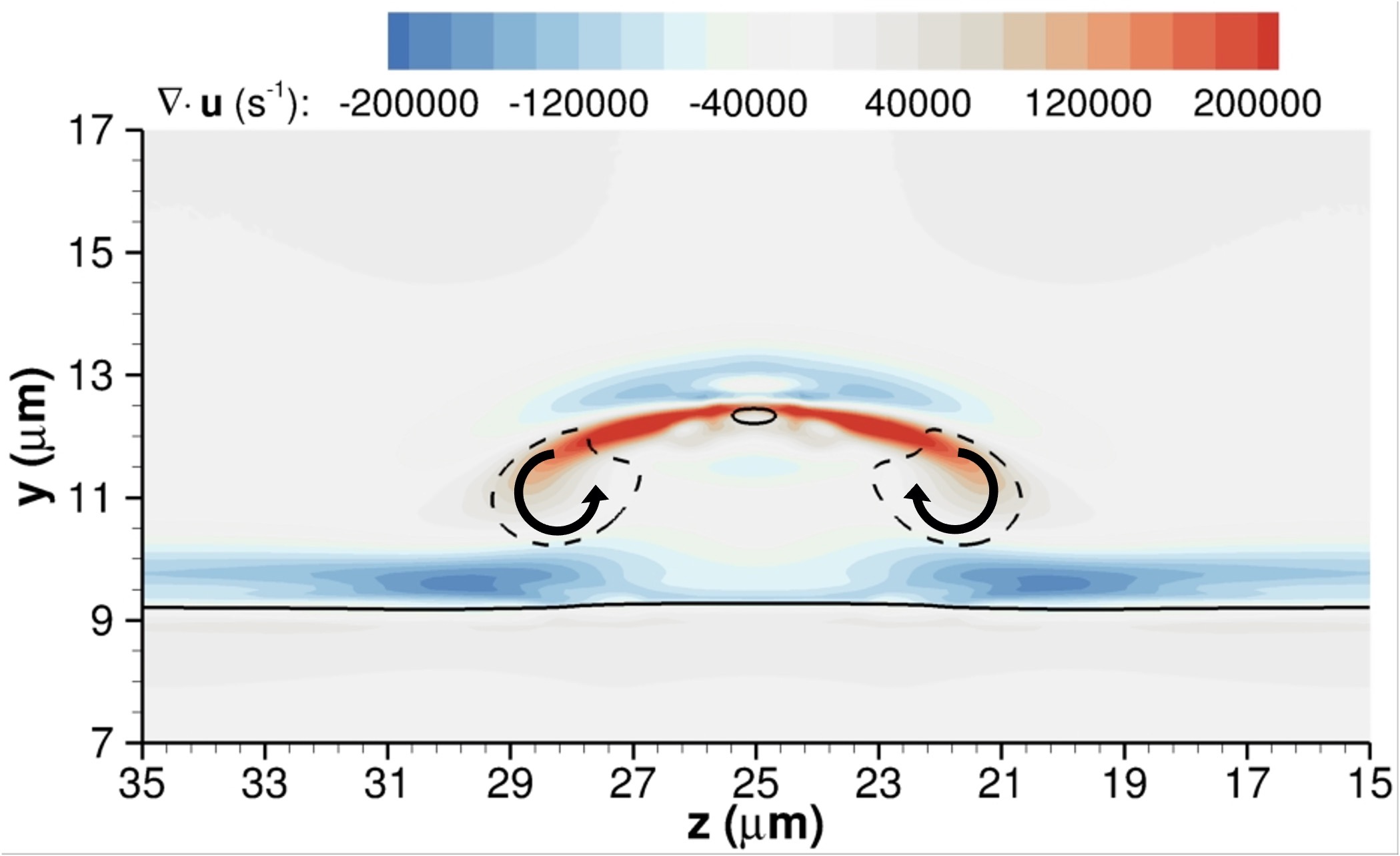}
  \caption{} 
  \label{subfig:Fig21c}
\end{subfigure}%
\caption{Density in the gas phase, relative pressure, \(p_{rel}=p-p_{amb}\), and volume dilatation rate during the stretching of the initial hairpin vortex for case C1 at \(t^*=3\). The front view from a \(yz\) plane at \(x=16\) \(\mu\)m is shown. The interface is identified by the solid isocontour with \(C=0.5\) and the cut vortex structures are identified by the dashed isocontour with \(\lambda_{\rho,t}=-2.5\times 10^{15}\). (a) gas density, \(\rho\); (b) relative pressure, \(p-p_\text{amb}\); and (c) volume dilatation rate, \(\boldsymbol{\nabla\cdot u}\).}
\label{fig:Fig21}
\end{figure}

All the terms in (\ref{eqn:vorticityX}) contribute to increasing the magnitude of the \(\omega_x\) observed in figure~\ref{subfig:Fig19a} between the liquid surface and the hairpin vortex. Vortex stretching and tilting still dominate, but the contributions from volume dilatation and baroclinicity cannot be neglected. Figure~\ref{fig:Fig21} shows the gas-phase density, the pressure field relative to the ambient pressure, and the divergence of the velocity field or volume dilatation rate. This snapshot's flow patterns reveal the vorticity generation mechanisms that occur throughout the computations in the variable-density transcritical flow. Note the local pressure minima inside the vortex pins in figure~\ref{subfig:Fig21b}, which reflects the rationale behind the \(\lambda_\rho\) vortex identification method. Moreover, the snapshot captures the interfacial pressure jump due to surface tension in the high curvature region near the lobe's tip. \par

The sign of the streamwise component of the baroclinic torque, \(\frac{1}{\rho^2}\bigg(\frac{\partial \rho}{\partial y}\frac{\partial p}{\partial z} - \frac{\partial \rho}{\partial z}\frac{\partial p}{\partial y}\bigg)\), can be justified by looking at the density and pressure fields in figure~\ref{fig:Fig21}. In general, the two-dimensional pressure gradient around the vortex pins (i.e., pressure gradient on the illustrated slice) is oriented radially outwards from each local pressure minimum. In contrast, the two-dimensional density gradient not only depends on the density jump across the interface, but also on the temperature and composition variations in the gas phase. Thus, the baroclinic term in (\ref{eqn:vorticityX}) changes sign multiple times around the vortex pins, increasing and decreasing \(\omega_x\) locally. For example, the sign of the baroclinic term in the region under the vortex pin centred around \(z=22\) \(\mu\)m in figure~\ref{subfig:Fig19d} is explained as follows. For \(z<22\) \(\mu\)m, the baroclinic term is positive since \(\frac{\partial\rho}{\partial y}<0\), \(\frac{\partial p}{\partial z}<0\), \(\frac{\partial\rho}{\partial z}>0\), and \(\frac{\partial p}{\partial y}<0\). That is, both \(\frac{\partial \rho}{\partial y}\frac{\partial p}{\partial z}>0\) and \(-\frac{\partial \rho}{\partial z}\frac{\partial p}{\partial y}>0\) are positive and increase \(\omega_x\) as depicted in figure~\ref{subfig:Fig19a}. Then, the baroclinic term becomes negative for \(z>22\) \(\mu\)m because of the two-dimensional pressure gradient. \(\frac{\partial\rho}{\partial y}<0\), \(\frac{\partial\rho}{\partial z}>0\), and \(\frac{\partial p}{\partial y}<0\) remain unchanged, but \(\frac{\partial p}{\partial z}>0\). This change results in \(\frac{\partial \rho}{\partial y}\frac{\partial p}{\partial z}<0\) and \(-\frac{\partial \rho}{\partial z}\frac{\partial p}{\partial y}>0\), where the negative term dominates because the transverse density gradient and the spanwise pressure gradient are larger in magnitude than the spanwise density gradient and the transverse pressure gradient, respectively. \par 

In contrast, the compressible vortex stretching sign is dictated by the local volume dilatation rate. If the fluid expands, the rotation of a fluid element is distributed over a larger volume, decreasing the vorticity magnitude. On the other hand, fluid compression increases vorticity. Here, the combination of the Stefan flow generated around the vaporising or condensing interface and the vortical flow determine the sign of \(\boldsymbol{\nabla \cdot u}\). Typically, the evaporation of the fuel results in a volume expansion near the interface. However, the transcritical environment introduces various changes in fluid behaviour. As discussed in section~\ref{subsec:lobe_bending}, the enhanced oxygen dissolution at 150 bar may result in net condensation. Therefore, volume compression occurs despite the evaporation of the fuel. Additionally, density variations in the gas phase are more significant, and the increase in fuel vapour results in higher gas densities that could limit the expansive behaviour of evaporating flows. \par 

Away from the recirculation region caused by the vortex pins, figure~\ref{subfig:Fig21c} reveals the local volume compression above the liquid surface, where weaker evaporation of the fuel results in net condensation. In contrast, the volume dilatation around the vortex structures behaves differently. On the one hand, the entrainment of the hot gas enhances the evaporation of the fuel, increasing the gas mixture density below the vortex pins. Thus, the lighter gas flows into a denser gas region, undergoing compression. On the other hand, the gas phase expands above the vortex pins. Fuel evaporation is stronger near the lobe's tip, substantially increasing the gas density in the surroundings. However, the vortical motion rapidly advects the gas into fluid regions of lower density. \par 

\begin{figure}
\centering
\begin{subfigure}{0.5\textwidth}
  \centering
  \includegraphics[width=1.0\linewidth]{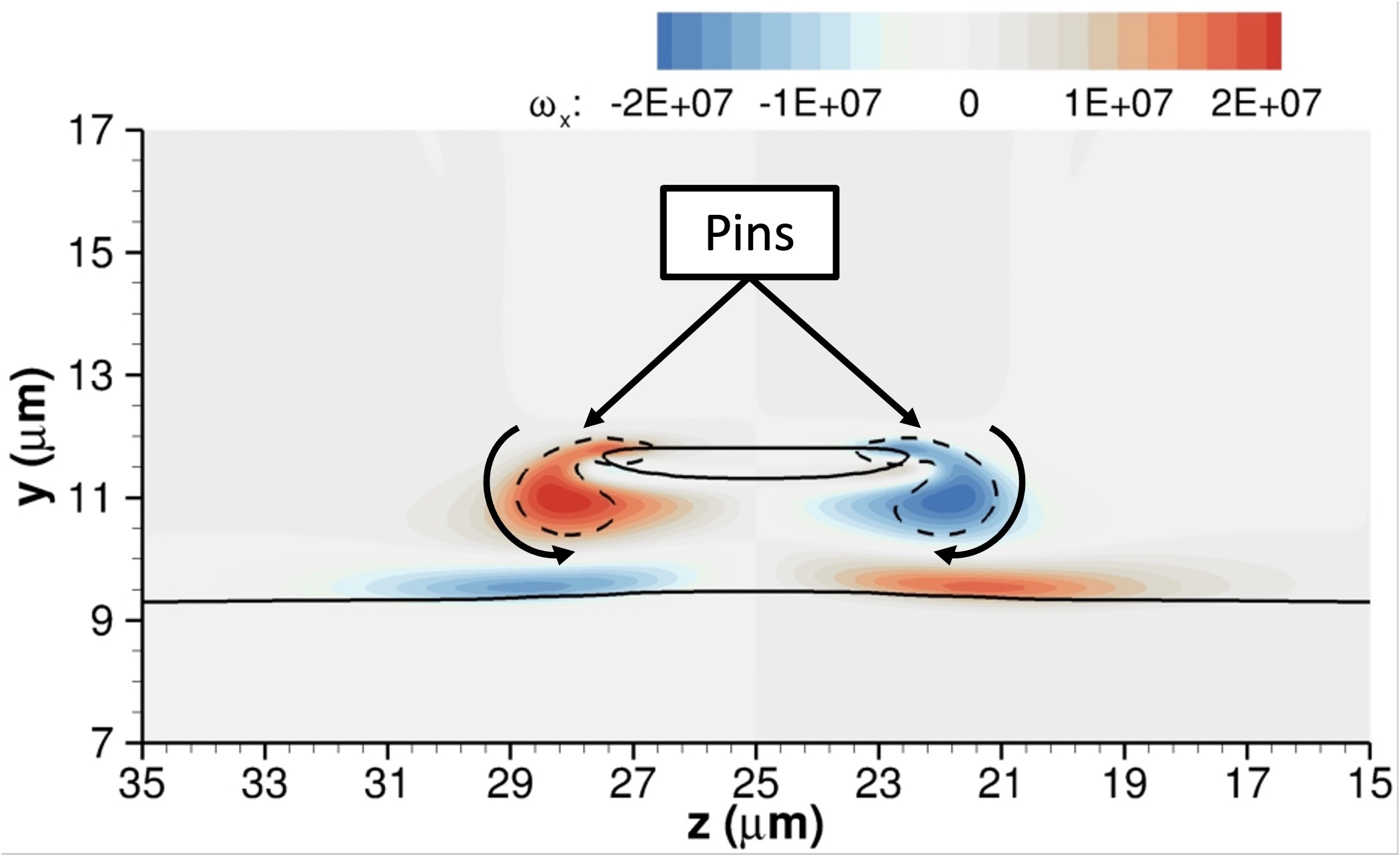}
  \caption{} 
  \label{subfig:Fig22a}
\end{subfigure}%
\begin{subfigure}{0.5\textwidth}
  \centering
  \includegraphics[width=1.0\linewidth]{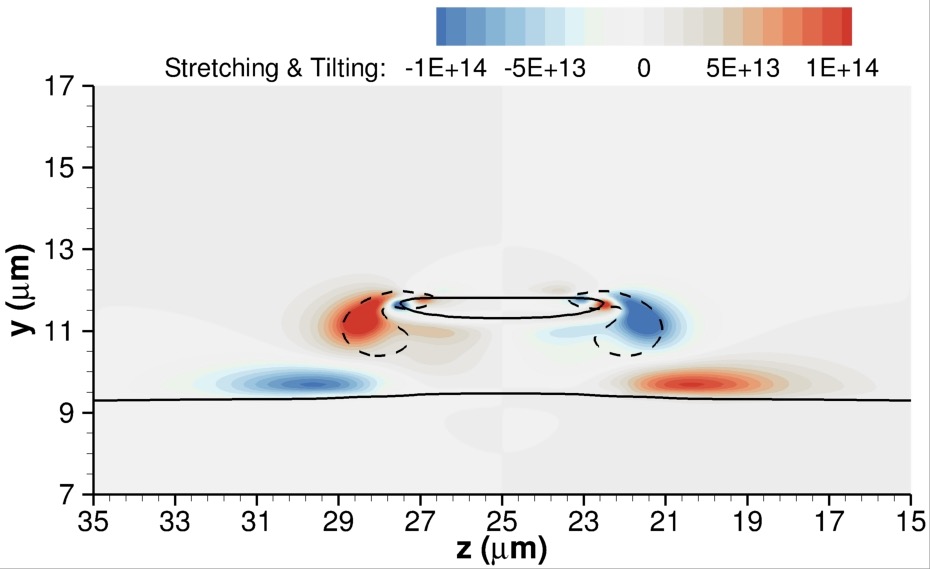}
  \caption{} 
  \label{subfig:Fig22b}
\end{subfigure}%
\\
\begin{subfigure}{0.5\textwidth}
  \centering
  \includegraphics[width=1.0\linewidth]{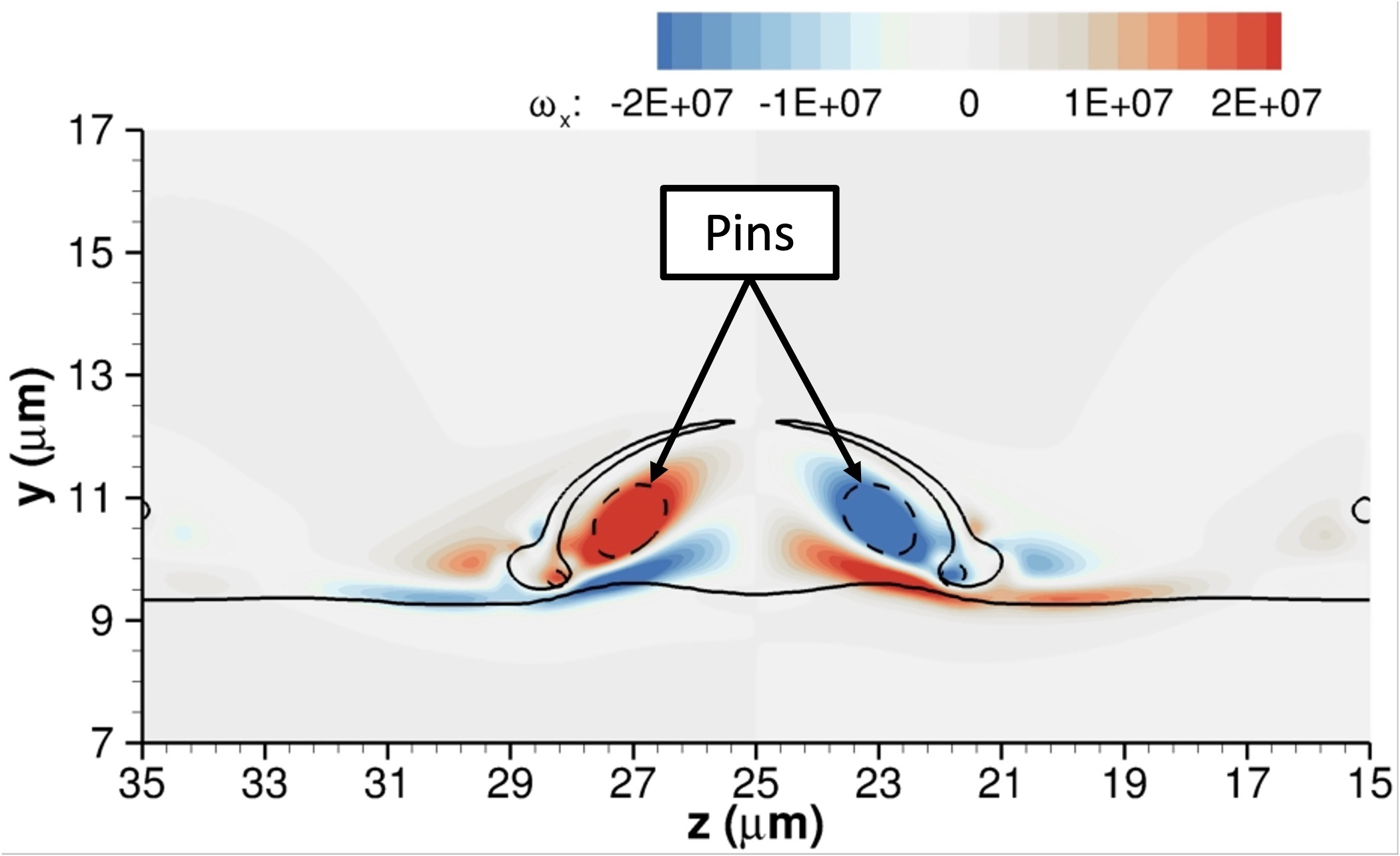}
  \caption{} 
  \label{subfig:Fig22c}
\end{subfigure}%
\begin{subfigure}{0.5\textwidth}
  \centering
  \includegraphics[width=1.0\linewidth]{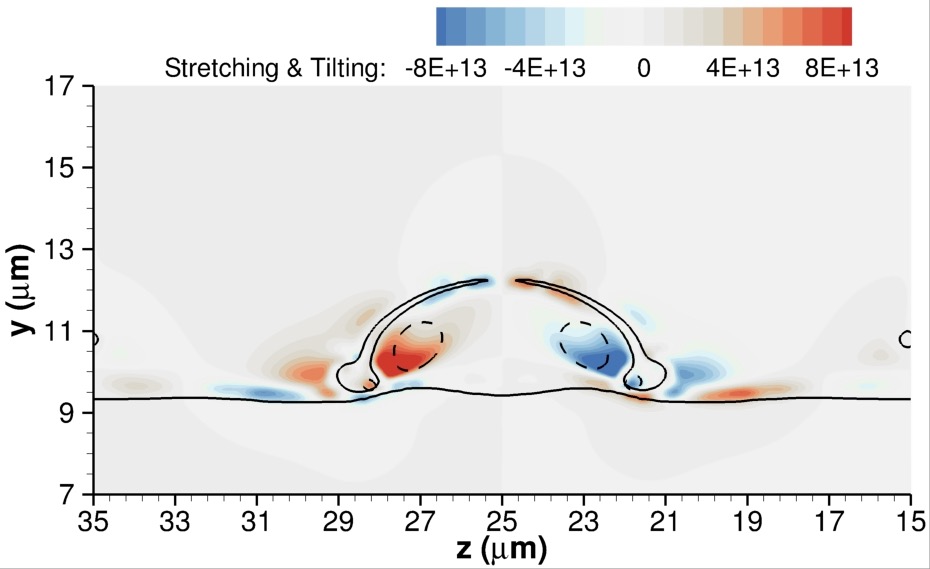}
  \caption{} 
  \label{subfig:Fig22d}
\end{subfigure}%
\caption{\(\omega_x\) generation terms during the stretching of the initial hairpin vortex for case C2. The contours of the different terms are shown in \(yz\) planes at various \(x\) locations and times. The interface is identified by the solid isocontour with \(C=0.5\), and the cut vortex structures are identified by the dashed isocontour with \(\lambda_{\rho,t}=-9\times 10^{15}\). (a) \(\omega_x\) at \(x=12\) \(\mu\)m and \(t^*=2.5\); (b) vortex stretching and tilting at \(x=12\) \(\mu\)m and \(t^*=2.5\); (c) \(\omega_x\) at \(x=28\) \(\mu\)m and \(t^*=4\); and (d) vortex stretching and tilting at \(x=28\) \(\mu\)m and \(t^*=4\).}
\label{fig:Fig22}
\end{figure}

The described \(\omega_x\) generation mechanisms for the stretching of the initial hairpin vortex do not change significantly across the analysed configurations that show lobe stretching, bending and perforation. Other deformation mechanisms show deviations in the evolution of the initial hairpin vortex. For example, figure~\ref{fig:Fig9} describing the lobe corrugation mechanism for case C2 shows a stronger stretching of the hairpin legs. Figure~\ref{fig:Fig22} depicts, for case C2, the streamwise vorticity and the combined contribution of vortex stretching and tilting in the generation of \(\omega_x\). Two different \(yz\) planes are shown: one at \(x=12\) \(\mu\)m and \(t^*=2.5\) and the other at \(x=28\) \(\mu\)m at \(t^*=4\). During the coupled deformation process between lobe and hairpin described in section~\ref{subsec:lobe_crest_corrugation}, \(\omega_x\) generation occurs inside the vortex pins, increasing the strength of each vortex. Here, \(\dot{\omega}_{x\rightarrow x}\) is the main vorticity generation mechanism, having a similar magnitude as the vortex tilting terms that dominate in other flow regions. The behaviour of the compressible vortex stretching and the baroclinic terms is very similar to what has been described previously. \par 

The importance of the coupled deformation between the liquid surface and the hairpin vortex is revealed by comparing cases B1 and C2, both having a gas freestream velocity \(u_G=50\) m/s. Case B1 shows a weaker contribution of \(\dot{\omega}_{x\rightarrow x}\) in \(\omega_x\) generation, and the hairpin deformation is consistent with the previous discussion of case C1. Not surprisingly, both cases B1 and C1 involve the same early deformation mechanism. Arguably, the term \(\frac{\partial u}{\partial x}\) has a similar strength in cases B1 and C2. Thus, the amplified contribution of \(\dot{\omega}_{x\rightarrow x}\) in case C2 must come from an \(\omega_x\) increase in the vortex pins. As described in section~\ref{subsec:lobe_crest_corrugation}, the lobe wraps around the vortex pins, enhancing the gas entrainment under the lobe as the gaps between the sides of the lobe and the liquid surface are reduced. This deformation process results in the strengthening of the \(\omega_x\) (see figures~\ref{fig:Fig10} or \ref{fig:Fig22}). \par

\subsubsection{The secondary roller vortex}
\label{subsubsec:rollers}

New roller vortices form during the jet development, particularly in front of the main perturbation wave as it grows (see figure~\ref{fig:Fig13}). These vortical structures are expected to emerge from the stretching of the vortex sheet between the liquid wave and the faster-moving gas. Along a sharp wave edge (i.e., a sharp flow turn), the vortex sheet separates and is rapidly affected by the gas entrainment underneath the liquid. Thence, a roller vortex forms. However, the formation of smaller roller vortices occurs without such a clear mechanism. For example, figure~\ref{fig:Fig9} shows that during the early lobe deformation in case C2, a small secondary roller vortex is observed on top of the hairpin vortex pins. No clear flow feature seems to explain its occurrence. Here, we describe the nature of the strengthening of the spanwise vorticity in that region, leading to the formation of the vortex. \par 

\begin{figure}
\centering
\begin{subfigure}{0.5\textwidth}
  \centering
  \includegraphics[width=0.9\linewidth]{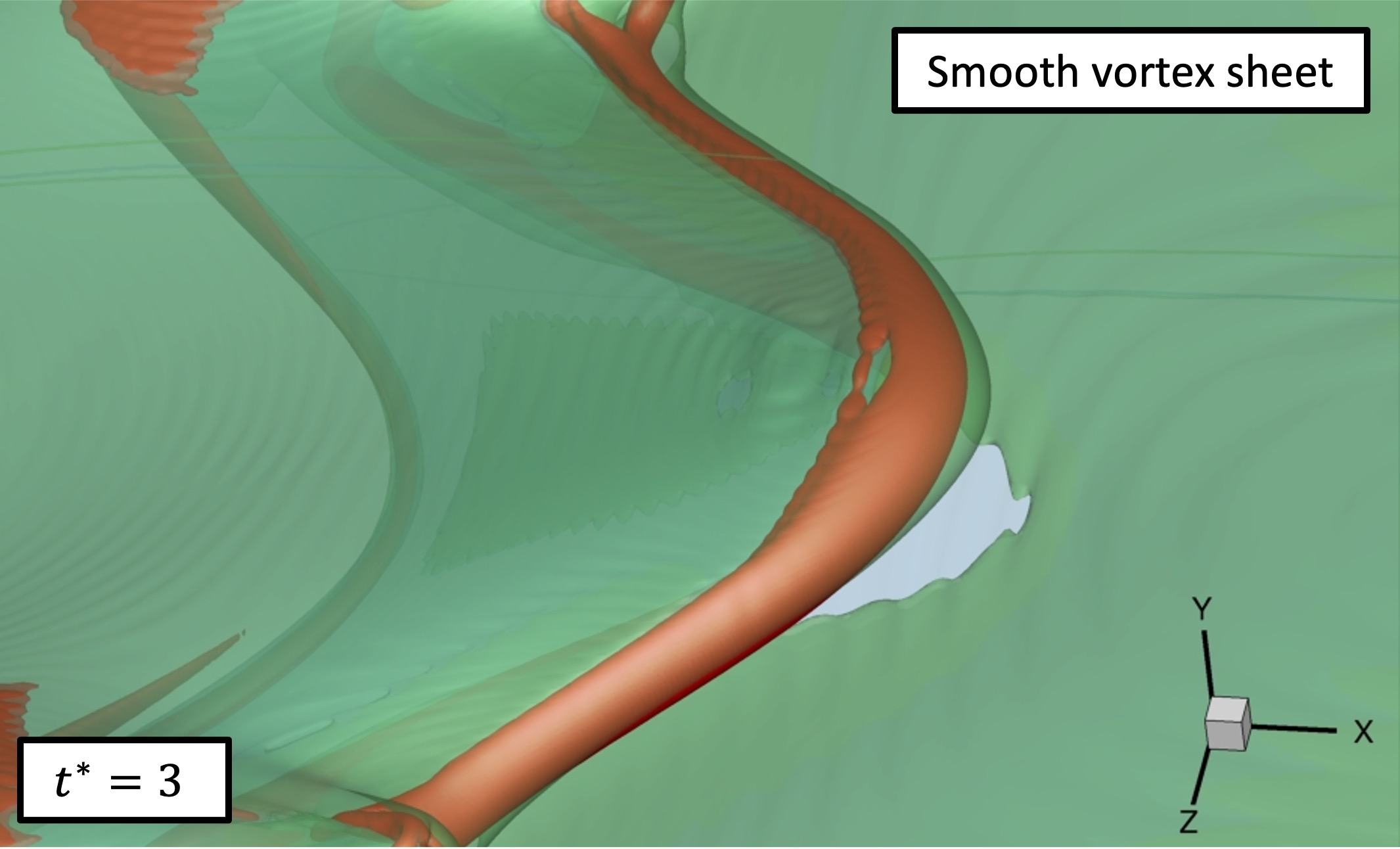}
  \caption{} 
  \label{subfig:Fig23a}
\end{subfigure}%
\begin{subfigure}{0.5\textwidth}
  \centering
  \includegraphics[width=0.9\linewidth]{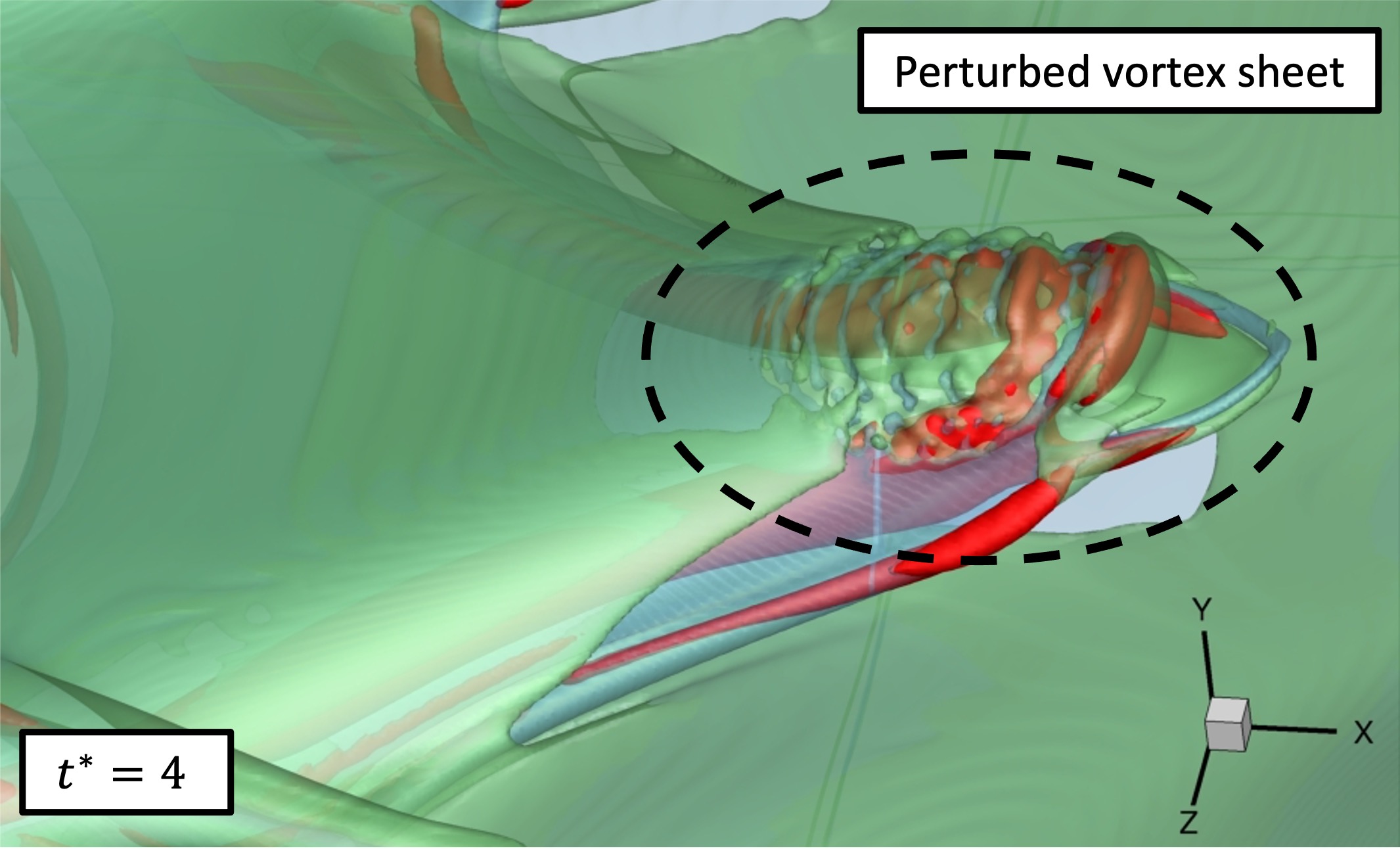}
  \caption{} 
  \label{subfig:Fig23b}
\end{subfigure}%
\caption{Destabilisation of the vortex sheet caused by the lobe bursting in case C2. The liquid surface is identified by the translucid blue isosurface with \(C=0.5\), the vortex structures are identified by the red isosurface with \(\lambda_{\rho,t}=-9\times 10^{15}\), and the vortex sheet is identified by the translucid green isosurface with \(\omega_z=-1.5\times10^{7}\) s\(^{-1}\).}
\label{fig:Fig23}
\end{figure}

The vortex formation mechanism originates in the destabilisation of the vortex sheet between liquid and gas. Figure~\ref{fig:Fig23} depicts the deformation of the early lobe as it corrugates between \(t^*=3\) and \(t^*=4\), showing the vortex sheet (represented by the green isosurface with \(\omega_z=-1.5\times10^{7}\) s\(^{-1}\)) and the hairpin vortex (red isosurface with \(\lambda_{\rho,t}=-9\times 10^{15}\)). The vortex sheet forms due to the shear between the two phases, and it follows the liquid surface along with the stretching and deformation of the lobe (i.e., the vortex sheet is also stretched and deformed). At \(t^*=3\), no distinctive perturbation is seen. However, once the lobe bursts into small ligaments and droplets, the vortex sheet detaches from the liquid surface and is suddenly perturbed by the increased multi-phase turbulence. This triggers a destabilisation of the vortex sheet and the formation of the secondary roller vortex, as explained in the following lines. \par 

\begin{figure}
\centering
\begin{subfigure}{0.5\textwidth}
  \centering
  \includegraphics[width=1.0\linewidth]{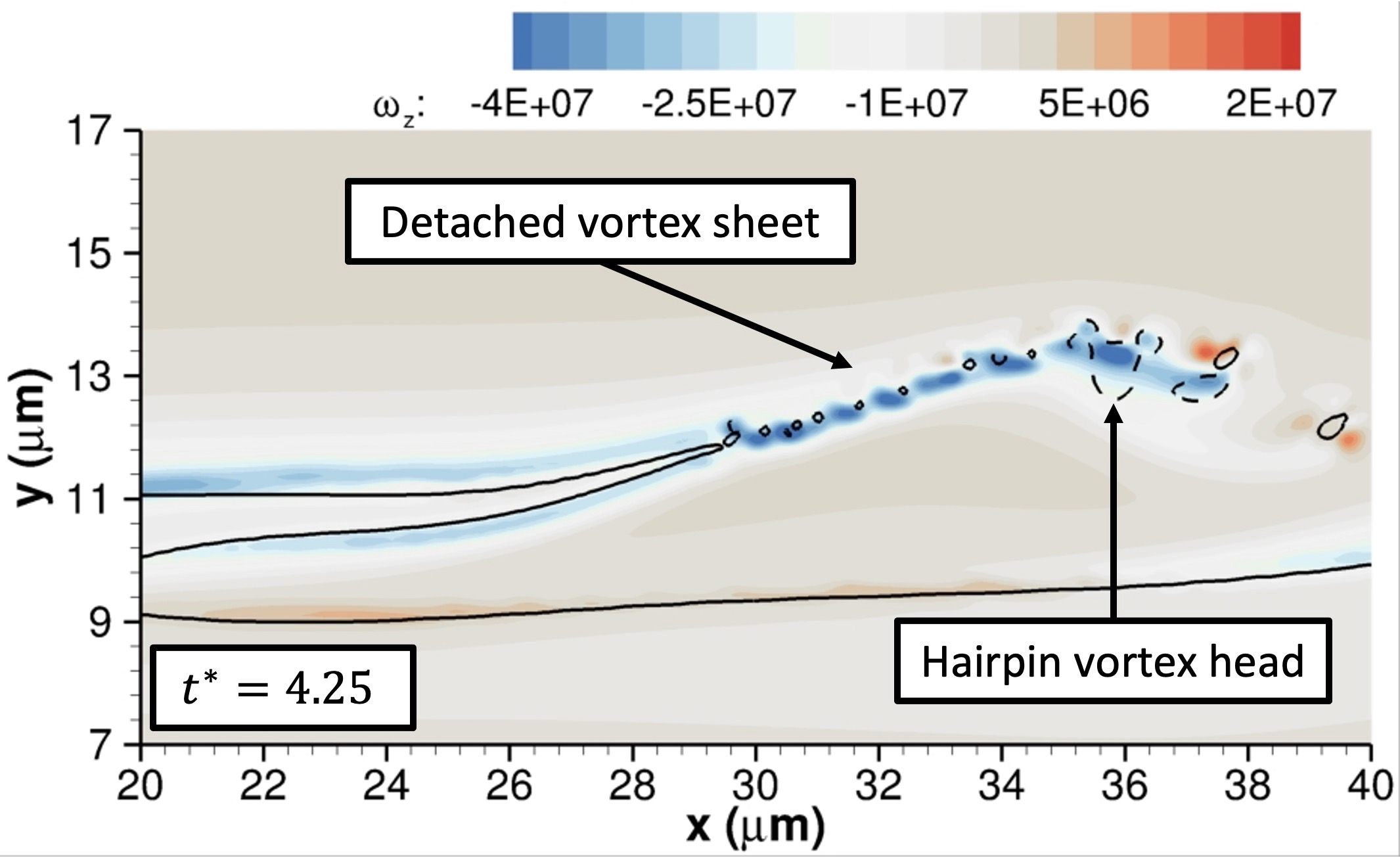}
  \label{subfig:Fig24a}
\end{subfigure}%
\begin{subfigure}{0.5\textwidth}
  \centering
  \includegraphics[width=0.9\linewidth]{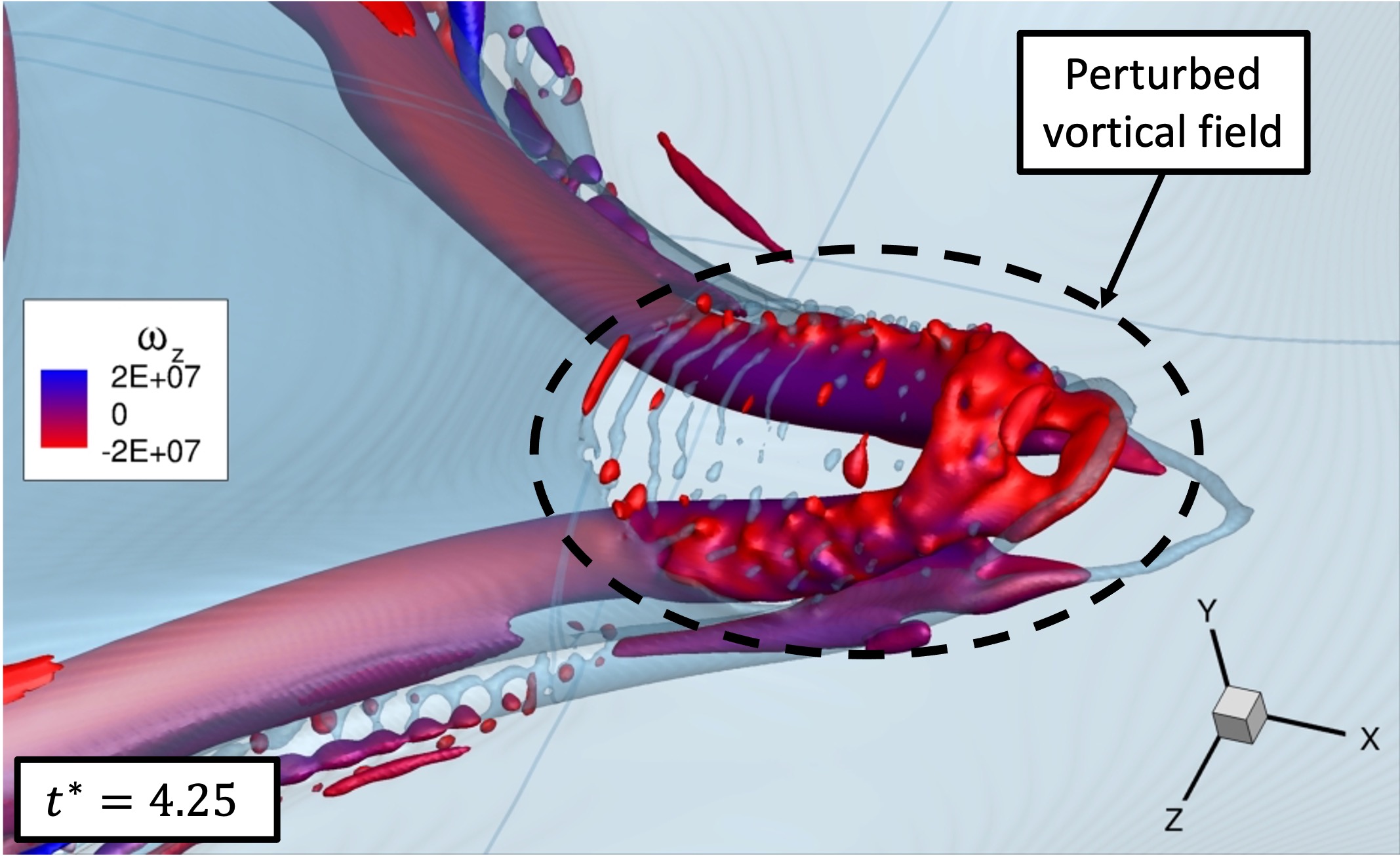}
  \label{subfig:Fig24b}
\end{subfigure}%
\\[-2ex]
\begin{subfigure}{0.5\textwidth}
  \centering
  \includegraphics[width=1.0\linewidth]{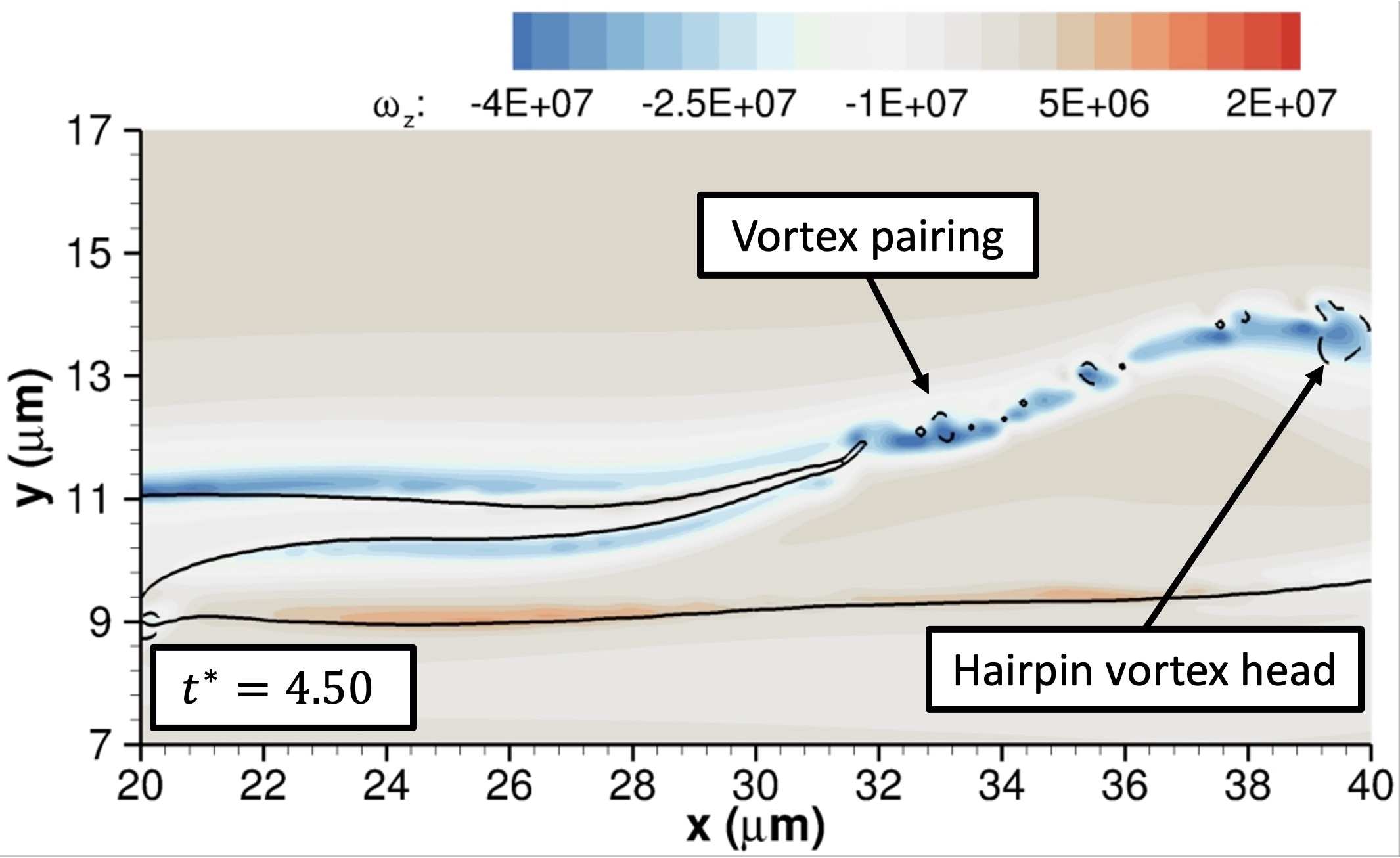}
  \label{subfig:Fig24c}
\end{subfigure}%
\begin{subfigure}{0.5\textwidth}
  \centering
  \includegraphics[width=0.9\linewidth]{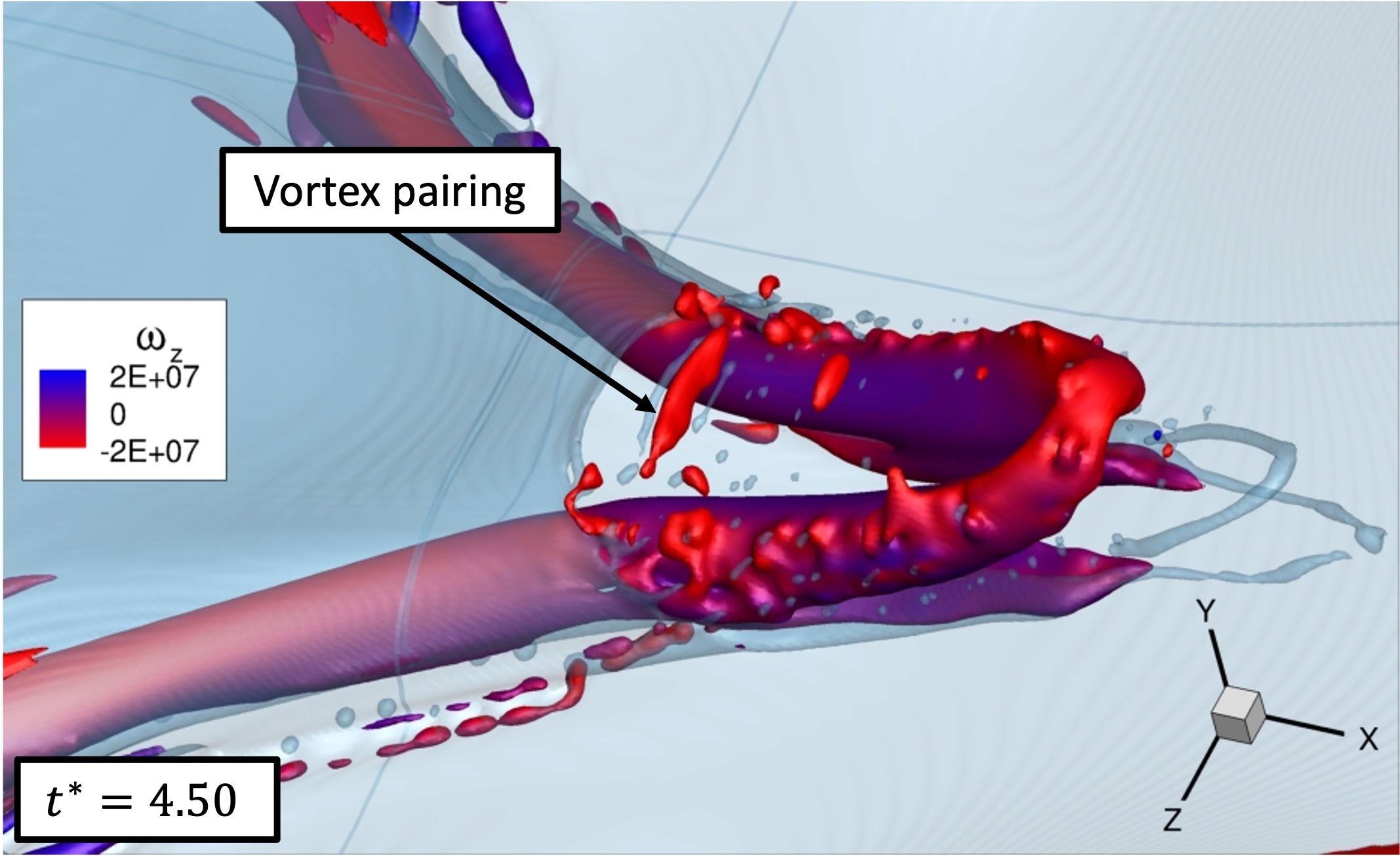}
  \label{subfig:Fig24d}
\end{subfigure}%
\\[-2ex]
\begin{subfigure}{0.5\textwidth}
  \centering
  \includegraphics[width=1.0\linewidth]{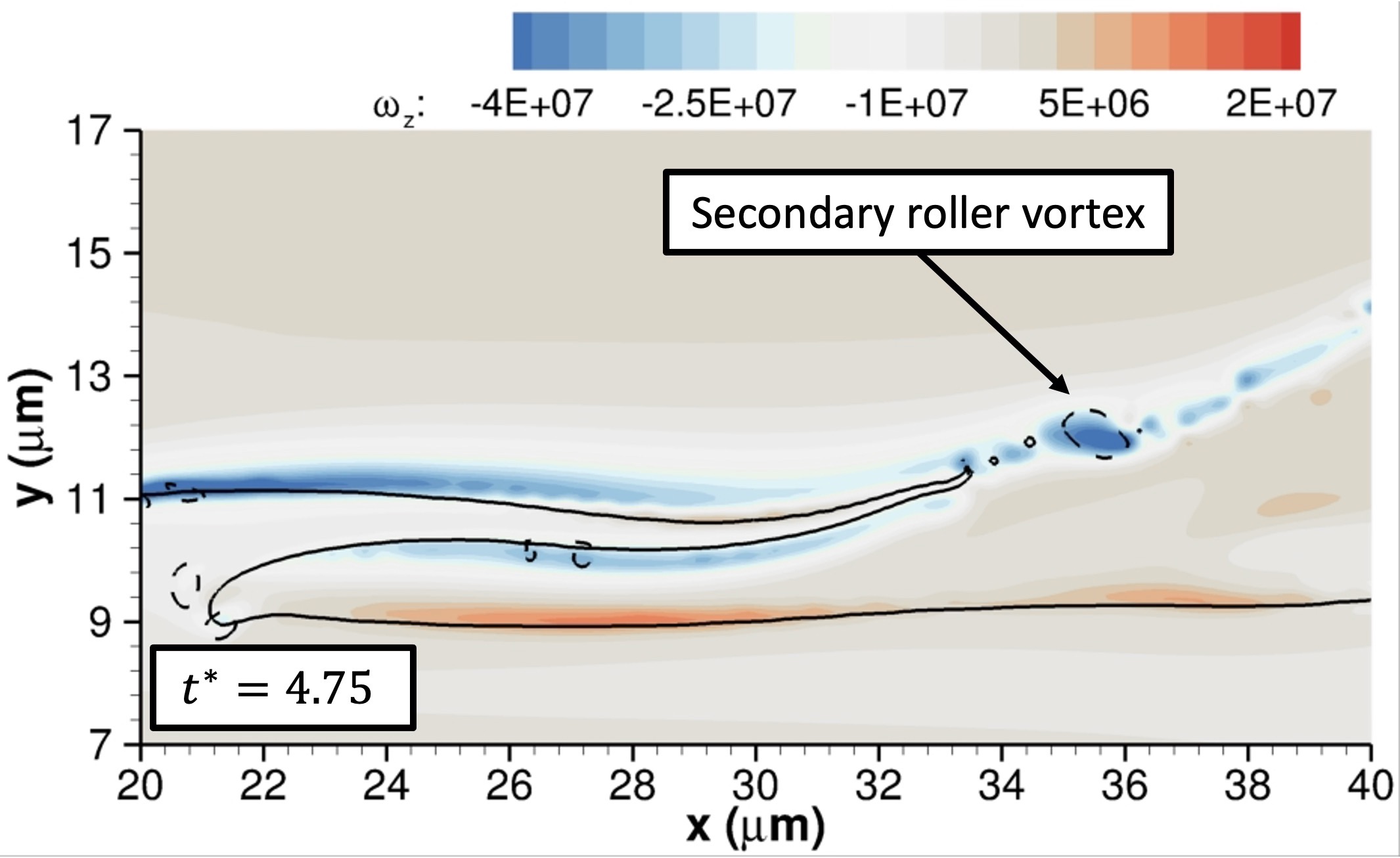}
  \label{subfig:Fig24e}
\end{subfigure}%
\begin{subfigure}{0.5\textwidth}
  \centering
  \includegraphics[width=0.9\linewidth]{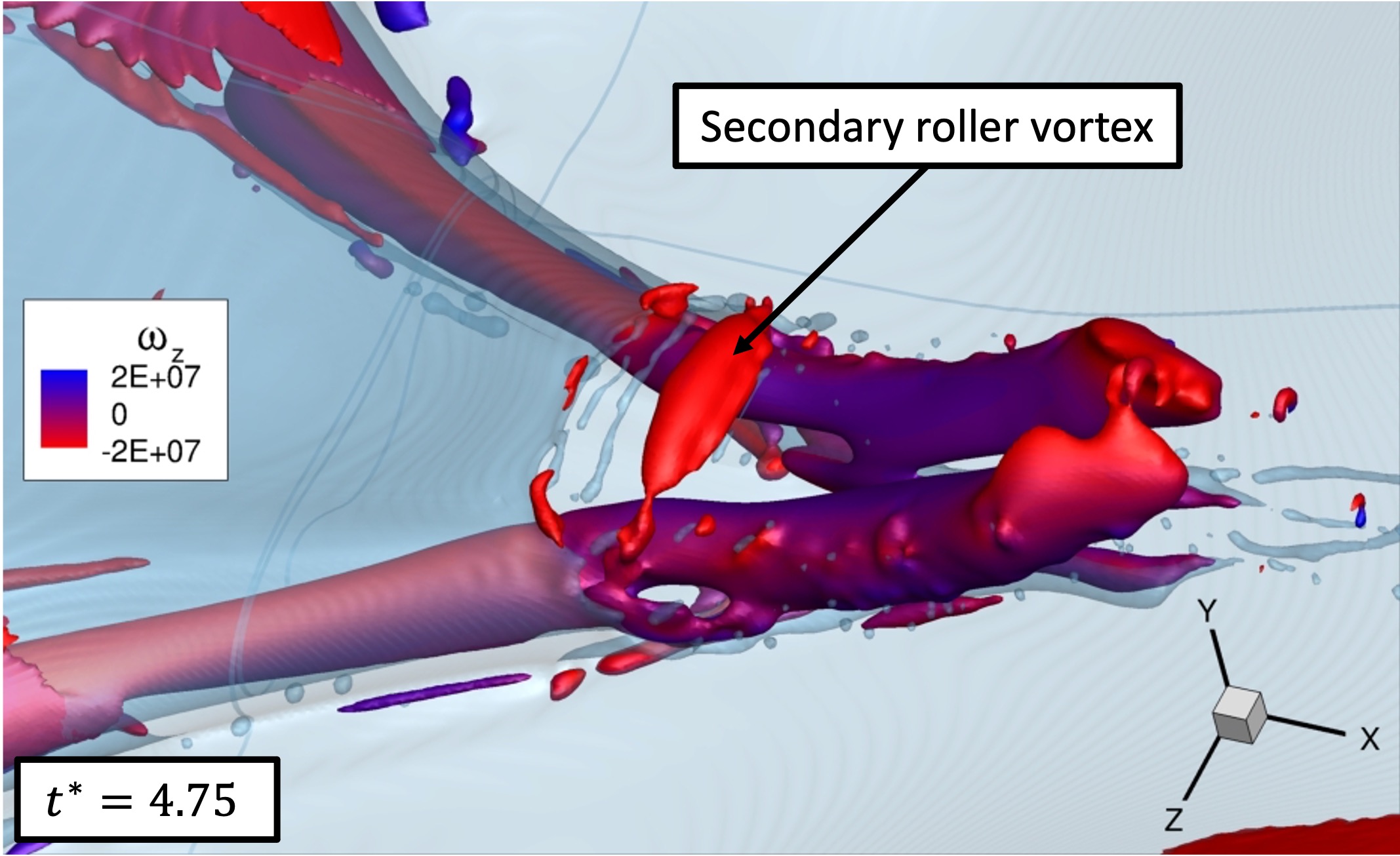}
  \label{subfig:Fig24f}
\end{subfigure}%
\\[-2ex]
\begin{subfigure}{0.5\textwidth}
  \centering
  \includegraphics[width=1.0\linewidth]{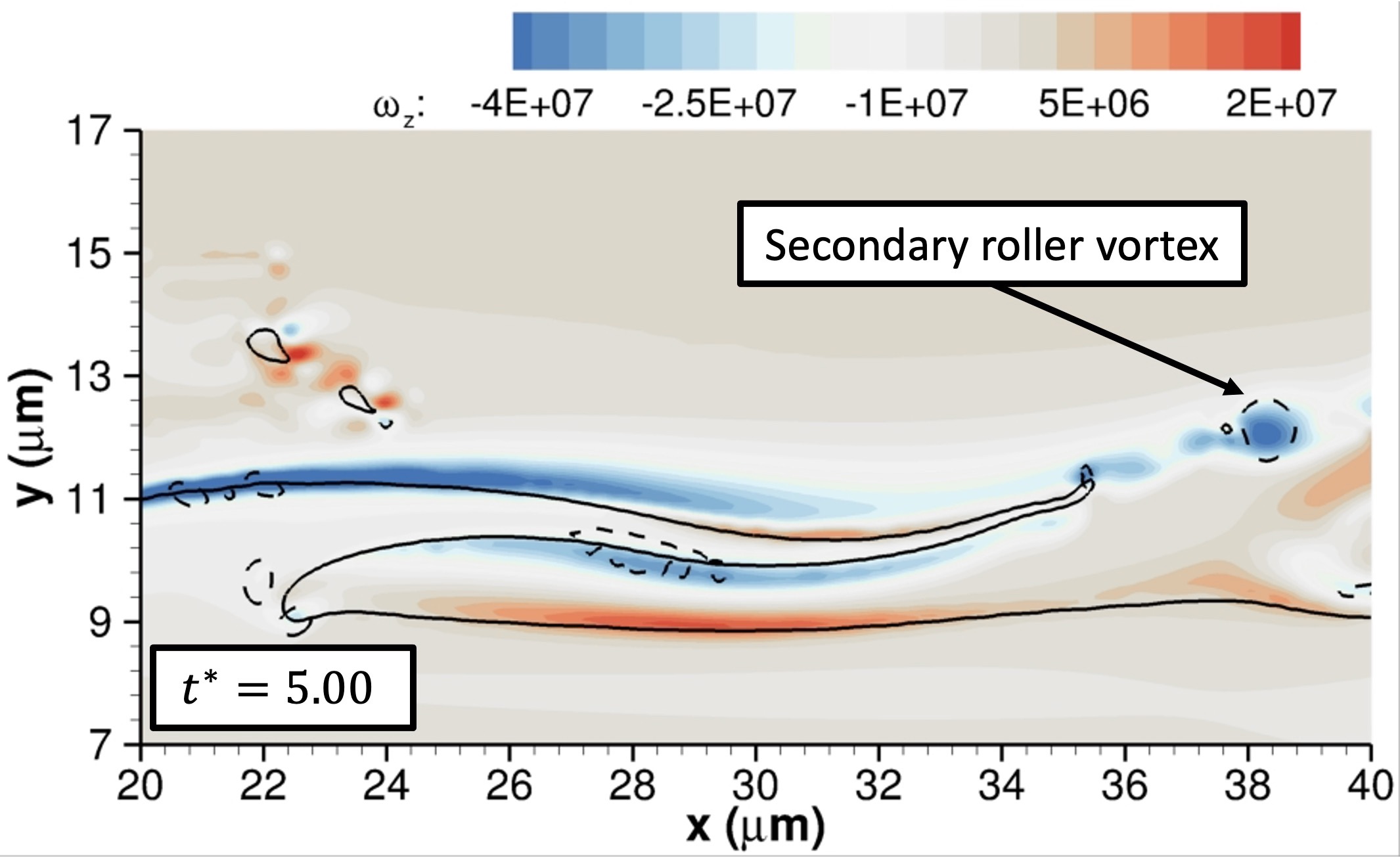}
  \label{subfig:Fig24g}
\end{subfigure}%
\begin{subfigure}{0.5\textwidth}
  \centering
  \includegraphics[width=0.9\linewidth]{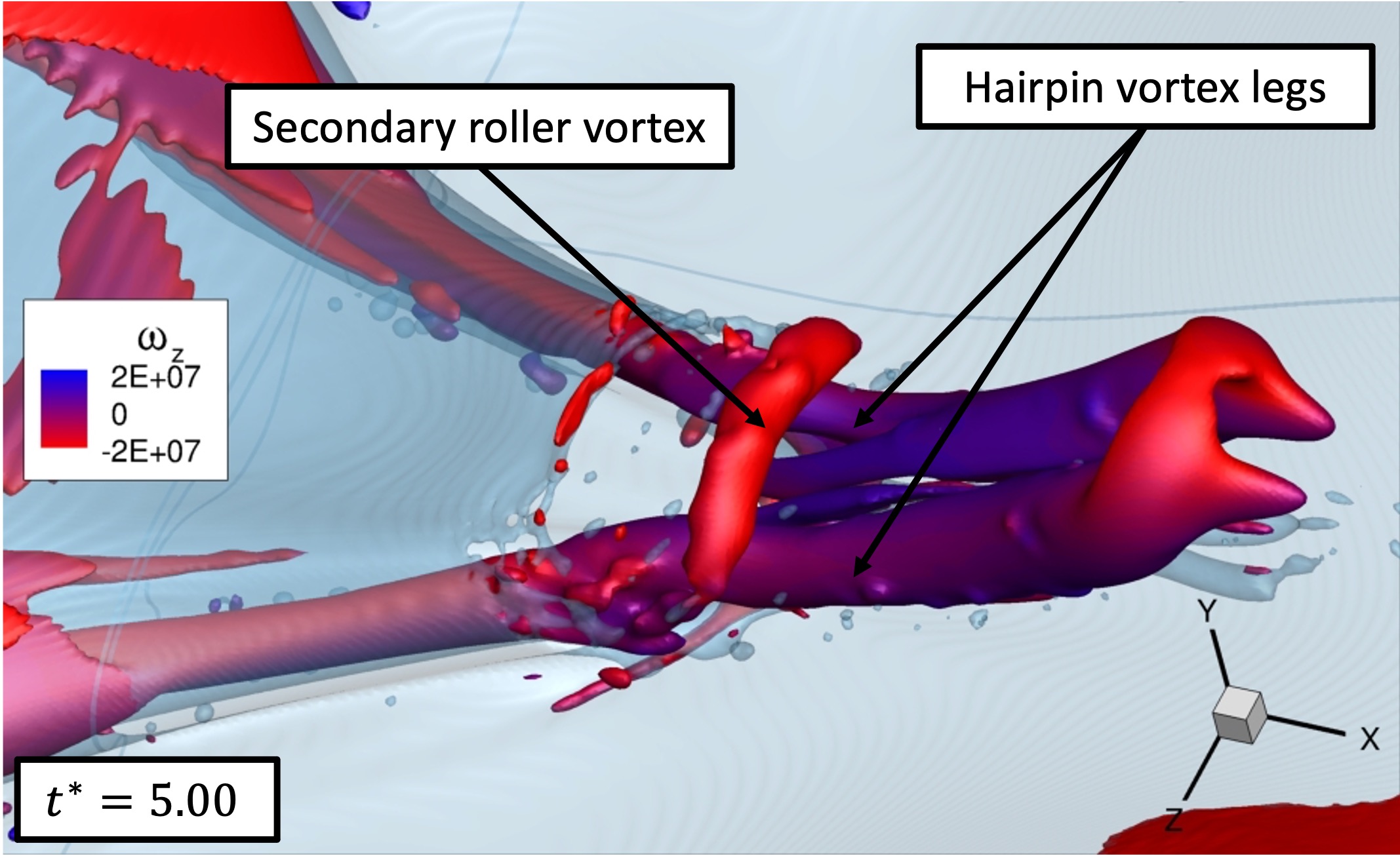}
  \label{subfig:Fig24h}
\end{subfigure}%
\caption{Vortex pairing mechanism resulting in the formation of the secondary roller vortex in case C2. Contours of \(\omega_z\) are shown in an \(xy\) plane at \(z=25\) \(\mu\)m. The liquid surface is identified by the solid isocontour or the translucid blue isosurface with \(C=0.5\), and the vortex structures are identified by the dashed isocontour or isosurface with \(\lambda_{\rho,t}=-5\times 10^{15}\).}
\label{fig:Fig24}
\end{figure}

Figure~\ref{fig:Fig24} shows the evolution of the vortex sheet between \(t^*=4.25\) and \(t^*=5\) after the lobe bursts. The contours of \(\omega_z\) are shown in an \(xy\) plane at \(z=25\) \(\mu\)m, together with a three-dimensional view of the liquid surface and the vortex structures. To enhance the visualisation of the vortices in this figure, \(\lambda_{\rho,t}=-5\times 10^{15}\) instead of \(\lambda_{\rho,t}=-9\times 10^{15}\) used in previous figures for case C2 and the vortex isosurface is coloured with \(\omega_z\). Pockets of stronger spanwise vorticity occur in the detached vortex sheet, as seen at \(t^*=4.25\), arguably due to the interactions between ligaments and droplets. Over time, these vortical regions come together via mutual induction and subsequent pairing/merging, and some small roller vortices can be observed at \(t^*=4.5\). In particular, one of these regions of vortex pairing grows considerably between \(t^*=4.25\) and \(t^*=5\), resulting in the formation of the secondary roller vortex seen in figure~\ref{fig:Fig9}. Moreover, the strengthening of \(\omega_z\) and the vortex pairing are aided by the induced flow coming from the hairpin vortex pins or legs. Once the lobe bursts, the vortex sheet is not only perturbed by the smaller liquid structures, but it is also under the direct influence of the hairpin vortex. Before, the presence of the liquid lobe between the vortex pins and the vortex sheet has a shielding effect. \par 

\begin{figure}
\centering
\begin{subfigure}{0.5\textwidth}
  \centering
  \includegraphics[width=1.0\linewidth]{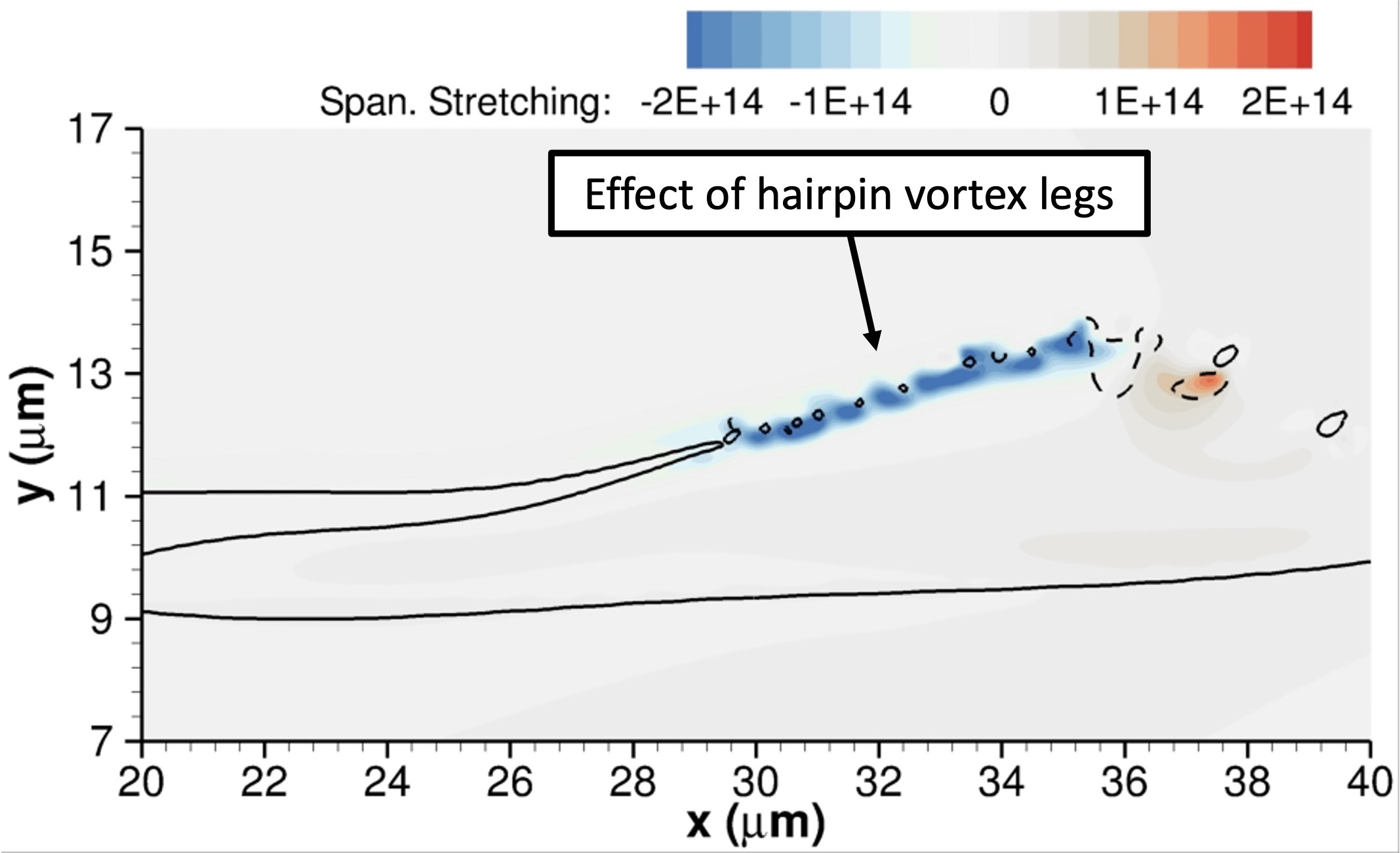}
  \caption{} 
  \label{subfig:Fig25a}
\end{subfigure}%
\begin{subfigure}{0.5\textwidth}
  \centering
  \includegraphics[width=0.9\linewidth]{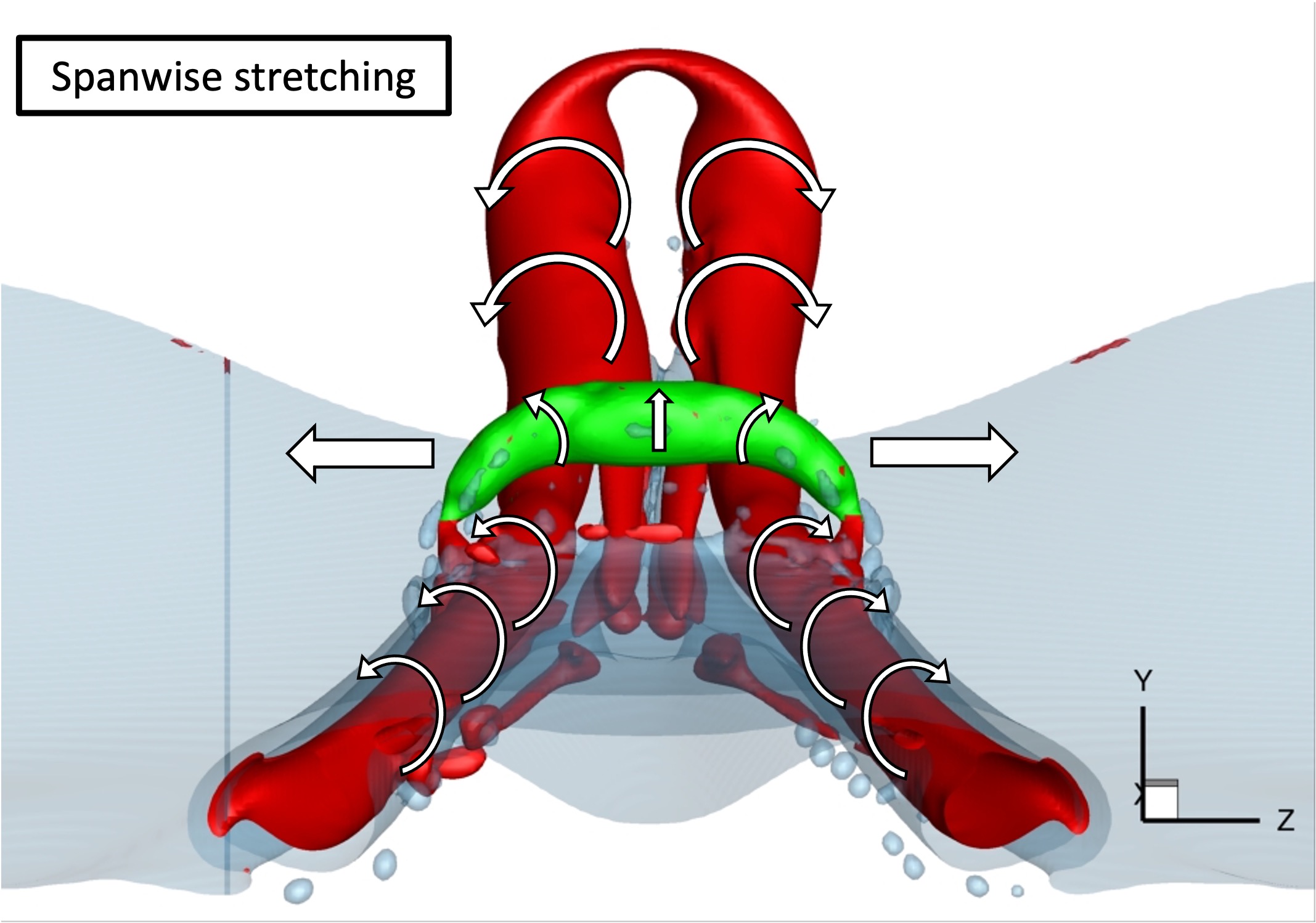}
  \caption{} 
  \label{subfig:Fig25b}
\end{subfigure}%
\caption{Spanwise vortex stretching affecting the formation of the secondary roller vortex in case C2. The liquid surface is identified by the solid isocontour or the translucid blue isosurface with \(C=0.5\), and the vortex structures are identified by the dashed isocontour or isosurface with \(\lambda_{\rho,t}=-5\times 10^{15}\). (a) \(\dot{\omega}_{z\rightarrow z}\) in the \(xy\) plane at \(z=25\) \(\mu\)m and \(t^*=4.25\); and (b) induced flow patterns from the hairpin vortex at \(t^*=5\).}
\label{fig:Fig25}
\end{figure}

Figure~\ref{subfig:Fig25a} manifests the implications of this relative position between the vortex sheet and the hairpin vortex. The contours of \(\dot{\omega}_{z\rightarrow z}\) from (\ref{eqn:vorticityZ}) are shown in the same \(xy\) plane from figure~\ref{fig:Fig24} at \(t^*=4.25\). Due to the presence of \(\omega_z\) and a spanwise stretching (i.e., \(\partial w/\partial z>0\)), there is \(\omega_z\) generation that tends to strengthen the small vortices during the pairing process. Other terms in (\ref{eqn:vorticityZ}) have a negligible effect in the detached vortex sheet. As seen in figure~\ref{subfig:Fig25b}, the hairpin legs (red vortex) are responsible for the spanwise stretching of the perturbed vortex sheet. Note that \(\lambda_{\rho,t}=-5\times 10^{15}\) in this figure too. Moreover, the secondary roller vortex (green) wraps around the hairpin vortex pins, similar to how the liquid phase wraps around the vortical structures. \par 

The mechanisms described in this section resemble the findings in previous works. The initial destabilisation or bursting of the vortex sheet is tightly coupled to the coiling of vortex lines and the turbulence cascade~\citep{stout2021genesis}, here involving the multi-phase flow. The later vortex pairing, emergence of the secondary roller vortex, and wrapping around the hairpin legs is a clear example of vortex reconnection~\citep{hussain2011mechanics,yao2022vortex,stout2022sadhana}. \par

\subsubsection{Streamwise vortex alignment in the layered flow}
\label{subsubsec:layers}

Lastly, we look at the initial stages of the vortex alignment in the streamwise direction inside the layered flow. As seen in figure~\ref{subfig:Fig14c}, which depicts a late snapshot of case C1 at \(t^*=15\) once various liquid layers have formed, some of the major structures contained in between the layers are aligned with \(x\). In section~\ref{subsec:layering}, we describe how the vorticity within the layered flow primarily comes from the various roller vortices that form during the growth of the main perturbation wave. Other minor vortical structures also emerge in the layered flow, such as rim vortices or small roller vortices. \par 

\begin{figure}
\centering
\begin{subfigure}{0.5\textwidth}
  \centering
  \includegraphics[width=1.0\linewidth]{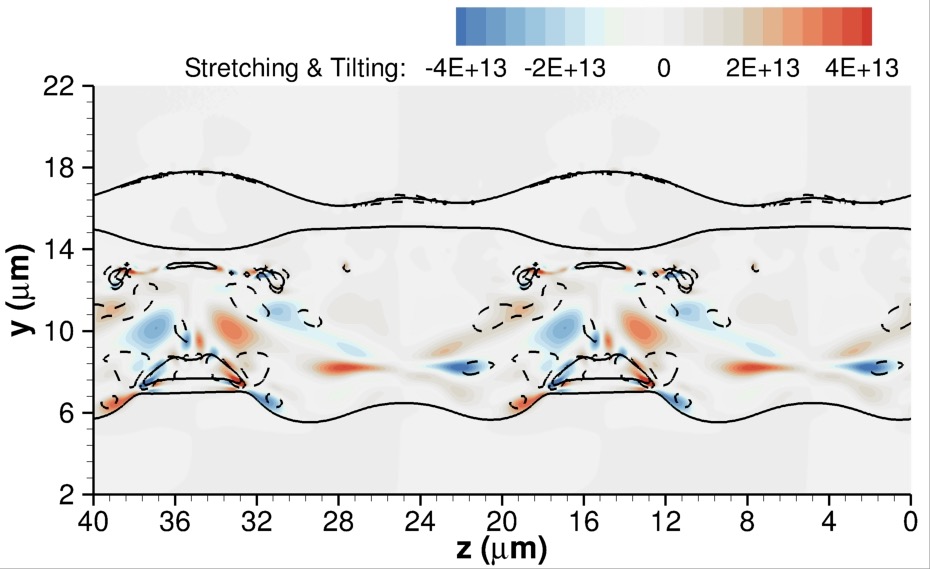}
  \caption{} 
  \label{subfig:Fig26a}
\end{subfigure}%
\begin{subfigure}{0.5\textwidth}
  \centering
  \includegraphics[width=1.0\linewidth]{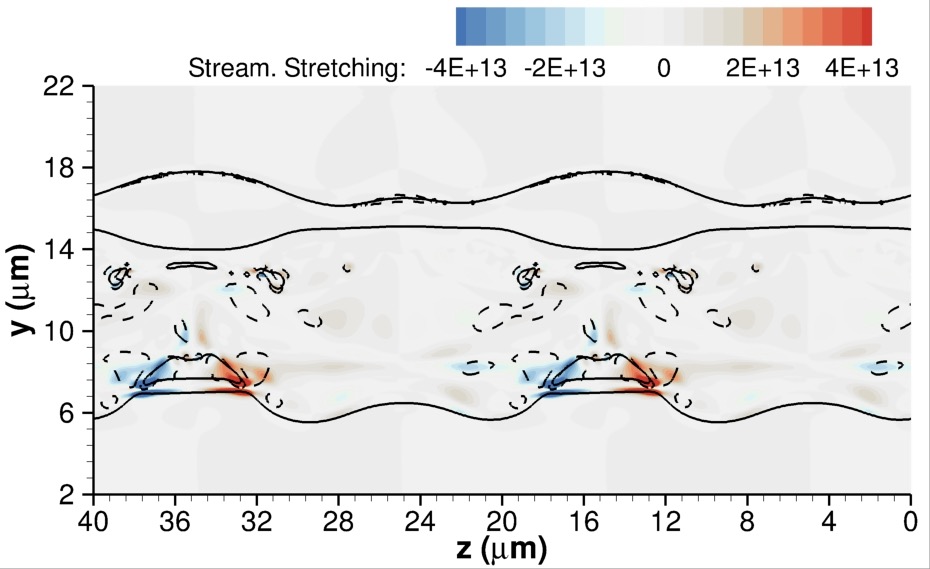}
  \caption{} 
  \label{subfig:Fig26b}
\end{subfigure}%
\\
\begin{subfigure}{0.5\textwidth}
  \centering
  \includegraphics[width=1.0\linewidth]{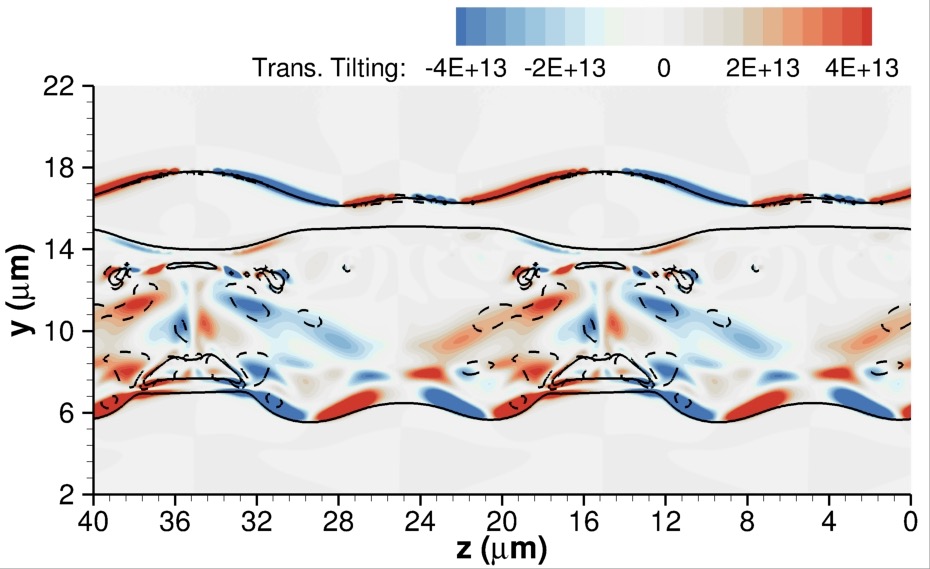}
  \caption{} 
  \label{subfig:Fig26c}
\end{subfigure}%
\begin{subfigure}{0.5\textwidth}
  \centering
  \includegraphics[width=1.0\linewidth]{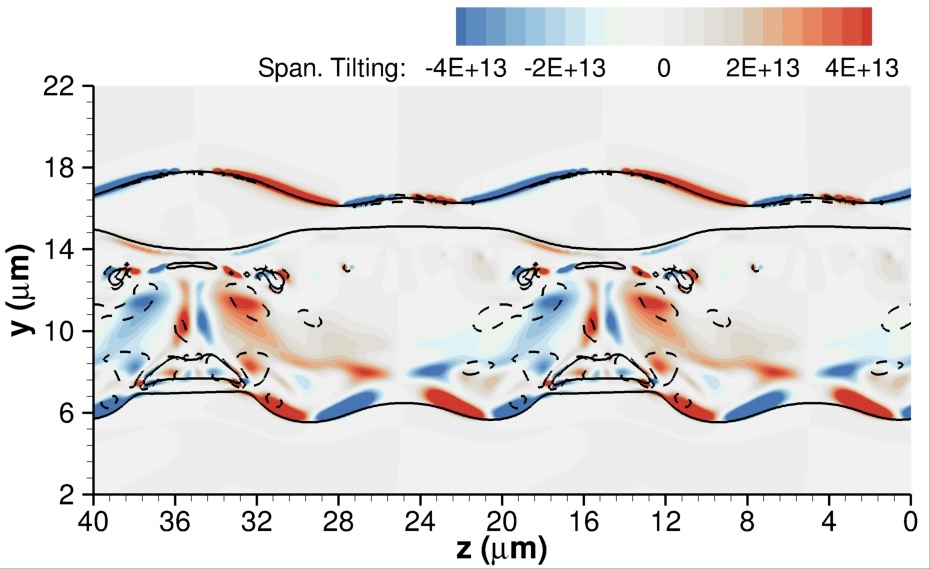}
  \caption{} 
  \label{subfig:Fig26d}
\end{subfigure}%
\\
\begin{subfigure}{0.5\textwidth}
  \centering
  \includegraphics[width=1.0\linewidth]{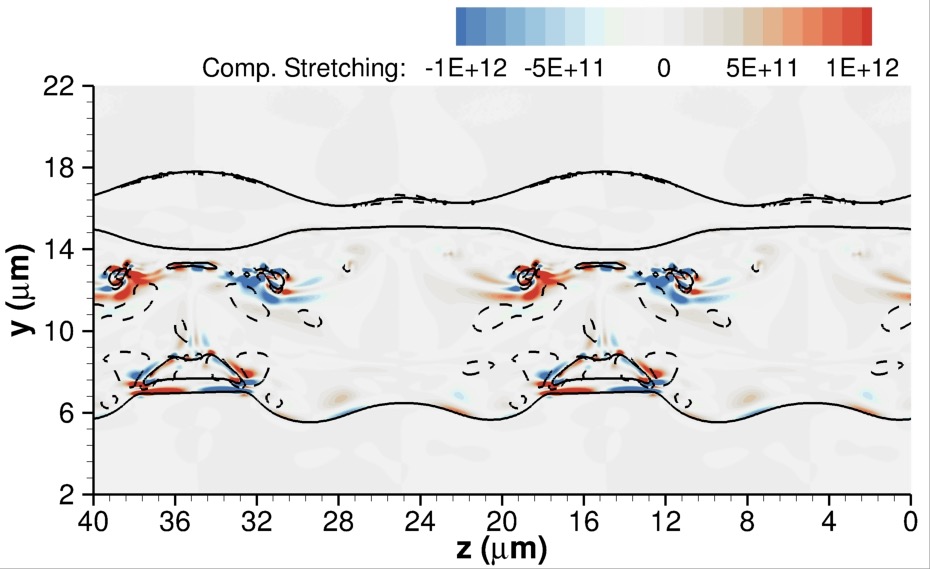}
  \caption{} 
  \label{subfig:Fig26e}
\end{subfigure}%
\begin{subfigure}{0.5\textwidth}
  \centering
  \includegraphics[width=1.0\linewidth]{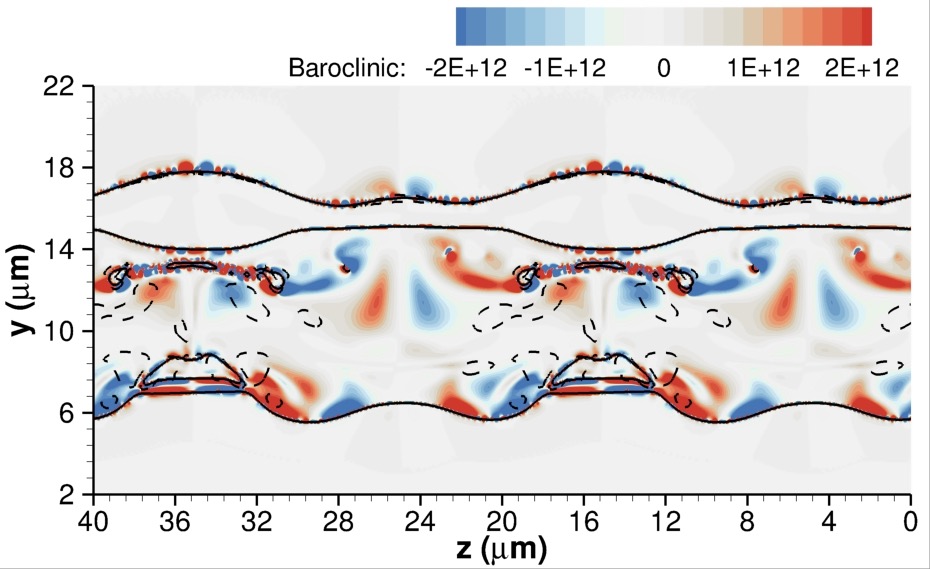}
  \caption{} 
  \label{subfig:Fig26f}
\end{subfigure}%
\caption{\(\omega_x\) generation terms in between the liquid layers for case C1 at \(t^*=8.4\). The front view from a \(yz\) plane at \(x=16\) \(\mu\)m is shown. The interface is identified by the solid isocontour with \(C=0.5\) and the cut vortex structures are identified by the dashed isocontour with \(\lambda_{\rho,t}=-2.5\times 10^{15}\). (a) vortex stretching and tilting; (b) \(\dot{\omega}_{x\rightarrow x}\); (c) \(\dot{\omega}_{y\rightarrow x}\); (d) \(\dot{\omega}_{z\rightarrow x}\); (e) compressible stretching; and (f) baroclinic term.}
\label{fig:Fig26}
\end{figure}

\begin{figure}
\centering
\begin{subfigure}{0.5\textwidth}
  \centering
  \includegraphics[width=1.0\linewidth]{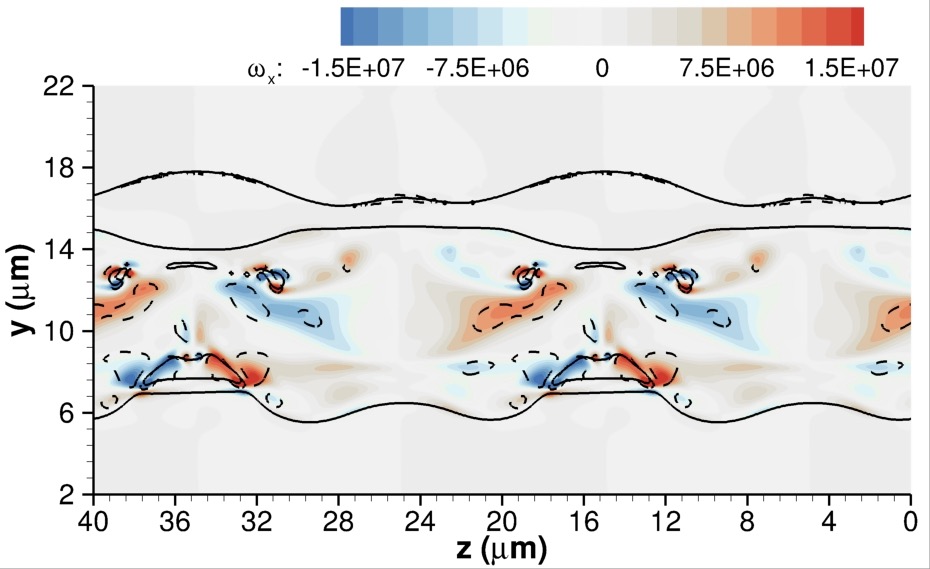}
  \caption{} 
  \label{subfig:Fig27a}
\end{subfigure}%
\begin{subfigure}{0.5\textwidth}
  \centering
  \includegraphics[width=1.0\linewidth]{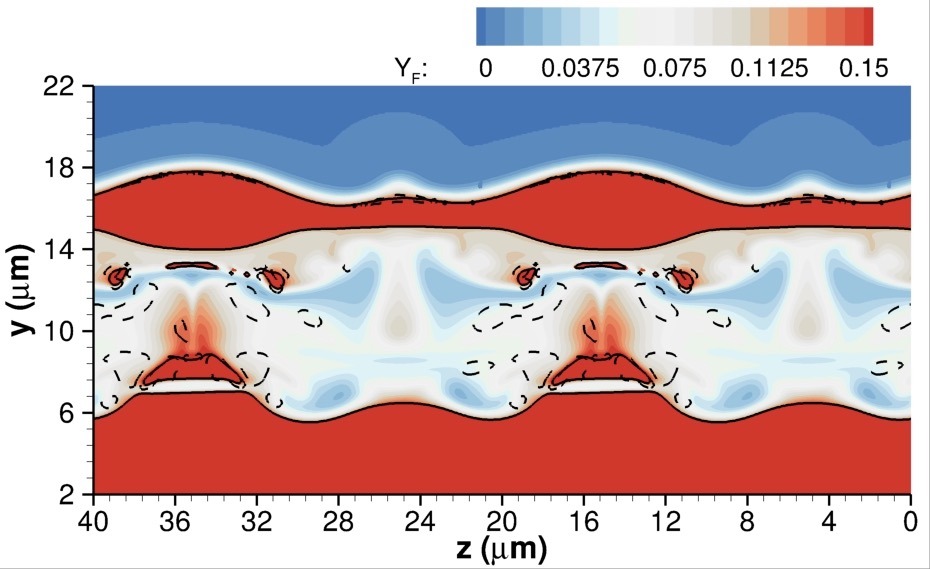}
  \caption{} 
  \label{subfig:Fig27b}
\end{subfigure}%
\\
\begin{subfigure}{0.5\textwidth}
  \centering
  \includegraphics[width=1.0\linewidth]{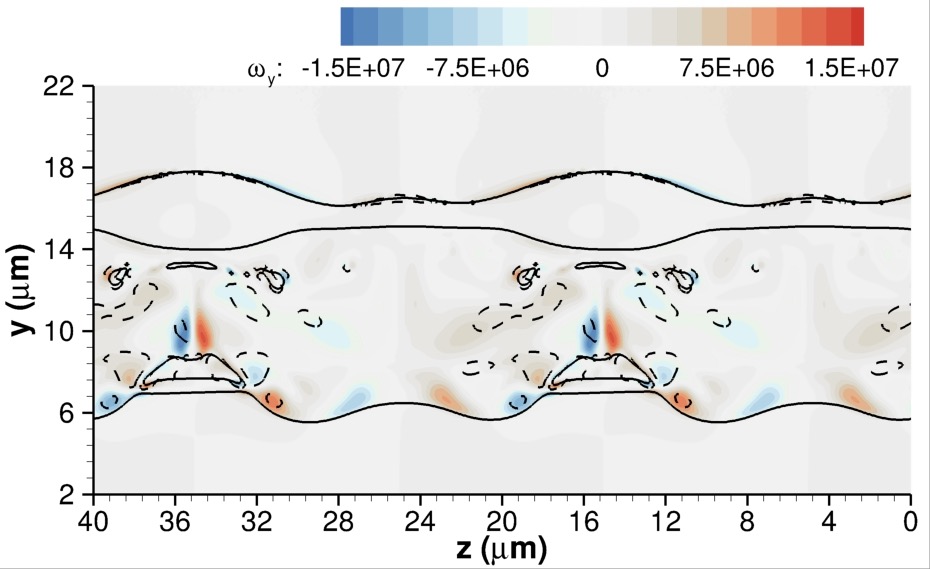}
  \caption{} 
  \label{subfig:Fig27c}
\end{subfigure}%
\begin{subfigure}{0.5\textwidth}
  \centering
  \includegraphics[width=1.0\linewidth]{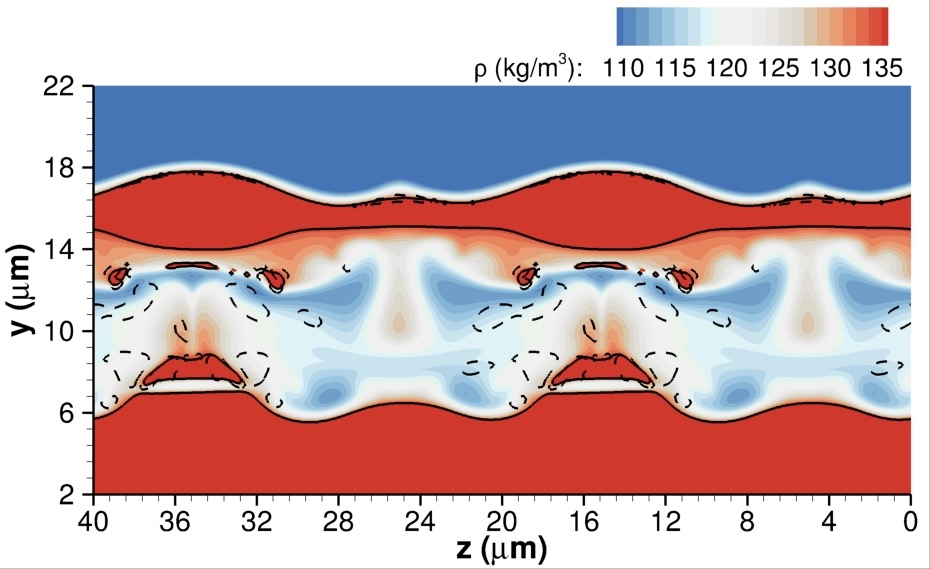}
  \caption{} 
  \label{subfig:Fig27d}
\end{subfigure}%
\\
\begin{subfigure}{0.5\textwidth}
  \centering
  \includegraphics[width=1.0\linewidth]{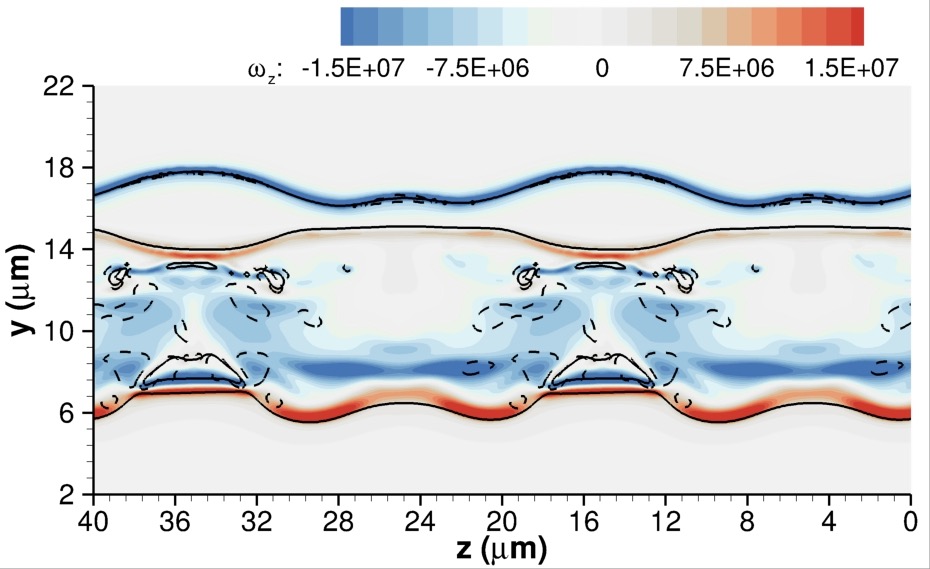}
  \caption{} 
  \label{subfig:Fig27e}
\end{subfigure}%
\begin{subfigure}{0.5\textwidth}
  \centering
  \includegraphics[width=1.0\linewidth]{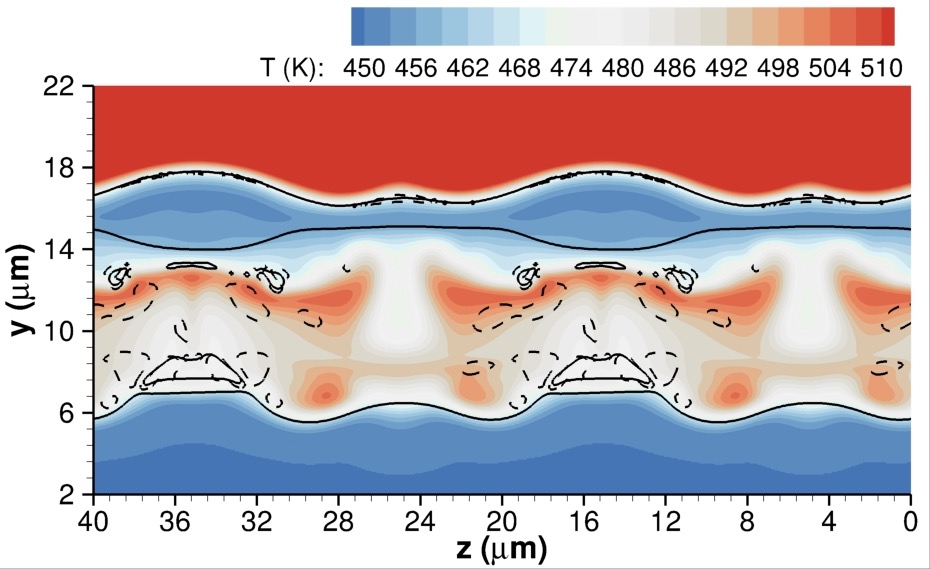}
  \caption{} 
  \label{subfig:Fig27f}
\end{subfigure}%
\caption{Vorticity components, fuel vapour mass fraction, gas density and temperature in between the liquid layers for case C1 at \(t^*=8.4\). The front view from a \(yz\) plane at \(x=16\) \(\mu\)m is shown. The interface is identified by the solid isocontour with \(C=0.5\) and the cut vortex structures are identified by the dashed isocontour with \(\lambda_{\rho,t}=-2.5\times 10^{15}\). (a) \(\omega_x\); (b) fuel vapour mass fraction, \(Y_F\); (c) \(\omega_y\); (d) gas density, \(\rho\); (e) \(\omega_z\); and (f) temperature, \(T\).}
\label{fig:Fig27}
\end{figure}

Figures~\ref{fig:Fig26} and \ref{fig:Fig27} present the contour plots of different variables in a \(yz\) plane at \(x=16\) \(\mu\)m and \(t^*=8.4\) for case C1. Figure~\ref{fig:Fig26} shows the various terms in (\ref{eqn:vorticityX}), and figure~\ref{fig:Fig27} shows the vorticity components, the fuel vapour mass fraction, gas density, and temperature. This non-dimensional time corresponds to the snapshot in figure~\ref{subfig:Fig14a}, where the remnants of the roller vortex R2 are found underneath the liquid. At \(x=16\) \(\mu\)m, the \(yz\) plane cuts through the region where most of the vortex structures are. Despite being an early snapshot of the layered flow, the results in figure~\ref{fig:Fig26} provide relevant details of the \(\omega_x\) generation mechanisms as the liquid layers form. Of course, other events impact the evolution of the vortical structures, such as the entrainment of the roller vortex R3, overlapping layers, and frequent tearing and hole formation. \par 

The contribution of vortex stretching and tilting to \(\omega_x\) generation mainly comes from \(\dot{\omega}_{y\rightarrow x}\) and \(\dot{\omega}_{z\rightarrow x}\). As reported in section~\ref{subsubsec:hairpin}, these two terms partially cancel each other. Note that \(\dot{\omega}_{x\rightarrow x}\) has a weak contribution except in a region above the liquid core near the symmetry plane of the computations. Vorticity is predominantly aligned in the spanwise direction coming from the roller vortex R2, but streamwise vorticity is not negligible (see figures~\ref{subfig:Fig27a} and~\ref{subfig:Fig27e}). Thus, \(\dot{\omega}_{x\rightarrow x}\) is not a major source of \(\omega_x\), at least at this stage, because of a negligible \(\partial u /\partial x\). In contrast to previous sections~\ref{subsubsec:hairpin} and~\ref{subsubsec:rollers}, the contributions from variable-density effects become more important in the layered flow. Despite being an order of magnitude lower, compressible vortex stretching and baroclinicity affect flow regions where vortex stretching and tilting are weak. Therefore, these terms can be substantial drivers of \(\omega_x\) and the overall vortex deformation. \par 

The increased importance of compressible terms is not unexpected. A closer look to the variations of composition, temperature and density in the gas phase is given in figures~\ref{subfig:Fig27b},~\ref{subfig:Fig27d} and~\ref{subfig:Fig27f}. Multiple fuel-lean pockets are observed with higher temperatures and lower densities. These regions exist because of the entrainment of hotter ambient gas, as detailed in section~\ref{subsec:layering} and visualised in figures~\ref{fig:Fig13} to~\ref{fig:Fig16}. This increased complexity in the scalar mixing patterns results in a coupled mixing-deformation mechanism where vorticity defines the scalar mixing, which, in turn, affects vorticity generation and the vortex deformation due to the variable-density and compressible effects seen in (\ref{eqn:vorticity}). Scalar mixing patterns are also affected by vorticity at low ambient pressures. Nevertheless, the gas mixture density variations are often negligible at low pressures~\citep{poblador2018transient,poblador2021selfsimilar}, and a relevant two-way coupling may not exist. Further analysis of this feature is beyond the scope of this work. Future studies of transcritical atomisation should address three key aspects together: (a) the liquid-gas mixing process (i.e., atomisation), (b) the mechanisms of scalar mixing (e.g., fuel mixing), and (c) the coupling with vorticity dynamics. \par

\section{Summary and Conclusions}
\label{sec:conclusions}

The vortex dynamics analysis using the vortex identification method \(\lambda_\rho\)~\citep{yao2018toward} reveals the interaction between deforming vortical structures and the liquid surface. It explains the deformation patterns identified in the temporal study of a transcritical planar liquid \textit{n}-decane jet submerged into a faster and hotter gaseous oxygen stream presented by PS. Three main deformation mechanisms have been studied: the lobe stretching, bending and perforation mechanism, the lobe and crest corrugation mechanism, and the layering of liquid sheets. The early lobe deformation is driven by the interaction between the liquid lobe and the initial roller vortex, which deforms into a hairpin. The heating of the liquid phase and the enhanced ambient gas dissolution at supercritical pressures generate significant variations of fluid properties across and along the interface. In particular, a liquid mixture with reduced density and gas-like viscosity is observed, and surface tension decreases rapidly. Thus, the vortical motion in the gas phase easily perturbs the liquid, which explains the lobe stretching, bending and corrugation mechanisms at relatively low velocities. During the growth of the primary perturbation, various strong roller vortices form and break up toward smaller structures. These move underneath the wave at very high pressures (100 bar and 150 bar), triggering layering of the liquid into sheets. This process traps most of the vorticity generated during the early stages of liquid-gas mixing. It limits the formation of ligaments and droplets above the two-phase mixture, which tends to uniformise the flow (PS). Nonetheless, the wave roll-up causes the entrainment of hotter liquid structures and hotter gas into a region of cooler temperatures, effectively mixing fluid regions with substantially different properties and generating fuel-rich or fuel-lean gas blobs. These gaseous regions of distinct fuel composition also exist outside the layered flow, resulting from the non-uniform fuel vaporisation and breakup of ligaments and droplets; thus, they might impact the combustion process. Their evolution is defined by the fluid dynamics and not only the high mass diffusivity characteristic of supercritical fluids. \par

Consistent with the literature, the alignment of vorticity with the streamwise direction explains the formation of three-dimensional perturbations responsible for atomisation. A look at the mechanisms for vorticity generation reveals that vortex stretching and tilting are the major contributors. Compressible vortex stretching and baroclinicity contribute to the vorticity generation everywhere in the mixing regions but are, in general, an order of magnitude lower. Moreover, variable-density and inertial terms induce different effects. Some regions show vortex stretching and tilting increasing vorticity, while compressible vortex stretching reduces it. \par 

Regardless, variable-density terms become important in the layered flow due to the large differences in mixture composition, temperature, and density. These observations highlight the strong two-way coupling between scalar mixing patterns and the development of vortical structures in the gaseous phase. Vorticity is responsible for the species and thermal mixing, modifying the mixture density and, in turn, the vorticity generation via compressible stretching and baroclinicity. This feature directly results from the transcritical environment since the variations in fluid properties are enhanced at high pressures. At low ambient pressures and similar low-Mach number conditions, density changes in the gas phase are negligible despite the vaporisation of the fuel. Thus, only the scalar mixing is strongly affected by vorticity. Future studies of transcritical atomisation must address the liquid-gas mixing, the species and thermal mixing within each phase, and the coupling with vorticity dynamics to fully understand the physical mechanisms driving the liquid jet destabilisation and atomisation. \par

\section*{Declaration of Interests}

The authors report no conflict of interest.

\section*{Acknowledgements}

The authors are grateful for the support of the NSF grant with Award Number 1803833 and Dr. Ron Joslin as Scientific Officer. This work utilised the infrastructure for high-performance and high-throughput computing, research data storage and analysis, and scientific software tool integration built, operated, and updated by the Research Cyberinfrastructure Center (RCIC) at the University of California, Irvine (UCI). The RCIC provides cluster-based systems, application software, and scalable storage to directly support the UCI research community (https://rcic.uci.edu). Additional computational resources included \textit{Stampede2} at the University of Texas at Austin, \textit{Bridges2} at the University of Pittsburgh, and \textit{Expanse} at the University of California San Diego, under the Extreme Science and Engineering Discovery Environment (XSEDE), which was supported by the NSF grant number ACI-1548562. \par

\newpage
\typeout{}
\bibliography{journal_bib}

\end{document}